\pdfoutput=1
\documentclass[cernpreprint, atlasdraft=false, texlive=2023, UKenglish, texmf, orcidlogo]{atlasdoc}
 
\usepackage{atlaspackage}
\usepackage{atlasbiblatex}
 
\usepackage[jetetmiss=true,process=true]{atlasphysics}

\graphicspath{{logos/}{figures/}}
 
\usepackage{ANA-TOPQ-2020-20-PAPER-defs}
\usepackage{longtable}
\usepackage{tablefootnote}
\usepackage{pdflscape}
\usepackage{multirow}
\usepackage{array}
\newcolumntype{P}[1]{>{\centering\arraybackslash}p{#1}}
\usepackage{changepage}
\usepackage{alphalph,etoolbox}

\AtlasTitle{Inclusive and differential cross-section measurements of $t\bar{t}Z$ production in $pp$ collisions at $\sqrt{s}=13$~TeV with the ATLAS detector, including EFT and spin-correlation interpretations}

\AtlasAbstract{
Measurements of both the inclusive and differential production cross sections of a top-quark--top-antiquark pair in association with a $Z$ boson ($t\bar{t}Z$) are presented.
Final states with two, three or four isolated leptons (electrons or muons) are targeted.
The measurements use the data recorded by the ATLAS detector in $pp$ collisions at $\sqrt{s}=13$~TeV at the Large Hadron Collider during the years 2015--2018, corresponding to an integrated luminosity of $140$~fb$^{-1}$.
The inclusive cross section is measured to be $\sigma_{t\bar{t}Z}= 0.86 \pm 0.04~\mathrm{(stat.)} \pm 0.04~\mathrm{(syst.)}~$pb and found to be in agreement with the most advanced Standard Model predictions.
The differential measurements are presented as a function of a number of observables that probe the kinematics of the $t\bar{t}Z$ system.
Both the absolute and normalised differential cross-section measurements are performed at particle level and parton level for specific fiducial volumes, and are compared with NLO+NNLL theoretical predictions.
The results are interpreted in the framework of Standard Model effective field theory and used to set limits on a large number of dimension-6 operators involving the top quark.
The first measurement of spin correlations in $t\bar{t}Z$ events is presented: the results are in agreement with the Standard Model expectations, and the null hypothesis of no spin correlations is disfavoured with a significance of $1.8$ standard deviations.
}

\AtlasRefCode{TOPQ-2020-20}
 
\PreprintIdNumber{CERN-EP-2023-252}

\AtlasJournal{JHEP}
\AtlasJournalRef{\JHEP 07 (2024) 163}
\AtlasDOI{10.1007/JHEP07(2024)163}

\hypersetup{pdftitle={ATLAS document},pdfauthor={The ATLAS Collaboration}}
 
\begin{document}
 
\maketitle
 
\tableofcontents

\section{Introduction}

The production of a top-quark--top-antiquark pair~(\ttbar) in association with a \Zboson boson is, according to the Standard Model (SM), a rare process in proton--proton (\pp) collisions at the LHC and a source of various multilepton final states. Given its high mass of approximately 173~\GeV~\cite{ATLAS:TopMassCombination}, and thus its large Yukawa coupling to the Higgs boson, the top quark plays a special role in electroweak~(EW) physics.
The coupling of the top quark to the \Zboson boson is not yet well constrained by available data and its value can be significantly altered by \enquote{beyond the Standard Model} (BSM) physics processes~\cite{Chivukula:1992ap,Chivukula:1994mn,Hagiwara:1995jx,Mahanta:1996ng,Mahanta:1996qe,Perelstein:2005ka}.
Precise measurements of the inclusive and differential cross sections of \ttbar production in association with a $Z$ boson, denoted by \ttZ, are thus of particular interest.
Furthermore, the \ttZ process is an irreducible background in other rare-top analyses, such as in four-top production~\cite{TOPQ-2021-08,CMS-TOP-22-013-custom}, as well as in several searches for BSM phenomena, such as in supersymmetric models~\cite{SUSY-2016-20,EXOT-2016-16,SUSY-2018-09,CMS-SUS-16-035}.
Also, measurements of important SM processes, such as \ttbar production in association with a Higgs boson~\cite{HIGG-2017-02,HIGG-2018-13,CMS-HIG-19-008-custom} or single top-quark production in association with a \Zboson boson~\cite{TOPQ-2018-01,CMS-TOP-20-010-custom} are affected by \ttZ background.
 
The most accurate theoretical prediction of the \ttZ cross section is at full next-to-leading order (NLO)~\cite{deFlorian:2016spz,Frixione:2015zaa}, including EW corrections.
Recently, these corrections were supplemented with a resummation of soft gluon corrections carried out at next-to-next-to-leading-logarithm (NNLL) accuracy and matched to the existing NLO results (NLO+NNLL), as reported in Ref.~\cite{Kulesza:2019}. The predicted value at $\sqrt{s}= 13$~\TeV~\cite{deFlorian:2016spz} is:
\begin{equation*}
\sigma_{\ttZ} = 0.863\,^{+0.073}_{-0.085}\, (\mathrm{scale}) \pm 0.028 \, (\mathrm{PDF}\oplus\mathrm{\alphas})\, \mathrm{pb}.\\
\end{equation*}
 
Predictions of \ttZ differential cross sections at NLO+NNLL accuracy, including EW corrections, were calculated in Ref.~\cite{Broggio:2019ewu} and include those as functions of the rapidity of the top quark, the transverse momenta \pt of the top (anti-top) quark and the \Zboson boson, and the invariant masses of the \ttbar and \ttZ systems.
 
The first differential cross-section measurement in \ttZ production at the LHC was performed by the CMS Collaboration, using its 2016--2017 dataset from 13~\TeV \pp collisions, which corresponds to 77.5~\ifb.
The cross section was measured as a function of two variables in final states with three or four leptons~\cite{ttZDifferentialCMS}.
Both the absolute and normalised differential cross sections were presented and compared with NLO+NNLL theoretical predictions.
In addition, an inclusive cross-section measurement was performed, yielding $\sigma_{\ttZ} = 0.95 \pm 0.05\ \mathrm{(stat.)} \pm 0.06\ \mathrm{(syst.)}\,$pb.
 
In Ref.~\cite{TOPQ-2018-08}, the ATLAS Collaboration reported the first measurements of \ttZ differential cross sections using its full dataset from \RunTwo of the LHC.
The inclusive cross section was also extracted in the three- and four-lepton channels and measured to be $\sigma_{\ttZ}=0.99 \pm 0.05\ \mathrm{(stat.)} \pm 0.08\ \mathrm{(syst.)}\,$pb.
The results from the ATLAS and CMS collaborations are compatible with the SM prediction~\cite{Kulesza:2019}, and with each other.
 
This paper presents an extended and refined measurement of the \ttZ cross section in the multilepton final states, using the full dataset collected by the ATLAS experiment during \RunTwo with the LHC.
An additional final state is considered, targeting the all-hadronic decay of the \ttbar system.
The precision of the cross-section measurements is enhanced by making use of improved calibrations of physics objects and the total integrated luminosity, together with reduced experimental uncertainties, while the estimation of theoretical and modelling uncertainties benefits from recent measurements of key background processes and from more accurate Monte Carlo event generators and fixed-order phenomenological calculations.
The cross section for \ttZ production is extracted by performing a profile-likelihood fit simultaneously in the targeted analysis regions, with the signal normalisation as the parameter of interest.
A similar profile-likelihood approach is employed to unfold the data to particle level and parton level, measuring absolute and normalised differential cross sections.
The improved treatment of background effects, more detailed models of systematic uncertainties, and
unfolding several detector-level selections simultaneously to the same fiducial volume provide a result that is more precise and robust than the one presented in Ref.~\cite{TOPQ-2018-08}.
The extracted inclusive \ttZ cross sections and selected normalised differential kinematic distributions at particle level and parton level are then used to constrain the dimension-6 effective field theory (EFT) operators relevant to the $t$--$Z$ interaction in the context of the SM effective field theory (SMEFT)~\cite{Brivio:2017vri,Brivio:2020onw}.
A further interpretation of the experimental results is given in terms of coefficients of the spin density matrix, never before measured for the \ttZ process~\cite{Ravina:2021kpr}.


\section{ATLAS detector}

The ATLAS detector~\cite{PERF-2007-01} at the LHC covers nearly the entire solid angle around the collision point.\footnote{\AtlasCoordFootnote}
It consists of an inner tracking detector surrounded by a thin superconducting solenoid, electromagnetic and hadron calorimeters,
and a muon spectrometer incorporating three large superconducting air-core toroidal magnets.
 
The inner-detector system (ID) is immersed in a \qty{2}{\tesla} axial magnetic field
and provides charged-particle tracking in the range \(|\eta| < 2.5\).
The high-granularity silicon pixel detector covers the vertex region and typically provides four measurements per track,
the first hit normally being in the insertable B-layer installed before \RunTwo~\cite{ATLAS-TDR-19,PIX-2018-001}.
It is followed by the silicon microstrip tracker, which usually provides eight measurements per track.
These silicon detectors are complemented by the transition radiation tracker (TRT),
which enables radially extended track reconstruction up to \(|\eta| = 2.0\).
The TRT also provides electron identification information
based on the fraction of hits (typically 30 in total) above a higher energy-deposit threshold corresponding to transition radiation.
 
The calorimeter system covers the pseudorapidity range \(|\eta| < 4.9\).
Within the region \(|\eta|< 3.2\), electromagnetic calorimetry is provided by barrel and
endcap high-granularity lead/liquid-argon (LAr) calorimeters,
with an additional thin LAr presampler covering \(|\eta| < 1.8\)
to correct for energy loss in material upstream of the calorimeters.
Hadron calorimetry is provided by the steel/scintillator-tile calorimeter,
segmented into three barrel structures within \(|\eta| < 1.7\), and two copper/LAr hadron endcap calorimeters.
The solid angle coverage is completed with forward copper/LAr and tungsten/LAr calorimeter modules
optimised for electromagnetic and hadronic energy measurements respectively.
 
The muon spectrometer (MS) comprises separate trigger and
high-precision tracking chambers measuring the deflection of muons in a magnetic field generated by the superconducting air-core toroidal magnets.
The field integral of the toroids ranges between \num{2.0} and \qty{6.0}Tm
across most of the detector.
Three layers of precision chambers, each consisting of layers of monitored drift tubes, cover the region \(|\eta| < 2.7\),
complemented by cathode-strip chambers in the forward region, where the background is highest.
The muon trigger system covers the range \(|\eta| < 2.4\) with resistive-plate chambers in the barrel, and thin-gap chambers in the endcap regions.
 
Interesting events are selected by the first-level trigger system implemented in custom hardware,
followed by selections made by algorithms implemented in software in the high-level trigger~\cite{TRIG-2016-01}.
The first-level trigger accepts events from the \qty{40}{\MHz} bunch crossings at a rate below \qty{100}{\kHz},
which the high-level trigger reduces in order to record events to disk at about \qty{1}{\kHz}.
 
An extensive software suite~\cite{ATL-SOFT-PUB-2021-001} is used in data simulation, in the reconstruction
and analysis of real and simulated data, in detector operations, and in the trigger and data acquisition
systems of the experiment.


\section{Data and simulated event samples}
\label{sec:samples}

This analysis uses the full \RunTwo dataset collected by ATLAS in 13~\TeV \pp collisions during 2015--2018 and corresponding to an integrated luminosity of \lumi~\ifb~\cite{DAPR-2021-01-custom}.
Only events recorded when LHC beams were stable and all ATLAS detector systems were operational are selected. The uncertainty in the total integrated luminosity is  0.83\% \cite{DAPR-2021-01-custom}, obtained using the LUCID-2 detector \cite{LUCID2} for the primary luminosity measurements, complemented by measurements using the inner detector and calorimeters. The average number of interactions per bunch crossing ranged from \num{0.5} to around 80, with a mean of \num{33.7}.
 
The data were collected using a combination of single-electron and single-muon triggers, with requirements
on the identification, isolation, and \pt of the leptons to maintain high efficiency across the full momentum range
while controlling the trigger rates~\cite{TRIG-2016-01,TRIG-2018-05,TRIG-2018-01}. For electrons the trigger thresholds were \pt = 26, 60 and 140~\GeV,
whereas for muons the thresholds were \pt = 26 and 50~\GeV.\footnote{Lower \pt thresholds of 24~\GeV and 120~\GeV for electrons and 20 GeV for muons were applied for 2015 data.}
Isolation requirements
were applied to the triggers with the lowest \pt thresholds of 26~\GeV for both electrons and muons~\cite{ATL-DAQ-PUB-2016-001,ATL-DAQ-PUB-2017-001,ATL-DAQ-PUB-2018-002}.
 
Simulated Monte Carlo (MC) samples are used to model the signal and the prompt SM background.
The effect of multiple interactions in the same and neighbouring bunch crossings (pile-up) was modelled by overlaying each simulated hard-scattering event with inelastic \pp events generated with \PYTHIA[8.186]~\cite{Sjostrand:2007gs} using the \NNPDF[2.3lo] set of parton distribution functions (PDF)~\cite{Ball:2012cx-custom} and the A3 set of tuned parameters~\cite{ATL-PHYS-PUB-2016-017}.
Separate MC production campaigns were used to model the different pile-up distributions observed in data during 2015/16, 2017 and 2018.
The simulated samples were reweighted to reproduce the observed distribution of the average number of collisions per bunch crossing in each data-taking period.
 
The simulation of detector effects was performed with either a full ATLAS detector simulation~\cite{SOFT-2010-01} based on the \GEANT~\cite{Geant} framework or a fast simulation (\textsc{AtlFast~II}) using a parameterisation of the performance of the electromagnetic and hadronic calorimeters and \GEANT for the other detector components~\cite{ATL-PHYS-PUB-2010-013,ATL-SOFT-PUB-2018-002}. The full simulation was used for most processes, while the fast simulation was used only for the nominal prediction of the \ttH process and alternative modelling samples for various other processes.
 
The associated production of \ttbar with a leptonically decaying \Zboson boson was modelled using the \MGNLO[2.8.1]~\cite{Alwall:2014hca} generator, which provided matrix elements at NLO in the strong coupling constant \alphas with the \NNPDF[3.0nlo]~\cite{Ball:2014uwa-custom} PDF set.
The $\gamma^{*}\to\ell^+\ell^-$ contribution and $Z/\gamma^{*}$ interference are taken into account, down to 5~\GeV in dilepton invariant mass.
The functional form of the renormalisation and factorisation scales (\muR, \muF) was set to the default scale $0.5 \times \sum_i \sqrt{m^2_i+p^2_{\textrm{T},i}}$, where the sum runs over all the particles generated from the matrix element calculation.
Top quarks were decayed at leading order (LO), using \MADSPIN~\cite{Frixione:2007zp,Artoisenet:2012st} to preserve all spin correlations.
The top-quark mass was set to 172.5~\GeV in all MC samples.
The events were interfaced with \PYTHIA[8.244]~\cite{Sjostrand:2014zea} for the simulation of the parton shower,
fragmentation, hadronisation, and underlying event, using the A14 set of tuned parameters\footnote{Tuning refers to the process of optimising the parameters of the MC to produce a reasonable description of measured observables.}
~\cite{ATL-PHYS-PUB-2014-021} and the \NNPDF[2.3lo] PDF set.
The decays of bottom and charm hadrons were simulated using the \EVTGEN[1.7.0] program~\cite{Lange:2001uf}.
 
To estimate theoretical uncertainties in the signal prediction, several alternative \ttZ MC samples are considered.
These include a sample generated with the same \MGNLO version as the nominal sample, but interfaced to \HERWIG[7.2.1]~\cite{Bahr:2008pv,Bellm:2015jjp} for the simulation of the parton shower (using the default angle-ordered shower model).
Two additional samples with the same settings as the nominal \ttZ sample, except for upward and downward variations of the Var3c parameter in the A14 tune, are used to evaluate uncertainties associated with the modelling of initial-state radiation (ISR), following an approach similar to the one described in Ref.~\cite{ATL-PHYS-PUB-2016-005}.
The Var3c variation corresponds to a variation of \alphas for ISR in the A14 tune.
 
Alternative samples generated with \SHERPA are used for comparisons with the unfolded differential distributions.
A \ttZ sample was produced with the \SHERPA[2.2.1]~\cite{Bothmann:2019yzt} generator at NLO accuracy.
Another sample was produced with a newer version of the same generator, \SHERPA[2.2.11], together with the \MEPSatNLO matching algorithm~\cite{Catani:2001cc,Hoeche:2009rj,Hoeche:2012yf}, which performed the multi-leg merging of up to three additional partons with the parton shower at LO, with a merging scale of 30~\GeV.
In both cases, a dynamic renormalisation scale defined similarly to that of the nominal \ttZ samples was used.
These samples also include off-shell effects down to $5~\GeV$ in the invariant mass of the lepton pair.
The default \SHERPA parton shower was used along with the \NNPDF[3.0nnlo] PDF set.
 
Events featuring the production of a \ttbar pair in association with a SM Higgs boson with a mass of 125~\GeV\ (\ttH) were generated using NLO matrix elements in \MGNLO[2.6.0] with the \NNPDF[3.0nlo] PDF set.
These events were showered with \PYTHIA[8.230] using the A14 tune.
 
For \ttbar production with a \Wboson boson (\ttW), the \SHERPA[2.2.10] generator and default parton shower were used at NLO accuracy in QCD, with \MEPSatNLO performing multi-leg merging of up to one additional parton at NLO and up to two additional partons at LO, with a merging scale of 30~\GeV.
Additionally, a LO QCD sample also generated with \SHERPA[2.2.10] but for the $\ttW j$ final state is used to model additional electroweak corrections to \ttW production.
In both \ttW samples, the \NNPDF[3.0nnlo] PDF set was used.
 
The production of single top quarks in association with a \Zboson boson~(\tZq) was modelled with the \MGNLO[2.9.5] generator at NLO, while their production in association with a \Wboson boson and \Zboson boson (\tWZ) used \MGNLO[2.2.2] at NLO, both with the \NNPDF[3.0nlo] PDF set.
The \tZq events were interfaced with \PYTHIA[8.245], and the \tWZ events with \PYTHIA[8.212], using the A14 tune and the \NNPDF[2.3lo] PDF set.
The \tZq sample was simulated in the four-flavour scheme (thus including an additional $b$-quark in the final state) and normalised to a cross section obtained in the five-flavour scheme.
It also includes off-shell effects down to $5~\GeV$ in dilepton invariant mass.
 
Uncertainties in modelling the \tZq process are estimated similarly to the case of \ttZ: the Var3c parameter of the A14 tune is varied upwards and downwards, and a separate sample is considered in which the generated events are interfaced to the \HERWIG[7.2.1] parton shower algorithm instead of \PYTHIA[8.245].
At NLO in QCD, the Feynman diagrams of the \tWZ process include contributions such as $gg\to tWZb$, which may also feature a second intermediate top resonance and thus interfere with the signal \ttZ process.
The nominal \tWZ sample follows the \enquote{diagram removal 1} (DR1) scheme described in Ref.~\cite{Demartin:2016axk} and ignores any Feynman diagrams containing two resonant top quarks.
An alternative sample was generated within the DR2 scheme, which additionally considers the interference terms (at the level of squared amplitudes) between single- and double-resonant \ttbar production.
This alternative sample is used to set an uncertainty on the modelling of the \tWZ process, as described in Section~\ref{subsubsec:syst_theory_backgrounds}.
 
The production of \ttbar background events was modelled with the \POWHEGBOX[v2] generator \cite{Alioli:2010xd} at NLO with the \NNPDF[3.0nlo] PDF set and the damping factor \hdamp\footnote{The \hdamp parameter is a resummation damping factor and one of the parameters that controls the matching of \POWHEG matrix elements to the parton shower and thus effectively regulates the high-\pT radiation against which the \ttbar system recoils.} set to 1.5 times the top-quark mass.
The events were interfaced with \PYTHIA[8.230], which used the A14 tune and \NNPDF[2.3lo] PDF set.
The top-quark decays were modelled at LO, while decays of bottom and charm hadrons were simulated with \EVTGEN[1.2.0].
 
Several alternative samples are used to evaluate theoretical uncertainties in the modelling of the \ttbar events.
These include a separate sample from the same \POWHEG generator version as above but where the \hdamp parameter was increased to 3.0 times the top-quark mass, a sample where the events generated in \POWHEG were interfaced to \HERWIG[7.0.4] for parton showering, and a sample from a different matrix element generator, \MGNLO[2.3.3], interfaced with \PYTHIA[8.230].
 
Diboson processes producing either three charged leptons and one neutrino or four charged leptons ($WZ$+jets or $ZZ$+jets, respectively) were simulated using the \SHERPA[2.2.2] generator.
In this set-up, multiple matrix elements were matched and merged with the \SHERPA parton shower based on the Catani--Seymour dipole factorisation scheme~\cite{Gleisberg:2008fv,Schumann:2007mg} using the \MEPSatNLO prescription~\cite{Hoeche:2011fd,Hoeche:2012yf,Catani:2001cc,Hoeche:2009rj}.
Virtual QCD corrections for NLO-accurate matrix elements were provided by the \OPENLOOPS library~\cite{Cascioli:2011va,Denner:2016kdg}.
Samples were generated using the \NNPDF[3.0nnlo] PDF set, along with the dedicated set of tuned parton shower parameters developed by the \SHERPA authors.
The $WZ/ZZ$+jets events with no or one additional parton were simulated at NLO, whereas events with two or three additional partons were simulated at LO precision.
 
The production of events with a \Wboson or \Zboson boson and multiple jets ($V+$jets) was simulated with the \SHERPA[2.2.1] generator using NLO-accurate matrix elements for up to two jets, and LO-accurate ones for up to four jets, calculated with the Comix~\cite{Gleisberg:2008fv} and \OPENLOOPS~\cite{Buccioni:2019sur,Cascioli:2011va,Denner:2016kdg} libraries.
They were matched with the \SHERPA parton shower using the \MEPSatNLO prescription.
The \NNPDF[3.0nnlo] set of PDFs was used and the samples are normalised to next-to-next-to-leading-order (NNLO) predictions~\cite{Anastasiou:2003ds}.
 
Events from both the diboson and $V+$jets processes are separated into light-, $b$- and $c$-flavour components, depending
on whether the MC event record has a $b$- or $c$-hadron in any of the selected jets.
 
MC samples featuring Higgs boson production in association with a \Wboson or \Zboson boson were generated with \PYTHIA[8.186] using the A14 tune and the \NNPDF[2.3lo] PDF set.
Triple top-quark production (\ttt) and the production of a \ttbar pair with two \Wboson bosons~(\ttWW) were simulated at LO using \MADGRAPH[2.2.2] interfaced to \PYTHIA[8.186] with the A14 tune and the \NNPDF[2.3lo] PDF set.
Four top-quark production (\tttt) was simulated at NLO using \MGNLO[2.3.3] interfaced to \PYTHIA[8.230] with the A14 tune and the \NNPDF[3.1nlo] PDF set; an alternative sample used the \HERWIG[7.04] parton shower instead.
Processes with three heavy gauge bosons ($WWW$, $WWZ$, $WZZ$ and $ZZZ$) yielding up to six final-state leptons were simulated with \SHERPA[2.2.2] and the \NNPDF[3.0nlo] PDF set.
Final states with no additional partons were calculated at NLO, whereas final states with one, two or three additional partons were calculated at LO.
 
\begin{table}[!htb]
\footnotesize
\caption{Versions of the generator, parton shower and PDF used for the nominal MC samples and reference cross sections used in the analysis. Whenever a reference is not indicated, the cross section is taken directly from the MC set-up described in the text.}
\label{tab:mc-samples}
\def\arraystretch{1.3}
\begin{center}
\begin{tabular}{lcccc}
\toprule
Process & Generator & Parton shower & PDF & Reference cross section [fb]\\
\midrule
\ttZ & \MGNLO[2.8.1] & \PYTHIA[8.244] & \NNPDF[3.0nlo] & 876~\cite{deFlorian:2016spz,ATL-PHYS-PUB-2016-005} \\
\ttH & \MGNLO[2.6.0] & \PYTHIA[8.230] & \NNPDF[3.0nlo] & 507~\cite{deFlorian:2016spz} \\
\ttW/$\ttW j$ & \SHERPA[2.2.10] & \SHERPA[2.2.10] & \NNPDF[3.0nnlo] & 722~\cite{Frederix:2021agh} \\
\tZq & \MGNLO[2.9.5] & \PYTHIA[8.245] & \NNPDF[3.0nlo] & 38.7 \\
\tWZ & \MGNLO[2.2.2] & \PYTHIA[8.212] & \NNPDF[2.3lo] & 16.1 \\
\ttbar & \POWHEGBOX[v2] & \PYTHIA[8.230] & \NNPDF[3.0nlo] & 87\;\!700~\cite{LHCTopWGttbarXsec} \\
$WZ$+jets/$ZZ$+jets & \SHERPA[2.2.2] & \SHERPA[2.2.2] & \NNPDF[3.0nnlo] & 7\;\!330 \\
$V$+jets & \SHERPA[2.2.1] & \SHERPA[2.2.1] & \NNPDF[3.0nnlo] & $6\;\!250\times 10^3$~\cite{Anastasiou:2003ds} \\
\tttt & \MGNLO[2.3.3] & \PYTHIA[8.230] & \NNPDF[3.1nlo] & 12.0~\cite{Frederix:2017wme} \\
\ttt & \MADGRAPH[2.2.2] & \PYTHIA[8.186] & \NNPDF[2.3lo] & 1.64 \\
$VH$ & \PYTHIA[8.186] & \PYTHIA[8.186] & \NNPDF[2.3lo] & 2\;\!250~\cite{Ciccolini:2003jy,Brein:2003wg,Brein:2011vx,Altenkamp:2012sx,Denner:2014cla,Brein:2012ne,Harlander:2014wda} \\
$VVV$ & \SHERPA[2.2.2] & \SHERPA[2.2.2] & \NNPDF[3.0nlo] &  13.7 \\
\bottomrule
\end{tabular}
\end{center}
\end{table}
 
The versions of the generator, parton shower and PDF used for each of the nominal MC samples, as well as the reference cross sections used to normalise the samples, are given in Table~\ref{tab:mc-samples}.
 
For the SMEFT interpretation, additional MC samples were produced at LO in QCD for the \ttZ and \tZq processes, using the \MADGRAPH[2.9.3] generator and \PYTHIA[8.245] parton shower (with the default A14 tune settings).
They rely on the SMEFTsim\,3.0~\cite{Brivio:2020onw} UFO model~\cite{Degrande:2011ua} implemented in \MADGRAPH with FeynRules~\cite{Alloul:2013bka}, in the $m_W$ electroweak input scheme \cite{Brivio:2021yjb} with the top-flavour restrictions (in the five-flavour scheme).
The nominal events were generated according to the SM, and the \MADGRAPH reweighting module was used to compute a large number of alternative event weights corresponding to the inclusion of dimension-6 EFT vertices and propagators in the production Feynman diagrams.
These internal weights can then be used to extract the dependence of various observables (including the cross section) on more than 30 different EFT operators related to \ttbar production, the $t$--$Z$ vertex and the off-shell $\ttbar\ell\ell$ vertex, as described in Section~\ref{sec:smeft}.
Since the EFT couplings cannot run in \MADGRAPH, the renormalisation and factorisation scales are kept fixed at $\mu=\sum_i m_i$ (where $i$ runs over the massive final-state resonances).
The EFT samples were passed through the fast detector simulation (using \textsc{AtlFast~II}) and the events reconstructed in order to assess their impact on the unfolding (efficiency and acceptance corrections).


\section{Object identification and event reconstruction}
\label{sec:reconstruction}

Electron candidates are reconstructed from clusters of energy deposits in the electromagnetic calorimeter that are matched to a track in the ID. They are required to
satisfy $\pt > \SI{7}{\GeV}$, $\abseta < 2.47$ and a \enquote{Medium} likelihood-based identification requirement~\cite{PERF-2017-01,EGAM-2018-01}. Electron candidates
are excluded if their calorimeter clusters lie within the transition region between the barrel and endcaps of the electromagnetic calorimeter, $1.37 < \abseta < 1.52$, to reduce the contribution from fake electrons.
The track associated with the electron must pass the requirements $|\zzsth| < \SI{0.5}{mm}$ and $|d_{0}| / \sigma(d_{0}) < 5$, where $z_{0}$ describes the longitudinal
impact parameter relative to the reconstructed primary vertex,\footnote{The primary vertex must have at least two associated tracks with $\pt > \SI{500}{\MeV}$ and is defined as the vertex with the highest scalar sum of the squared transverse
momenta of such tracks.} \dzero is the transverse impact parameter relative to the beam axis, and $\sigma(d_{0})$ is the uncertainty
in \dzero.
Furthermore, a requirement on the electron isolation, corresponding to the PLVLoose\footnote{This algorithm is a multivariate discriminant similar to the non-prompt lepton BDT detailed in Ref.~\cite{HIGG-2017-02}, but retrained on updated input variables, including more sophisticated $b$-tagging algorithms.} isolation working point (WP) is applied to identify \enquote{signal} electrons; \enquote{baseline} electrons have no isolation requirement.
 
Muon candidates are reconstructed from MS tracks matched to ID tracks in the pseudorapidity range of $\abseta < 2.5$. They must satisfy $\pt > \SI{7}{\GeV}$ along with the
\enquote{Medium} identification requirements defined in Refs.~\cite{MUON-2022-01-custom,MUON-2018-03}. The latter impose requirements on the number of hits in the different ID and MS subsystems
and on the significance of the charge-to-momentum ratio $q/p$. In addition, the track associated with the muon candidate must have $|\zzsth| < \SI{0.5}{mm}$ and
$|d_{0}| / \sigma(d_{0}) < 3$.
As is the case for electrons, the baseline muons have no isolation requirements, whereas the muons selected for the analysis must pass the PLVLoose isolation WP.
The lepton trigger, reconstruction and selection efficiencies from simulation receive small corrections derived from measurements of $Z\to\ell\ell$ and $\jpsi\to\mu\mu$ events in the data~\cite{TRIG-2018-01,TRIG-2018-05,MUON-2018-03,EGAM-2018-01}.
 
Jets are reconstructed using the \antikt jet algorithm~\cite{Cacciari:2008gp}  as implemented in the \textsc{FastJet} package~\cite{Fastjet}, with the radius parameter $R$ set to $0.4$ and particle-flow objects~\cite{PERF-2015-09} as input.
The jets are calibrated by applying a jet energy scale derived from 13~\TeV\ data and simulation~\cite{JETM-2018-05-custom}.
The jets are kept only if they have $\pt > 25$~\GeV\ and are inside a pseudorapidity range of $\abseta < 2.5$.
The jet-vertex tagger~(JVT)~\cite{PERF-2014-03} algorithm is employed in order to mitigate pile-up effects in jets with $\pt < 60$~\GeV, applying the \enquote{Tight} WP.
 
Jets containing a $b$-hadron, referred to as $b$-jets, are identified with the DL1r $b$-tagging algorithm~\cite{FTAG-2018-01,FTAG-2019-07-custom}.
A WP corresponding to $85\%$ efficiency\footnote{The $b$-tagging efficiency is determined with respect to generator-level $b$-jets with $\pt>20$~\GeV and $\abseta < 2.5$ in \ttbar MC simulations.} is used for most preselections in the analysis.
Exclusive bins of $b$-tagging discriminant values corresponding to different $b$-jet identification efficiencies are also used, as pseudo-continuous $b$-tagging (PCBT).
This allows different calibrated $b$-tagging WPs to be used in defining selections targeting specific signal or background processes, referred to as regions.
 
The missing transverse momentum is defined as the negative vector sum of the transverse momenta of all selected and calibrated physics objects.
Low-momentum tracks from the primary vertex that are not associated with any of the reconstructed physics objects described previously are also included as a `soft term' in the
calculation~\cite{PERF-2016-07}. The magnitude of the missing transverse momentum vector is denoted by \met.
 
Ambiguities can arise from the independent reconstruction of electron, muon and jet candidates in the detector. A sequential procedure~(overlap removal) is applied to
resolve these ambiguities and thus prevent double counting of physics objects. It is applied to signal electrons, muons, and jets as follows. If an electron candidate and a muon candidate share a track, the electron candidate is removed. The jet candidate closest to a remaining electron candidate is removed if they are less than a distance
$\DeltaR_{y,\phi} =\sqrt{(\Delta y)^2+(\Delta \phi)^2} = 0.2$ apart, where $y$ is the jet's rapidity.
If the electron--jet separation is between \num{0.2} and \num{0.4}, the electron candidate is removed. If the $\DeltaR_{y,\phi}$ between any
remaining jet and a muon candidate is less than \num{0.4}, the muon candidate is removed if the jet has more than two associated tracks, otherwise the jet is discarded.
 
In the differential measurements of the \ttZ cross section, two definitions of particles in the MC generator-level event record are considered: parton level and particle level.
While the former leads to a very inclusive phase-space for ease of comparison with fixed-order matrix element calculations, the latter is used to build a fiducial volume much closer to that of the detector-level analysis.
 
Parton-level objects are obtained from the MC generation history of the \ttZ system.
The top~(anti-top) quarks and \Zboson bosons are selected as the last instances of these particles in this \enquote{truth} record, after radiation but immediately before their \tWb or \Zll decay, respectively.
The leptons originating from \Wboson and \Zboson bosons are selected as the first instances, immediately following the decay of the parent boson.
 
Particle level refers to a collection of objects which are considered stable in the MC simulation ($\tau\ge30\,\mathrm{ps}$) but without any simulation of the interaction of these particles with the detector components or any additional \pp interactions.
Unlike for parton-level objects, the hadronisation of the quarks is included.
Particle-level leptons are selected as leptons originating from the decay of a \Wboson~or \Zboson~boson.
The four-momentum of an electron or muon is summed with the four-momenta of all radiated photons within a cone of size $\DeltaR = 0.1$ around its direction, excluding photons from hadron decays.
The parents of the electrons or muons are required not to be a hadron or quark ($u$,$d$,$s$,$c$,$b$).
The particle-level jets are reconstructed with the \antikt algorithm with a radius parameter of $R = 0.4$, using all stable particles except for the selected electrons, muons, and photons used in the definition of the selected leptons, and neutrinos originating from the \Zboson boson or \Wboson bosons.
A small-$R$ jet is considered a $b$-jet if it is ghost-matched~\cite{Cacciari:2008gn,FTAG-2018-01} to a $b$-hadron with $\pt > 5$~GeV.
The particle-level missing transverse momentum is defined as the vector sum of the transverse momenta of all neutrinos found in the simulation history of the event, excluding those originating from hadron decays.


\section{Analysis strategy}

 
The signal regions (SRs) used in this analysis are designed to offer the highest possible purity of \ttZ events, as well as to provide yields sufficient for a differential measurement of the \ttZ cross section.
The analysis is performed in three orthogonal channels, distinguished by lepton multiplicity.
A multivariate analysis (MVA) approach is employed in each channel to better discriminate between the signal and background processes.
An improvement on the previous measurement~\cite{TOPQ-2018-08}, the MVA has the largest impact on the dilepton (2$\ell$) and trilepton ($3\ell$) channels that have large background contributions, whereas the tetralepton (4$\ell$) channel only receives a modest enhancement due to its already excellent signal purity at the preselection level.
The MVA input variables consist mostly of kinematic variables for individual objects such as jets and leptons,
$b$-tagging information for jets, and kinematics of the reconstructed top quarks and $Z$ boson. The exact list of the variables
can be found in the Appendix, in Tables~\ref{tab:2l_variable_definitions},~\ref{tab:3l_variable_definitions}~and~\ref{tab:4l_variable_definitions}.

Neural networks are used in the three channels, and these are trained using the Keras~\cite{chollet2015keras} backend of Tensorflow~\cite{tensorflow2015-whitepaper}. In all cases, the Adam optimiser is used for training.
In the 3$\ell$ channel, the categorical cross-entropy loss is minimised.  The networks in the 2$\ell$ and 4$\ell$ channels employ binary cross-entropy as the loss function. These networks cater to binary classification scenarios, albeit with varying objectives. To ensure optimal performance, the hyperparameters of all networks are tuned using a grid search, which systematically varies the number of layers, activation functions, and regularisation techniques such as batch normalisation and dropout.
Additionally, K-folding techniques~\cite{Kfolding} are employed to enable comprehensive training and evaluation with the entire set of MC simulations. This approach ensures that the networks are trained and tested on statistically independent subsets of the MC simulations.

\FloatBarrier

\subsection{Dilepton signal regions}
 
The dilepton channel targets \ttZ events where the \ttbar system decays hadronically, while an opposite-sign same-flavour (OSSF) pair of leptons originates from the \Zboson boson. Events are required to have exactly one pair of OSSF leptons.
The invariant mass of the lepton pair is required to be within $10~\GeV$ of the \Zboson-boson mass \cite{PDG}, with the two leptons required to have transverse momenta of at least $30~\GeV$ and $15~\GeV$ respectively.
 
The $2\ell$OS channel generally suffers from a low signal-to-background ratio, with the \ttbar and $Z$+jets processes (both characterised by the presence of two prompt\footnote{The term \enquote{prompt} refers to leptons which are directly produced by the hard-scatter process or by the decays of heavy resonances such as \Wboson, \Zboson or Higgs bosons.} leptons) constituting major backgrounds.
Three signal regions are defined, based on jet and $b$-tagged jet (at $77\%$ $b$-tagging efficiency) multiplicities
($N_\text{jets}$ and $N_{b\text{-tagged jets}@77\%}$ respectively).
The splitting is motivated by the different background compositions and the fact that it is not possible to fully reconstruct the $t\bar{t}$ system in events with only five jets.
SR-2$\ell$-5j2b requires exactly five jets, of which at least two must be $b$-tagged; SR-2$\ell$-6j2b SR provides its jet-inclusive complement (at least six jets).
SR-2$\ell$-6j1b, inclusive in jet multiplicity (at least six jets) but requiring exactly one $b$-tagged jet, targets events with the appropriate jet multiplicity for tree-level \ttZ events but with one non-identified $b$-tagged jet.
All dileptonic preselections and SR selections are summarised in Table~\ref{tab:2l-selection}.
 
To improve the discrimination between the \ttbar and $Z+$jets background processes and the \ttZ signal, one deep neural network (DNN) is trained for each signal region on events passing the corresponding selection.
The categorisation of events into signal regions based on jet and $b$-tagged-jet multiplicities allows the DNNs to be tuned on different background compositions and signal-to-background ratios, enhancing the overall performance.
This is particularly needed here, given the much larger background contributions than in the other two analysis channels.
All DNNs are constructed as binary classification networks (with \ttZ as signal and all other processes as background) and the distributions of the DNN outputs are used directly in the inclusive measurement and are not employed in the definition of the signal regions.
Details of the variables used in the training of the DNN are given in Table~\ref{tab:2l_variable_definitions} in the Appendix.
 
\begin{table}[!htb]
\footnotesize
\caption{Definition of the dilepton signal regions.}
\label{tab:2l-selection}
\def\arraystretch{1.3}
\begin{center}
\begin{tabular}{lccc}
\toprule
Variable & \multicolumn{3}{c}{\textbf{Preselection}}\\
\midrule
$N_\ell~\left(\ell=e,\mu\right)$ & \multicolumn{3}{c}{$=2$}\\
& \multicolumn{3}{c}{$=1$ OSSF lepton pair with $\lvert~\mZone -m_Z\rvert<10~\GeV\ $}\\
$\pt\left(\ell_1,\ell_2\right)$ & \multicolumn{3}{c}{$> 30,~15~\GeV\ $}\\
\midrule
& \textbf{SR-2$\ell$-5j2b} & \textbf{SR-2$\ell$-6j1b} & \textbf{SR-2$\ell$-6j2b} \\
$N_\text{jets}\left(\pt>25~\GeV\right)$ & $=5$ & $\ge 6$ & $\ge 6$\\
$N_{b\text{-tagged jets}@77\%}$ & $\ge 2$ & $=1$ & $\ge 2$\\
\bottomrule
\end{tabular}
\end{center}
\end{table}
 
\subsection{Trilepton signal regions}
 
A trilepton preselection is defined by requiring exactly three signal leptons,
and their transverse momenta must be higher than 27, 20 and 15~\GeV.
Amongst these three leptons, the OSSF pair with invariant mass closest to the \Zboson-boson mass is considered to originate from the \Zboson-boson decay, and its invariant mass (\mZone) has to be within $10~\GeV$ of the \Zboson-boson mass.
Furthermore, all OSSF lepton combinations are required to have $m_\text{OSSF}>10~\GeV$ to remove contributions arising from low-mass resonances, which are not included in the simulation.
At least three jets are required, of which at least one has to be $b$-tagged (with $85\%$ tagging efficiency).
All trileptonic preselections and SR selections are summarised in Table~\ref{tab:3l-selection}.
 
A 3-class DNN is trained to identify \ttZ, \tZq and diboson events amongst those kept after the 3$\ell$ preselection is applied.
Table~\ref{tab:3l_variable_definitions} in the Appendix details the variables used in the DNN.
The trilepton phase-space after preselection is partitioned into three signal regions labelled SR-3$\ell$-ttZ, SR-3$\ell$-tZq and SR-3$\ell$-WZ, according to the largest output of the three decision nodes. These selections are summarised in Table~\ref{tab:3l-selection}.
While the \tZq and \WZb contributions are largest in SR-3$\ell$-tZq and SR-3$\ell$-WZ, respectively, these regions still contain a non-negligible number of \ttZ signal events.
The three SRs are mutually exclusive by construction and together fill the entire phase-space after preselection.
A tighter $b$-tagging requirement (at least one $b$-tagged jet at $60\%$ efficiency) is then additionally applied in SR-3$\ell$-WZ
to efficiently suppress the contributions from the lighter-flavour \WZl\footnote{In this context, we refer to the light-flavour quarks $u$, $d$, and $s$ by the label $l$, not to be confused with the label $\ell$ reserved for leptons.} and \WZc backgrounds,
retaining only the \WZb component since it is an important background in the other two SRs also.
 
\begin{table}[!htb]
\footnotesize
\caption{Definition of the trilepton signal regions.}
\label{tab:3l-selection}
\def\arraystretch{1.3}
\begin{center}
\begin{tabular}{lccc}
\toprule
Variable & \multicolumn{3}{c}{\textbf{Preselection}}\\
\midrule
$N_\ell~\left(\ell=e,\mu\right)$ & \multicolumn{3}{c}{$=3$}\\
& \multicolumn{3}{c}{$\ge 1$ OSSF lepton pair with $\lvert~\mZone -m_Z\rvert<10~\GeV\ $}\\
& \multicolumn{3}{c}{for all OSSF combinations: $m_\text{OSSF}>10~\GeV\ $}\\
$\pt\left(\ell_1,\ell_2,\ell_3\right)$ & \multicolumn{3}{c}{$> 27,~20,~15~\GeV\ $}\\
$N_\text{jets}\left(\pt>25~\GeV\right)$ & \multicolumn{3}{c}{$\ge 3$}\\
$N_{b\text{-tagged jets}}$ & \multicolumn{3}{c}{$\ge 1 @ 85\%$}\\
\midrule
& \textbf{SR-3$\ell$-ttZ} & \textbf{SR-3$\ell$-tZq} & \textbf{SR-3$\ell$-WZ} \\
DNN-tZq output & $<0.43$ & $\ge 0.43$ & --- \\
DNN-WZ output & $<0.27$ & $<0.27$ & $\ge 0.27$ \\
$N_{b\text{-tagged jets}}$ & --- & --- & $\ge 1 @ 60\%$\\
\bottomrule
\end{tabular}
\end{center}
\end{table}
 
\FloatBarrier
 
\subsection{Tetralepton signal regions}
 
The tetralepton channel is defined by the requiring exactly four leptons, of which at least one must have transverse momentum greater than $27~\GeV$ and two must form an OSSF pair with invariant mass within $20~\GeV$ of the \Zboson-boson mass.
The sum of the lepton charges is required to be zero.
Low-mass dilepton resonances are removed by requiring the invariant mass of all OSSF pairs to be greater than $10~\GeV$.
The OSSF pair with invariant mass closest to the \Zboson-boson mass is taken to be the \Zboson candidate; selected events can then be further categorised according to the flavour of the non-\Zboson lepton pair, into a same-flavour (SF) or different-flavour (DF) signal region.
As a result, the $ZZ+$jets background mostly populates the SF region.
Additionally, at least two jets, including at least one $b$-tagged jet (with $85\%$ tagging efficiency), are required.
All tetraleptonic preselections and SR selections are summarised in Table~\ref{tab:4l-selection}.
 
To achieve better signal sensitivity, a DNN is trained in both signal regions to discriminate between the \ttZ signal and the
background processes, with the input variables listed in Table~\ref{tab:4l_variable_definitions} of the Appendix.
Events featuring two pairs of same-flavour leptons are particularly sensitive to contributions from the $ZZ+$jets background process;
they can be removed from the SR by applying a selection requirement on the DNN output.
The SR definitions are summarised in Table~\ref{tab:4l-selection}.
 
\begin{table}[!htb]
\footnotesize
\caption{Definition of the tetralepton signal regions.}
\label{tab:4l-selection}
\def\arraystretch{1.3}
\begin{center}
\begin{tabular}{lcc}
\toprule
Variable & \multicolumn{2}{c}{\textbf{Preselection}}\\
\midrule
$N_\ell~\left(\ell=e,\mu\right)$ & \multicolumn{2}{c}{$=4$}\\
& \multicolumn{2}{c}{$\ge 1$ OSSF lepton pair with $\lvert~\mZone -m_Z\rvert<20~\GeV\ $}\\
& \multicolumn{2}{c}{for all OSSF combinations: $m_\text{OSSF}>10~\GeV\ $}\\
$\pt\left(\ell_1,\ell_2,\ell_3,\ell_4\right)$ & \multicolumn{2}{c}{$> 27,~7,~7,~7~\GeV\ $}\\
The sum of lepton charges & \multicolumn{2}{c}{$=0$} \\
$N_\text{jets}\left(\pt>25~\GeV\right)$ & \multicolumn{2}{c}{$\ge 2$}\\
$N_{b\text{-tagged jets}}$ & \multicolumn{2}{c}{$\ge 1 @ 85\%$}\\
\midrule
& \textbf{SR-4$\ell$-SF} & \textbf{SR-4$\ell$-DF}\\
$\ell\ell^{\text{non-}Z}$ & $e^{+}e^{-}$ or $\mu^{+}\mu^{-}$ & $e^{\pm}\mu^{\mp}$\\
DNN output & $\ge 0.4$ & --- \\
\bottomrule
\end{tabular}
\end{center}
\end{table}
 
\subsection{Particle- \& parton-level selections}
\label{sec:particlepartondefs}
 
The particle- and parton-level selections for the $3\ell$ and $4\ell$ channels used for the differential measurements are summarised in Table~\ref{tab:fiducial-volumes}.
The particle-level fiducial regions are constructed to closely follow the detector-level regions, using the particle-level objects defined in Section~\ref{sec:reconstruction} and with at least one OSSF lepton pair within $\pm 10~\GeV$ of the \Zboson-boson mass.
The parton-level fiducial volumes are defined by the \ttbar decays: semileptonic ($e,\mu$+jets only) in the $3\ell$ channel, and dileptonic ($e^+e^-,e^\pm\mu^\mp,\mu^+\mu^-$ only) in the $4\ell$ channel.
The \Zboson boson is required to decay dileptonically via $Z\to e^+e^-,\mu^+\mu^-$.
Events featuring $\tau$-leptons which originate directly from either the \Zboson boson or the \Wboson bosons from the \ttbar system are removed from the parton-level fiducial volume and are not considered in the unfolding, regardless of their subsequent decay.
The differential variables are reconstructed from the top quarks after final-state radiation, immediately prior to their decays.
The invariant mass of the two leptons from the \Zboson-boson decay is required to be within $\pm 15~\GeV$ of the \Zboson-boson mass; this widening of the mass window at parton level diminishes the impact of reconstruction and acceptance efficiency uncertainties on the unfolding procedure.
 
\begin{table}[!htb]
\footnotesize
\caption{Definition of the fiducial volumes at particle level and parton level. Leptons refer exclusively to electrons and muons; they are dressed with additional photons at particle level, but not at parton level.}
\label{tab:fiducial-volumes}
\def\arraystretch{1.3}
\begin{center}
\begin{tabular}{cc}
\toprule
\multicolumn{2}{c}{\textbf{Particle-level selection}}\\
$3\ell$  channel & $4\ell$ channel\\
\midrule
Exactly 3 leptons, with $\pT(\ell_1,\ell_2,\ell_3)>27,20,15~\GeV$ & Exactly four leptons, with  $\pT(\ell_1,\ell_2,\ell_3,\ell_4)>27,7,7,7~\GeV$ \\
The sum of charges is $\pm 1$ & The sum of charges is $=0$ \\
At least 3 jets, with $\pT >25~\GeV$ & At least 2 jets, with $\pT >25~\GeV$ \\
\multicolumn{2}{c}{At least 1 $b$-jet (jet ghost-matched to a $b$-hadron)}\\
\multicolumn{2}{c}{At least one OSSF lepton pair, with $\lvert m_{\ell\ell}-m_Z\rvert<10~\GeV$}\\
\addlinespace[0.3em]
\toprule
\multicolumn{2}{c}{\textbf{Parton-level selection}}\\
$3\ell$  channel & $4\ell$ channel\\
\midrule
$\ttbar\to e^\pm/\mu^\pm$ + jets & $\ttbar\to e^\pm\mu^\mp/e^\pm e^\mp/\mu^\pm\mu^\mp$ \\
\multicolumn{2}{c}{$Z\to e^\pm e^\mp/\mu^\pm\mu^\mp$}\\
\multicolumn{2}{c}{$\lvert m_{\ell\ell}-m_Z\rvert<15~\GeV$}\\
\bottomrule
\end{tabular}
\end{center}
\end{table}
 
\subsection{Top-quark reconstruction}
\label{subsec:top_reco}
 
Several different approaches for the kinematic reconstruction of either the \ttbar system or the single (anti-)top quarks are used in this measurement and described in the following, tailored to the characteristics of the various \ttZ decay channels considered in this analysis. A brief summary for each channel is given in the following.
 
\subsubsection{\twolos reconstruction}
In the \twolos channel, two methods are employed to reconstruct the \ttbar system.
The outputs of both of these algorithms are used in the construction of variables that are subsequently used in the training of neural networks since they provide complementary information that can be used to discriminate between the signal and background processes.
 
The first method, referred to as the \textit{multi-hypothesis hadronic top/$W$ reconstruction method},
targets the fully hadronically decaying \ttbar system associated with the signal process,
taking into account several hypotheses for the numbers of available and missing top-quark final-state particles.
At tree level, six jets from the fully hadronic decay of the \ttbar system are part of the \twolos \ttZ signature;
however, due to the jet energy resolution and coverage of the detector, some jets matched to a final-state quark
may not be reconstructed. Five different scenarios\footnote{Considered scenarios: jets from one $W$ boson are present, jets from two $W$ bosons are present, jets from one top quark are present, jets one top quark and one $W$ boson are present, and jets both top quarks are present.}
are considered for the reconstruction procedure,
depending on the numbers of hadronically decaying \Wboson bosons and top quarks that can be reconstructed,
each giving an output weight. For each hypothesis, all jets-to-quarks assignments are tested and the probability of
the hypothesis and combination being correct is calculated from known distributions of two-jet or three-jet invariant mass originating
from the $W$ bosons or top quarks, respectively. The final weight for each hypothesis is the probability of the most likely jet permutation,
and this permutation is considered correct for a given hypothesis.
 
An alternative approach attempts  to reconstruct the all-hadronic \ttbar system through the use of SPANet (Symmetry Preserving Attention Network), an attention-based neural network originally designed for the reconstruction of all-hadronic \ttbar events \cite{Shmakov:2021qdz}.
The network was trained for the dilepton \ttZ topology using both the nominal and alternative \ttZ sample events, required to pass the 2$\ell$OS selection from Table \ref{tab:2l-selection} and have at least six jets (at least one $b$-tagged). One or both top quarks have to be correctly matched, where correct jet assignments are found by matching detector-level objects to parton level.
Inputs are the kinematic and $b$-tagging information of all jets present in the event, as well as the correct jet assignments.
The network predicts the jet assignments for the top and anti-top quarks.
It assigns them correctly in $\approx$56$\%$ of the events where jets from the top quarks are present.
The transverse momentum of the reconstructed all-hadronic \ttbar system, $p_{\mathrm{T}}^{t\bar{t}}$, is then used as an input to the MVA discriminant in the 2$\ell$OS channel.
 
\subsubsection{$3\ell$ reconstruction}
The full reconstruction of the \ttbar system in $3\ell$ events is performed by first reconstructing the leptonic-side top quark\footnote{The term leptonic-side top quark is taken to mean the top or anti-top quark from the \ttbar pair for which the \Wboson-boson decays via $\Wboson\to\ell\nu$, and similarly $\Wboson\to q'\bar{q}$ for the term hadronic-side.} and subsequently reconstructing the hadronic-side top quark.
For the leptonic-side the \MET is attributed to the neutrino from the associated \Wboson-boson decay.
The neutrino momentum in the $z$ direction ($p_{\nu z}$) can be determined from a quadratic equation constrained by the SM \Wboson-boson mass, resulting in up to two distinct solutions, which are both considered.
In the few cases where no real solution exists, the neutrino $\pT$ is adjusted such that the quadratic determinant becomes zero and a single solution for $p_{\nu z}$ exists.
Each \Wboson~boson candidate is then paired with the nearest (in $\DeltaR$) $b$-tagged jet.
The most likely top-quark candidate is determined from the $p(m_{b\ell\nu})$ probability density distribution obtained from MC \ttZ simulations.
 
The hadronic-side reconstruction builds top-quark candidates from jet pairs compatible with a \Wboson~boson, and a $b$-tagged jet.
Since the $b$-tagged jet on the leptonic-side has already been determined, the remaining jet with the highest $b$-tagging score is assigned to the hadronic-side top quark.
From all the other jets, the two most compatible with having originated from a \Wboson~boson are determined via interpolation of $m_{jj}$ with the reference distributions used in the multi-hypothesis reconstruction method.
 
The exact same top reconstruction algorithms are applied at particle level, using the two highest-\pT jets that were ghost-matched to $b$-hadrons to define the $b$-candidates.
 
\subsubsection{$4\ell$ reconstruction}
In this channel a full kinematic reconstruction of the \ttZ system is performed employing the Two Neutrino Scanning Method (\twovSM), improving on the previous analysis~\cite{TOPQ-2018-08} where it was reconstructed in the transverse plane only.
Values of the azimuthal angle and pseudorapidity of either neutrino are tested by systematically scanning the $\eta$--$\phi$ space and, with the set of the respective values at each point in the $\eta$--$\phi$ space of the two neutrinos, the \ttbar signature is constructed from the information about the two leptons which are not associated with the \Zboson-boson decay and the two jets which have the highest $b$-tagging score.
 
Kinematic constraints from reference distributions are used to create a single output weight, \wvsm, for each of the hypotheses and then the combination with the largest weight is selected as the reconstructed dileptonic \ttbar system.
This output weight shows high discrimination power between \ttZ and dileptonic \ttbar events, and therefore can be used as a discriminating variable for MVA training.
 
At particle level, a pseudo-top-quark reconstruction algorithm is employed. First, the two leading-\pT neutrinos (from the MC truth record) are matched to the two charged leptons left in the event after the \Zboson boson candidate is determined.
From the two possible lepton--neutrino pairings, the one that yields an invariant mass closest to the \Wboson-boson mass is retained.
The two highest-\pT jets that are ghost-matched to $b$-hadrons define the $b$-candidates; when only one such jet exists in the event, the second $b$-candidate is taken to be the leading-\pT jet amongst those left available.
As before, both pairings of $b$- and $W$-candidates are considered, and the one that yields an invariant mass closest to the MC top-quark mass ($172.5\,\GeV$) is used to define the reconstructed top and anti-top quarks.


\section{Background estimation}

Several processes can lead to background contamination in the signal regions. The contributions from SM processes featuring the production of two, three or four prompt
leptons is discussed in Section~\ref{subsec:prompt_lepton_background}, whereas the estimation of backgrounds from processes where at least one of the reconstructed leptons originates from a non-prompt process is explained in Section~\ref{subsec:fake_lepton_background}.
 
\subsection{Prompt-lepton backgrounds}
\label{subsec:prompt_lepton_background}
 
\subsubsection{Prompt backgrounds in $2\ell$OS regions}
\label{subsec:DD_ttbar}
 
The opposite-sign dilepton channel is dominated by two large, prompt contributions: dileptonic \ttbar and $Z+$jets.
The former enters primarily in the regions where the two $b$-jets can be tagged and reconstructed; it makes up  $35\%$ of the total expected event yields in the SR-$2\ell$-5j2b and SR-$2\ell$-6j2b signal regions, but only $10\%$ in SR-$2\ell$-6j1b.
The $Z+$jets process, on the other hand, contributes around $80\%$ of the event yield in the SR-$2\ell$-6j1b signal region; this decreases to 55\%--60\% in the other two regions.
Since the modelling of $Z+$jets in high jet-multiplicity regions can be problematic, especially when involving heavy-flavour jets, it is important to correct the predictions obtained from MC simulations with data: the normalisation of the \Zb and \Zc components are therefore obtained in data, simultaneously with the extraction of the signal strength in the combined inclusive fit described in Section~\ref{sec:results_inclusive}.
 
To better model the dileptonic \ttbar process, where also a number of additional jets are present due to the selection, a fully data-driven approach which relies on the high \ttbar purity of an $e\mu$ selection is preferred instead.
Selection criteria are further applied to replicate those of the signal regions, as defined in Table~\ref{tab:2l-selection},
and therefore limit the extrapolation between regions only to the change in lepton flavour to an OS different-flavour (DF)
lepton pair ($e^\pm\mu^\mp$ in the regions used for the data-driven \ttbar estimation, and $e^\pm e^\mp/\mu^\pm\mu^\mp$ in the SRs).
These requirements are summarised in Table~\ref{tab:vr-ttbar-selection} below.
The estimation of the \ttbar background in the signal regions uses the distributions of the DNN output in data in the regions 2$\ell$-$e\mu$-6j1b, 2$\ell$-$e\mu$-6j2b and 2$\ell$-$e\mu$-5j2b.
 
\begin{table}[!htb]
\footnotesize
\caption{Definition of the dilepton regions used for data-driven estimation of the \ttbar background.}
\label{tab:vr-ttbar-selection}
\def\arraystretch{1.3}
\begin{center}
\begin{tabular}{lccc}
\toprule
Variable & \multicolumn{3}{c}{\textbf{Preselection}}\\
\midrule
$N_\ell~\left(\ell=e,\mu\right)$ & \multicolumn{3}{c}{$=2$}\\
& \multicolumn{3}{c}{$=1$ OSDF lepton pair with $\lvert \mZone -m_Z\rvert<10~\GeV\ $}\\
$\pt\left(\ell_1,\ell_2\right)$ & \multicolumn{3}{c}{$> 30,~15~\GeV\ $}\\
\midrule
& \textbf{2$\ell$-$e\mu$-5j2b} & \textbf{2$\ell$-$e\mu$-6j1b} & \textbf{2$\ell$-$e\mu$-6j2b} \\
$N_\text{jets}\left(\pt>25~\GeV\right)$ & $=5$ & $\ge 6$ & $\ge 6$\\
$N_{b\text{-tagged jets}@77\%}$ & $\ge 2$ & $=1$ & $\ge 2$\\
\bottomrule
\end{tabular}
\end{center}
\end{table}
 
To be able to use the distribution of the DNN output from data $e\mu$ events in the $\ell\ell$ signal regions, the different acceptances and efficiencies need to be considered, and the non-\ttbar background must be taken into account.
The MC prediction for all non-\ttbar background is first subtracted from the distribution of the DNN output in the $e\mu$ data.
Then, the following correction factor is applied to the resulting data $e\mu$ distributions of the DNN output\footnote{Since the DNN is not sensitive to the flavour of the leptons (see Table~\ref{tab:2l_variable_definitions}), no $e\mu\to\ell\ell$ correction to the shape of the DNN output is needed.}:
\begin{equation*}
C_{\ttbar} = \dfrac{N_{\ttbar}^{\ell\ell}}{N_{\ttbar}^{e\mu}},
\end{equation*}
where $N_{\ttbar}^{\ell\ell}$ and $N_{\ttbar}^{e\mu}$ are the numbers of expected \ttbar events (from MC predictions) after the $\ell\ell$ selection in the SRs and the $e\mu$ selection in the regions for the data-driven \ttbar estimation, respectively.
The total uncertainty in this number is derived by including both the MC statistical error and the differences between alternative MC predictions, obtained by comparing the nominal value of the ratio with the values obtained by using a different \hdamp parameter value, a different showering algorithm (\HERWIG[7.0.4]) and a different matrix element generator (\MGNLO[2.3.3]).
The correction factors obtained in the three regions are found to agree very closely, and therefore an average factor of $0.982\pm0.009$ is used to apply the $e\mu \ \rightarrow \ \ell\ell$ correction in all dilepton signal regions. In addition to this uncertainty, both the statistical uncertainty related to the Poisson fluctuations of the data and the statistical uncertainty of the subtracted MC backgrounds are taken into account bin-by-bin for this background in all regions.
The distribution of the $b$-tagged-jet multiplicity in 2$\ell$-$e\mu$-6j2b, illustrating the mismodelling of the $t\bar{t}$ background
in high jet-multiplicity regions and highlighting the need to use the data-driven approach, is shown in Figure~\ref{fig:2L-vr-control_1}(a).
The distributions of the DNN output, which are used in the data-driven estimation of the
\ttbar background in the 2$\ell$OS signal regions, are shown in Figures~\ref{fig:2L-vr-control_1}(b)--\ref{fig:2L-vr-control_1}(d).
 
\begin{figure}[!htb]
\centering
\subfloat[]{\includegraphics[width=0.25\textwidth]{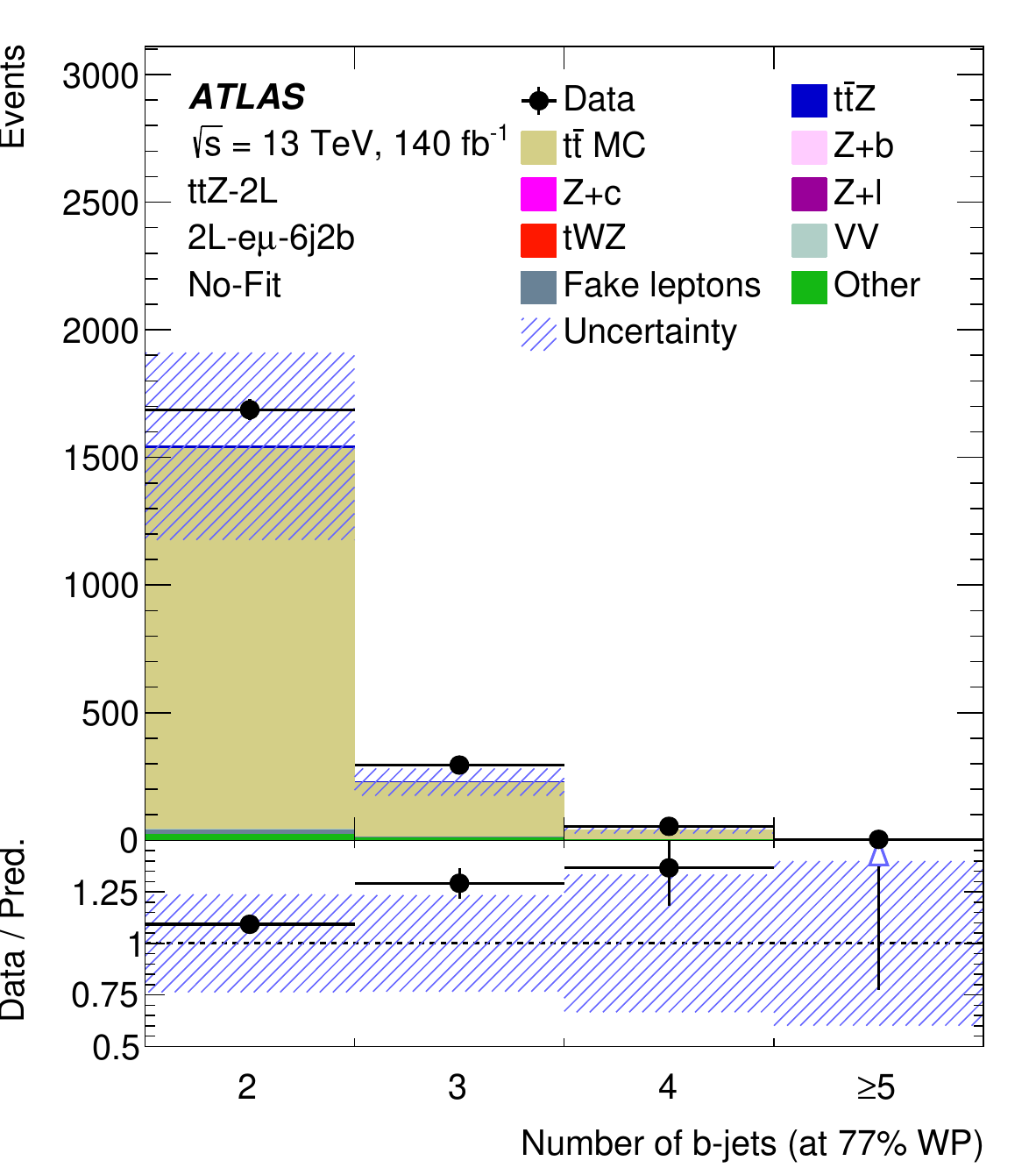}}
\subfloat[]{\includegraphics[width=0.25\textwidth]{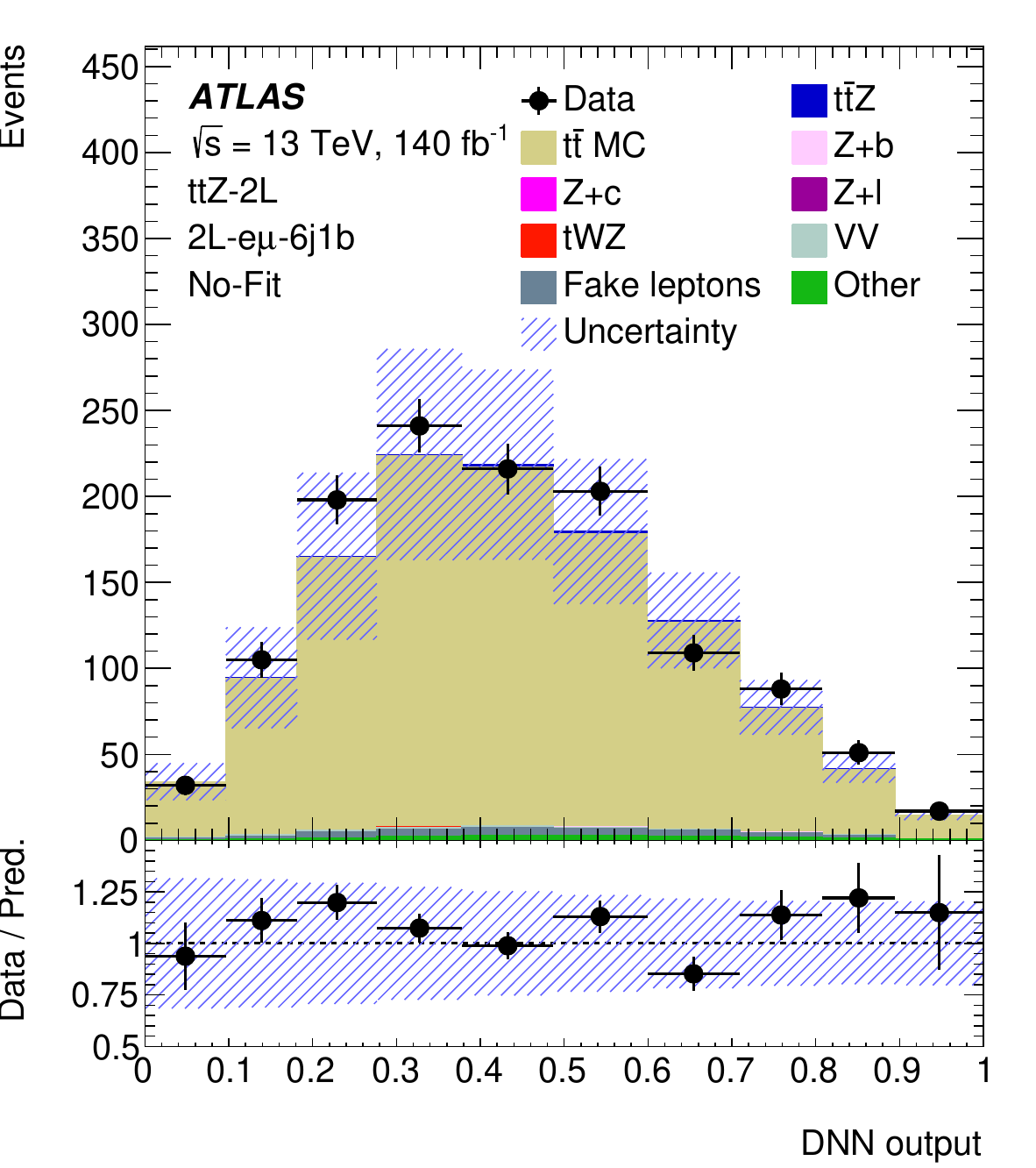}}
\subfloat[]{\includegraphics[width=0.25\textwidth]{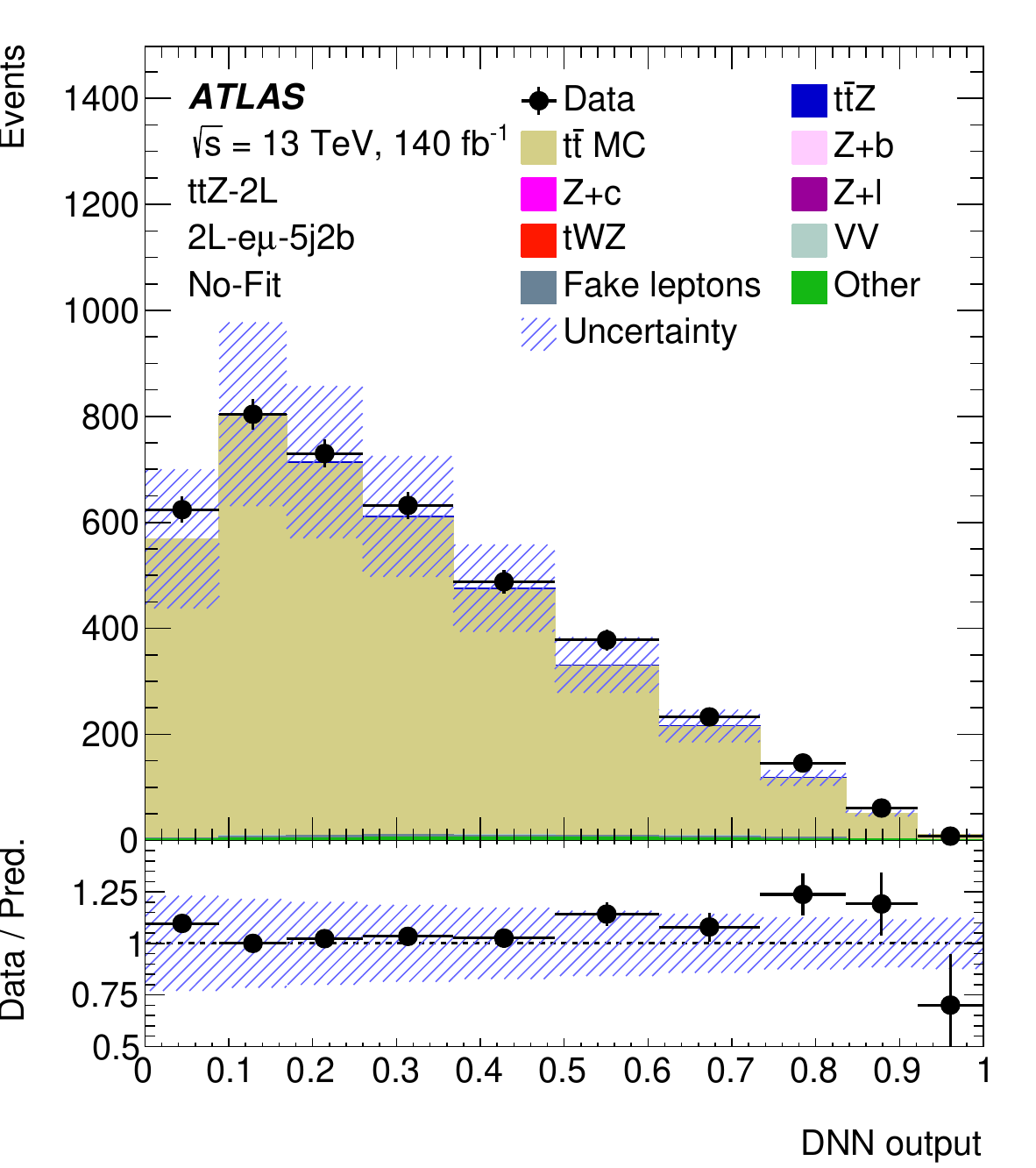}}
\subfloat[]{\includegraphics[width=0.25\textwidth]{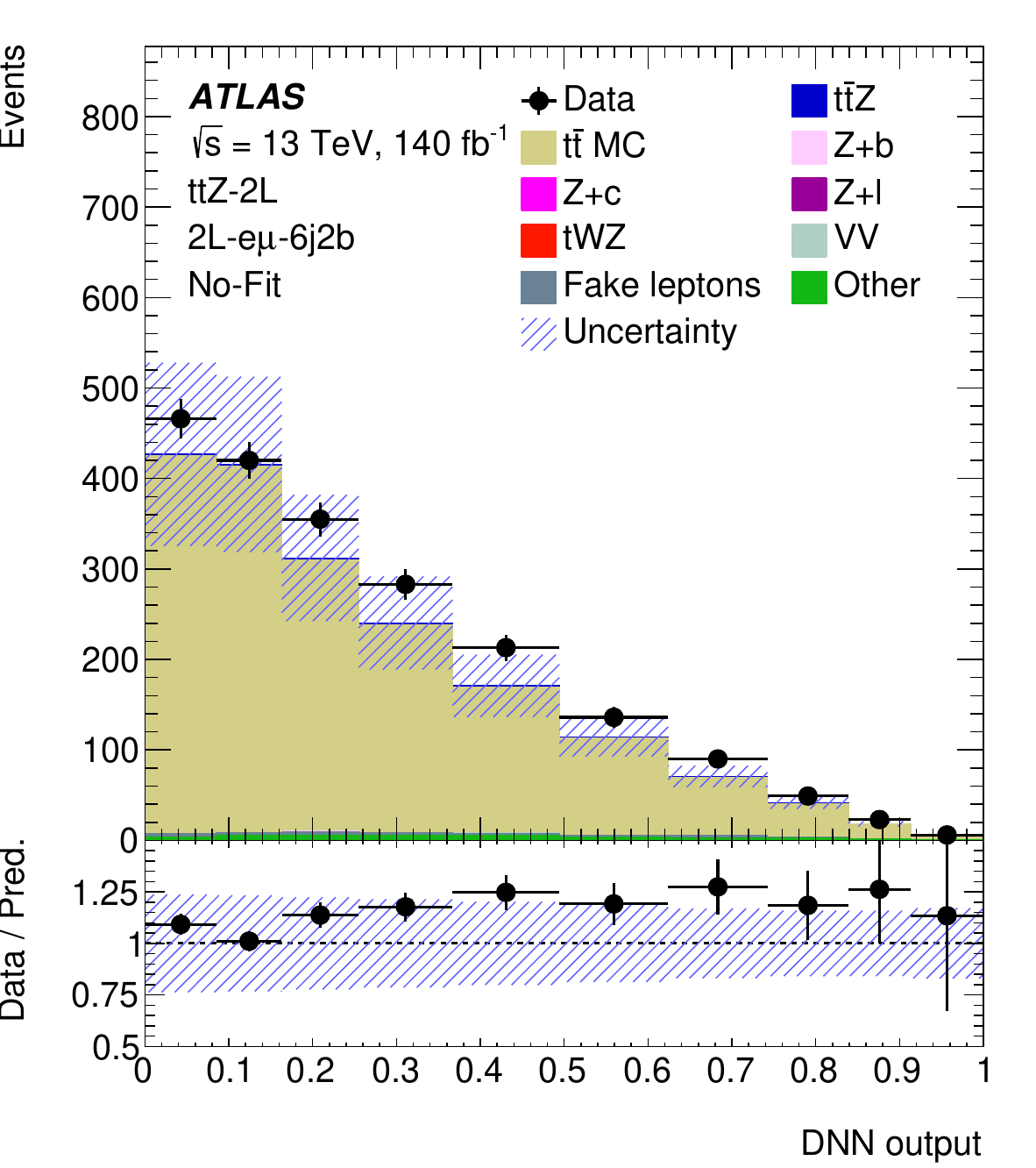}}
\caption{Distributions of (a) the number of $b$-tagged jets in 2$\ell$-$e\mu$-6j2b, and the DNN output in (b) 2$\ell$-$e\mu$-6j1b,
(c) 2$\ell$-$e\mu$-5j2b and (d) 2$\ell$-$e\mu$-6j2b.
The shaded band corresponds to the total uncertainty (systematic and statistical) of the total pre-fit SM prediction.
The lower panel shows the ratio of the data to the SM prediction. The last bin includes the overflow.
These regions are not included in the fit and are only used for the fully data-driven estimation of the \ttbar background in the dilepton signal regions.
}
\label{fig:2L-vr-control_1}
\end{figure}

\subsubsection{Prompt backgrounds in $3\ell$ regions}
 
The dominant background processes in the trilepton signal regions are $WZ+$jets (with $WZ\to\ell\ell\ell\nu$) and $tZq$ production.
The heavy-flavour components of $WZ+$jets, in particular \WZb, is most relevant, accounting for ${\sim}5\%$ of the predicted event yield in SR-3$\ell$-ttZ, ${\sim}8\%$ in SR-3$\ell$-tZq and ${\sim}31\%$ in SR-3$\ell$-WZ.
The \WZc and \WZl contributions are roughly 3 and 20 times smaller, respectively.
The \tZq background is most relevant in SR-3$\ell$-tZq, where it makes up ${\sim}22\%$ of the predicted event yield -- while the \ttZ signal process is still twice as large.
Since the \tZq process cannot be completely separated from \ttZ, it is kept fixed to its best SM prediction and an appropriate set of normalisation and shape uncertainties are considered.
The \WZb background, on the other hand, can be normalised to data because of its large contribution compared to other processes in SR-3$\ell$-WZ.
Another relevant process is $tWZ$, an irreducible singly resonant background to \ttZ.
Due to its kinematic properties being so close to those of the signal process, no efficient discrimination can be obtained from the neural network used to define the $3\ell$ signal regions: the $tWZ$ process is therefore fixed to its SM prediction and yields a uniform 6\%--10\% contribution across the various DNN discriminant bins.
 
\subsubsection{Prompt backgrounds in $4\ell$ regions}
 
The main background in the tetralepton channel is $ZZ+$jets, with $ZZ\to\ell^+\ell^-\ell^{'+}\ell^{'-}$.
This background mainly affects the same-flavour signal region, but can also contribute in a minor way to the different-flavour region through $Z\to\tau^+\tau^-\to e^\pm \mu^\mp \nu_{\tau^+}\nu_{\tau^-}\nu_{e^\mp}\nu_{\mu^\pm}$.
As in the trilepton channel, the \tWZ process is a significant irreducible background, contributing between $8\%$ and $10\%$ of the total event yield in each of the signal regions.
Other rare processes, such as $VH$ or \ttH, typically contribute ${\lesssim}2\%$.
The contribution from the $ZZ+$jets process in the same-flavour signal region is about $13\%$, mostly from the \ZZb component: it is therefore useful to design a dedicated control region to normalise this component in data.
The definition of CR-4$\ell$-ZZ is given in Table~\ref{tab:cr-zz-selection} below.
It is similar to that of SR-4$\ell$-SF (see Table~\ref{tab:4l-selection}), but relies on an inverted cut on the DNN discriminant to ensure orthogonality.
To suppress contributions from the \ZZl and \ZZc components in the extraction of the \ZZb normalisation (each contributing approximately a third of the event yield), the control region is split into two bins based on the PCBT bin value of the $b$-tagged jet with the highest PCBT value.
The first bin contains events for which it is tagged at the 85\%, 77\% and 70\% WPs, while the second bin corresponds to the tightest, i.e.\ 60\%, WP.
The second bin is dominated by the \ZZb background, and there is non-negligible \ZZl and \ZZc contamination in the first bin.
Apart from the \ZZb background, all other prompt background processes in the $4\ell$ channel are kept fixed to their best SM prediction.
 
\begin{table}[!htb]
\footnotesize
\caption{Definition of the tetralepton control region.}
\label{tab:cr-zz-selection}
\def\arraystretch{1.3}
\begin{center}
\begin{tabular}{lc}
\toprule
Variable & \textbf{Preselection}\\
\midrule
$N_\ell~\left(\ell=e,\mu\right)$ & $=4$\\
& $\ge 1$ OSSF lepton pair with $\lvert \mZone -m_Z\rvert<20~\GeV\ $\\
& for all OSSF combinations: $m_\text{OSSF}>10~\GeV\ $\\
$\pt\left(\ell_1,\ell_2,\ell_3,\ell_4\right)$ & $> 27,~7,~7,~7~\GeV\ $\\
The sum of lepton charges & $=0$ \\
$N_\text{jets}\left(\pt>25~\GeV\right)$ & $\ge 2$\\
$N_{b\text{-tagged jets}}$ & $\ge 1 @ 85\%$\\
\midrule
& \textbf{CR-4$\ell$-ZZ}\\
$\ell\ell^{\text{non-}Z}$ & $e^{+}e^{-}$ or $\mu^{+}\mu^{-}$\\
DNN-SF output & $< 0.4$ \\
\bottomrule
\end{tabular}
\end{center}
\end{table}
 
\FloatBarrier
\subsection{Background from non-prompt/fake leptons}
\label{subsec:fake_lepton_background}
 
Fake or non-prompt leptons are objects unintentionally misidentified as prompt leptons.
They can originate from various sources including meson decays, photon conversions or light jets accidentally creating lepton-like detector signatures. In the signal regions the typical source of non-prompt leptons\footnote{Hereafter referred to as \enquote{fake leptons} or \enquote{fakes}.} is the semileptonic decay of heavy-flavour hadrons, mainly from \ttbar and $Z+$jets processes. While the impact of fake leptons is negligible in the $2\ell$OS channel, the contribution can reach $12\%\,(5\%)$ of the expected yield in the $3\ell\,(4\ell)$ SRs.
 
Due to the small number of fake leptons in the dilepton channel, the MC estimate of the fake-lepton background is assigned a conservative 50$\%$ normalisation uncertainty.
To estimate the contribution of fake leptons in the trilepton and tetralepton signal regions, a semi-data-driven method is used, called the \enquote{template fit method}, since unlike in the $2\ell$OS regions the jet multiplicities in the $3\ell$ and $4\ell$ regions are lower and kinematic variables are generally well modelled by the MC. The template fit method relies on using the data to normalise dedicated MC fake-lepton templates built from MC truth-record information about the origin and flavour of the fake leptons. Four distinct MC templates are defined for the major sources of fakes in this analysis: electrons from heavy-flavour sources (\enquote{F-e-HF}), electrons from other sources (\enquote{F-e-Other}), muons from heavy-flavour sources (\enquote{F-$\mu$-HF}), and any other fake leptons that do not belong to any of the other sets or that come from events containing multiple fake leptons (\enquote{F-Other}).
Only this last category is not normalised in data, and receives instead a $50\%$ normalisation uncertainty.
 
To remain as kinematically close as possible to the $3\ell$ signal regions, where the fakes contribution is most important, a set of trilepton control regions are designed. These employ the same lepton \pt, jet multiplicity and $b$-tagged jet multiplicity requirements as in the SRs.
To ensure orthogonality with the SRs defined in Table~\ref{tab:3l-selection}, exactly one lepton must fail to satisfy the identification and isolation requirements applied to signal leptons -- this lepton is said to be \enquote{loose}.
The heavy-flavour fake components can be isolated by defining \ttbar-enriched control regions: any event with an OSSF pair of leptons is vetoed, and the \enquote{loose} lepton is required to be part of the same-sign pair of leptons.
The flavour of the \enquote{loose} lepton is then used to categorise events as CR-\ttbar-e or CR-\ttbar-$\mu$.
On the other hand, the \enquote{F-e-Other} component can be obtained from a \Zboson-like selection (CR-Z-e), requiring exactly three electrons of which two form an OS pair, and vetoing events with \met greater than $80~\GeV$.
These selection criteria are summarised in Table~\ref{tab:3l-fake-cr-selection}.
 
\begin{table}[!htb]
\footnotesize
\caption{Definition of the trilepton fakes control regions.}
\label{tab:3l-fake-cr-selection}
\def\arraystretch{1.3}
\begin{center}
\begin{tabular}{lccc}
\toprule
Variable & \multicolumn{3}{c}{\textbf{Preselection}}\\
\midrule
$N_\ell~\left(\ell=e,\mu\right)$ & \multicolumn{3}{c}{$=3$ (of which $=1$ loose non-tight)}\\
$\pt\left(\ell_1,\ell_2,\ell_3\right)$ & \multicolumn{3}{c}{$> 27,~20,~15~\GeV\ $}\\
Sum of lepton charges & \multicolumn{3}{c}{$\pm 1$}\\
$N_\text{jets}\left(\pt>25~\GeV\right)$ & \multicolumn{3}{c}{$\ge 3$}\\
$N_{b\text{-tagged jets}}$ & \multicolumn{3}{c}{$\ge 1 @ 85\%$}\\
\midrule
& \textbf{CR-\ttbar-e} & \textbf{CR-\ttbar-$\mu$} & \textbf{CR-Z-e} \\
Lepton flavours & no OSSF pair & no OSSF pair & OSSF pair\\
& (loose lepton is an electron) & (loose lepton is a muon) & (exactly 3 electrons)\\
\met & --- & --- & $<80~\GeV$ \\
\bottomrule
\end{tabular}
\end{center}
\end{table}
 
The extraction of the three fake normalisation factors (\nFakesElHF, \nFakesElOther and \nFakesMuHF) is first performed independently of the inclusive combined fit, in order to determine an additional uncertainty due to non-closure of key kinematic distributions in the fake-lepton CRs. This uncertainty, found to be $20\%$ for fake electrons and $10\%$ for fake muons, is applied to the fake templates in the SRs and later used in the combined fits.
In CR-\ttbar-e and CR-\ttbar-$\mu$, the overall event yields are used in the template fit, while in CR-Z-e, the distribution of the transverse mass of the trailing lepton and the missing transverse momentum is fitted in six bins.
These distributions are displayed in Figure~\ref{fig:3L-fake-cr-postfit}, with the corresponding event yields shown in Table~\ref{tab:3L_fake_yields_postfit}, after the inclusive combined fit to data.
The fitted normalisations of the fake lepton backgrounds from the combined fit are shown in Table~\ref{tab:norm_factors_inclusive}
and are consistent with the values obtained from the fit to the fake lepton control regions only.

\begin{figure}[!htb]
\centering
\subfloat[]{\includegraphics[width=0.32\textwidth]{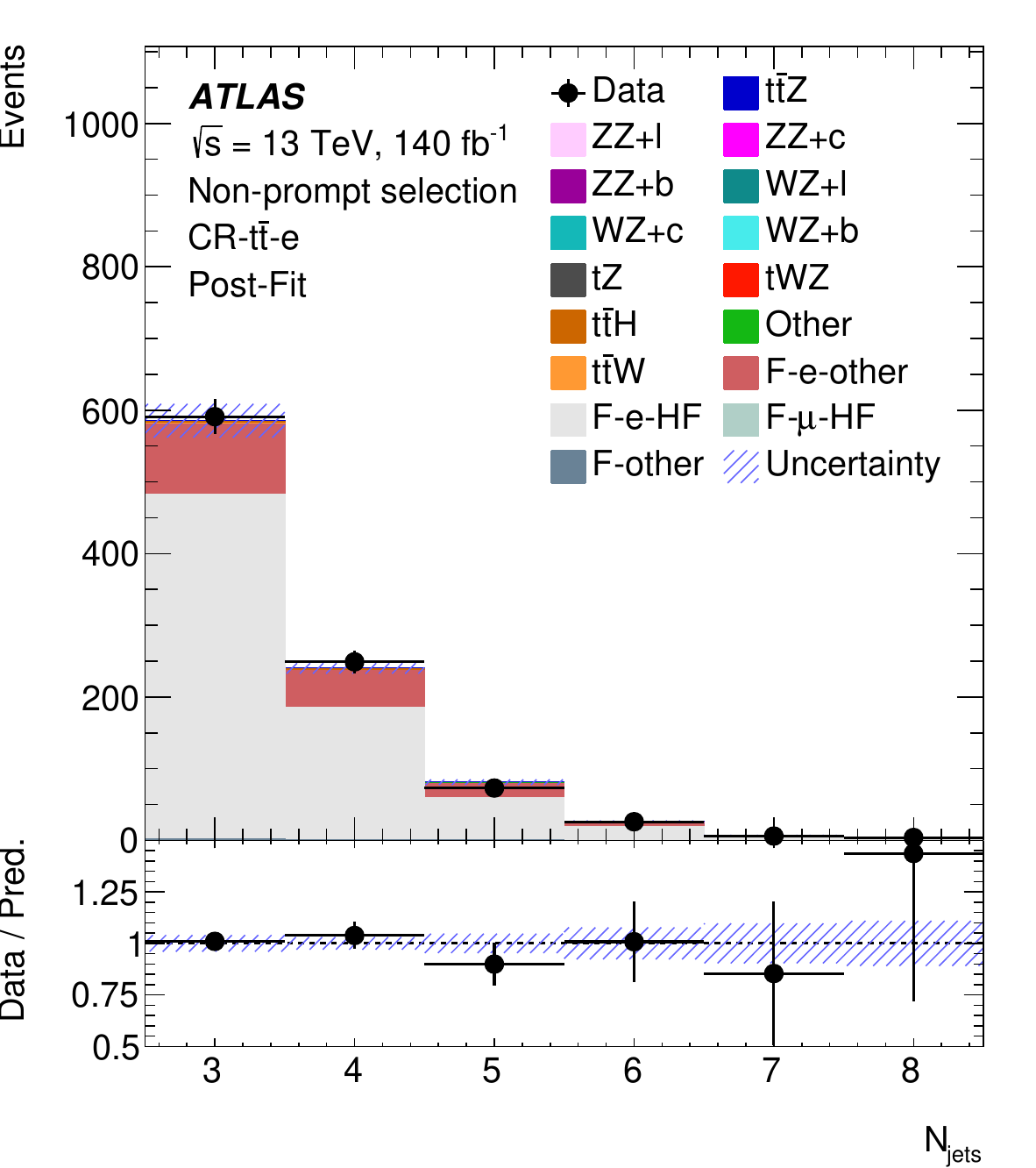}}
\subfloat[]{\includegraphics[width=0.32\textwidth]{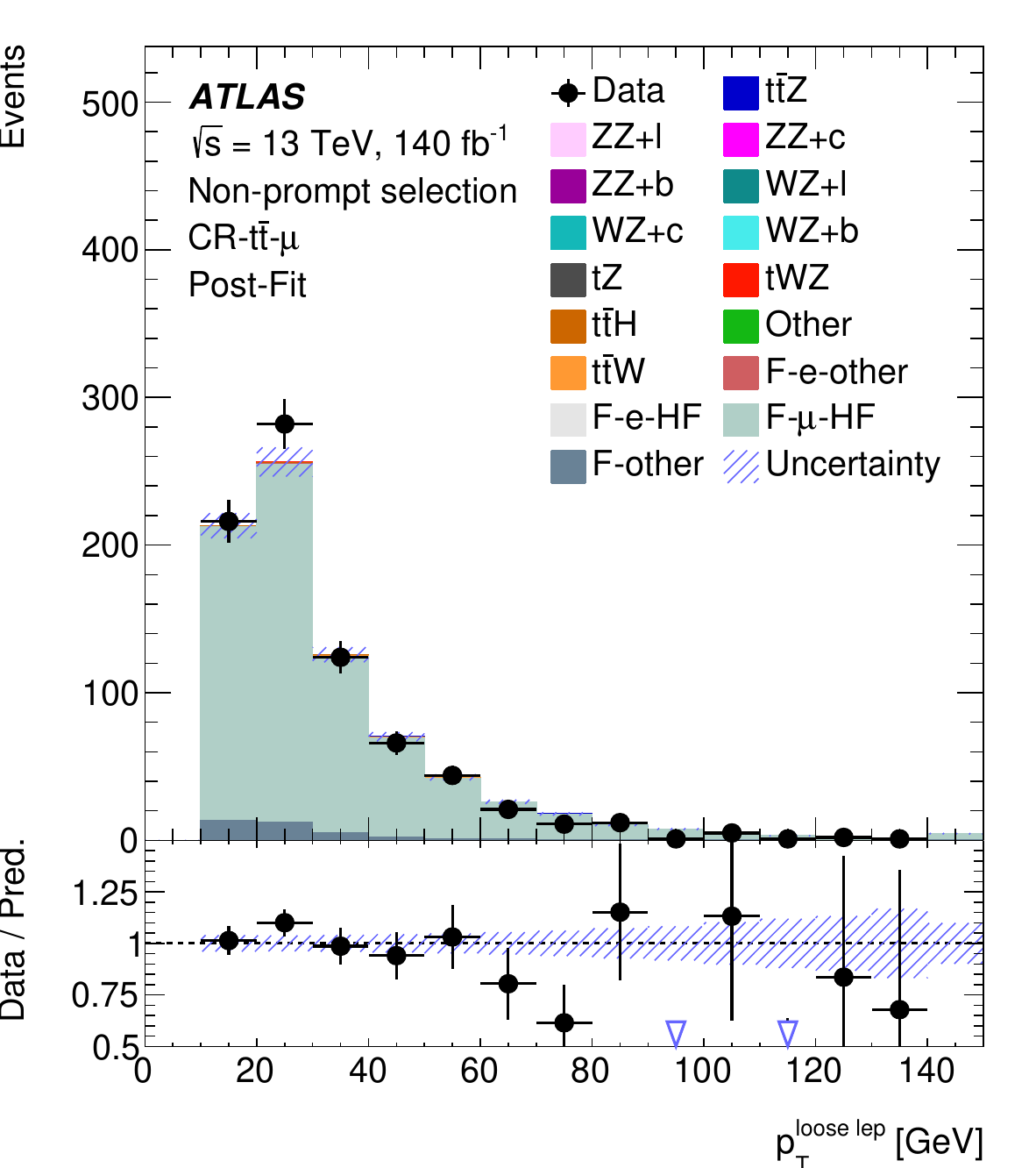}}
\subfloat[]{\includegraphics[width=0.32\textwidth]{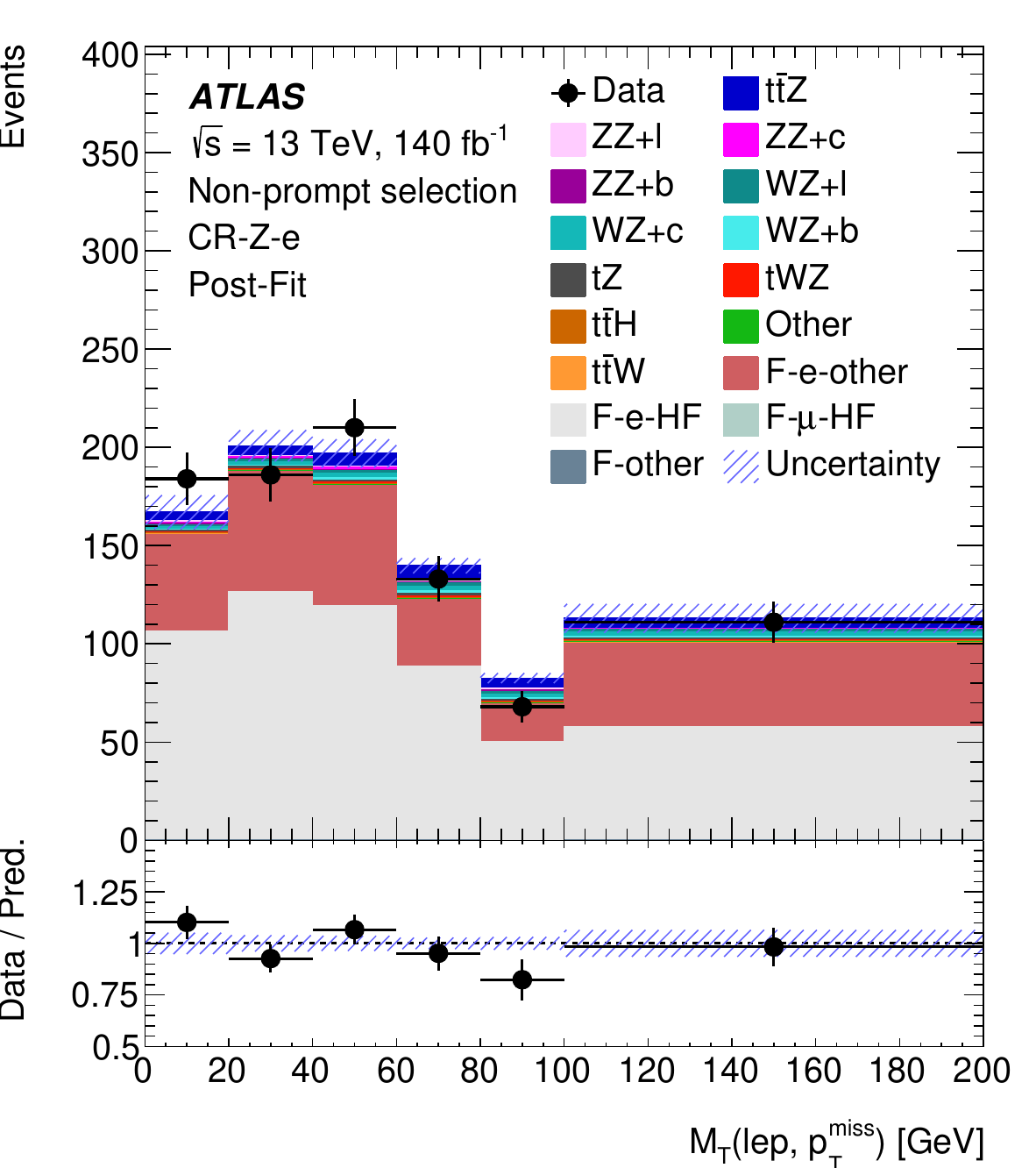}}
\caption{Post-fit distributions of (a) number of jets in CR-\ttbar-e, (b) transverse momentum of the loose lepton in CR-\ttbar-$\mu$, and of (c) the transverse mass of the trailing lepton and the missing transverse momentum in CR-Z-e. The shaded band corresponds to the total uncertainty (systematic and statistical) of the total SM prediction. The lower panel shows the ratio of the data to the SM prediction. The last bin includes the overflow. The distributions (a) and (b) are not used in the fit, instead just an overall event yield in each of these regions is fitted.}
\label{fig:3L-fake-cr-postfit}
\end{figure}
 
\begin{table}[!htb]
\footnotesize
\caption{Observed and expected event yields in the fake-lepton control regions, obtained for an integrated luminosity of \lumi~\ifb after the combined fit to data.
The indicated errors include the Monte Carlo statistical uncertainty as well as all other systematic uncertainties discussed in Section~\ref{sec:systematics}.
A dash (---) indicates event yields smaller than 0.1. Pre-fit yields are reported in Table~\ref{tab:3L_fake_yields_prefit}.}
\label{tab:3L_fake_yields_postfit}
\def\arraystretch{1.3}
\begin{center}
\begin{tabular}{
l
r@{\(\,\pm\,\)}r
r@{\(\,\pm\,\)}r
r@{\(\,\pm\,\)}r
}
\toprule
& \multicolumn{2}{c}{CR-\ttbar-e} & \multicolumn{2}{c}{CR-\ttbar-$\mu$} & \multicolumn{2}{c}{CR-Z-e} \\
\midrule
$t\bar{t}Z$               &  \numRF{2.53459}{3} & \numRF{0.212202}{2} & \numRF{0.711603}{2} & \numRF{0.121359}{2} & \numRF{33.4595}{3}\pho & \numRF{1.39367}{2}\pho \\
\ZZl                      &  \multicolumn{2}{c}{---}         &  \multicolumn{2}{c}{---}       &  \numRF{3.77002}{2}\pho & \numRF{1.14974}{2}\pho \\
\ZZc                      &  \multicolumn{2}{c}{---}         &  \multicolumn{2}{c}{---}      &  \numRF{2.98729}{3} & \numRF{0.953839}{2} \\
\ZZb                      &  \multicolumn{2}{c}{---}        &  \multicolumn{2}{c}{---}        &  \numRF{3.45602}{2}\pho & \numRF{1.78155}{2}\pho \\
\WZl                      &  \numRF{0.287849}{2} & \numRF{0.127478}{2} & \multicolumn{2}{c}{---} & \numRF{7.83794}{2}\pho & \numRF{3.08669}{2}\pho \\
\WZc                      &  \numRF{0.362393}{2} & \numRF{0.134557}{2} & \multicolumn{2}{c}{---} & \numRF{11.2105}{3}\pho & \numRF{4.32305}{2}\pho \\
\WZb                      &  \numRF{0.162643}{2} & \numRF{0.10471}{2} & \multicolumn{2}{c}{---} & \numRF{5.7941}{2}\pho & \numRF{3.29707}{2}\pho \\
\tZq                      &  \numRF{0.213257}{2} & \numRF{0.0521185}{1} & \multicolumn{2}{c}{---} & \numRF{6.54786}{3} & \numRF{0.994654}{2} \\
\tWZ                      &  \numRF{0.260953}{2} & \numRF{0.0448714}{1} & \multicolumn{2}{c}{---} & \numRF{3.81252}{3} & \numRF{0.412437}{2} \\
\ttW                      &  \numRF{2.92355}{2}\pho & \numRF{1.45873}{2}\pho & \numRF{1.50113}{3} & \numRF{0.751444}{2} & \numRF{1.41269}{3} & \numRF{0.705499}{2} \\
$t\bar{t}H$               &  \numRF{4.46172}{3} & \numRF{0.384417}{2} & \numRF{2.30892}{3} & \numRF{0.198353}{2} & \numRF{3.20194}{3} & \numRF{0.288297}{2} \\
Other                     &  \numRF{1.19869}{2}\pho & \numRF{0.538635}{2} & \numRF{0.71652}{2} & \numRF{0.322265}{2} & \numRF{0.517154}{2} & \numRF{0.232626}{2} \\
F-e-Other                 &  \numRF{176.993}{3}\phdoo & \numRF{50.4674}{2}\phdoo & \multicolumn{2}{c}{---} & \numRF{267.131}{3}\phdoo & \numRF{72.4677}{2}\phdoo \\
F-e-HF                    &  \numRF{749.441}{3}\phdoo & \numRF{70.3103}{2}\phdoo & \multicolumn{2}{c}{---} & \numRF{548.462}{3}\phdoo & \numRF{55.8972}{2}\phdoo \\
F-$\mu$-HF                &  \numRF{0.254887}{2} & \numRF{0.0181981}{1} & \numRF{744.156}{3}\phdoo & \numRF{31.4308}{2}\phdoo & \multicolumn{2}{c}{---} \\
F-Other                   &  \numRF{3.01592}{2}\pho & \numRF{1.35098}{2}\pho & \numRF{36.3742}{2}\phdoo & \numRF{16.2006}{2}\phdoo & \numRF{1.03191}{3} & \numRF{0.518964}{2} \\
\midrule
Total                     &  \numRF{942.567}{3}\phdoo & \numRF{30.4194}{2}\phdoo & \numRF{786.214}{3}\phdoo & \numRF{28.0839}{2}\phdoo & \numRF{900.793}{3}\phdoo & \numRF{27.9046}{2}\phdoo \\
\midrule
Data                      &  \multicolumn{2}{l}{\num{949}\phdoo}                      &  \multicolumn{2}{l}{\num{786}\phdoo}                      &  \multicolumn{2}{l}{\num{892}\phdoo} \\
\bottomrule
\end{tabular}
\end{center}
\end{table}
 
\FloatBarrier


\section{Systematic uncertainties}
\label{sec:systematics}

The signal and background predictions are affected by several sources of experimental and theoretical systematic uncertainty.
These are considered for both the inclusive and differential measurements presented in Sections~\ref{sec:results_inclusive} and~\ref{sec:results_differential}.
The uncertainties can be classified into the different categories which are described in the following subsections.
 
\subsection{Detector-related uncertainties}
 
The uncertainty in the combined 2015--2018 integrated luminosity is 0.83\%~\cite{DAPR-2021-01-custom}.
This systematic uncertainty affects all processes determined from MC simulations for which the normalisation is not extracted from data.
 
The uncertainty in the reweighting of the MC pile-up distribution to match the distribution in data is evaluated by varying the pile-up correction factors and has a small impact on both the inclusive and differential results.
 
Uncertainties associated with the lepton selection arise from the trigger, reconstruction, identification and isolation efficiencies, and the lepton momentum scale and resolution~\cite{PERF-2017-01,EGAM-2018-01,MUON-2022-01-custom,MUON-2018-03}. They each have impacts below 1\% on the \ttZ cross-section for the individual sources and have a total impact of 1\%--3\% on the inclusive cross-section measurements depending on the channel.
 
Uncertainties associated with the jet selection arise from the jet energy scale~(JES), the JVT requirement and the jet energy resolution~(JER). The JES and its uncertainties
are derived by combining information from test-beam data, collision data and simulation~\cite{JETM-2018-05-custom}.
The JER has been measured separately for data and MC simulation using two in situ techniques, 
and a systematic uncertainty is obtained by defining its square to be the difference of the squares of the jet energy resolution in data and simulation.
The uncertainties in the JER and from the JVT requirement increase at lower jet \pt.
 
The efficiency of the flavour-tagging algorithm is measured for each jet flavour using control samples in data and in simulation. From these
measurements, correction factors are derived to correct the tagging rates in the simulation. In the case of $b$-tagged jets, the correction factors and
their uncertainties are estimated from dileptonic \ttbar events in data~\cite{FTAG-2018-01}. For $c$-tagged jets, they are derived from
jets arising from \Wboson boson decays in \ttbar events~\cite{FTAG-2020-08}. For light-flavour jets, the correction factors are derived
using dijet events~\cite{FTAG-2019-02}. Sources of uncertainty affecting the $b$-, $c$- and light-flavour-tagging efficiencies are evaluated as a function of
jet \pt, including bin-to-bin correlations.
An additional uncertainty is assigned to account for the extrapolation of the $b$-tagging efficiency measurement from the \pt region used
to determine the correction factors to regions with higher \pt.
The impact of flavour-tagging uncertainties on the inclusive measurement is 1.7\% overall.
 
Uncertainties are assigned to the scale and resolution of the soft track component of the missing transverse momentum. They are derived from differences seen between data and MC simulation when measuring the \pt balance between the hard and soft \MET components~\cite{ATLAS-CONF-2018-023}.
 
\subsection{Signal modelling uncertainties}
To estimate the uncertainties related to missing higher-order effects, the renormalisation and factorisation scale parameters \muR and \muF are either both doubled or both halved in the matrix element calculation and the results of the two variations are compared with the nominal predictions.
Uncertainties in the PDF are evaluated by following the recommended \PDFforLHC prescription~\cite{Butterworth:2015oua} and include uncertainties related to the choice of PDF.
 
The uncertainties associated with the parton showering algorithm and the underlying event model\footnote{In the following, it is referred to as the \enquote{parton shower uncertainty}.} are evaluated by comparing the nominal samples, generated with \MGNLO interfaced to \PYTHIA[8], with equivalent samples interfaced to \HERWIG[7] instead.
Uncertainties related to the modelling of initial-state radiation are obtained by varying the Var3c parameter of the \PYTHIA A14 tune in dedicated alternative samples, and their impact is found to largely cover that of final state radiation uncertainties.
 
The modelling of the \ttZ signal process is cross-checked by comparing the nominal samples with alternative ones produced with the \SHERPA generator, at various levels of precision (see Section~\ref{sec:samples} for more details).
This comparison is not treated as an additional source of systematic uncertainty, as the shape differences between the nominal and \SHERPA setups are found to be contained within the uncertainty band corresponding to the \HERWIG and \tune{A14}~\tune{Var3c} variations described above.
 
\subsection{Background modelling uncertainties}
\label{subsubsec:syst_theory_backgrounds}
 
Uncertainties in the $WZ \rightarrow \ell\ell\ell\nu$ ($WZ$+jets) and $ZZ \rightarrow \ell\ell\ell\ell$ ($ZZ$+jets) backgrounds related to the CKKW matching scale, QSF parameter (resummation scale) and alternative recoil scheme are estimated through the use of alternative truth-level samples.
The renormalisation and factorisation scale uncertainties are evaluated simultaneously in both the matrix element calculation and the parton shower: the values of the scales are varied jointly by a factor of two and the results are compared with the nominal predictions.
Uncertainties related to the choice of PDF are evaluated following the recommended \PDFforLHC prescription and are derived by comparing the nominal value with those from the \CT[14] and \MMHT PDF sets.
Variations of \alphas in the nominal PDF are also included.
In addition, a normalisation uncertainty of $30\%$ is assigned to the \WZl and \WZc components of the $WZ+$jets background, evaluated from discrepancies found in comparisons between data and MC simulations.
Similarly, normalisation uncertainties of $10\%$ and $30\%$ are assigned to the \ZZl and \ZZc components of the $ZZ+$jets background.
 
A cross-section normalisation uncertainty of $14\%$ is assigned to the \tZq process, based on the dedicated ATLAS measurement presented in Ref.~\cite{TOPQ-2018-01}.
A parton shower uncertainty is obtained by comparing an alternative sample of \tZq events generated with \MGNLO interfaced to \HERWIG[7] with the nominal set-up (\MGNLO interfaced to \PYTHIA[8]).
As with \ttZ, variations of the Var3c parameter of the \PYTHIA  A14 tune, and of the matrix element factorisation and renormalisation scales, are considered.
Uncertainties related to the choice of PDF are evaluated following the \PDFforLHC prescriptions.
 
For the \tWZ background process, no parton shower uncertainty is considered, but instead the difference between samples generated with the DR1 and DR2 diagram removal schemes~\cite{Demartin:2016axk} is treated as a modelling uncertainty.
Furthermore, both the shape and normalisation components of this systematic uncertainty are considered: in the absence of higher-order theoretical calculations for the \tWZ process, a comparison of the cross sections obtained in the five-flavour scheme in the DR1 and DR2 set-ups leads to an overall 10\%--15\% normalisation uncertainty.
PDF and scale uncertainties are also taken into account, in the same way as for the other processes described above.
 
Only theoretical uncertainties in the normalisation of the cross section of the \ttH process are considered.
Following the NLO QCD+EWK calculation presented in Ref.~\cite{deFlorian:2016spz}, the scale uncertainty is taken to be $^{+5.8\%}_{-9.2\%}$ and the $\mathrm{PDF}\oplus\mathrm{\alphas}$ uncertainty is $\pm 3.6\%$.
 
Similarly to the diboson processes, variations of the CKKW matching scale and QSF parameter define independent modelling uncertainties for the $Z+$jets process.
Renormalisation and factorisation scales in both the matrix element calculation and the parton shower are raised and lowered by a factor of two, and PDF uncertainties are evaluated according to the \PDFforLHC prescription.
The \Zl component, the only one not extracted directly from a fit to data, is assigned a $10\%$ normalisation uncertainty~\cite{STDM-2016-01}.
The total impact of modelling and normalisation uncertainties on the \Zl component ranges from $32\%$ to $56\%$ in the $2\ell$OS signal regions.
 
The dileptonic \ttbar background in the $2\ell$OS channel is estimated from the data-driven approach described in Section~\ref{subsec:DD_ttbar}.
Modelling uncertainties only enter via the correction factor $C_{\ttbar}$.
The uncertainties on the final data-driven \ttbar template also account for the MC statistical uncertainties of the subtracted MC background templates.
 
For other minor background processes, such as $HV$, $VVV$, \ttW, $\ttWW$ or multi-top-quark (\ttt, \tttt) production, an overall normalisation uncertainty of $50\%$ is applied.
For \tttt, an additional parton shower uncertainty is considered, by comparing samples interfaced to either \PYTHIA[8] or \HERWIG[7].
These background components typically contribute ${\lesssim}1\%$ of the event yield in the signal regions.
As described in Section~\ref{subsec:fake_lepton_background}, the MC template for fake leptons, which cannot be normalised in data as part of the fake-factor method, is assigned a $50\%$ normalisation uncertainty.
Non-closure uncertainties are derived for the other fake templates.
 
A few channel-specific exceptions to the above-mentioned treatment of the theoretical systematic uncertainties of the background are also included.
In the 2$\ell$OS channel, the diboson background is significantly smaller than in the 3$\ell$ and 4$\ell$ channels.
Splitting it into three flavor components and using all aforementioned mentioned uncertainties would lead to large MC statistical uncertainties in those templates,
so all diboson events are treated as one background and are assigned a conservative 50\% uncertainty.
In the 3$\ell$ channel, the \ZZb background is not negligible, but still not large enough to measure its normalisation directly in data, as is done in the 4$\ell$ channel.
A $50\%$ normalisation uncertainty is assigned instead.


\section{Results of the inclusive cross-section measurement}
\label{sec:results_inclusive}

The \ttZ cross section is first measured separately in each channel.
The final result is then obtained by simultaneously fitting all three channels.
 
The fit is based on the profile-likelihood technique~\cite{Cowan:2010js}, with a likelihood function defined as a product of Poisson
probability functions given by the observed event yields in the signal and control regions.
The signal strength $\mu_{\ttZ}$, defined as the ratio of the observed \ttZ cross section to the cross section predicted by the Monte Carlo simulation, and normalisations of some of the backgrounds (specified later in this section) are treated as free parameters of the fit.
Systematic uncertainties described in Section~\ref{sec:systematics} are introduced using additional nuisance parameters with Gaussian constraints.
None of the uncertainties are found to be significantly constrained or pulled in the fit.
 
Since the signal MC samples with leptonic decay of the $Z$ boson also contain a contribution from $\gamma^{*} \rightarrow \ell^{+}\ell^{-}$,
the total cross section has to be corrected in order to remove the photon contribution.
In accordance with the previous ATLAS~\cite{TOPQ-2018-08} and CMS~\cite{ttZDifferentialCMS} measurements,
the inclusive cross-section fiducial volume is defined using the requirement that the invariant mass of the fermion pair originating
from $Z/\gamma^{*}$ must be close to the $Z$-boson mass: $70~\GeV < m_{f\kern-0.1em\bar{f}} <110~\GeV$.
The total cross section is thus corrected by the fraction of parton-level events with a fermion-pair mass in this mass window;
this fraction is found to be 94.5\%, with MC statistical and signal modelling uncertainties well below 0.1$\%$.
The cross section, after correcting by the scale factor, is 0.828~pb.
 
In the dilepton channel, the distributions of the DNN output are fitted in all three SRs. No additional control regions are used.
The binning is optimised to achieve the lowest possible $\mu_{t\bar{t}Z}$ uncertainty.
The event yields are shown in Table~\ref{tab:2L_incl_region_yields_postfit_comb} and the post-fit distributions of the DNN output
are shown in Figure~\ref{fig:incl-SRs-2L}. The free parameters of the dilepton fit are the signal strength and the
normalisations of the $Z+c$ and $Z+b$ backgrounds.
The $t\bar{t}$ background is fixed to its estimated value obtained from $e\mu$ events in data as described in Section~\ref{subsec:DD_ttbar}.
The fitted value of the cross section can be found in the
Table~\ref{tab:incl_cross_section_results}.
 
\begin{table}[!htb]
\footnotesize
\sisetup{group-minimum-digits=4}
\caption{Post-fit event yields in the dilepton signal regions, obtained for an integrated luminosity of \lumi~\ifb.
The values of the fitted parameters from the combined fit are used.
The data-driven approach described in Section~\ref{subsec:DD_ttbar} is used to estimate the $t\bar{t}$ background, denoted by $t\bar{t}$\,DD.
The indicated errors include the Monte Carlo statistical uncertainty as well as all other systematic uncertainties discussed in Section~\ref{sec:systematics}.
Because of rounding and correlations between systematic uncertainties, the values quoted for the total yield and its uncertainty may differ from the simple sum over all processes.}
\label{tab:2L_incl_region_yields_postfit_comb}
\def\arraystretch{1.3}
\begin{center}
\begin{tabular}{
l
r@{\(\,\pm\,\)}r
r@{\(\,\pm\,\)}r
r@{\(\,\pm\,\)}r
}
\toprule
& \multicolumn{2}{c}{SR-2$\ell$-5j2b} & \multicolumn{2}{c}{SR-2$\ell$-6j2b} & \multicolumn{2}{c}{SR-2$\ell$-6j1b} \\
\midrule
$t\bar{t}Z$   & \numRF{297.09 }{3}\phdo & \numRF{20.02 }{2}\phdoo & \numRF{442.54 }{3}\phdo & \numRF{27.36 }{2}\phdo & \numRF{305.15 }{3}\phdo & \numRF{28.39 }{2}\phdo \\
$t\bar{t}$ DD & \numRF{4001.02}{4}\phdo & \numRF{72.34 }{2}\phdoo & \numRF{1912.64}{4}\phdo & \numRF{45.39 }{2}\phdo & \numRF{1161.28}{4}\phdo & \numRF{35.21 }{2}\phdo \\
$Z+b$         & \numRF{5714.10}{3}\phdo & \numRF{169.49}{2}\phdoo & \numRF{2679.04}{3}\phdo & \numRF{111.67}{2}\phdo & \numRF{4833.12}{3}\phdo & \numRF{283.85}{2}\phdo \\
$Z+c$         & \numRF{348.93 }{3}\phdo & \numRF{95.27 }{2}\phdoo & \numRF{188.93 }{3}\phdo & \numRF{47.47 }{2}\phdo & \numRF{2020.02}{3}\phdo & \numRF{479.28}{2}\phdo \\
$Z+l$         & \numRF{59.05  }{2}\phdo & \numRF{25.06 }{2}\phdoo & \numRF{19.58  }{3}      & \numRF{8.10  }{2}      & \numRF{1019.30}{3}\phdo & \numRF{241.43}{2}\phdo \\
$tWZ$         & \numRF{23.16  }{3}      & \numRF{0.92  }{2}       & \numRF{34.40  }{3}      & \numRF{2.09  }{2}      & \numRF{40.16  }{3}      & \numRF{1.86  }{2} \\
Diboson       & \numRF{147.45 }{2}\phdo & \numRF{80.20 }{2}\phdoo & \numRF{94.99  }{2}\phdo & \numRF{51.94 }{2}\phdo & \numRF{337.83 }{2}\phdo & \numRF{184.62}{2}\phdo \\
Fake leptons  & \numRF{28.45  }{2}\phdo & \numRF{14.02 }{2}\phdoo & \numRF{18.57  }{3}      & \numRF{9.15  }{2}      & \numRF{25.35  }{2}\phdo & \numRF{12.49 }{2}\phdo \\
Other         & \numRF{55.13  }{2}\phdo & \numRF{24.76 }{2}\phdoo & \numRF{48.82  }{2}\phdo & \numRF{21.95 }{2}\phdo & \numRF{22.54  }{2}\phdo & \numRF{10.16 }{2}\phdo \\
\midrule
Total         & \numRF{10674.4}{3}\phdo & \numRF{102.88}{2}\phdoo & \numRF{5439.50}{4}\phdo & \numRF{68.32 }{2}\phdo & \numRF{9764.75}{3}\phdo & \numRF{105.01}{2}\phdo \\
\midrule
Data          & \multicolumn{2}{l}{\num{10702}}   & \multicolumn{2}{l}{\num{5435}}   & \multicolumn{2}{l}{\num{9737}} \\
\bottomrule
\end{tabular}
\end{center}
\end{table}
 
\begin{figure}[!htb]
\centering
\subfloat[]{\includegraphics[width=0.32\textwidth]{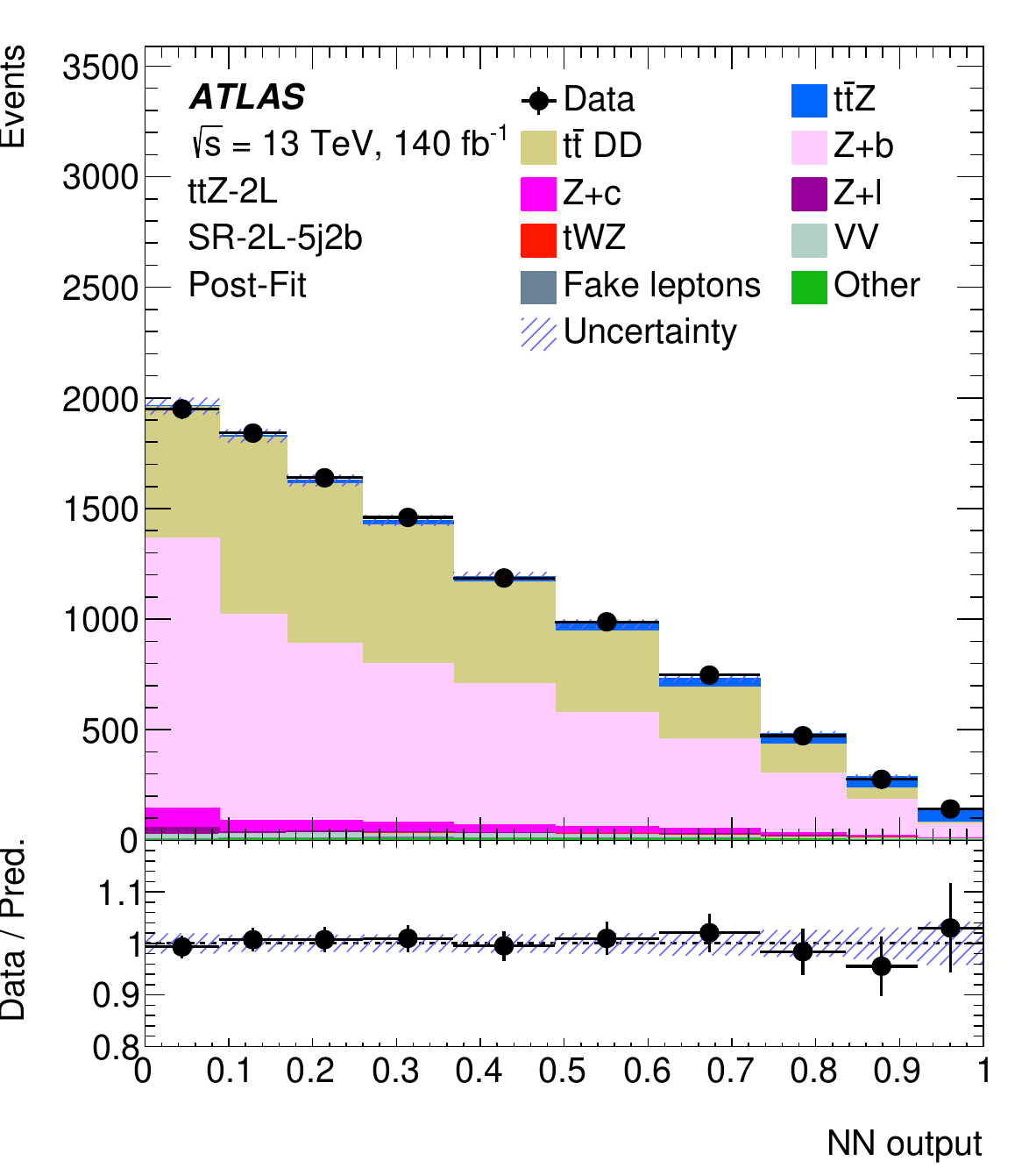}}
\subfloat[]{\includegraphics[width=0.32\textwidth]{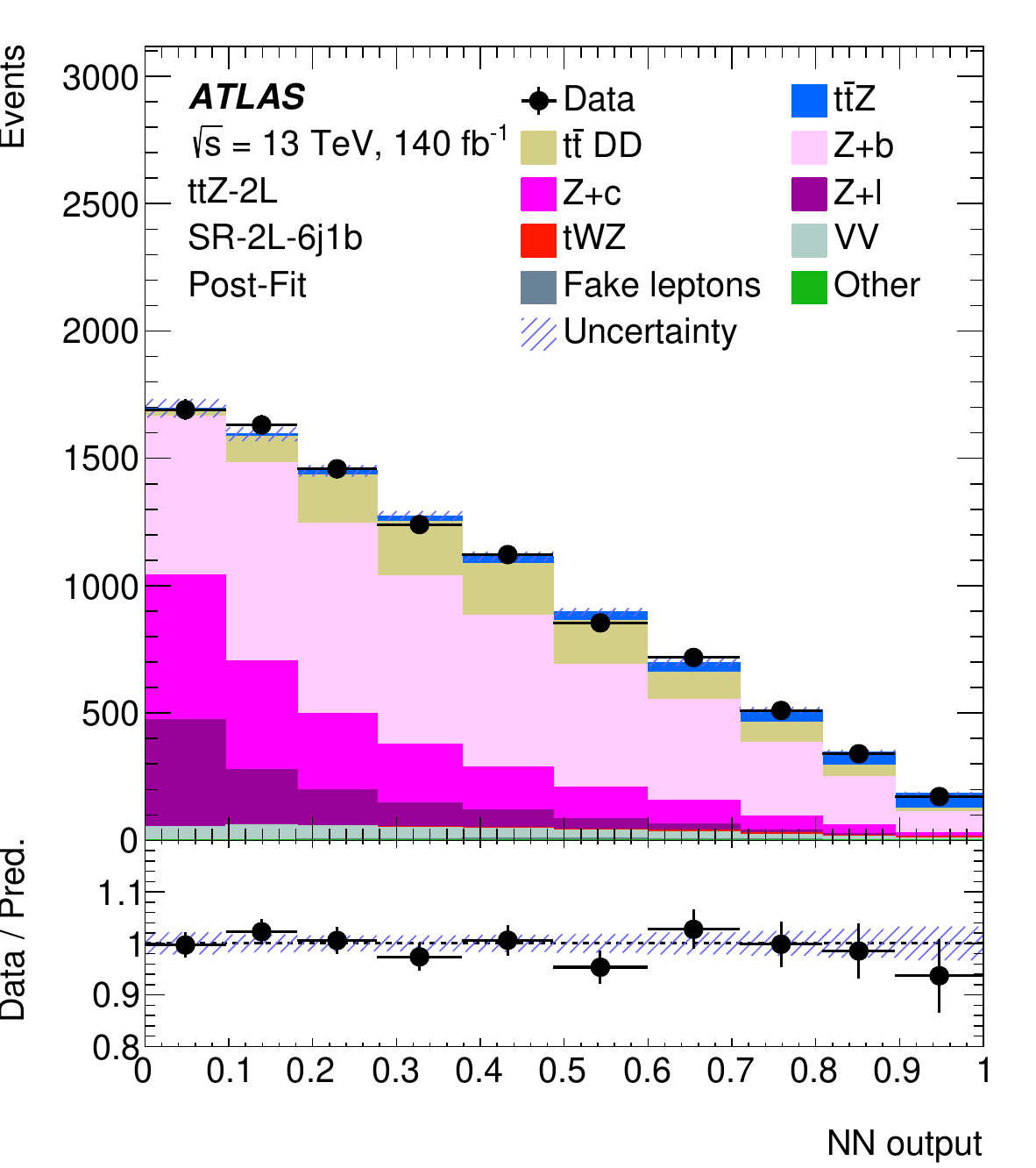}}
\subfloat[]{\includegraphics[width=0.32\textwidth]{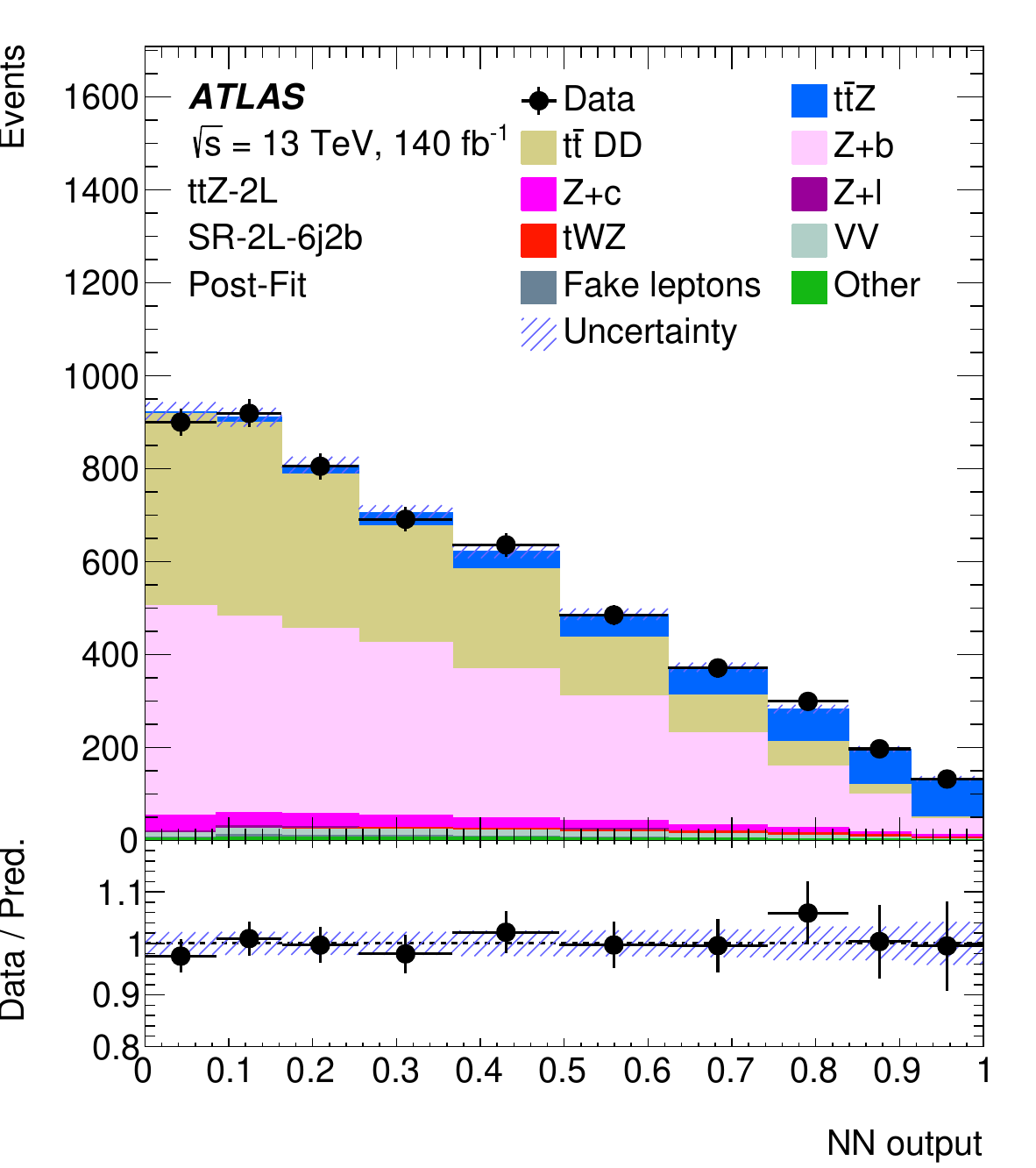}}
\caption{Distributions of the binary DNN output in the three dilepton signal regions used in the fit: (a) SR-2$\ell$-5j2b, (b) SR-2$\ell$-6j1b, and (c) SR-2$\ell$-6j2b. The data-driven approach described in Section~\ref{subsec:DD_ttbar} is used to estimate the \ttbar background, denoted by $t\bar{t}$\,DD. The fitted values of the signal strength, background normalisations and nuisance parameters were applied to the MC templates in the plots. The shaded band corresponds to the total uncertainty (systematic and statistical) of the total SM prediction. The lower panel shows the ratio of the data to the SM prediction.}
\label{fig:incl-SRs-2L}
\end{figure}

In the trilepton channel, the distribution of the DNN output, related to the probability of the event being a \ttZ signal event, is fitted in the SR-3$\ell$-ttZ and SR-3$\ell$-tZq regions.
The total number of events is fitted in the SR-3$\ell$-WZ region. The post-fit plots for these distributions are shown in Figure~\ref{fig:incl-SRs-3L}.
The event yields in the trilepton signal regions are shown in Table~\ref{tab:incl_region_yields_combined_postfit}.
In addition to the three trilepton signal regions, the three fake-lepton control regions, shown in Figure~\ref{fig:3L-fake-cr-postfit},
are used in the fit to extract the normalisations of the fake-lepton backgrounds.
The trilepton fit has five free parameters: the signal strength, normalisation of the $WZ+b$ background and normalisations of the three fake-lepton backgrounds.
The fitted value of the cross section can be found in Table~\ref{tab:incl_cross_section_results}.

\begin{table}[!htb]
\footnotesize
\sisetup{group-minimum-digits=4,}
\caption{Post-fit event yields in the trilepton signal regions and the tetralepton signal and control regions, obtained for an integrated luminosity of \lumi~\ifb.
The values of the fitted parameters from the combined fit are used.
The indicated errors include the Monte Carlo statistical uncertainty as well as all other systematic uncertainties discussed in Section~\ref{sec:systematics}.
Because of rounding and correlations between systematic uncertainties, the values quoted for the total yield and its uncertainty may differ from the simple sum over all processes.
A dash (---) indicates event yields smaller than 0.1.}
\label{tab:incl_region_yields_combined_postfit}
\def\arraystretch{1.3}
\begin{center}
\begin{tabular}{
l
r@{\(\,\pm\,\)}r
r@{\(\,\pm\,\)}r
r@{\(\,\pm\,\)}r
r@{\(\,\pm\,\)}r
r@{\(\,\pm\,\)}r
r@{\(\,\pm\,\)}r
}
\toprule
& \multicolumn{2}{c}{SR-3$\ell$-ttZ}     & \multicolumn{2}{c}{SR-3$\ell$-WZ}      & \multicolumn{2}{c}{SR-3$\ell$-tZq}     & \multicolumn{2}{c}{SR-4$\ell$-SF}        & \multicolumn{2}{c}{SR-4$\ell$-DF}        & \multicolumn{2}{c}{CR-4$\ell$-ZZ} \\
\midrule
\ttZ        & \numRF{440.549}{3}\phdoo & \numRF{21.0505}{2}\phdoo & \numRF{48.9892}{3}\pho & \numRF{3.66904}{2}\pho & \numRF{150.699}{3}\phdoo & \numRF{10.9678}{2}\phdoo & \numRF{49.4412}{3}\pho & \numRF{2.95592}{2}\pho & \numRF{51.0797}{3}\pho & \numRF{2.93847}{2}\pho & \numRF{2.35991}{3} & \numRF{0.227596}{2} \\
\ttW        & \numRF{4.30523}{2}\pho & \numRF{2.15328}{2}\pho & \numRF{2.18911}{2}\pho & \numRF{1.09577}{2}\pho & \numRF{5.27268}{2}\pho & \numRF{2.64129}{2}\pho & \multicolumn{2}{c}{---}             & \multicolumn{2}{c}{---}             & \multicolumn{2}{c}{---} \\
\ttH        & \numRF{11.8527}{3}\pho & \numRF{1.05713}{2}\pho & \numRF{1.43392}{3} & \numRF{0.134595}{2} & \numRF{6.69634}{3} & \numRF{0.572591}{2} & \numRF{2.79154}{3} & \numRF{0.238812}{2} & \numRF{2.82212}{3} & \numRF{0.242363}{2} & \numRF{0.320916}{2} & \numRF{0.0404033}{1} \\
\WZb        & \numRF{21.1047}{3}\pho & \numRF{7.40528}{2}\pho & \numRF{46.8442}{2}\phdoo & \numRF{16.275}{2}\phdoo & \numRF{27.0666}{3}\pho & \numRF{9.49298}{2}\pho & \multicolumn{2}{c}{---}             & \multicolumn{2}{c}{---}             & \multicolumn{2}{c}{---} \\
\WZc        & \numRF{8.94937}{2}\pho & \numRF{3.6089}{2}\pho & \numRF{12.2286}{3}\pho & \numRF{5.01716}{2}\pho & \numRF{11.1355}{3}\pho & \numRF{4.60547}{2}\pho & \multicolumn{2}{c}{---}             & \multicolumn{2}{c}{---}             & \multicolumn{2}{c}{---} \\
\WZl        & \numRF{1.19137}{3} & \numRF{0.52393}{2} & \numRF{1.69634}{3} & \numRF{0.7563}{2} & \numRF{1.81433}{3} & \numRF{0.797836}{2} & \multicolumn{2}{c}{---}             & \multicolumn{2}{c}{---}             & \multicolumn{2}{c}{---} \\
\ZZb        & \numRF{4.28825}{2}\pho & \numRF{2.52661}{2}\pho & \numRF{6.94379}{2}\pho & \numRF{4.01197}{2}\pho & \numRF{7.25599}{2}\pho & \numRF{4.21331 }{2}\pho & \numRF{7.53103}{2}\pho & \numRF{2.0392}{2}\pho & \numRF{0.457713}{2} & \numRF{0.122272}{2} & \numRF{26.6938}{3}\pho & \numRF{6.93884}{2}\pho \\
\ZZc        & \numRF{1.23042}{3} & \numRF{0.415198}{2} & \numRF{1.21665}{3} & \numRF{0.428336}{2} & \numRF{1.60989}{3} & \numRF{0.534124}{2} & \numRF{2.12593}{3} & \numRF{0.662127}{2} & \numRF{0.299369}{2} & \numRF{0.092099}{1} & \numRF{24.62}{3}\pho & \numRF{7.13643}{2}\pho \\
\ZZl        & \numRF{0.416802}{2} & \numRF{0.130645}{2} & \numRF{0.263483}{2} & \numRF{0.0925761}{1} & \numRF{0.527992}{2} & \numRF{0.15474}{2} & \numRF{0.828687}{2} & \numRF{0.238657}{2} & \numRF{0.337013}{2} & \numRF{0.088456}{1} & \numRF{22.5565}{3}\pho & \numRF{5.16099}{2}\pho \\
\tZq        & \numRF{20.7676}{3}\pho & \numRF{3.99084}{2}\pho & \numRF{13.1673}{3}\pho & \numRF{2.27621}{2}\pho & \numRF{99.1451}{2}\phdoo & \numRF{15.9948}{2}\phdoo & \multicolumn{2}{c}{---}             & \multicolumn{2}{c}{---}             & \multicolumn{2}{c}{---} \\
\tWZ        & \numRF{40.0497}{3}\pho & \numRF{7.55923}{2}\pho & \numRF{17.9746}{3}\pho & \numRF{4.20146}{2}\pho & \numRF{24.1946}{3}\pho & \numRF{2.98233}{2}\pho & \numRF{6.59925}{3} & \numRF{0.815642}{2} & \numRF{7.31955}{2}\pho & \numRF{1.181}{2}\pho & \numRF{0.6911}{2} & \numRF{0.0969607}{1} \\
\tttt       & \numRF{1.56017}{3} & \numRF{0.781096}{2} & \numRF{0.132041}{2} & \numRF{0.0673318}{1} & \numRF{0.268402}{2} & \numRF{0.135197}{2} & \multicolumn{2}{c}{---}             & \multicolumn{2}{c}{---}             & \multicolumn{2}{c}{---}\\
Other       & \numRF{1.33121}{3} & \numRF{0.606096}{2} & \numRF{1.40442}{3} & \numRF{0.633991}{2} & \numRF{0.393735}{2} & \numRF{0.188567}{2} & \numRF{0.554128}{2} & \numRF{0.249076}{2} & \numRF{1.12356}{3} & \numRF{0.519546}{2} & \numRF{0.54711}{2} & \numRF{0.25204}{2} \\
F-e-HF      & \numRF{4.59239}{2}\pho & \numRF{1.0159}{2}\pho & \numRF{3.89565}{3} & \numRF{0.867596}{2} & \numRF{11.9666}{3}\pho & \numRF{2.64159}{2}\pho & \numRF{0.282757}{2} & \numRF{0.0651239}{1} & \numRF{0.452665}{2} & \numRF{0.10189}{2} & \numRF{0.109649}{2} & \numRF{0.0265224}{1} \\
F-e-Other   & \numRF{7.75857}{2}\pho & \numRF{2.74507}{2}\pho & \numRF{7.27877}{2}\pho & \numRF{2.58853}{2}\pho & \numRF{15.185}{3}\pho & \numRF{5.40486}{2}\pho & \numRF{0.39269}{2} & \numRF{0.137994}{2} & \numRF{0.50177}{2} & \numRF{0.177327}{2} & \numRF{0.100932}{2} & \numRF{0.03572}{1} \\
F-$\mu$-HF      & \numRF{6.98385}{3} & \numRF{0.859344}{2} & \numRF{5.2717}{3} & \numRF{0.65966}{2} & \numRF{18.2327}{3}\pho & \numRF{2.23753}{2}\pho & \numRF{0.580684}{2} & \numRF{0.0734729}{1} & \numRF{0.616894}{2} & \numRF{0.0771338}{1} & \numRF{0.163473}{2} & \numRF{0.0203077}{1} \\
F-Other     & \numRF{2.79523}{2}\pho & \numRF{1.24812}{2}\pho & \numRF{2.69283}{2}\pho & \numRF{1.19522}{2}\pho & \numRF{4.4151}{2}\pho & \numRF{1.96246}{2}\pho & \numRF{0.901747}{2} & \numRF{0.402261}{2} & \numRF{1.65771}{3} & \numRF{0.74269}{2} & \numRF{0.33191}{2} & \numRF{0.148282}{2} \\
\midrule
Total     & \numRF{579.727}{3}\phdoo & \numRF{18.6618}{2}\phdoo & \numRF{173.623}{3}\phdoo & \numRF{12.5802}{2}\phdoo & \numRF{385.879}{3}\phdoo & \numRF{14.9478}{2}\phdoo & \numRF{72.0297}{3}\pho & \numRF{3.42735}{2}\pho & \numRF{66.668}{3}\pho & \numRF{3.04926}{2}\pho & \numRF{78.4954}{3}\pho & \numRF{8.00424}{2}\pho \\
\midrule
Data   & \multicolumn{2}{l}{\num{569}} & \multicolumn{2}{l}{\num{175}} & \multicolumn{2}{l}{\num{388}} & \multicolumn{2}{l}{\num{79}\phdo} & \multicolumn{2}{l}{\num{74}\phdo} & \multicolumn{2}{l}{\num{81}\phdo} \\
\bottomrule
\end{tabular}
\end{center}
\end{table}
 
\begin{figure}[!htb]
\centering
\subfloat[]{\includegraphics[width=0.32\textwidth]{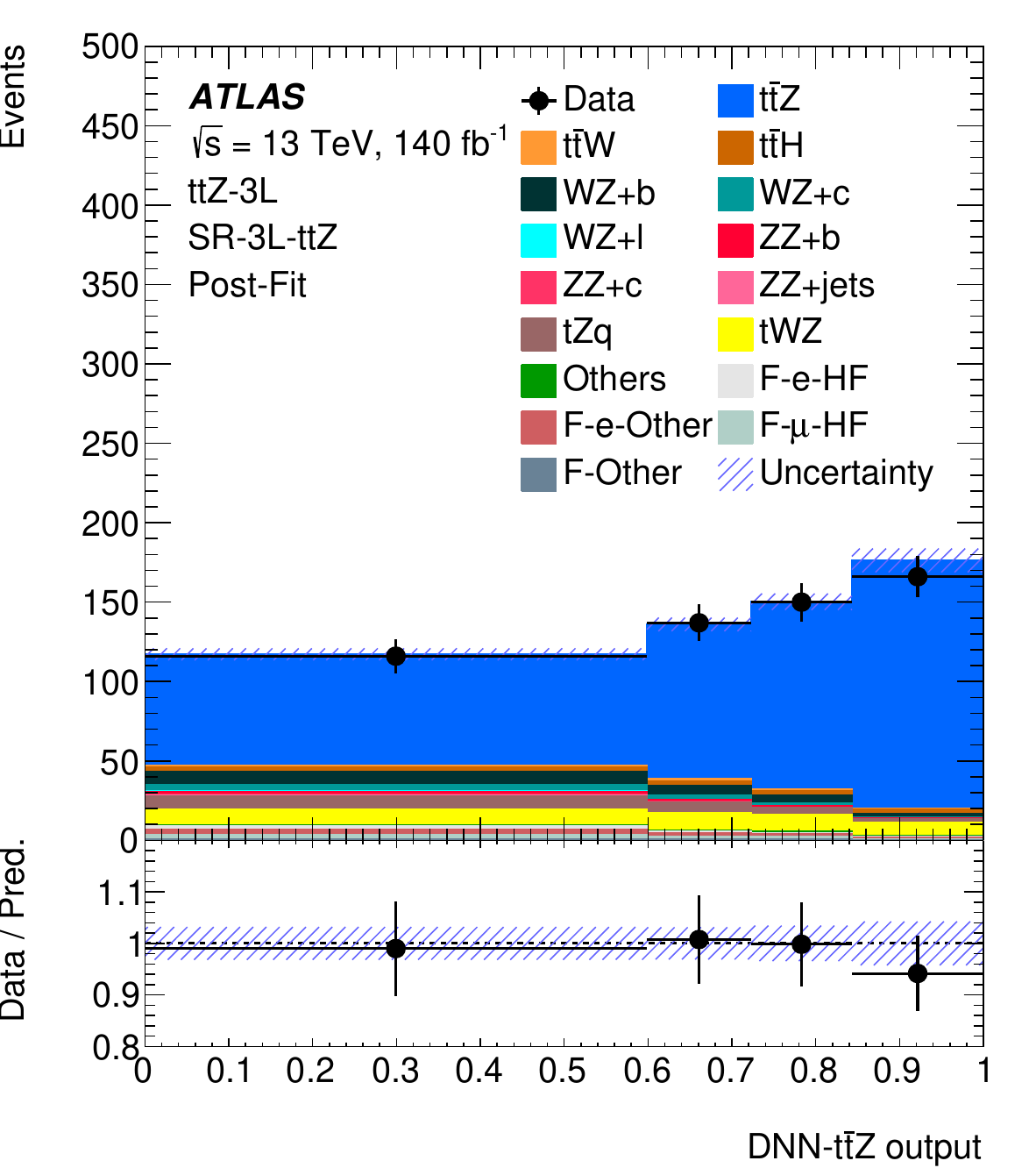}}
\subfloat[]{\includegraphics[width=0.32\textwidth]{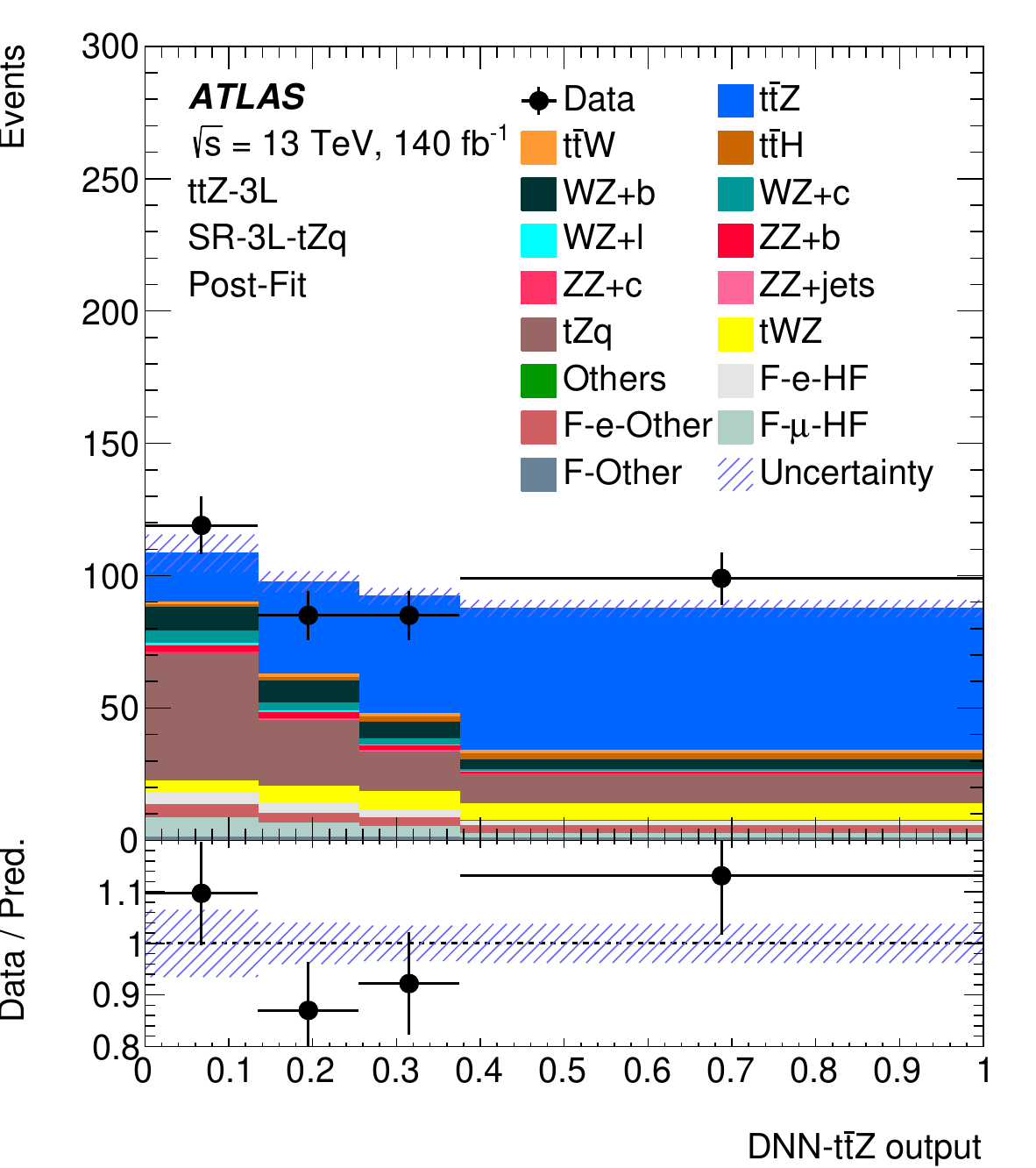}}
\subfloat[]{\includegraphics[width=0.32\textwidth]{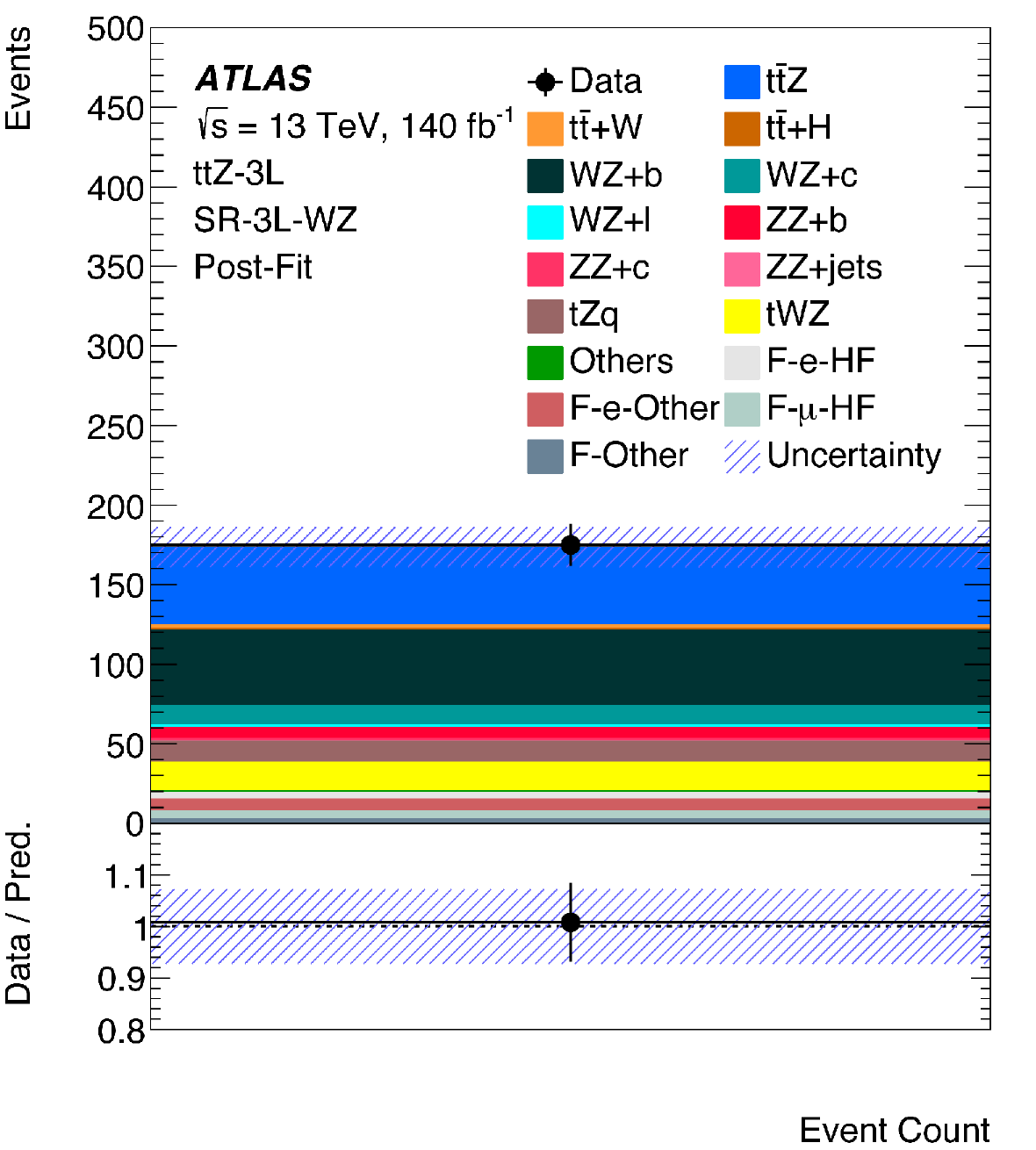}}
\caption{Distributions of the \ttZ-node of the multi-class DNN output and total event yield in the three trilepton signal regions used in the fit: (a) SR-3$\ell$-ttZ, (b) SR-3$\ell$-tZq, and (c) SR-3$\ell$-WZ. The fitted values of the signal strength, background normalisations and nuisance parameters were applied to the MC templates in the plots. The shaded band corresponds to the total uncertainty (systematic and statistical) of the total SM prediction. The lower panel shows the ratio of the data to the SM prediction.}
\label{fig:incl-SRs-3L}
\end{figure}
 
In the tetralepton channel, the distributions of the two DNN outputs are fitted in the SR-4$\ell$-SF and SR-4$\ell$-DF regions.
The distribution of the $b$-tagging score of the leading $b$-tagged jet is fitted in CR-4$\ell$-ZZ.
The post-fit plots for these distributions are shown in Figure~\ref{fig:incl-SRs-4L}.
The event yields in the tetralepton signal regions are shown in Table~\ref{tab:incl_region_yields_combined_postfit}.
In addition to the two tetralepton signal regions and $ZZ$ control region,
the three fake-lepton control regions, shown in Figure~\ref{fig:3L-fake-cr-postfit},
are used in the fit to extract the normalisations of the fake-lepton backgrounds.
The tetralepton fit has five free parameters: the signal strength, normalisation of the $ZZ+b$ background and normalisations of the three fake-lepton backgrounds.
The fitted value of the cross section can be found in the
Table~\ref{tab:incl_cross_section_results}.

\begin{figure}[!htb]
\centering
\subfloat[]{\includegraphics[width=0.32\textwidth]{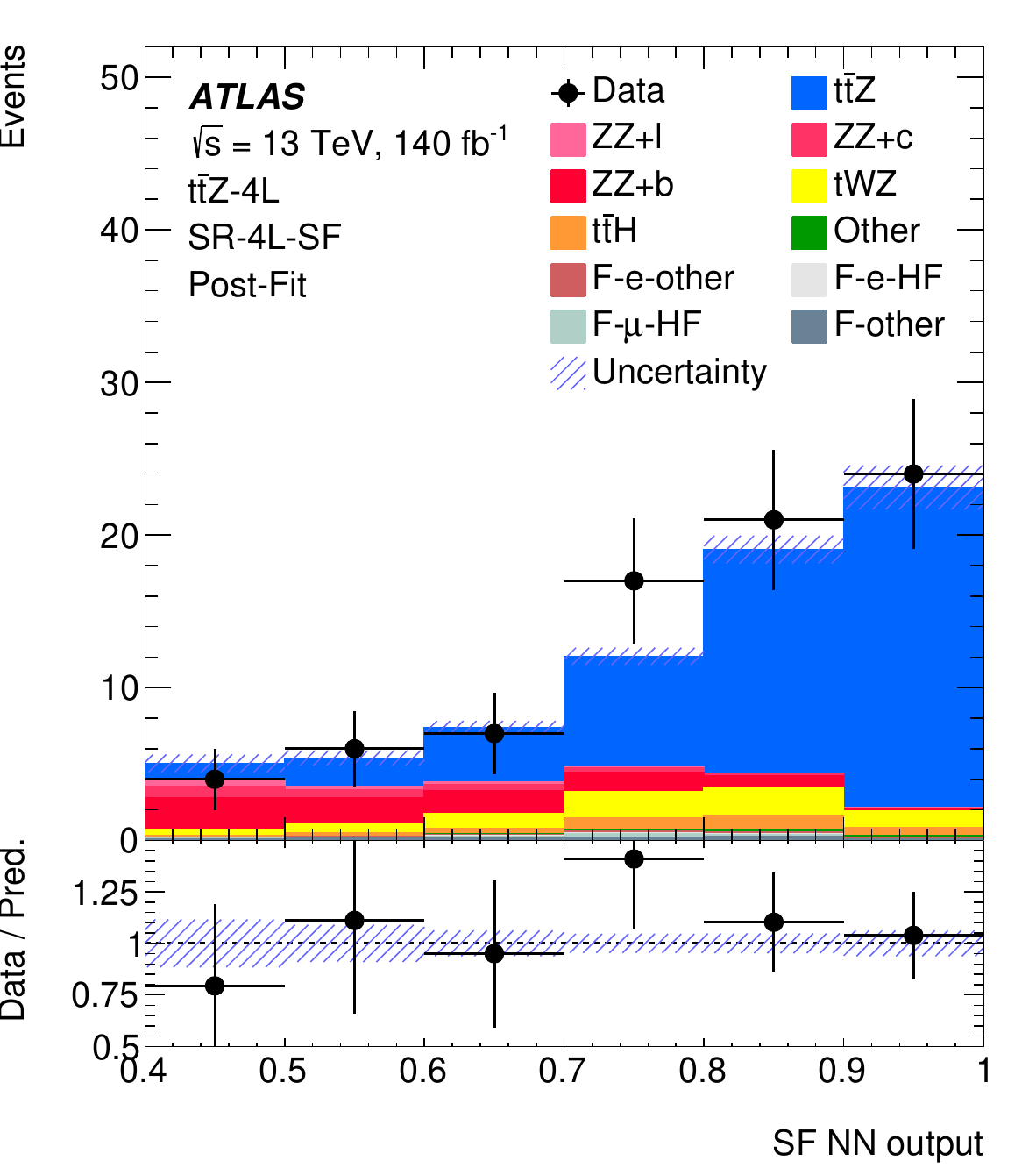}}
\subfloat[]{\includegraphics[width=0.32\textwidth]{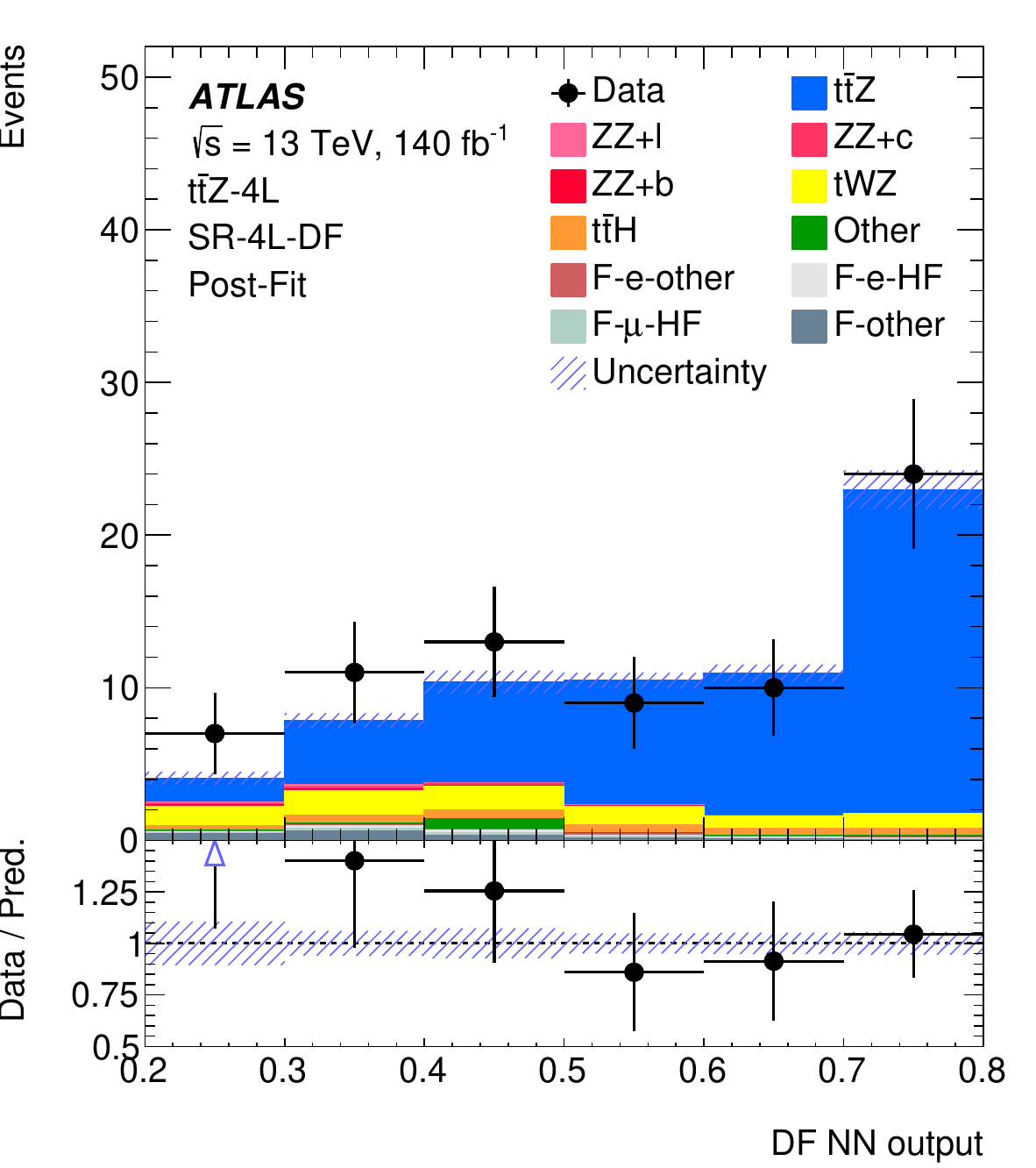}}
\subfloat[]{\includegraphics[width=0.32\textwidth]{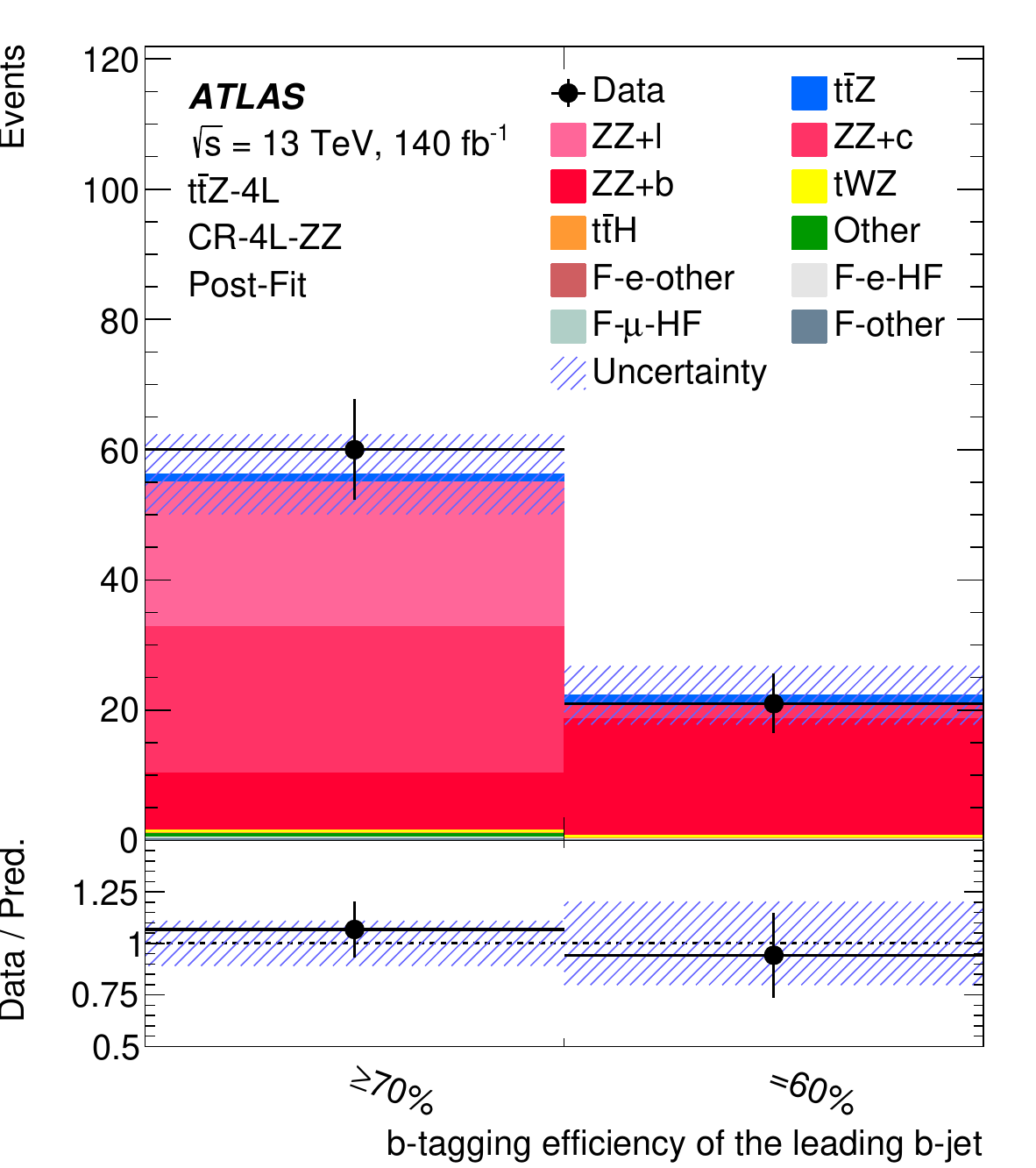}}
\caption{Distributions of the binary DNN outputs in the two tetralepton signal regions used in the fit, (a) SR-4$\ell$-SF and (b) SR-4$\ell$-DF, as well as of (c) the $b$-tagging efficiency of the leading $b$-tagged jet in the control region CR-4$\ell$-ZZ. The fitted values of the signal strength, background normalisations and nuisance parameters were applied to the MC templates in the plots. The shaded band corresponds to the total uncertainty (systematic and statistical) of the total SM prediction. The lower panel shows the ratio of the data to the SM prediction.}
\label{fig:incl-SRs-4L}
\end{figure}
 
The combined inclusive fit uses all of the aforementioned signal and control regions, i.e.\
three dilepton signal regions, three trilepton signal regions, two tetralepton signal regions, CR-4$\ell$-ZZ and three fake-lepton control regions.
The fitted value of the inclusive cross section is found to be $0.86 \pm 0.05\,\mathrm{pb}$, in good agreement with the theory prediction~\cite{Kulesza:2019} and with a relative precision of $6\%$.
For comparison, the previous result~\cite{TOPQ-2018-08} achieved a $10\%$ uncertainty for the \ttZ cross section, using the same dataset.
All free floating background normalisations are consistent with the SM predictions within their uncertainties.
The breakdown of the combination into the contributing channels can be found in Table~\ref{tab:incl_cross_section_results}.
A mild upwards fluctuation in the data is observed in the 4$\ell$ channel.
The $p$-value for compatibility of the cross section measured in the combination and the values fitted individually is 53\%,
indicating good agreement between the different analysis channels.
The values of the background normalisation factors obtained from the inclusive cross-section fit can be found in the
Table~\ref{tab:norm_factors_inclusive}.
A deviation from unity is observed only for \nZc, which compensates for the mismodelling of the $Z+c$ background in SR-2$\ell$-6j1b.
All other normalisation factors are consistent with their respective SM predictions.
No significant pulls or constraints are observed in the fit, except for the CKKW and QSF scales of the $Z+b$ background,
which are pulled by $\approx$1$\sigma$. These pulls are caused by the mismodelling of the $Z$+jets background
in SR-2$\ell$-5j2b and SR-2$\ell$-6j1b.
The effects of systematic uncertainties on the result are listed in Table~\ref{tab:combined_inclusive_grouped_impact_data}.
The dominant systematic uncertainties are those in the background normalisations and those related to jets and missing transverse momentum.
 
\begin{table}[!htb]
\footnotesize
\caption{Values of background normalisation factors measured in the combined inclusive fit. Both statistical and systematic uncertainties are considered.}
\def\arraystretch{1.4}
\begin{center}
\begin{tabular}{ll@{\,}l}
\toprule
Norm. factor & \multicolumn{2}{c}{Value} \\
\midrule
\nZZb           & \numRF{1.143}{2}\pho & $^{+\numRF{0.386}{1}}_{-\numRF{0.386}{1}}$ \\
\nWZb           & \numRF{0.905}{1}\pho & $^{+\numRF{0.399}{1}}_{-\numRF{0.399}{1}}$ \\
\nZb            & \numRF{1.078}{3}     & $^{+\numRF{0.110}{2}}_{-\numRF{0.101}{2}}$ \\
\nZc            & \numRF{0.612}{2}     & $^{+\numRF{0.225}{2}}_{-\numRF{0.195}{2}}$ \\
\nFakesElHF     & \numRF{0.892}{2}     & $^{+\numRF{0.092}{1}}_{-\numRF{0.092}{1}}$ \\
\nFakesElOther  & \numRF{1.193}{2}\pho & $^{+\numRF{0.360}{1}}_{-\numRF{0.360}{1}}$ \\
\nFakesMuHF     & \numRF{1.021}{3}     & $^{+\numRF{0.075}{1}}_{-\numRF{0.075}{1}}$ \\
\bottomrule
\end{tabular}
\end{center}
\label{tab:norm_factors_inclusive}
\end{table}
 
\begin{table}[!htb]
\footnotesize
\caption{Measured \ttZ cross sections obtained from the fits in the signal regions with different lepton multiplicities and corrected to parton level. The uncertainties include those from statistical and systematic sources.}
\def\arraystretch{1.3}
\begin{center}
\begin{tabular}{ll@{}l}
\toprule
Channel & \multicolumn{2}{c}{$\sigma_{t\bar{t}Z}$}\\
\midrule
Dilepton & $0.84 \pm 0.11\,\mathrm{pb}\,$ & $= 0.84 \pm 0.06\,(\text{stat.}) \pm 0.09\,(\text{syst.})\,\mathrm{pb}$\\
\addlinespace[0.5em]
Trilepton & $0.84 \pm 0.07\,\mathrm{pb}\,$ & $= 0.84 \pm 0.05\,(\text{stat.}) \pm 0.05\,(\text{syst.})\,\mathrm{pb}$\\
\addlinespace[0.5em]
Tetralepton & $0.97\,^{+\,0.13}_{-\,0.12}\,\mathrm{pb}\,$ & $= 0.97 \pm 0.11\,(\text{stat.}) \pm 0.05\,(\text{syst.})\,\mathrm{pb}$\\
\midrule
Combination $(2\ell,3\ell\,\&\,4\ell)$ & $0.86 \pm 0.05\,\mathrm{pb}\,$ & $= 0.86 \pm 0.04\,(\text{stat.}) \pm 0.04\,(\text{syst.})\,\mathrm{pb}$\\
\bottomrule
\end{tabular}
\end{center}
\label{tab:incl_cross_section_results}
\end{table}

\begin{table}[!htb]
\footnotesize
\caption{Grouped impact of systematic uncertainties in the combined inclusive fit to data. The uncertainties are symmetrised for presentation and grouped into the categories described in the text. The quadrature sum of the individual uncertainties is not equal to the total uncertainty because of correlations introduced by the fit.}
\label{tab:combined_inclusive_grouped_impact_data}
\def\arraystretch{1.3}
\begin{center}
\begin{tabular}{lc}
\toprule
Uncertainty Category & $\Delta\sigma_{\ttZ}/\sigma_{\ttZ}$ [\%]  \\
\midrule
Background normalisations & 2.0 \\
Jets and \met & 1.9 \\
$b$-tagging & 1.7 \\
\ttZ \muF and \muR scales & 1.6 \\
Leptons & 1.6 \\
$\Zboson$+jets modelling & 1.5 \\
\tWZ modelling & 1.1 \\
\ttZ showering & 1.0 \\
\ttZ A14 tune& 1.0 \\
Luminosity & 1.0 \\
Diboson modelling & 0.8 \\
\tZq modelling & 0.7 \\
PDF (signal \& backgrounds) & 0.6 \\
MC statistical  & 0.5 \\
Other backgrounds & 0.5 \\
Fake leptons  & 0.4 \\
Pile-up & 0.3 \\
Data-driven $t\bar{t}$ & 0.1 \\
\bottomrule
\end{tabular}
\end{center}
\end{table}

\FloatBarrier


\section{Unfolding and differential cross-section measurements}
\label{sec:results_differential}

 
The differential measurement is performed in the 3$\ell$ and 4$\ell$ channels,
using definitions of the fiducial volumes described in Section~\ref{sec:particlepartondefs}.
The dilepton channel is not used because of its high background contamination.
A profile-likelihood unfolding method is used, as described in Section~\ref{subsec:PLU}.
The observables that are unfolded and their binning are listed in Sections~\ref{subsec:differential-observables} and~\ref{subsec:binning-choice}, and the results are presented in Section~\ref{subsec:unfolded-results}.
 
\subsection{Profile-likelihood unfolding procedure}
\label{subsec:PLU}
 
A profile-likelihood unfolding (PLU) procedure is used to measure differential cross sections.
A likelihood model is built for the detector-level distribution by treating the contribution from each truth-level bin as a subcomponent of the detector-level signal.
The likelihood takes the form:
 
\begin{eqnarray}
L\left(\vec{n}|\vec{\mu}, \vec{\theta},\vec{\lambda}\right) = \prod_{r \in\text{reg.}}\prod_{i\in \substack{\text{det.}\\\text{bins}}} \mathrm{Pois}\left(n_{i,r}|L_{\mathrm{int}}\sum_{j\in \substack{\text{gen.}\\\text{bins}}}\mathcal{R}_{ij,r}(\vec{\theta})\mu_j\sigma_j^\mathrm{MC} + B_{i,r}(\vec{\theta},\vec{\lambda})\right)\times\prod_{k\in \text{NPs}}\mathrm{Gaus}\left(\theta_k\right)\times R(\vec{\mu}),
\label{eq:UnfoldingLikelihood}
\end{eqnarray}
 
where $i$ indicates the detector-level bin index, $j$ the truth-level bin index, $k$ the systematic uncertainty index, and the index $r$ runs over all control and signal regions.
The number of observed detector-level events in a given bin, $n_{i,r}$, is compared with the prediction obtained by folding the truth-level signal differential cross section $\sigma_j^\text{MC}$ through the response matrix $\mathcal{R}_{ij,r}$, appropriately normalised to the integrated luminosity $L_{\mathrm{int}}$, and adding the background contribution $B_{i,r}$, which depends on the background normalisation factors $\vec{\lambda}$.
The expected truth-level differential cross section $\vec{\sigma}$ is associated bin-wise with the normalisation factors $\vec{\mu}$, the parameters of interest of PLU.
To obtain a normalised differential distribution, the last element of $\vec{\mu}$ is reparameterised in terms of all the other ones and the integrated fiducial cross section.
 
The response matrix is defined as:
\begin{equation*}
\mathcal{R}_{ij} = \frac{1}{\alpha_i}\epsilon_{j}\mathcal{M}_{ij},
\end{equation*}
where the migration matrix $\mathcal{M}_{ij}$ quantifies the
bin-to-bin migrations of events from truth level to detector level due to resolution effects.
It is computed as:
\begin{equation*}
\mathcal{M}_{ij} = \frac{N_{ij}^{\text{det.}\,\cap\, \text{fid.}}}{N_j^{\text{det.}\,\cap\, \text{fid.}}}.
\end{equation*}
The superscript $\textrm{det.}\,\cap\, \text{fid.}$ indicates events that pass both the detector-level event selection and the fiducial selection. The numerator $N^{\textrm{det.}\,\cap\,\textrm{fid.}}_{ij}$ is the expected number of detector-level events in detector-level bin $i$ and truth-level bin $j$, while the denominator $N_j^{\text{det.}\,\cap\, \text{fid.}}$ is the expected number of detector-level events in truth-level bin $j$ summed over all detector-level bins.
 
The response matrix then receives acceptance corrections, $\alpha_i$, which define the fraction of events
that satisfy the detector-level selection and originate from
configurations outside the truth-level fiducial selection:
\begin{equation*}
\alpha_i= \frac{N_i^{\text{det.}\,\cap\, \text{fid.}}}{N_i^{\text{det.}}}.
\end{equation*}
 
Finally, efficiency corrections, $\epsilon_j$, are also applied
to account for events that satisfy the fiducial phase-space
selection criteria but are not reconstructed in the detector:
\begin{equation*}
\epsilon_j= \frac{N_j^{\text{det.}\,\cap\, \text{fid.}}}{N_j^\text{fid.}}.
\end{equation*}
Here, $N_i^\text{det.}$ and $N_j^\text{fid.}$ represent the expected number of events in the $i^{\mathrm{th}}$ and $j^{\mathrm{th}}$ bins of the detector-level and fiducial truth-level histograms, while $N_i^{\text{det.}\,\cap\, \text{fid.}}$ is the expected number of detector-level events in detector-level bin $i$ summed over all truth-level bins.
 
Systematic uncertainties, associated with the nuisance parameters (NPs) $\vec{\theta}$, enter the likelihood in two ways: in the background term $B_{i,r}(\vec{\theta},\vec{\lambda})$, and in the folded signal term via $\mathcal{R}_{ij,r}(\vec{\theta})$.
Each NP is subject to a Gaussian constraint in the likelihood fit.
Those accounting for the statistical uncertainties in each bin due to the limited size of the simulated event samples are assigned Poisson constraints instead, following the Beeston--Barlow \enquote{lite} method~\cite{Barlow:1993dm}.
 
The PLU procedure can be regularised via the additional constraint term $R(\vec{\mu})$ in Eq.~(\ref{eq:UnfoldingLikelihood}).
Specifically, a discretised second-derivative regularisation, also known as Tikhonov regularisation~\cite{Phillips1962ATF,Tikhonov63,Tikhonov77,Cowan:1998ji}, is used:
 
\begin{eqnarray*}
R(\vec{\mu}) =  \exp \left[ - \frac{\tau^2}{2} \sum_{i=2}^{i+1 < N_\text{bins}} \left( (\mu_{i} - \mu_{i-1}) - (\mu_{i+1} - \mu_{i}) \right)^{2} \right],
\end{eqnarray*}
 
where the sum runs over the bins of the unfolded distribution and $\tau$ is the regularisation parameter.
Its value depends on the observable, and is shown in Table~\ref{tab:variable_definitions}.
 
The aim of the regularisation is to suppress strong anti-correlations between the bins of the unfolded distribution,
arising from migrations between the bins. Higher values of the regularisation parameter $\tau$ lead to lower uncertainties.
However, too strong regularisation can lead to biased results and turns anti-correlations between the neighbouring bins into correlations.
In order to find a compromise between these two effects, the $\tau$ values were scanned with a step size of 0.1, starting from zero.
For each $\tau$, a pseudo-dataset generated according to the SM expectation values (so-called \enquote{Asimov dataset}) was unfolded and the global correlation factor
defined in Eq.~(\ref{eq:global_correlation}) was evaluated:
 
\begin{equation}
\rho = \Biggl \langle \sqrt{1 - (C_{ii}C_{ii}^{-1})^{-1}} \Biggr \rangle,
\label{eq:global_correlation}
\end{equation}
 
where $C_{ij}$ is the correlation matrix, $C_{ij}^{-1}$ its inverse, and the brackets represent the mean value.
The $\tau$ value where the global correlation reaches its minimum was chosen as the optimal value.
 
In order to verify that the unfolding is able to recover an alternative truth-level distribution, a number of stress tests are performed.
Motivated by the potential impact of EFT effects on the high-$\pT$ tail of the $\pT^{Z}$ distribution, this observable's distribution is linearly reweighted, with slopes ranging from $-0.25$ to 1.00, between 0 and 1~\TeV.
The corresponding event-wise correction factor, which is 1.0 at $\pT^{Z}=0$~\TeV and ranges from 0.75 to 2.0 at $\pT^{Z}=1$~\TeV, is used to reweight all of the other observables in this pseudo-EFT-effect stress test.
Additionally, data-driven stress tests are performed, where the histogram of the ratio of observed to expected signal
in the unfolded variable at detector level is fitted with a quadratic function and the value of the function is then used to apply the weight.
Another kind of stress test uses a histogram of the ratio of observed to expected signal as a function of the scalar sum of lepton transverse momenta, $\HT^{\ell}$,
which was found to be the most poorly modelled variable (lowest $p$-value at detector level), and the value of the ratio is used to apply the weight.
All the stress tests show that the unfolding is quite able to recover an alternative distribution, with an average $\chi^2/$ndf well below 0.05.

\subsection{Differential observables}
\label{subsec:differential-observables}
 
The optimal choice of unfolded observables was based on their physics importance and sensitivity to EFT operators.
A list of the observables used in the differential measurement can be found in Table~\ref{tab:variable_definitions}.

The observables are unfolded either in one of the channels, or in their combination, taken as the union of the fiducial volumes defined in Section~\ref{sec:particlepartondefs}.
The observables relying on a particular decay mode of the $t\bar{t}$ system are unfolded only in the corresponding channel.
Two observables, $N_{\mathrm{jets}}$ and $\HT^{\ell}$, were unfolded separately in each of the two channels,
as their distributions are expected to differ between the channels.
The observables not requiring a particular decay of the $t\bar{t}$ system are unfolded in the combined 3$\ell$ and 4$\ell$ channels.
Each of the observables is unfolded to both particle level and parton level, except for $N_{\mathrm{jets}}$, which is unfolded only to particle level.
For observables requiring hadronic top reconstruction, the migration matrices are somewhat non-diagonal ($\sim$40\%--70\% on diagonal, compared to purely leptonic observables with $>$90\% on the diagonal).
In these cases, unregularised unfolding would yield large fluctuations, so Tikhonov regularisation is applied.
 
\begin{table}[!htb]
\footnotesize
\caption{Summary of the variables used for the differential measurement, and the values of the regularisation parameter $\tau$ for particle level and parton level for the relevant variables.}
\label{tab:variable_definitions}
\def\arraystretch{1.3}
\begin{center}
\begin{tabular}{cccccp{0.5\textwidth}}
\toprule
& Variable                                                     & Regularisation & $\tau^{\mathrm{particle}}$ & $\tau^{\mathrm{parton}}$ &  Definition \\
\midrule
\multirow{8}{*}{\rotatebox{90}{$3\ell+4\ell$}} & $\pT^{Z}$     & No   & -   & -   &  Transverse momentum of the \Zboson boson  \\
& $\lvert y^{Z}\rvert$                                         & No   & -   & -   &  Absolute rapidity of the \Zboson boson  \\
& $\cos{\theta^*_Z}$                                           & No   & -   & -   &  Angle between the direction of the \Zboson boson in the detector reference frame and the direction of the negatively charged lepton in the rest frame of the \Zboson boson\\
& $\pT^{t}$                                                    & Yes  & 1.5 & 1.4 &  Transverse momentum of the top quark \\
& $\pT^{\ttbar}$                                               & Yes  & 1.6 & 1.5 &  Transverse momentum of the \ttbar system \\
& $\lvert\Delta\Phi({\ttbar},\Zboson)\rvert /\pi$              & Yes  & 2.4 & 2.1 &  Absolute azimuthal separation between the \Zboson boson and the \ttbar system \\
& $m^{\ttZ}$                                                   & Yes  & 1.5 & 1.6 &  Invariant mass of the \ttZ system\\
& $m^{\ttbar}$                                                 & Yes  & 1.5 & 1.4 &  Invariant mass of the \ttbar system\\
& $\lvert y^{\ttZ}\rvert$                                      & Yes  & 1.5 & 1.5 &  Absolute rapidity of the \ttZ system\\
\midrule
\multirow{5}{*}{\rotatebox{90}{$3\ell$}} & $\HT^{\ell}$        & No & - & - &  Sum of the transverse momenta of all the signal leptons \\
& $|\Delta\Phi(\Zboson,t_{\mathrm{lep}})| /\pi$                & No & - & - &  Absolute azimuthal separation between the \Zboson boson and the top (anti-top) quark featuring the \Wln decay\\
& $|\Delta y(\Zboson,t_{\mathrm{lep}})|$                       & No & - & - &  Absolute rapidity difference between the \Zboson boson and the top (anti-top) quark featuring the \Wln decay\\
& $\pT^{\ell,{\textrm{non-}}\Zboson}$                          & No & - & - &  Transverse momentum of the lepton that is not associated with the \Zboson boson\\
& $N_{\mathrm{jets}}$                                          & No & - & - &  Number of selected jets with $\pt > \SI{25}{GeV}$ and $\abseta < 2.5$ \\
\midrule
\multirow{3}{*}{\rotatebox{90}{$4\ell$}} & $\HT^{\ell}$        & No & - & - &  Sum of the transverse momenta of all the signal leptons \\
& $|\Delta\Phi(\ell^{+}_{t}, \ell^{-}_{\bar{t}})|/\pi$        & No & - & - &  Absolute azimuthal separation between the two leptons from the \ttbar system\\
& $N_{\mathrm{jets}}$                                          & No & - & - &  Number of selected jets with $\pt > \SI{25}{GeV}$ and $\abseta < 2.5$ \\
\bottomrule
\end{tabular}
\end{center}
\end{table}
 
\FloatBarrier

\subsection{Choice of binning}
\label{subsec:binning-choice}
 
The binning for all the observables is summarised in Table~\ref{tab:diff_observables_binning} in the Appendix.
The binning is optimised, starting from the requirement that the statistical uncertainty in all unfolded bins be
lower than 35\% for variables unfolded in the $4\ell$ channel only, and lower than 25\% for the 3$\ell$ and 3$\ell$+4$\ell$ variables.
Furthermore, diagonal elements of the migration matrix (MM) must exceed a chosen threshold (starting from a high value and iteratively decreasing it).
The first bin of the distribution is chosen to have the smallest possible range for which these two requirements are satisfied
(narrow bins lead to large migrations and high statistical uncertainties; widening the bin decreases the uncertainty and increases the diagonal elements of the migration matrix).
Once the range of the first bin is chosen, the next bin is optimised.
Given the two requirements, only a limited number of bins can be chosen.
The requirement on the minimum value of MM diagonal elements is gradually decreased to obtain binning for a higher number of bins.
The binning is optimised for a number of bins ranging from two to ten, using only the predicted statistical uncertainties and unfolding to particle level.
Then stress tests are performed for each choice of binning: the one with the largest number of bins that still passes all stress tests is selected.
The same binning is used for both particle level and parton level.
 
\subsection{Unfolded cross-section measurements}
\label{subsec:unfolded-results}
 
The full set of unfolded results for the transverse momentum of the $Z$ boson are presented in this section.
For brevity, only the absolute differential cross sections unfolded to particle level are shown for other observables.
All other differential results (including normalised distributions, and unfolding to parton level) are available in the Appendix.
 
Figure~\ref{fig:trilepton-asimov-unfolding-reco-migration-eff-acc-particle-ptz} shows the distribution of detector-level $\pT^{Z}$
in $3\ell$ signal regions, as well as the migration matrices, acceptances and efficiencies.
The acceptance is the fraction of the signal in a given bin of the detector-level distribution which passes the truth-level fiducial volume cuts.
The efficiency is defined as the fraction of the events from a given bin of the truth-level distribution passing the selection in the given
detector-level region. Figure~\ref{fig:tetralepton-asimov-unfolding-reco-migration-eff-acc-particle-ptz} shows similar plots for $4\ell$ signal regions.
The unfolded distributions in the combination of the $3\ell$ and $4\ell$ channels, as well as the nominal truth-level MC prediction, and alternative generator predictions
are shown in Figure~\ref{fig:combined-asimov-unfolding-result-ptz}.
The uncertainties of the unfolded distributions range approximately from 15$\%$ to 40$\%$.
In general, small uncertainties are observed for the variables that are reconstructed
only from leptons and unfolded in the combined 3$\ell$ and 4$\ell$ channels.
A large number of data events and small migrations between the bins lead to the small uncertainties.
On the other hand, the variables unfolded only in the 4$\ell$ channel suffer from a low number of events,
which results in relatively high statistical uncertainties.
The unfolding to particle level has lower uncertainties than unfolding to parton level, because of the more diagonal migration matrices,
and normalised unfolded distributions are also more precise than the absolute distributions,
because the normalisation significantly reduces the uncertainties.
The statistical uncertainty of the data is the dominant source of uncertainty in all distributions and bins.
Compared to the statistical uncertainty, the systematic uncertainties have a significantly smaller effect.
 
The compatibility of the unfolded distributions and the predictions was assessed by calculating $\chi^2$ values,
using the uncertainties of the unfolded distribution and their correlations. The corresponding $p$-values can be found in the Appendix, in Table~\ref{tab:unfolding_p_values_absolute} for absolute unfolded distributions and in Table~\ref{tab:unfolding_p_values_normalised} for normalised distributions.
The $p$-values indicate good agreement between the unfolded data and the prediction for most of the variables.
The lowest $p$-values, around 2$\%$, are observed for $\HT^{\ell}$; this is expected because this variable
was seen to be the most poorly modelled one (with the lowest $p$-values) at detector level.
The signal strengths obtained from the integrated cross sections of differential distributions are very close to each other for each channel (and combination),
having a standard deviation around 1~$\%$ and they are in a good agreement with the inclusive cross section measurement performed in these channels.
 
The absolute differential cross sections unfolded to particle level for observables defined in the 3$\ell$ and 4$\ell$ channels separately are shown in Figures~\ref{fig:3l-4l-unfolded-results-1} and~\ref{fig:3l-4l-unfolded-results-2}, while Figures~\ref{fig:combination-unfolded-results-1} and~\ref{fig:combination-unfolded-results-2} present those defined in the combination of the two channels.
All other results are available in the Appendix: Figures~\ref{fig:combined-observed-unfolding-result-yz1} and \ref{fig:combined-observed-unfolding-result-cos-theta-starZ} for the unregularised observables in the combined 3$\ell$ and 4$\ell$ channels, Figures~\ref{fig:combined-observed-unfolding-result-pT_top-particle}--\ref{fig:combined-observed-unfolding-result-y_ttZ-particle} for the regularised ones, Figures~\ref{fig:trilepton-observed-unfolding-result-ht_leptons_3L-particle}--\ref{fig:trilepton-observed-unfolding-result-n_jets_3L-particle} for those defined in the 3$\ell$ channel only, and Figures~\ref{fig:tetralepton-observed-unfolding-result-sum_pT_leptons}--\ref{fig:tetralepton-observed-unfolding-result-nJets} for those defined in the 4$\ell$ channel only.
 
\begin{figure}[!htb]
\centering
\subfloat[]{\includegraphics[width=0.285\textwidth]{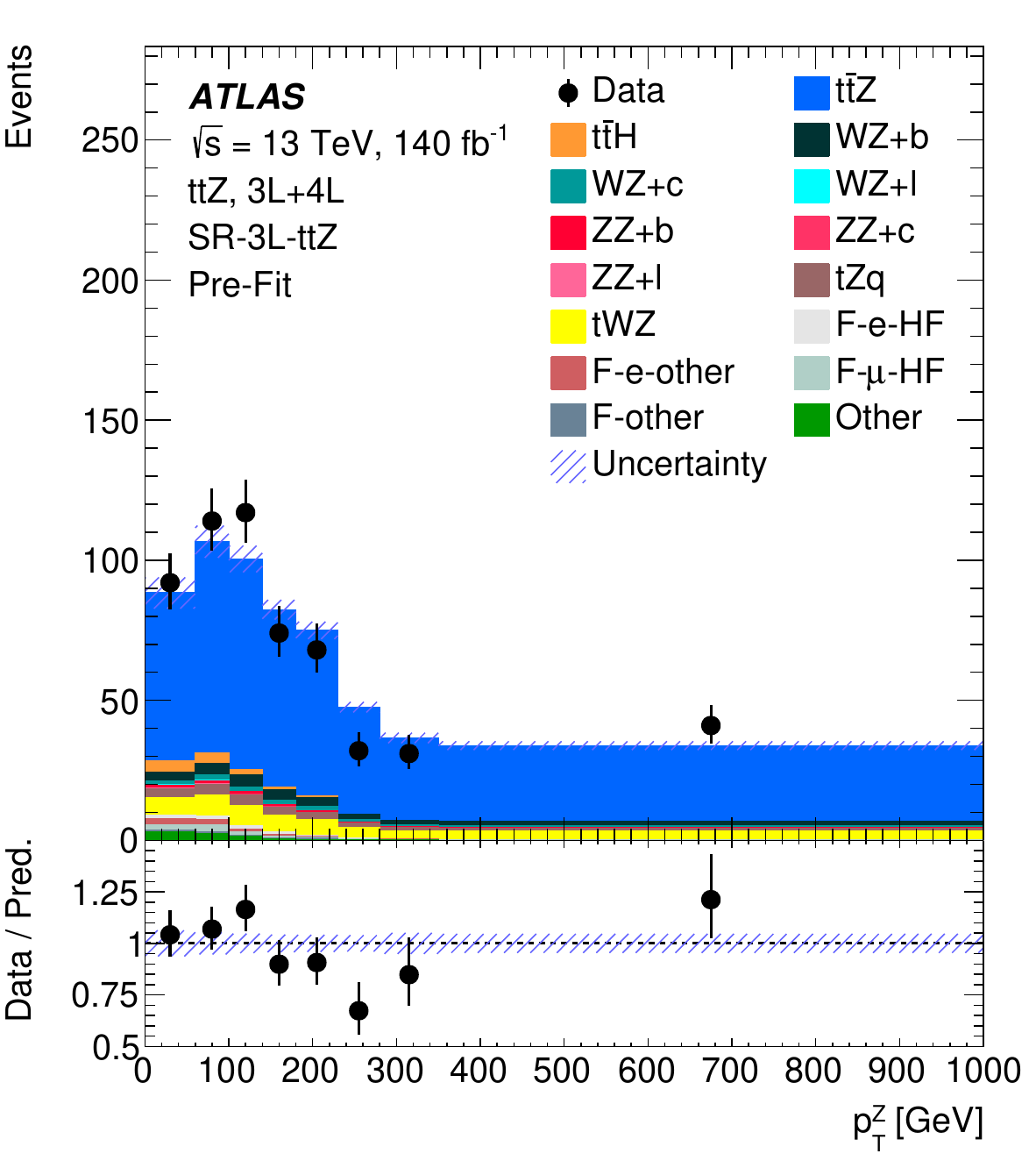}}
\subfloat[]{\includegraphics[width=0.285\textwidth]{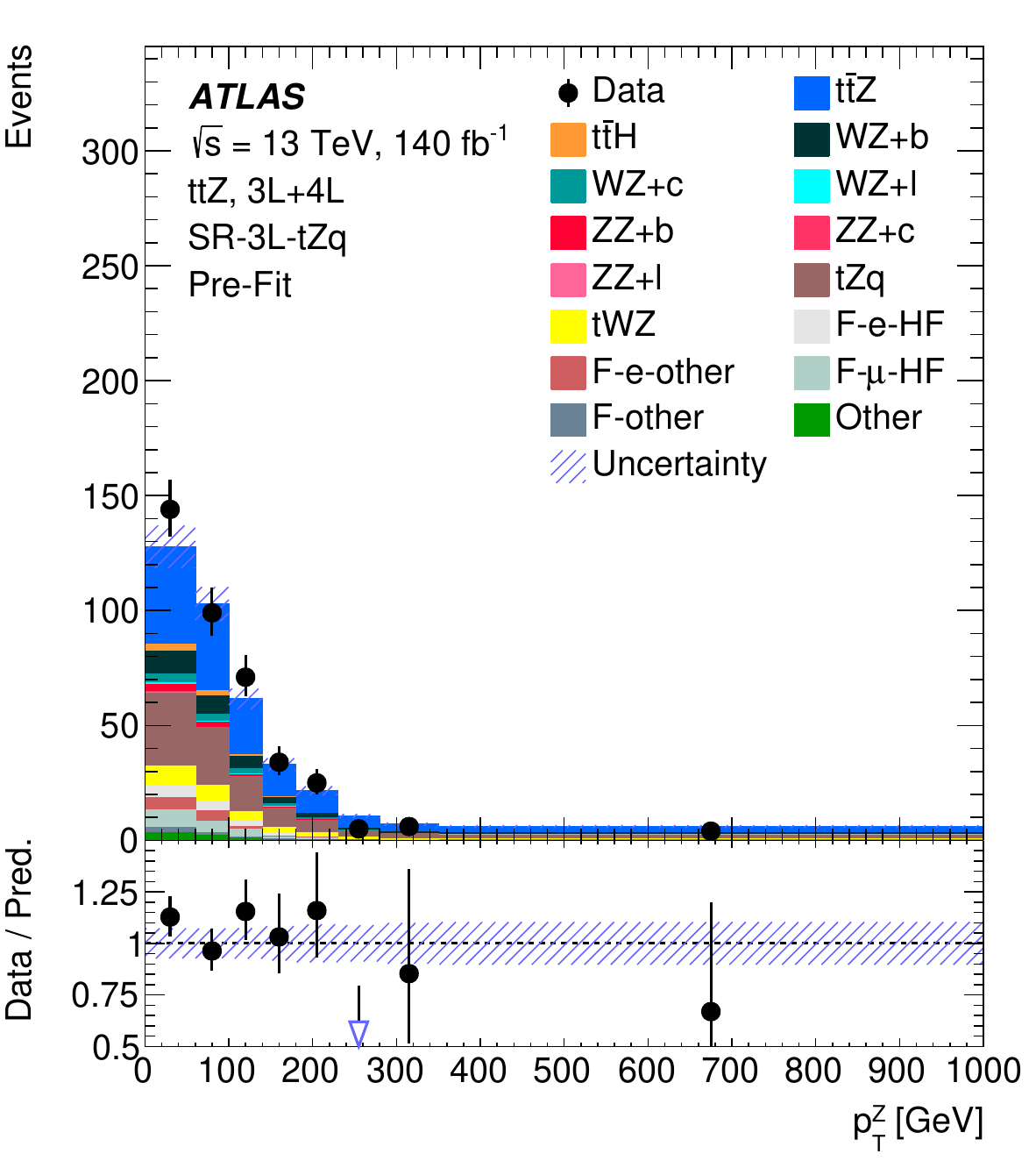}}
\subfloat[]{\includegraphics[width=0.285\textwidth]{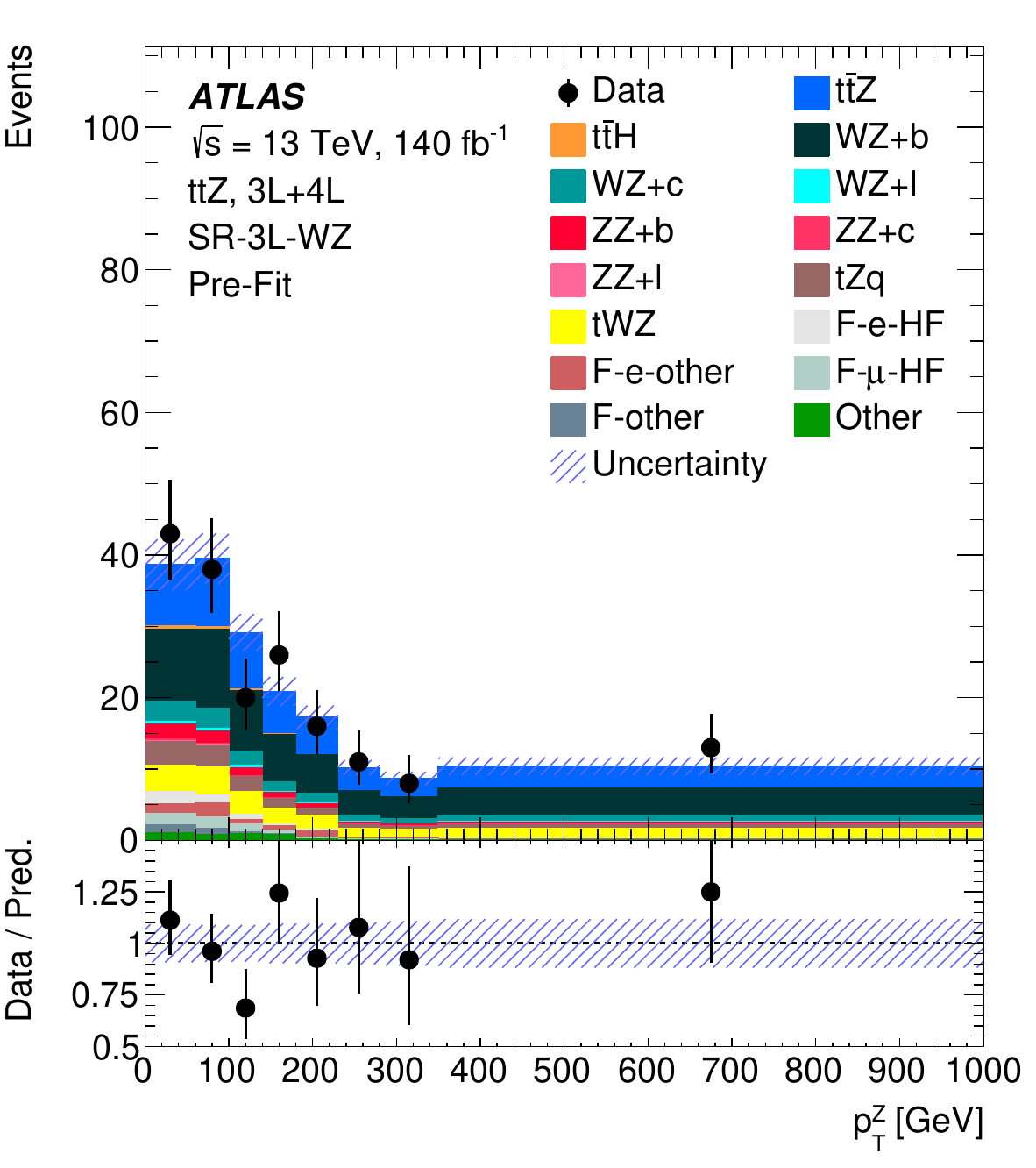}}\\
\subfloat[]{\includegraphics[width=0.285\textwidth]{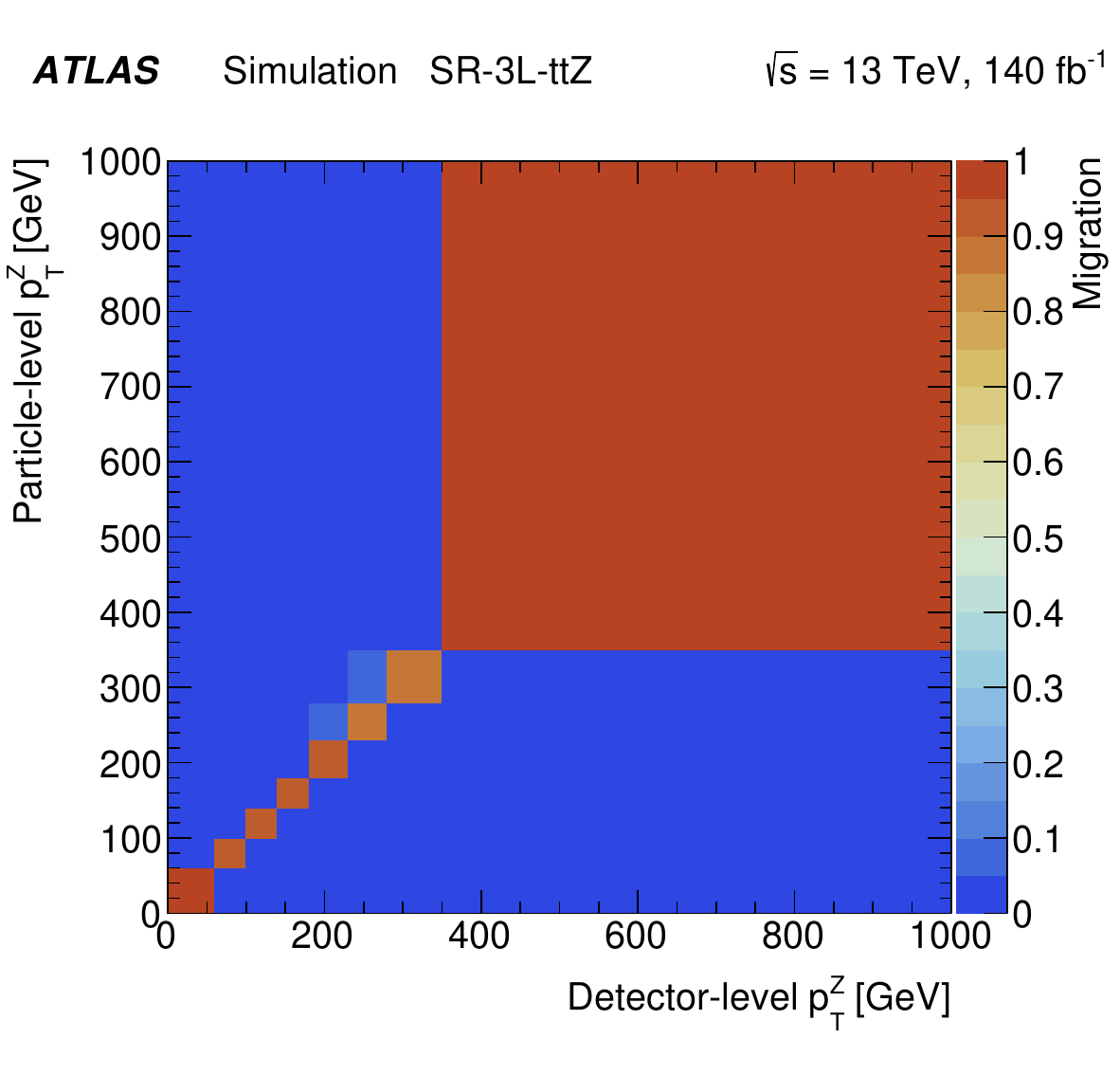}}
\subfloat[]{\includegraphics[width=0.285\textwidth]{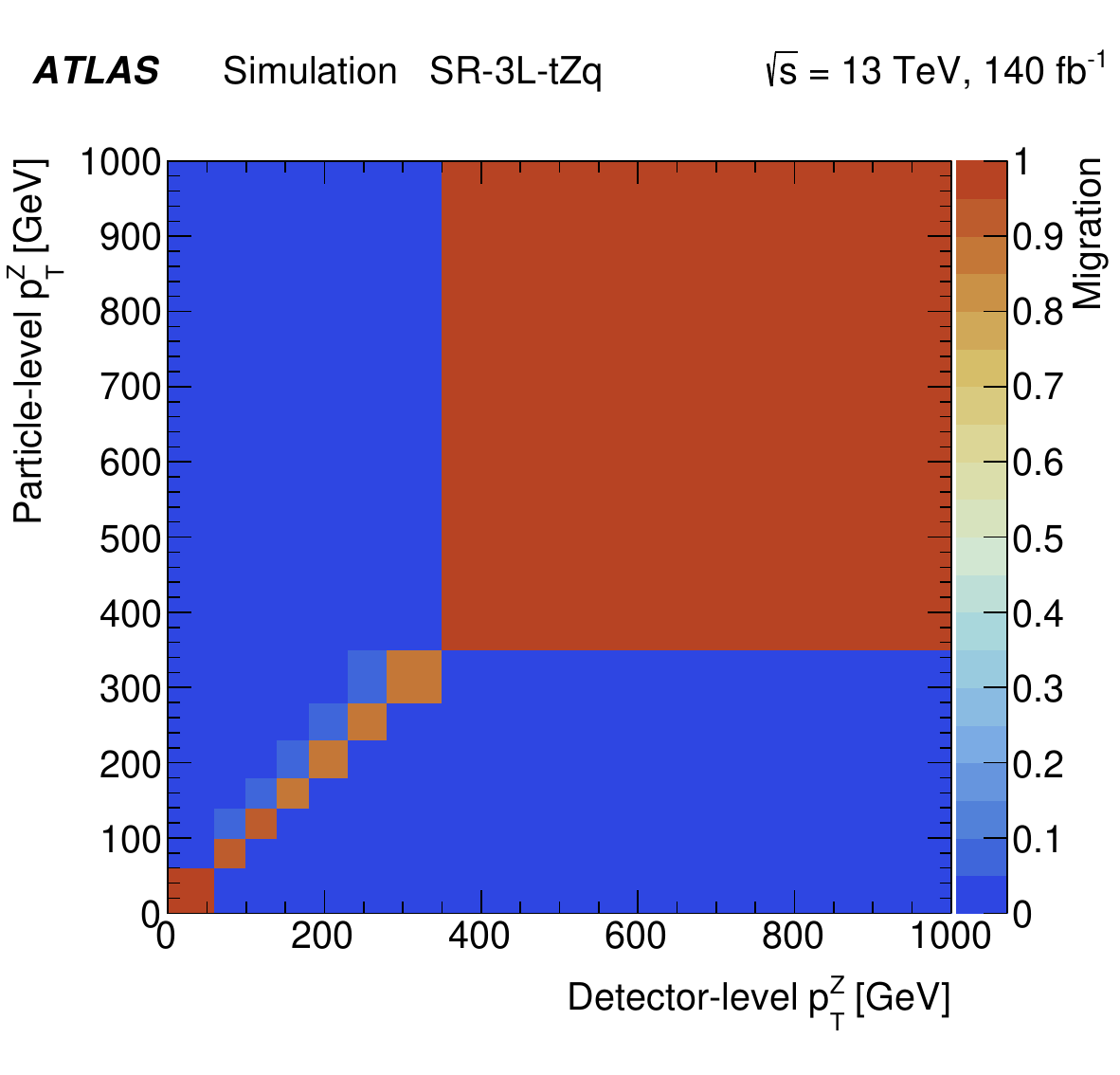}}
\subfloat[]{\includegraphics[width=0.285\textwidth]{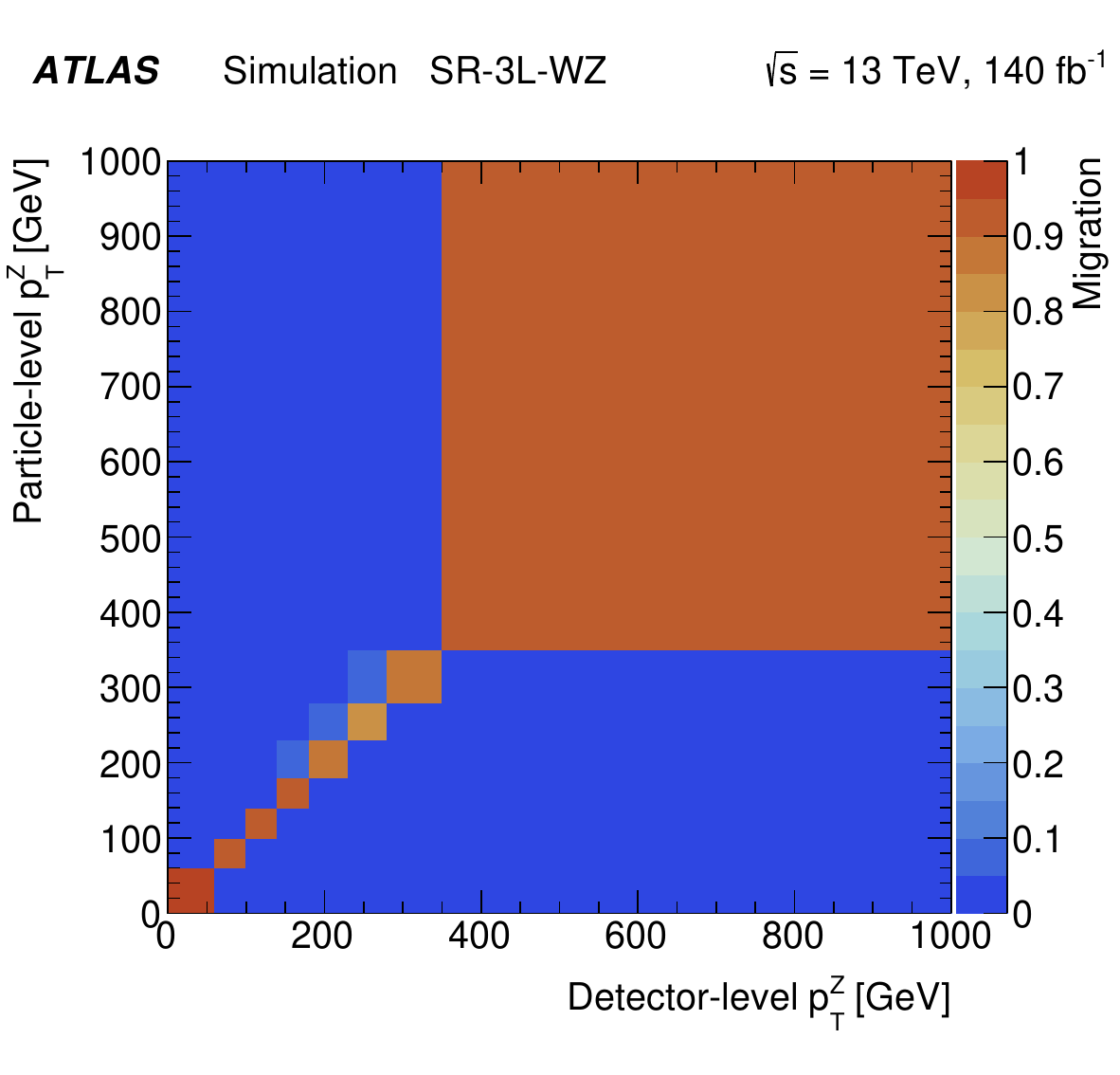}}\\
\subfloat[]{\includegraphics[width=0.285\textwidth]{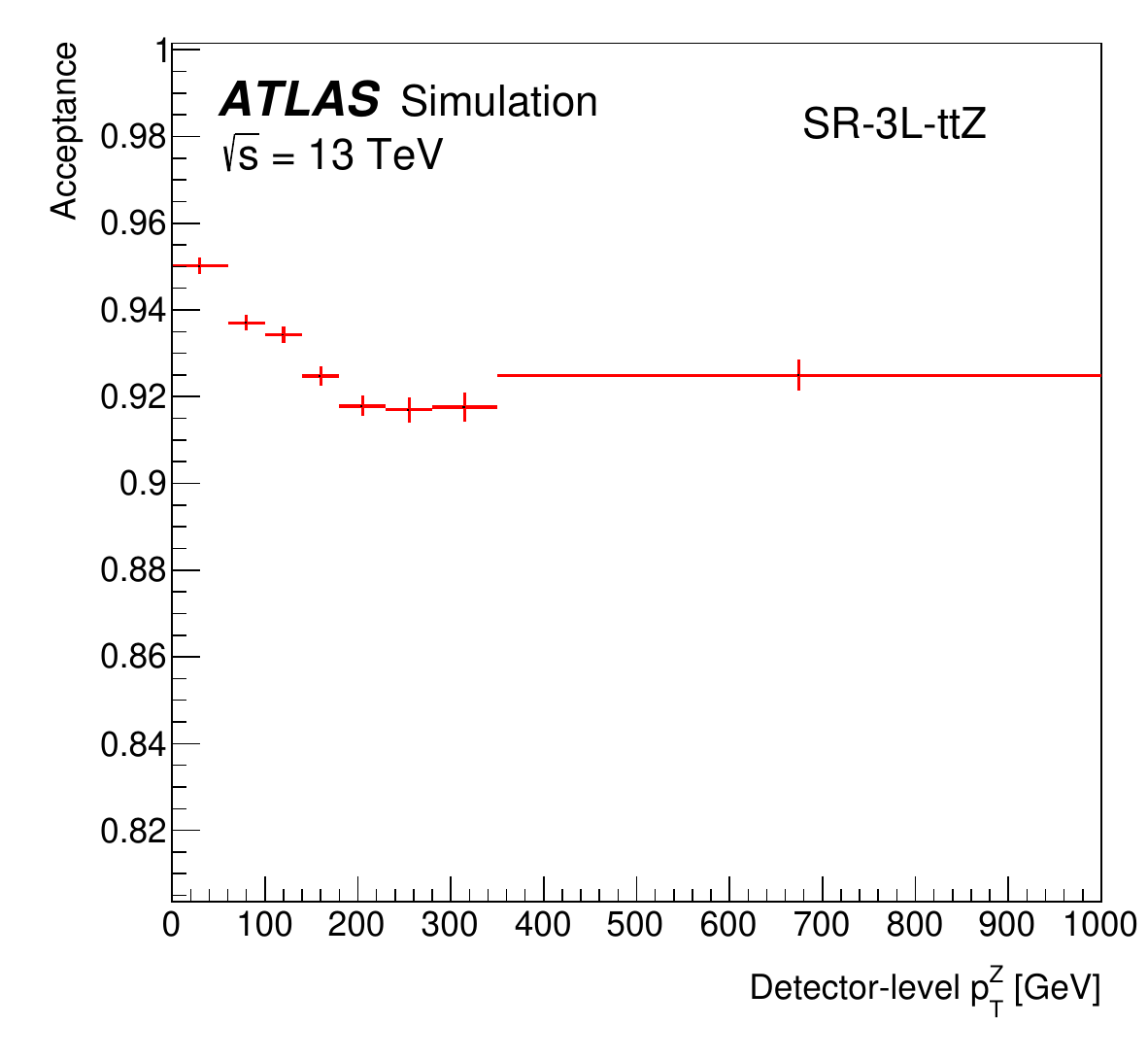}}
\subfloat[]{\includegraphics[width=0.285\textwidth]{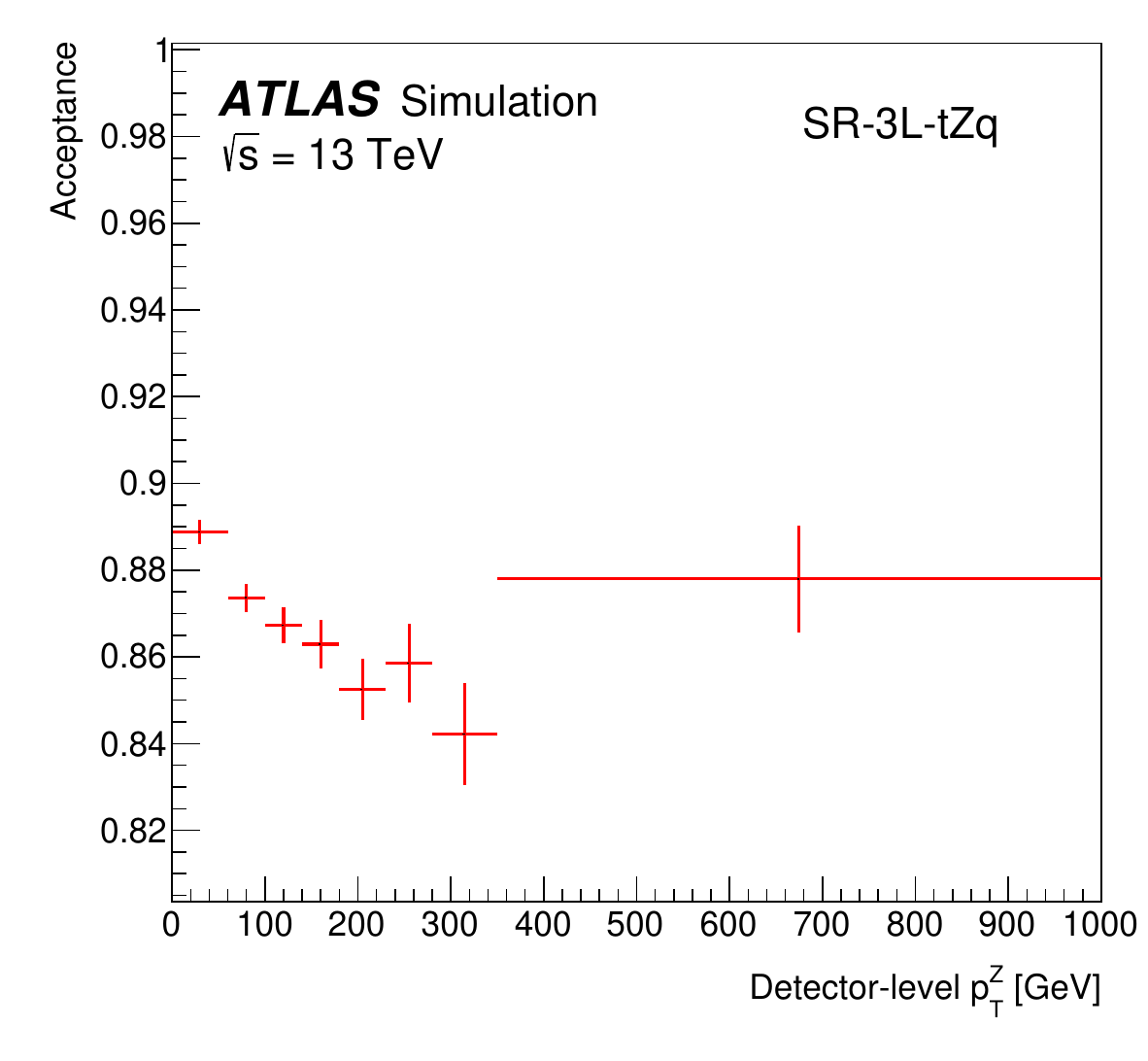}}
\subfloat[]{\includegraphics[width=0.285\textwidth]{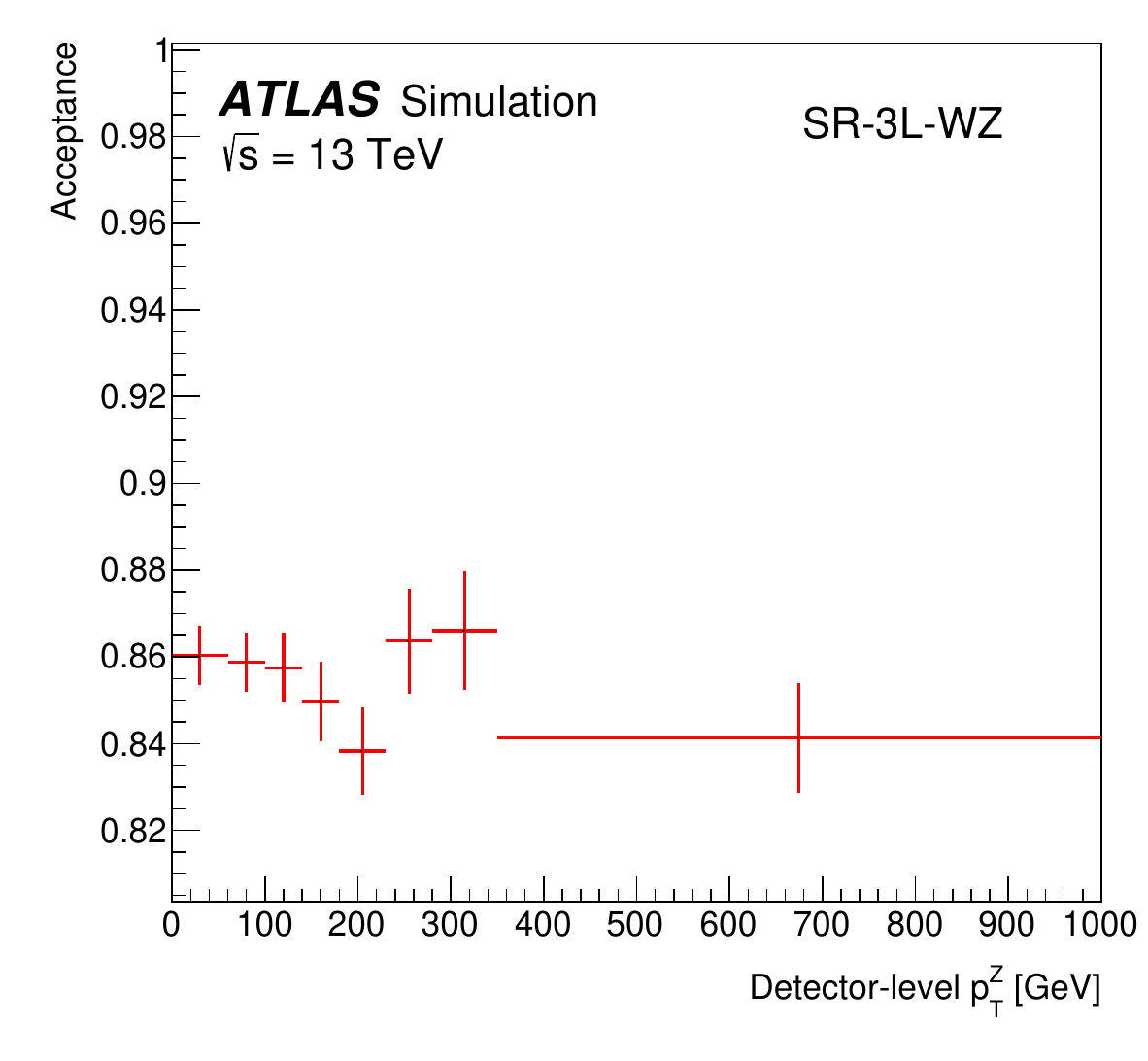}}\\
\subfloat[]{\includegraphics[width=0.285\textwidth]{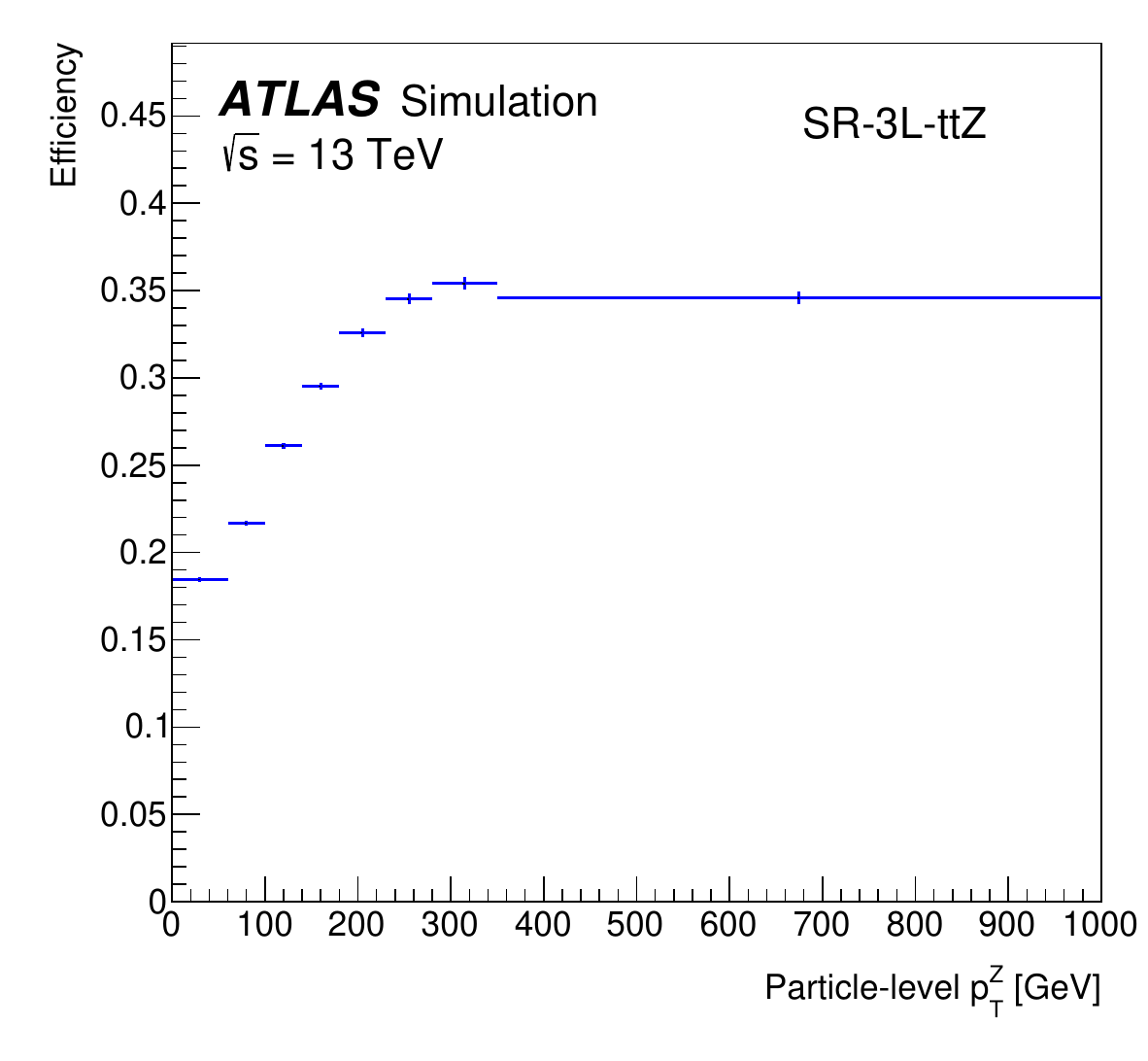}}
\subfloat[]{\includegraphics[width=0.285\textwidth]{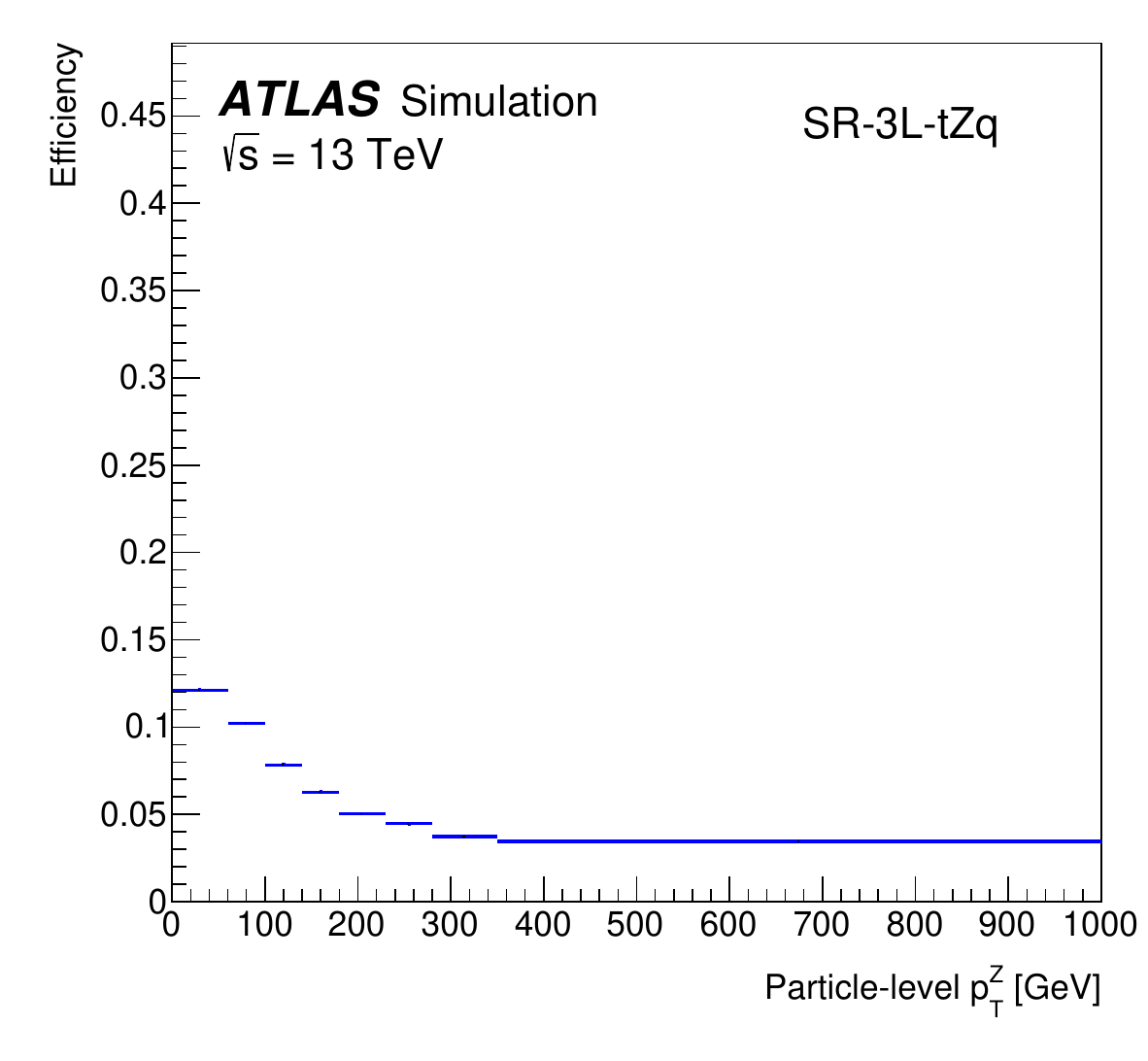}}
\subfloat[]{\includegraphics[width=0.285\textwidth]{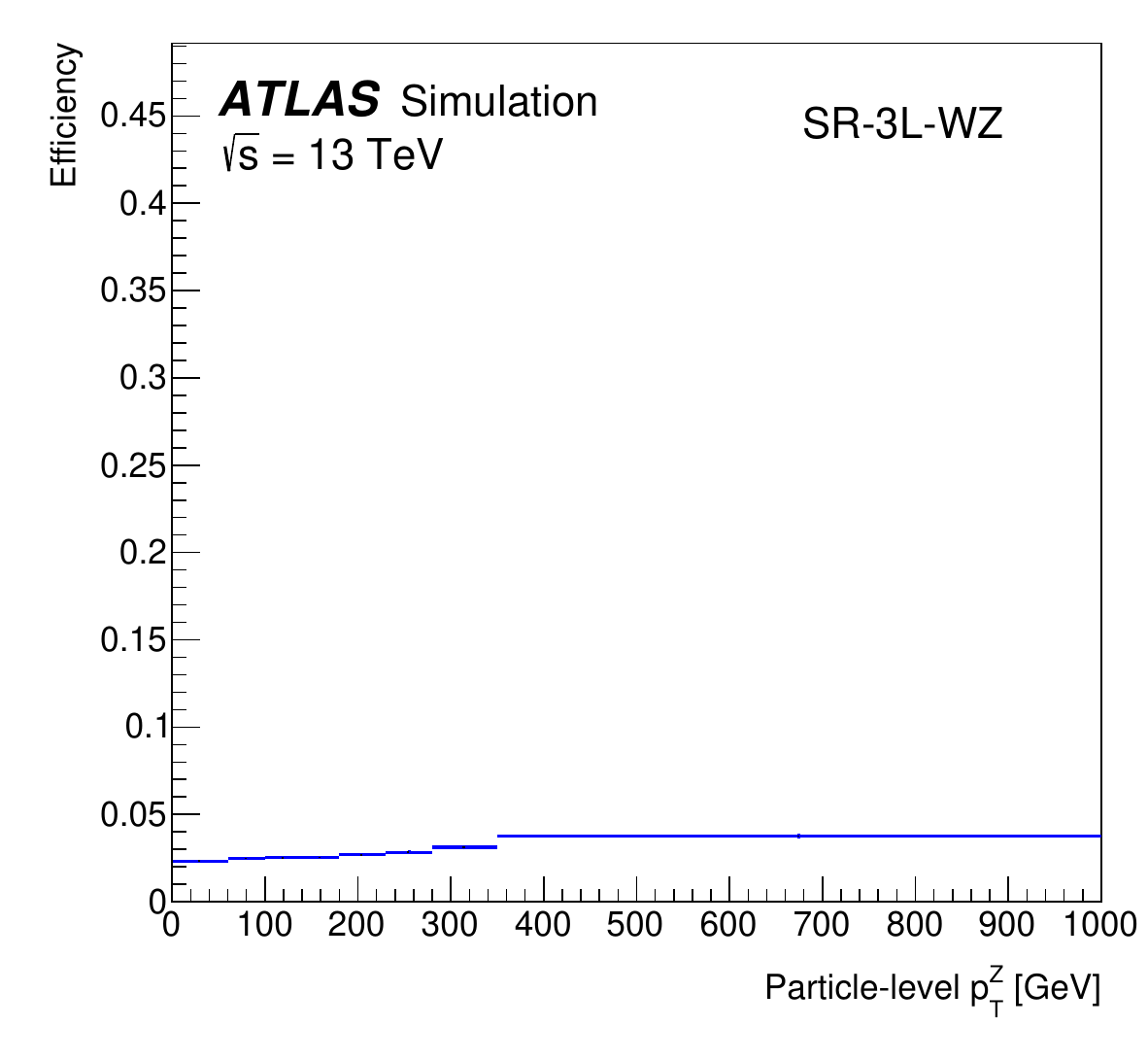}}
\caption{Detector-level distributions (a,b,c), together with migration matrices (d,e,f) and acceptance (g,h,i) and efficiency (j,k,l) histograms for the $\pT^{Z}$ observable in the trilepton channel regions: SR-$3\ell$-ttZ (a,d,g,j), SR-$3\ell$-tZq (b,e,h,k) and SR-$3\ell$-WZ (c,f,i,l). Migration matrices and corrections apply to the particle level.}
\label{fig:trilepton-asimov-unfolding-reco-migration-eff-acc-particle-ptz}
\end{figure}
 
\begin{figure}[!htb]
\centering
\subfloat[]{\includegraphics[width=0.285\textwidth]{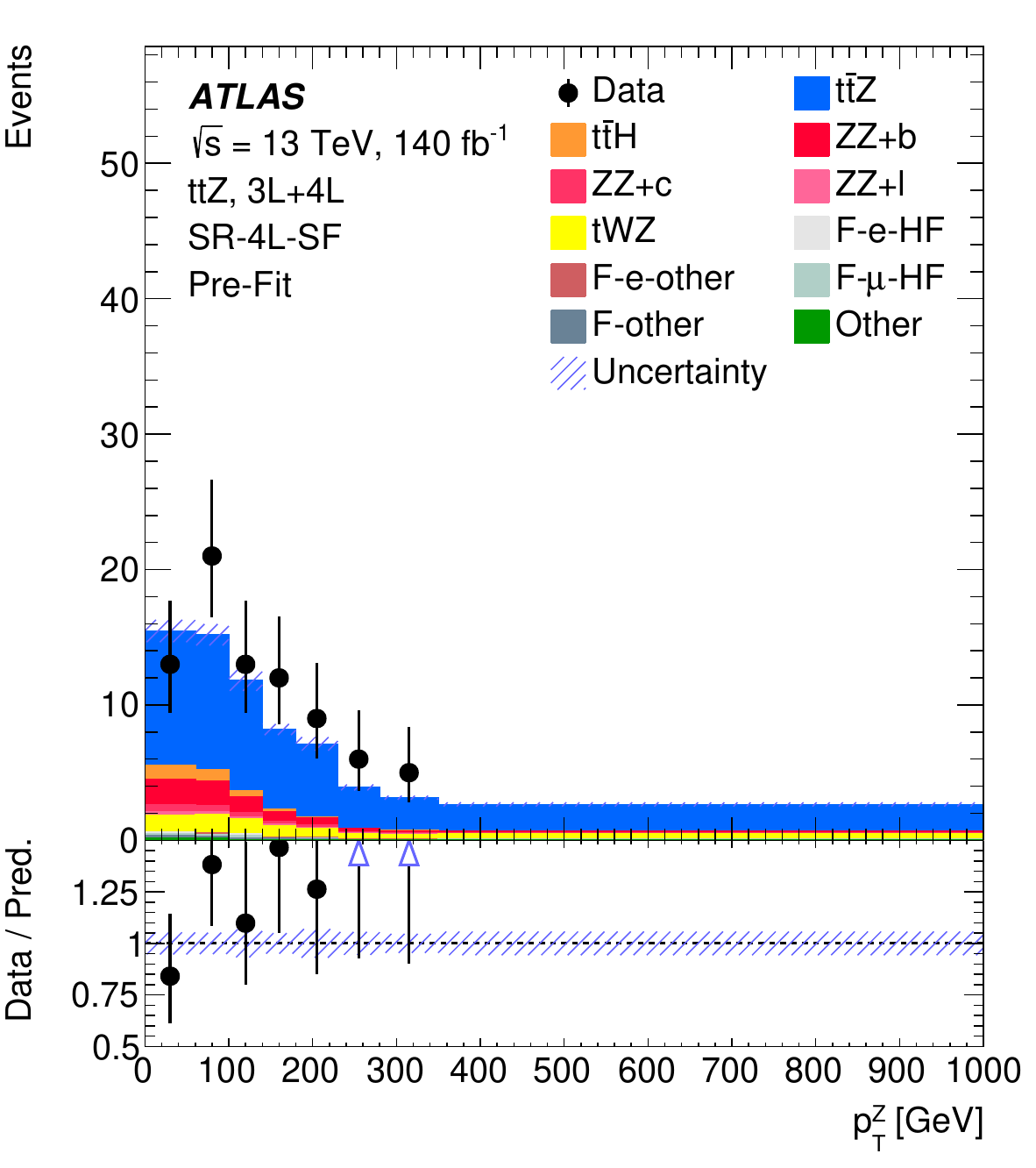}}
\subfloat[]{\includegraphics[width=0.285\textwidth]{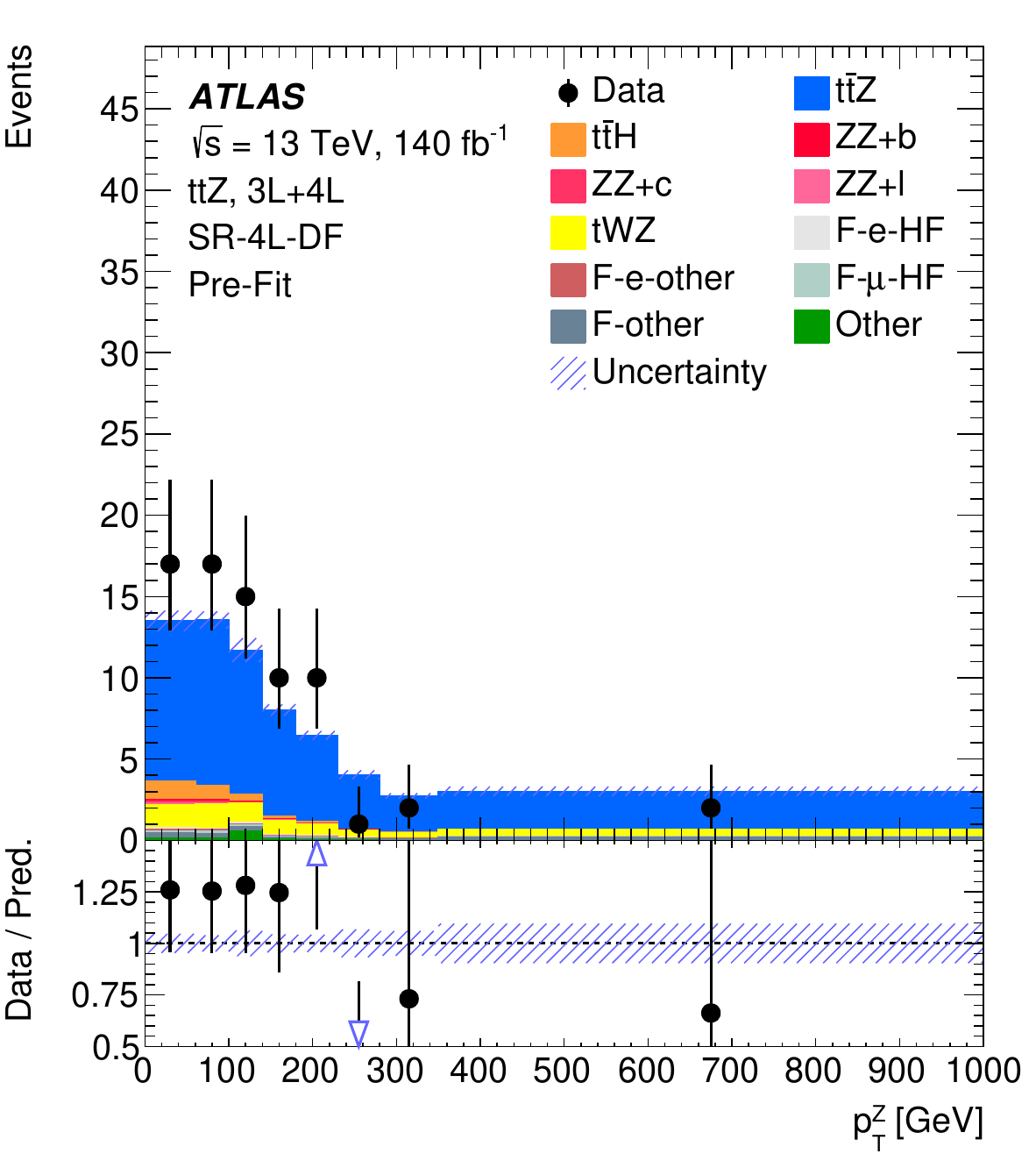}}
\subfloat[]{\includegraphics[width=0.285\textwidth]{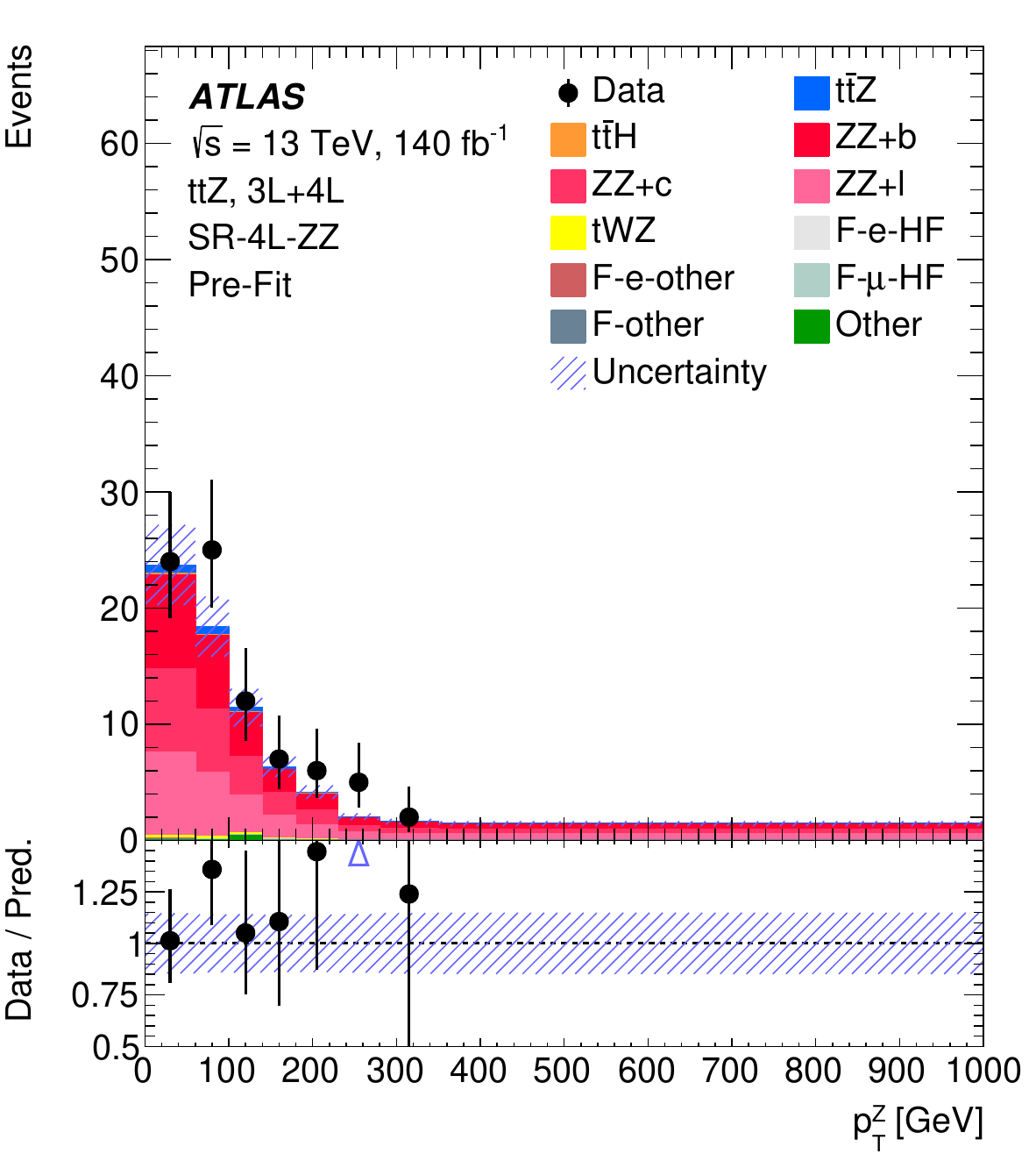}}\\
\subfloat[]{\includegraphics[width=0.285\textwidth]{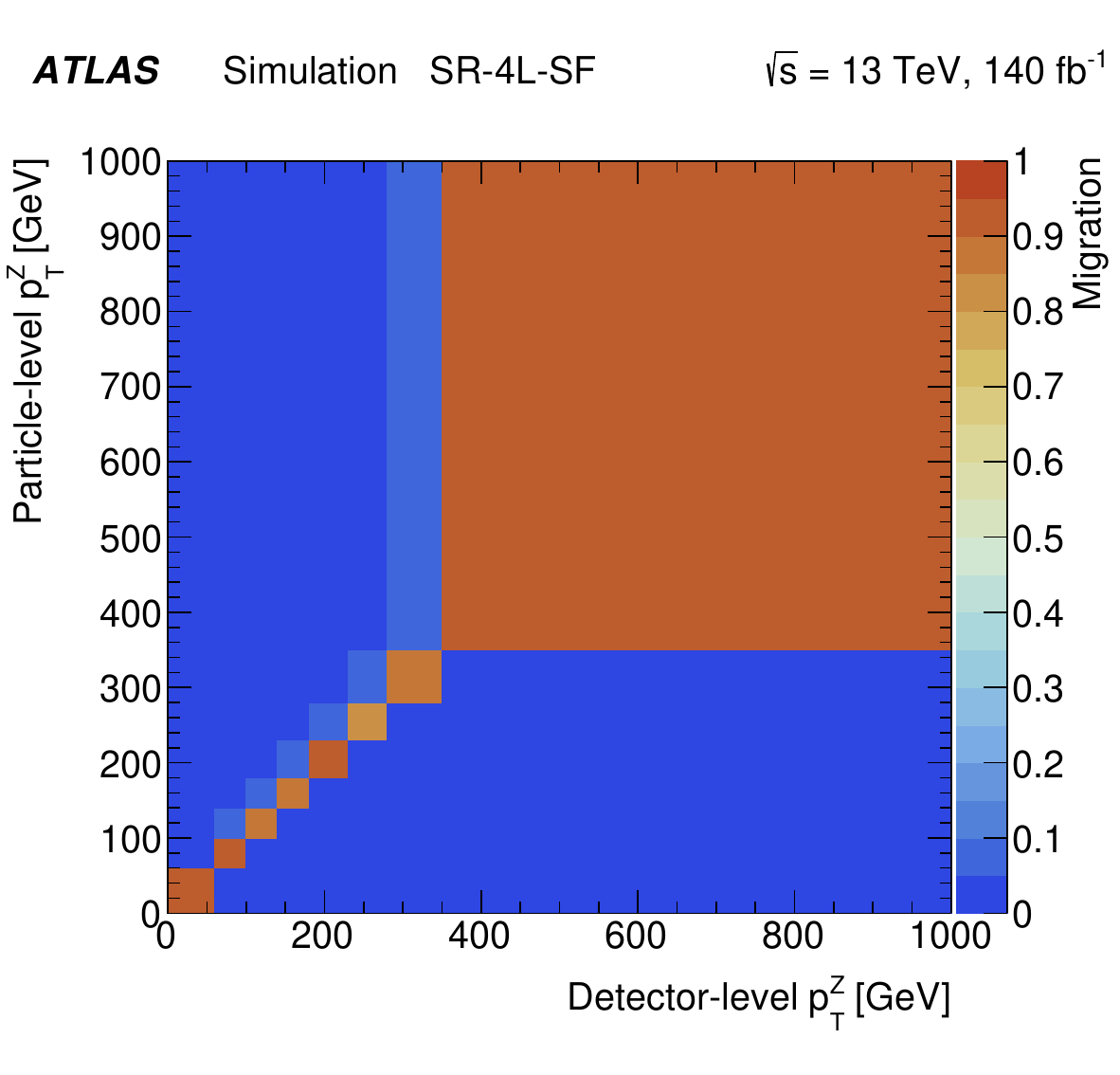}}
\subfloat[]{\includegraphics[width=0.285\textwidth]{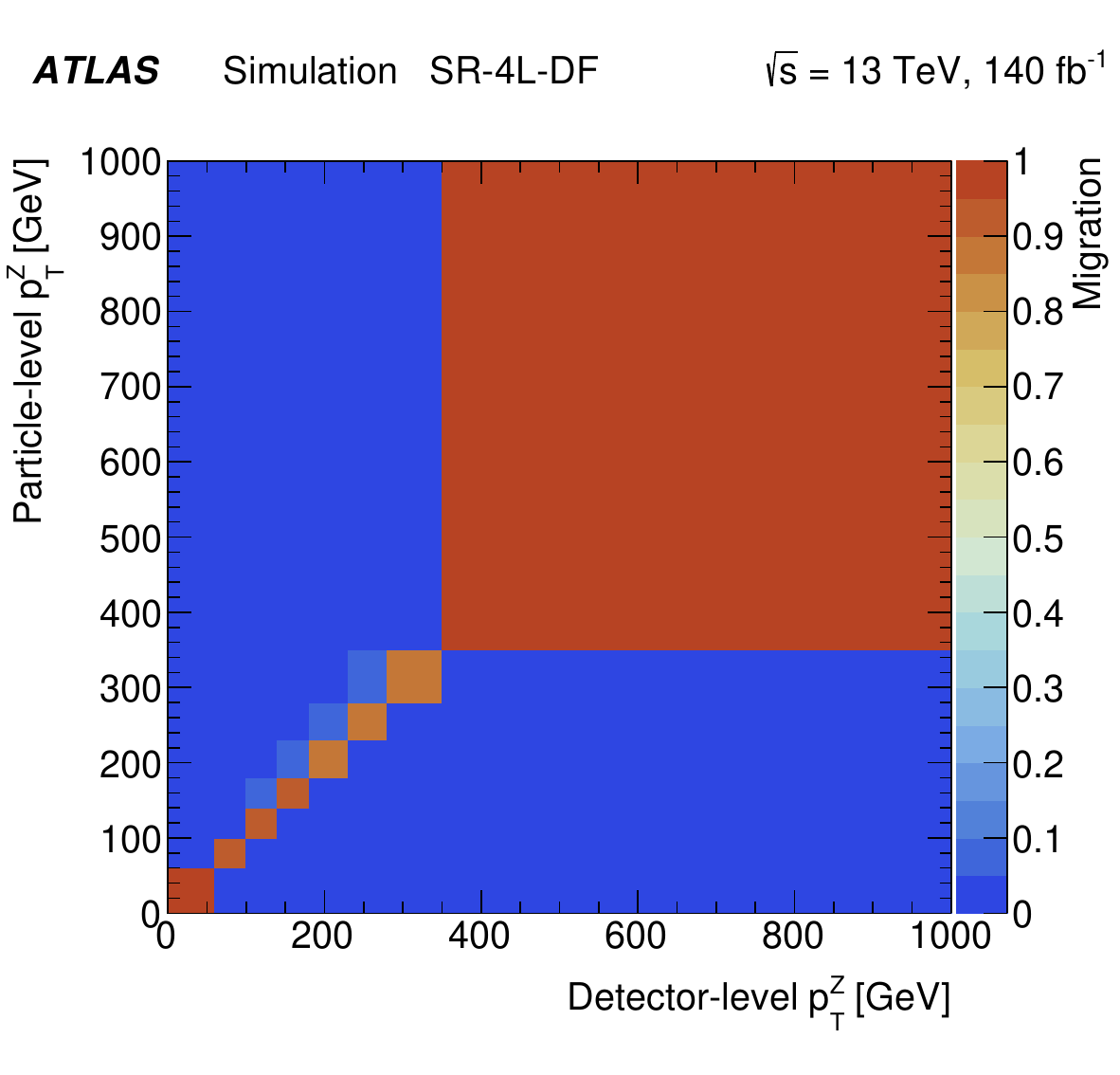}}
\subfloat[]{\includegraphics[width=0.285\textwidth]{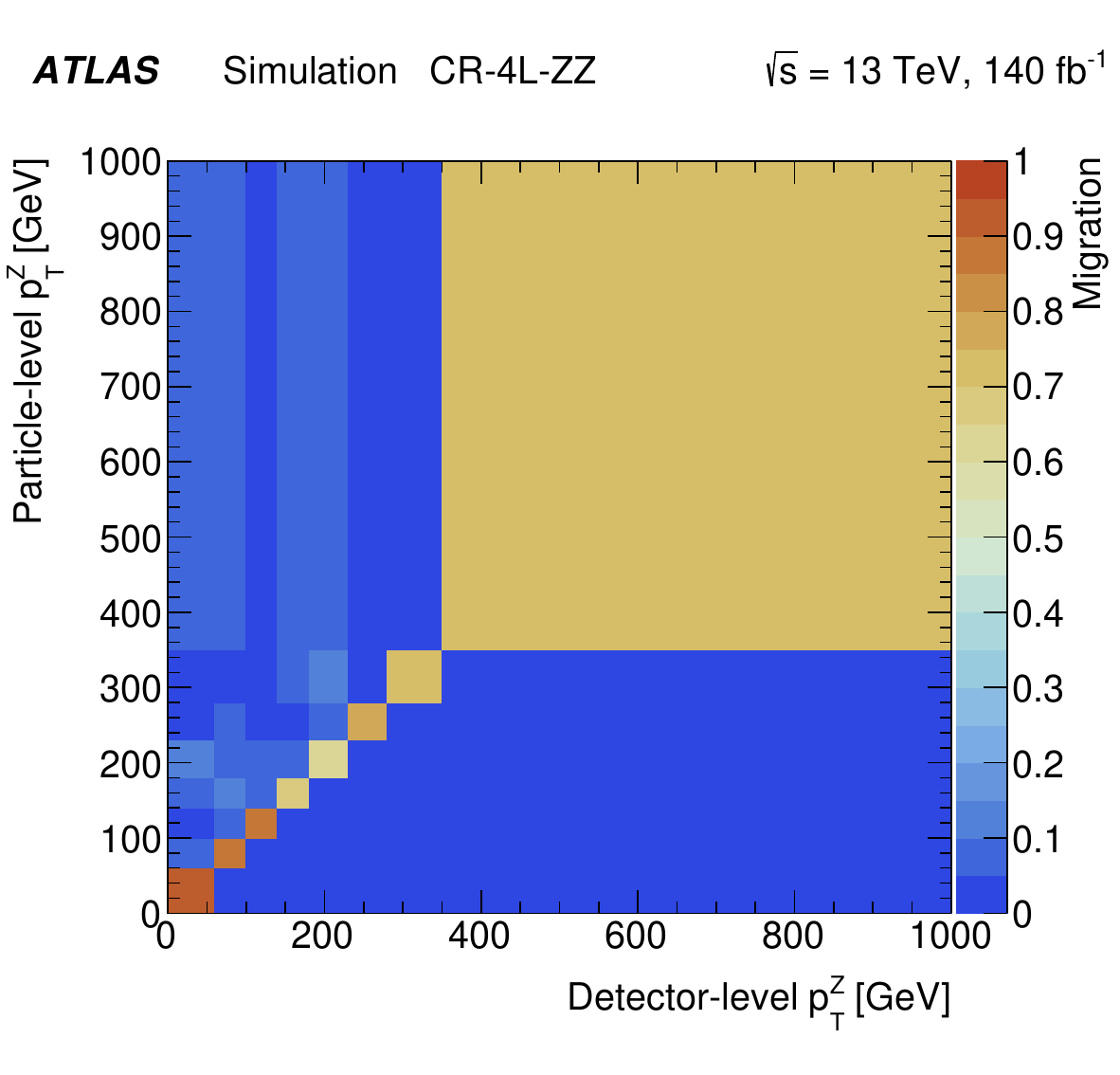}}\\
\subfloat[]{\includegraphics[width=0.285\textwidth]{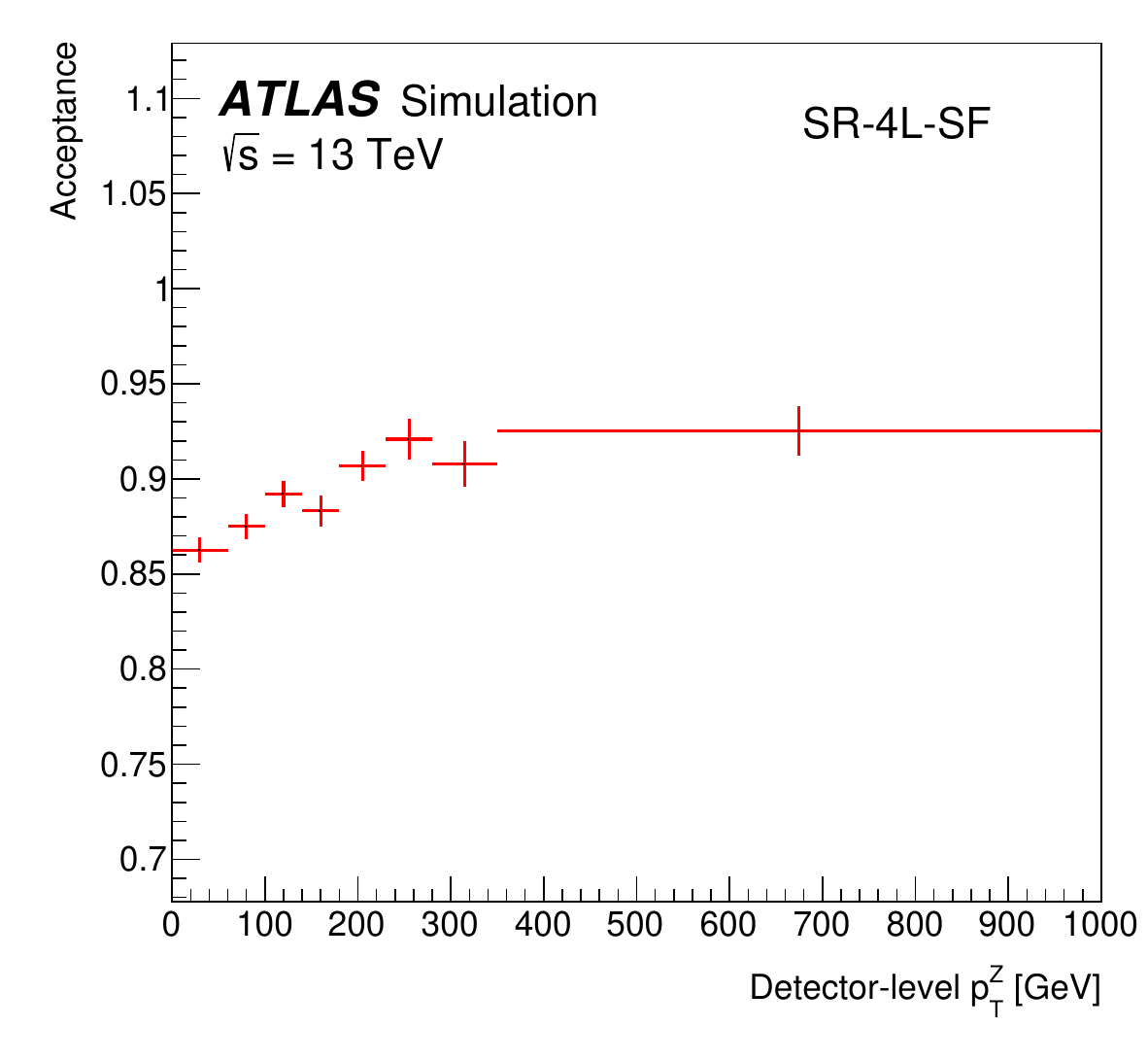}}
\subfloat[]{\includegraphics[width=0.285\textwidth]{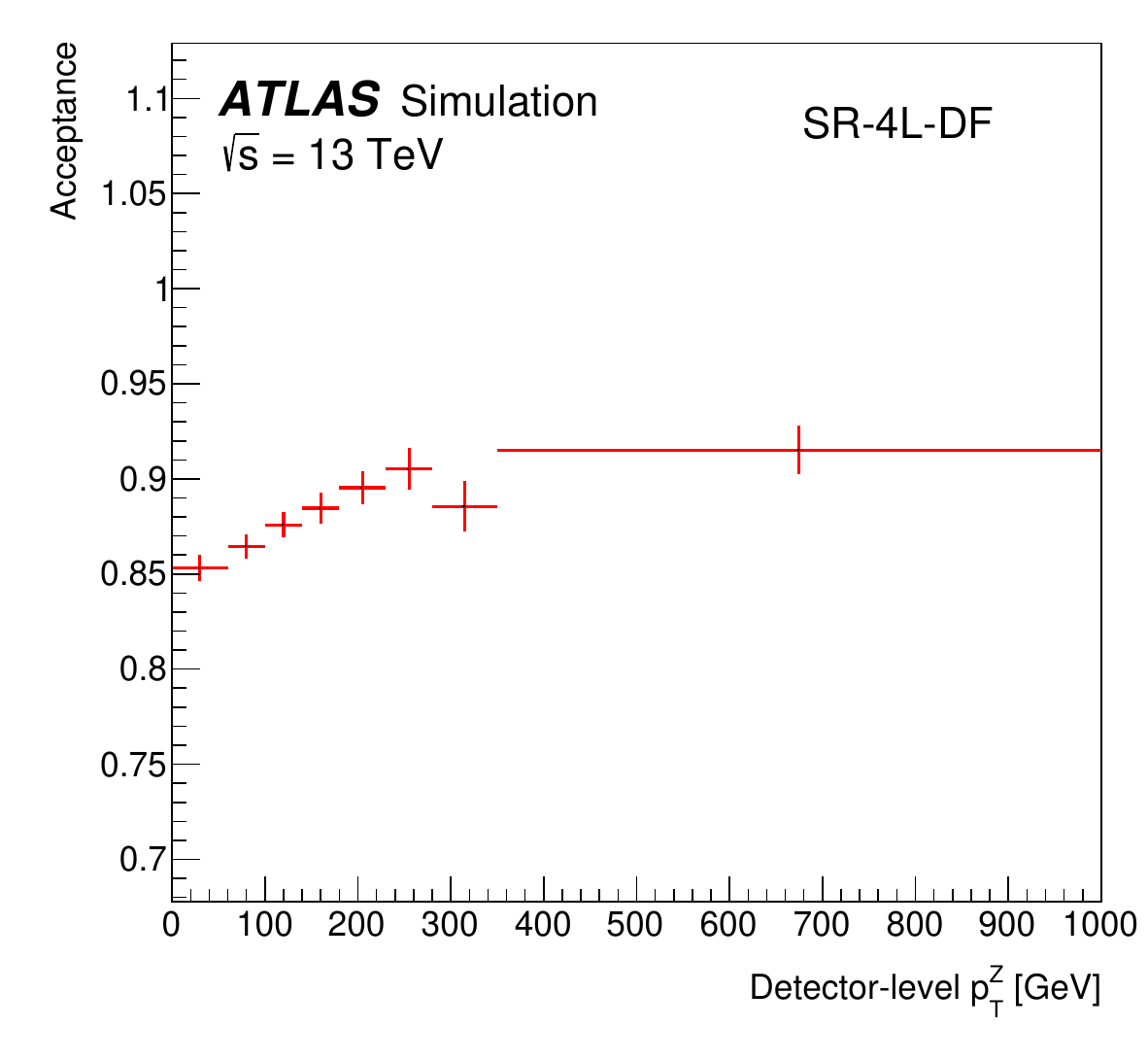}}
\subfloat[]{\includegraphics[width=0.285\textwidth]{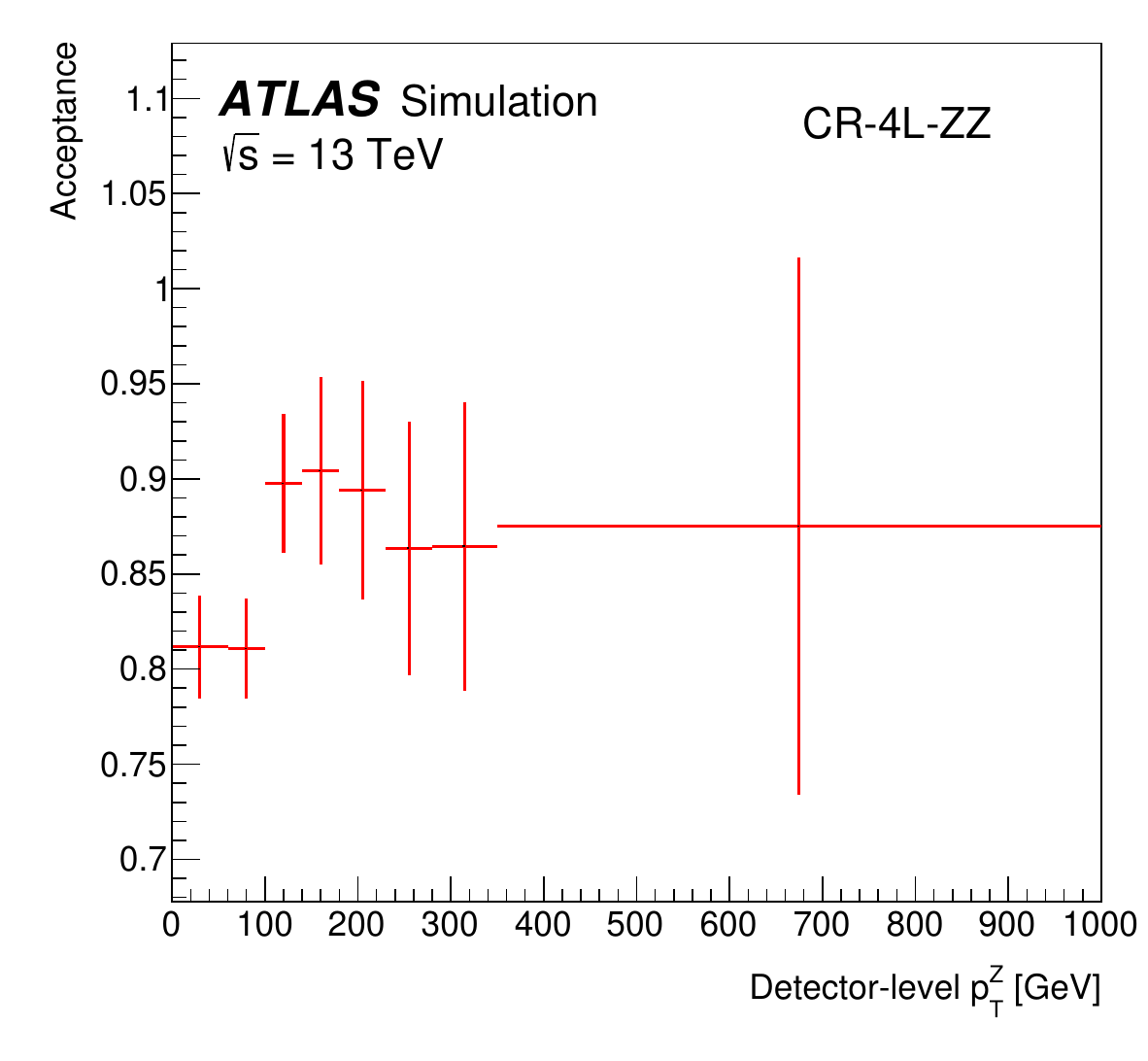}}\\
\subfloat[]{\includegraphics[width=0.285\textwidth]{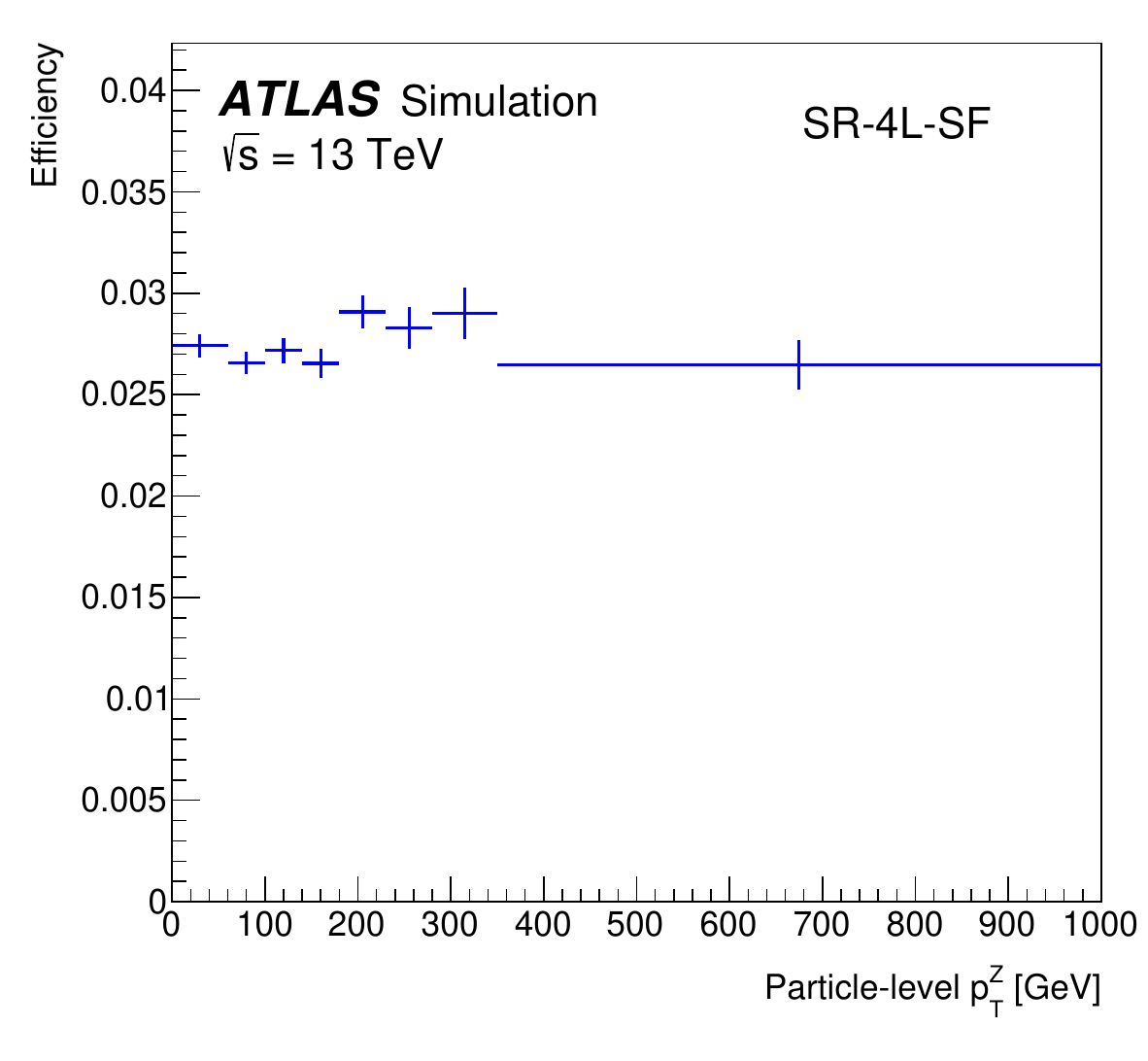}}
\subfloat[]{\includegraphics[width=0.285\textwidth]{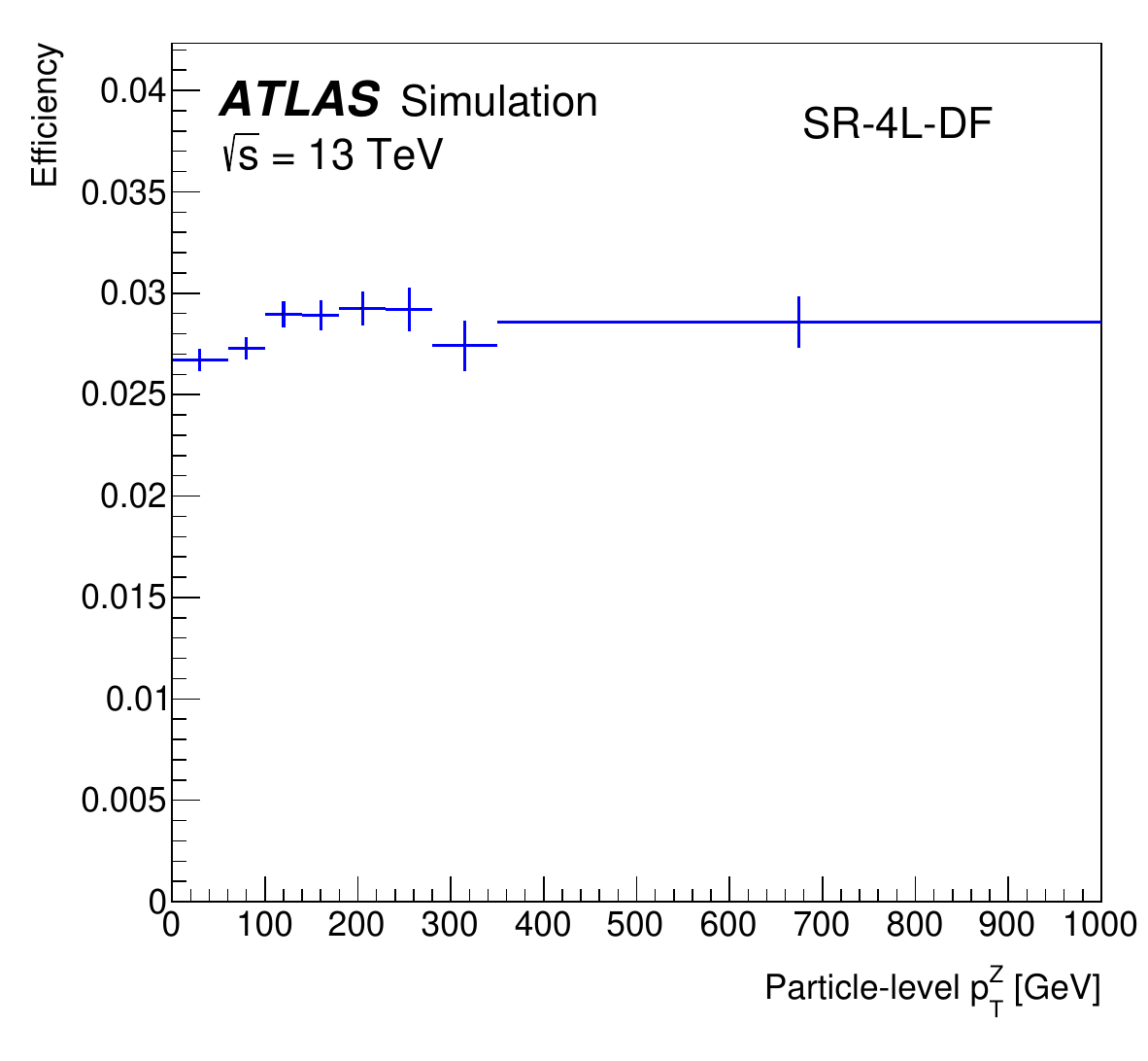}}
\subfloat[]{\includegraphics[width=0.285\textwidth]{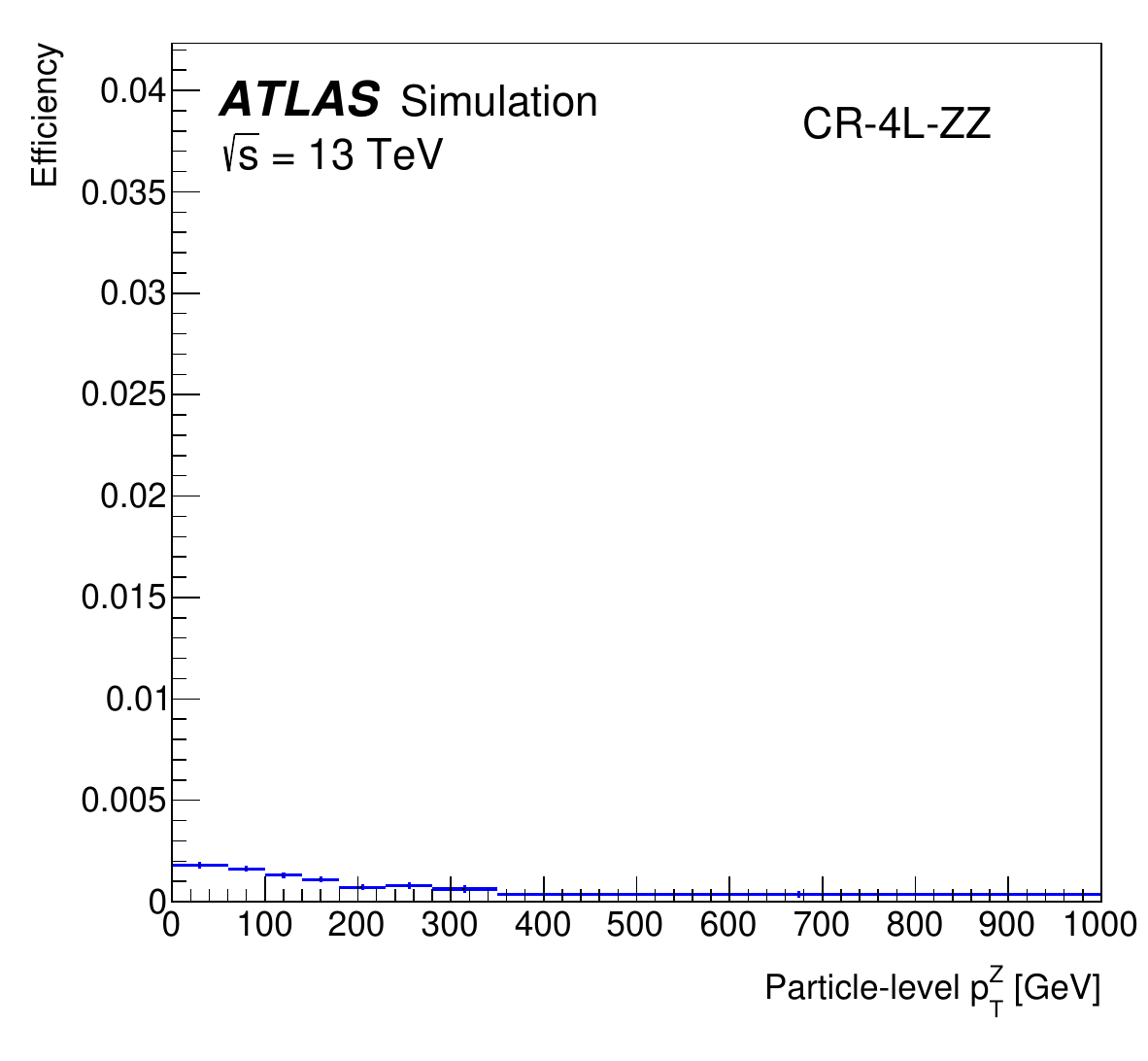}}
\caption{Detector-level distributions (a,b,c), together with migration matrices (d,e,f) and acceptance (g,h,i) and efficiency (j,k,l) histograms for the $\pT^{Z}$ observable in the tetralepton channel regions: SR-$4\ell$-SF (a,d,g,j), SR-$4\ell$-DF (b,e,h,k) and CR-$4\ell$-ZZ (c,f,i,l). Migration matrices and corrections apply to the particle level.}
\label{fig:tetralepton-asimov-unfolding-reco-migration-eff-acc-particle-ptz}
\end{figure}
 
\begin{figure}[!htb]
\centering
\subfloat[]{\includegraphics[width=0.46\textwidth]{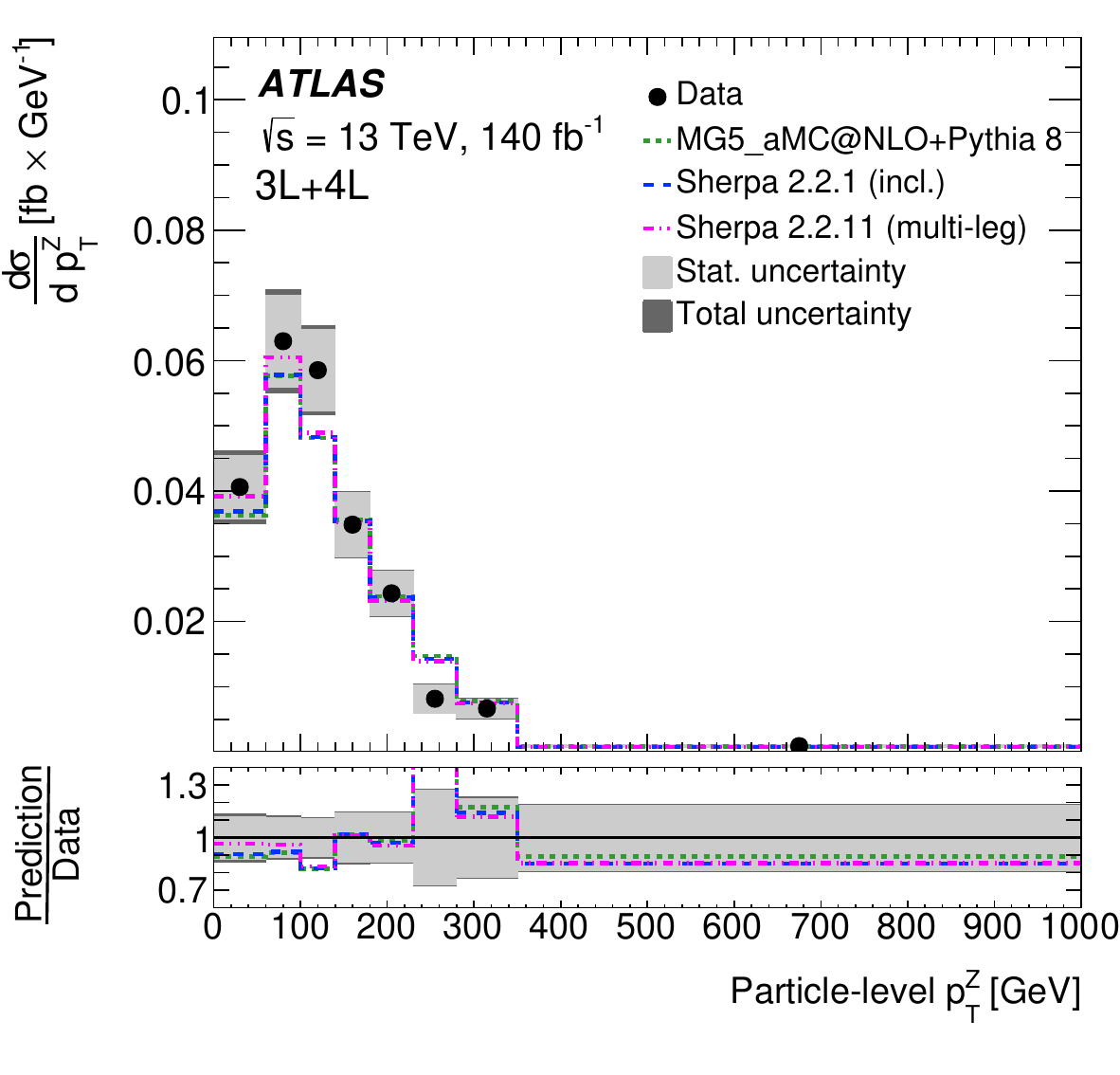}}
\hspace*{0.06\textwidth}
\subfloat[]{\includegraphics[width=0.46\textwidth]{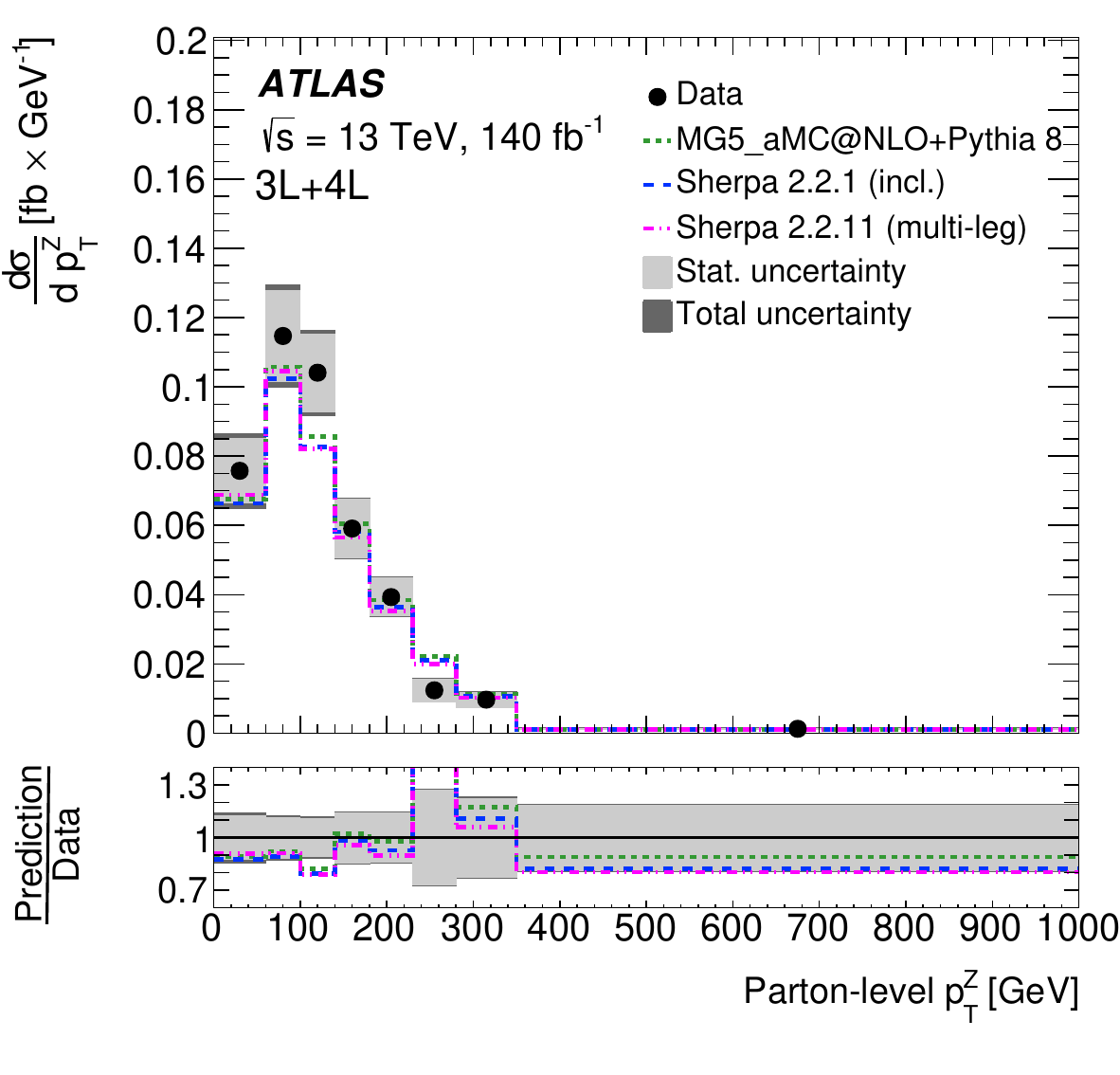}} \\
\subfloat[]{\includegraphics[width=0.46\textwidth]{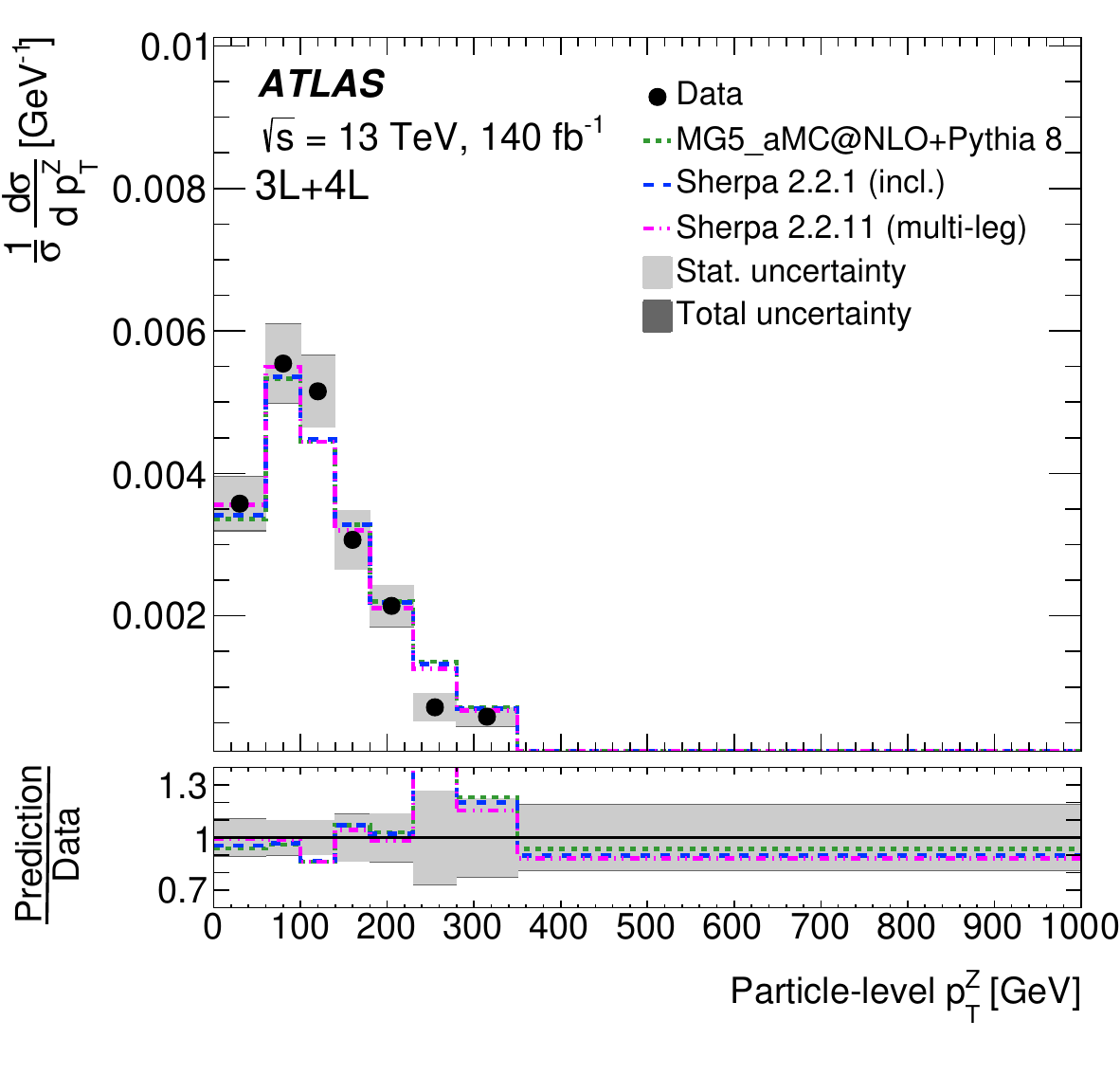}}
\hspace*{0.06\textwidth}
\subfloat[]{\includegraphics[width=0.46\textwidth]{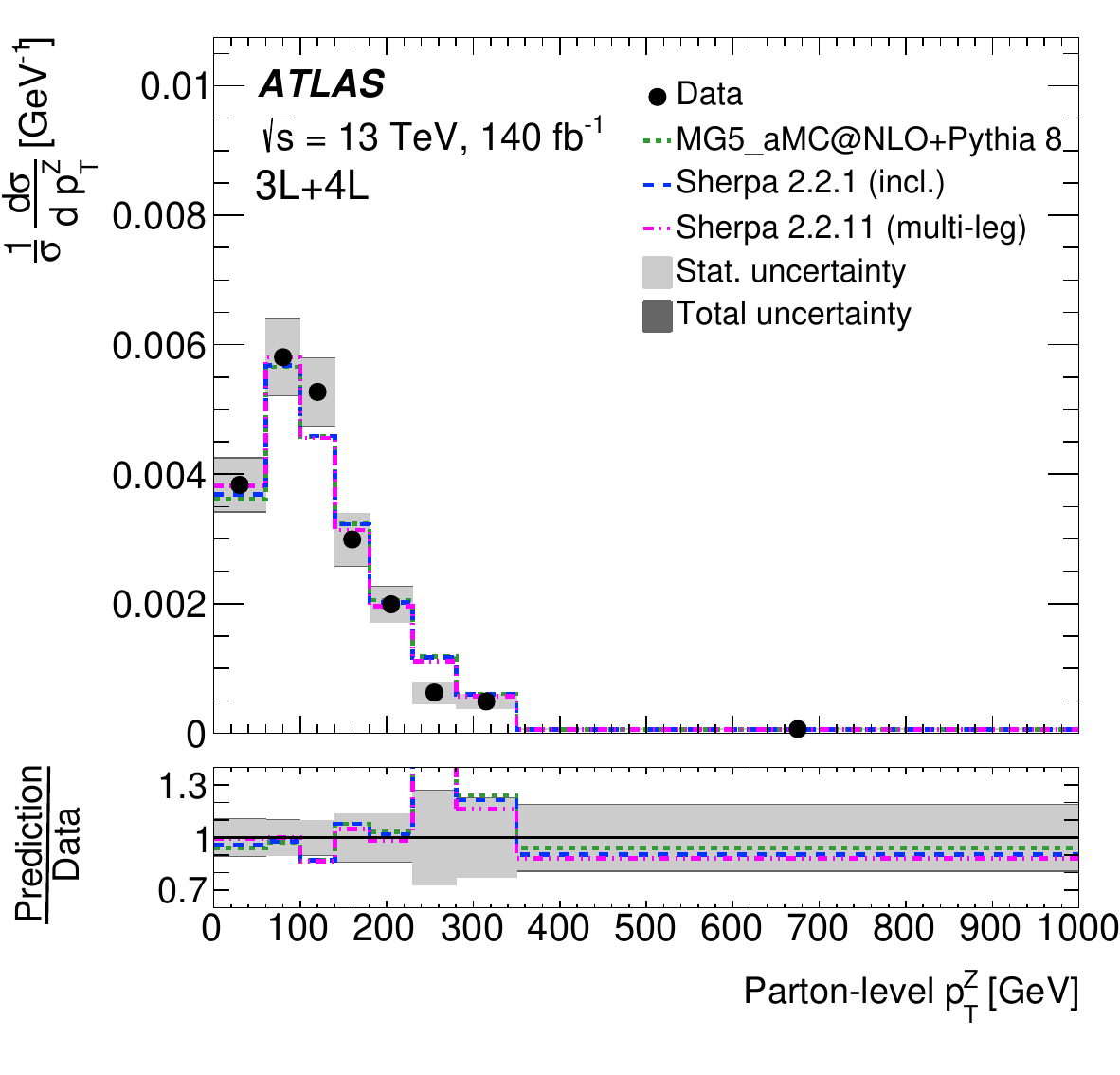}}
\caption{Differential cross-section measurements as a function of $\pT^{Z}$ observable in the combined $3\ell$ and $4\ell$ channels, (a,b) absolute and (c,d) normalised, unfolded to (a,c) particle level and (b,d) parton level. The dark grey band corresponds to the total uncertainty of the measurement; in some cases, it is almost fully covered by the light grey band, representing the dominant statistical uncertainty. Alternative generator predictions are overlaid as additional coloured lines.}
\label{fig:combined-asimov-unfolding-result-ptz}
\end{figure}
 
\begin{figure}[!htb]
\centering
\subfloat[]{\includegraphics[width=0.46\textwidth]{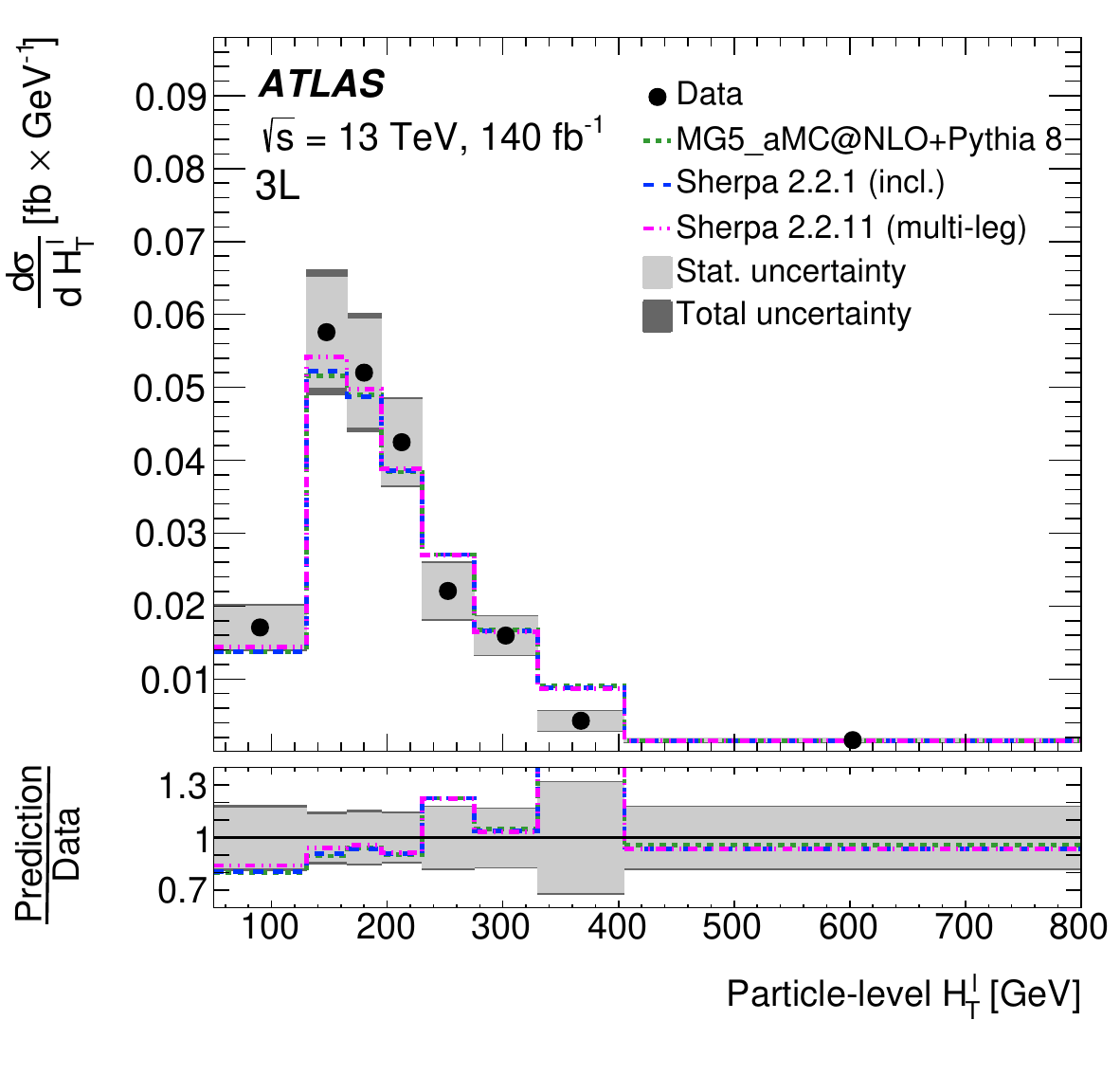}}
\hspace*{0.06\textwidth}
\subfloat[]{\includegraphics[width=0.46\textwidth]{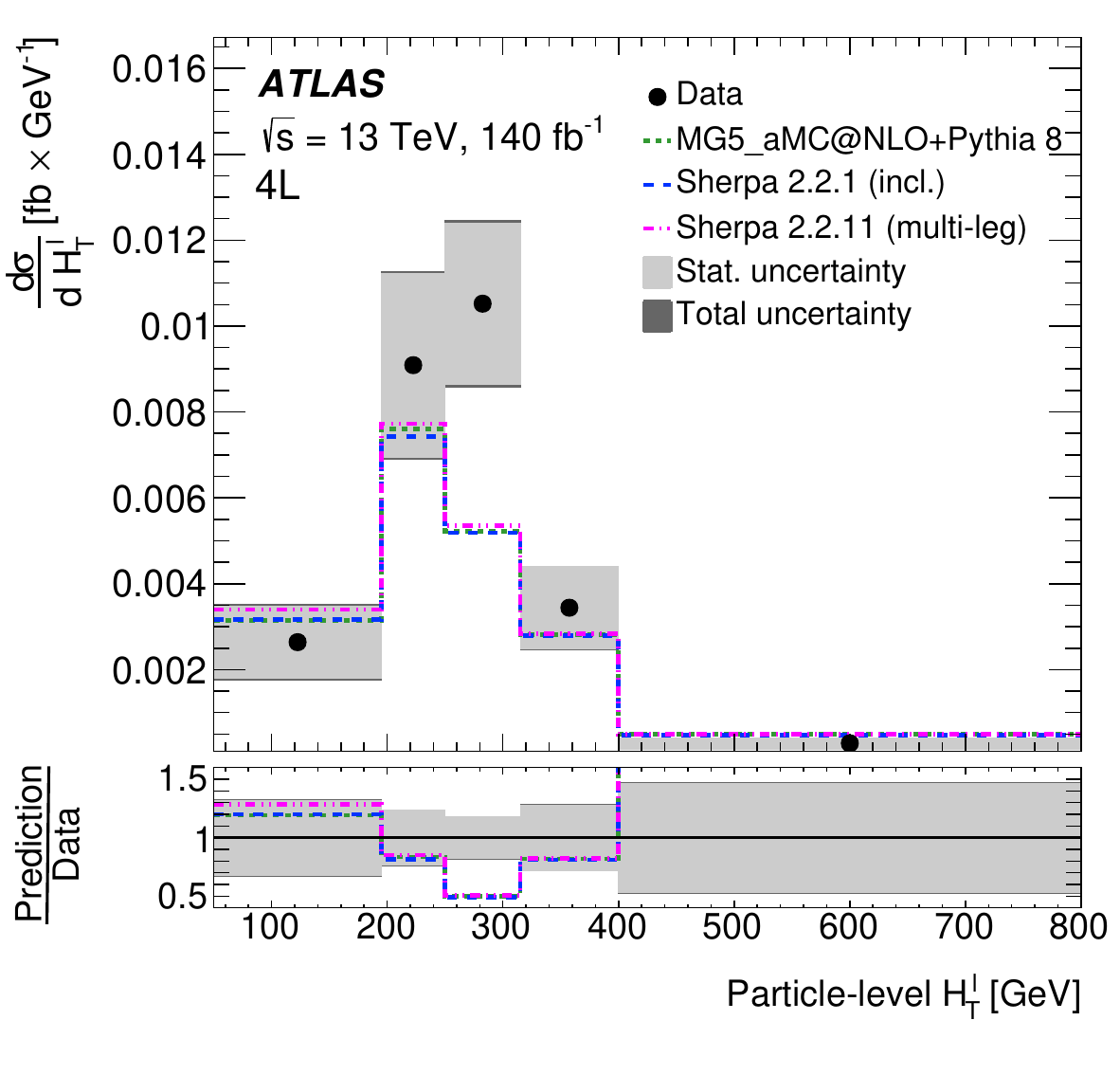}} \\
\subfloat[]{\includegraphics[width=0.46\textwidth]{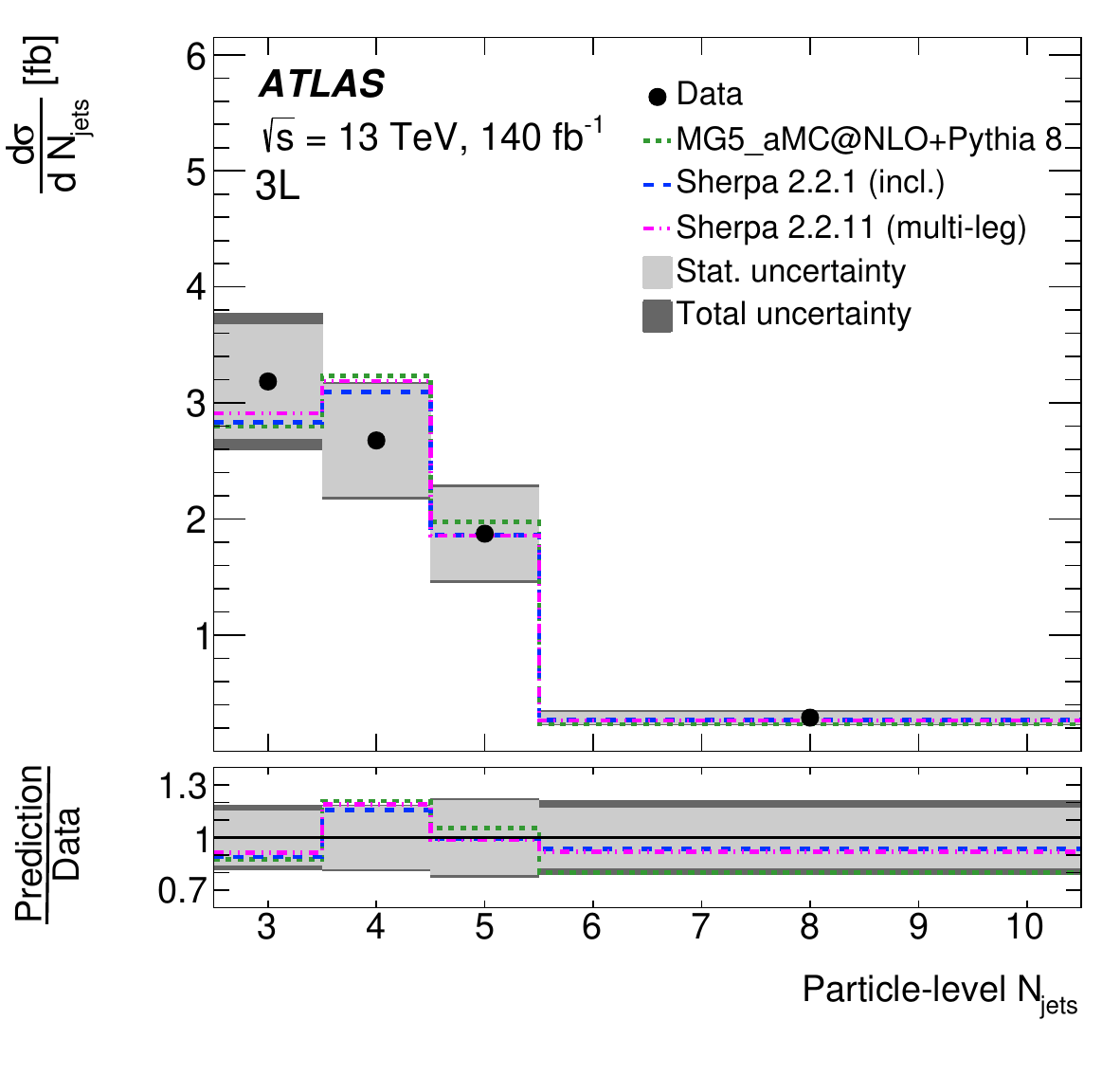}}
\hspace*{0.06\textwidth}
\subfloat[]{\includegraphics[width=0.46\textwidth]{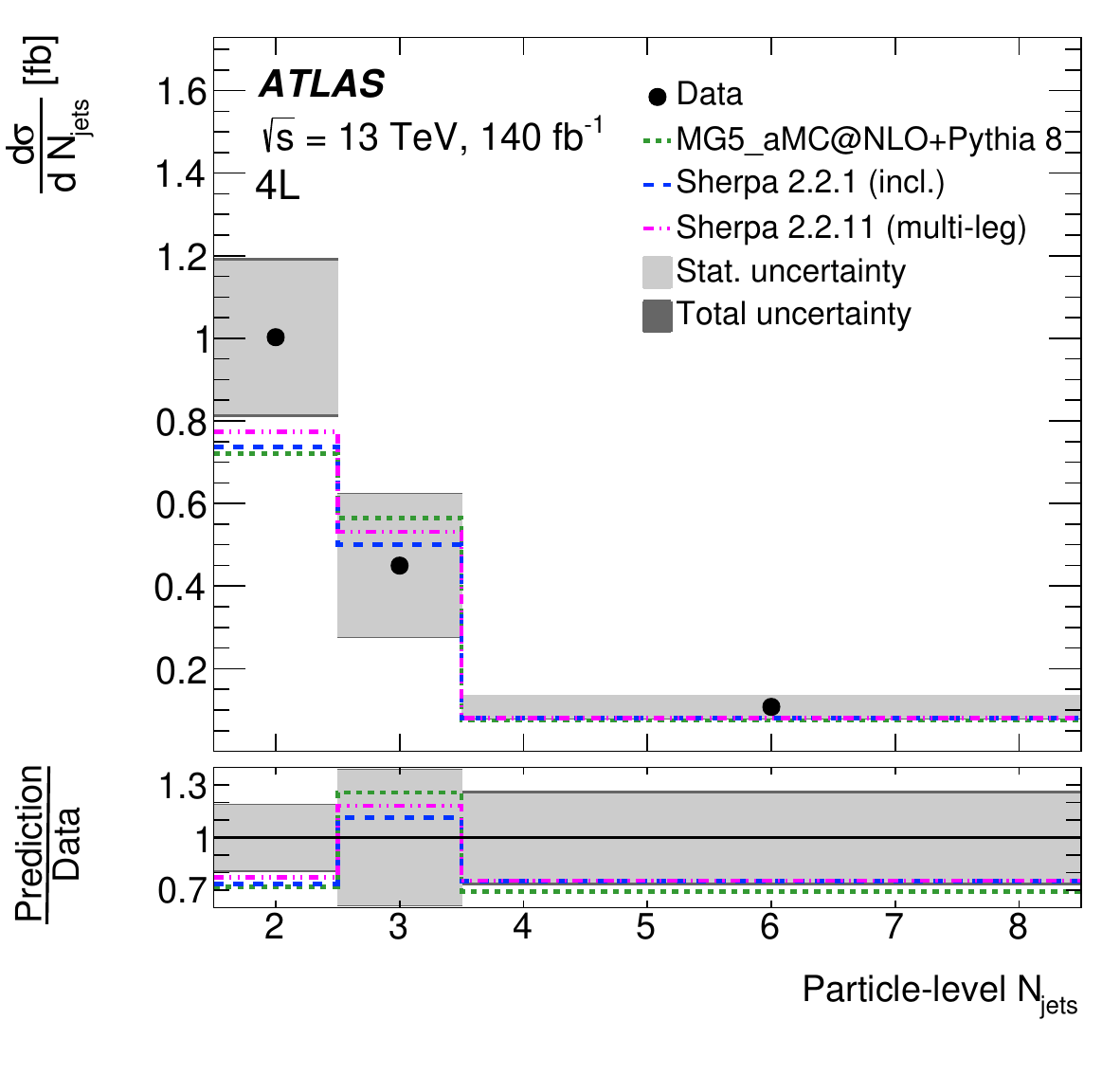}}
\caption{Absolute differential cross-section measurements, unfolded to particle level, as a function of $\HT^\ell$ in the (a) 3$\ell$ and (b) 4$\ell$ channels, and $N_{\mathrm{jets}}$ in the (c) 3$\ell$ and (d) 4$\ell$ channels. The dark grey band corresponds to the total uncertainty of the measurement; in some cases, it is almost fully covered by the light grey band, representing the dominant statistical uncertainty. Alternative generator predictions are overlaid as additional coloured lines.}
\label{fig:3l-4l-unfolded-results-1}
\end{figure}
 
\begin{figure}[!htb]
\centering
\subfloat[]{\includegraphics[width=0.46\textwidth]{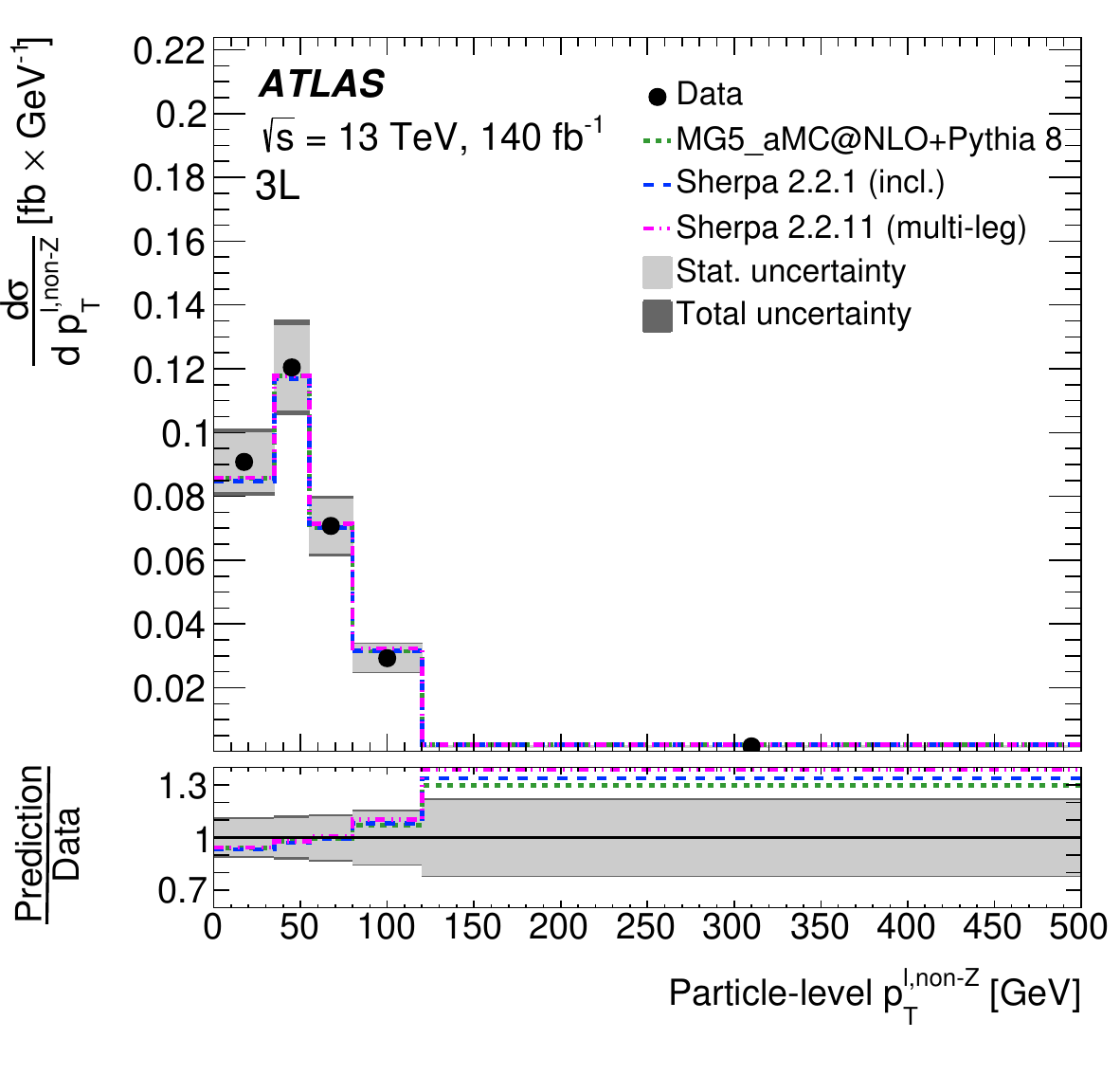}}
\hspace*{0.06\textwidth}
\subfloat[]{\includegraphics[width=0.46\textwidth]{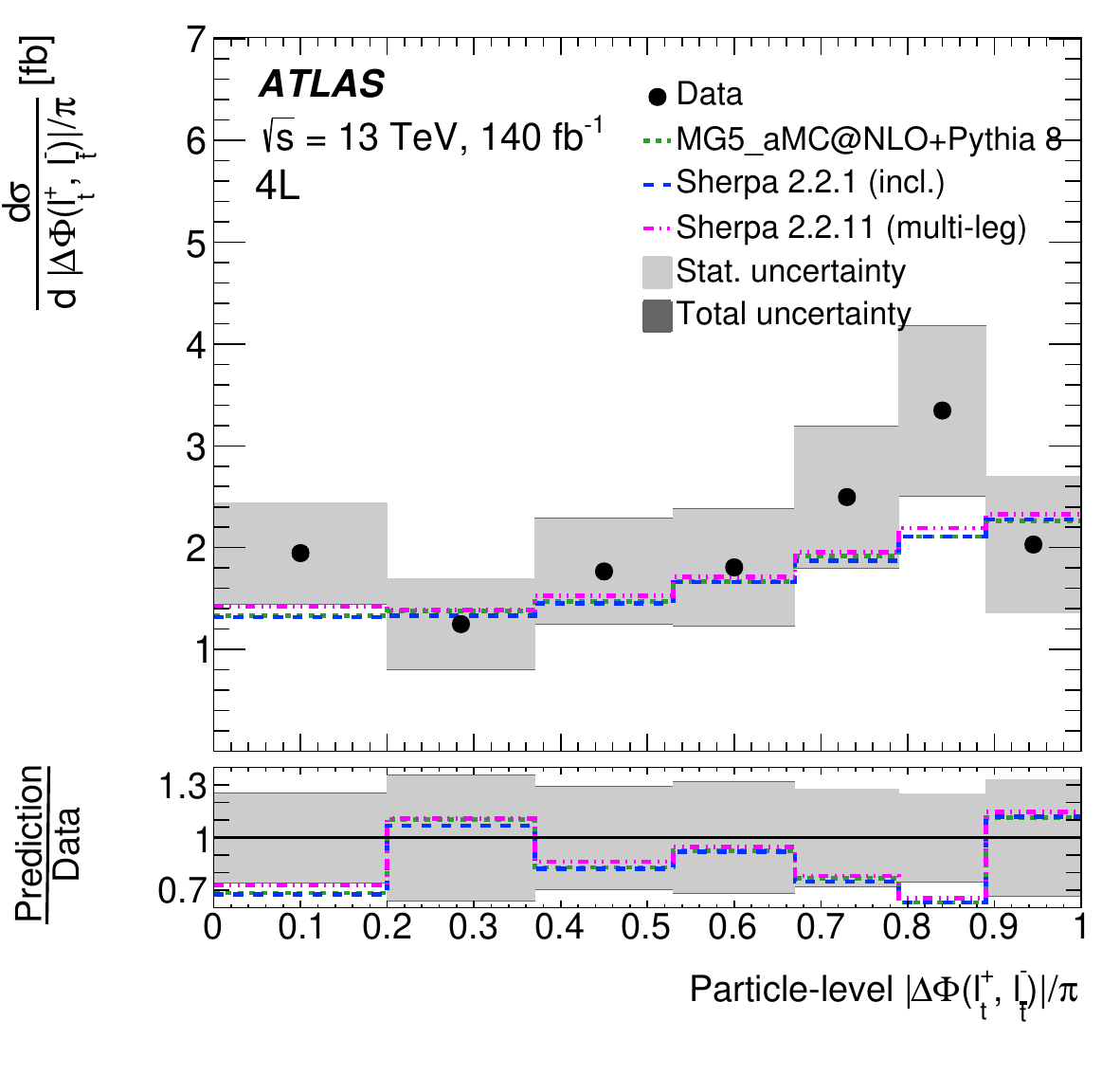}} \\
\subfloat[]{\includegraphics[width=0.46\textwidth]{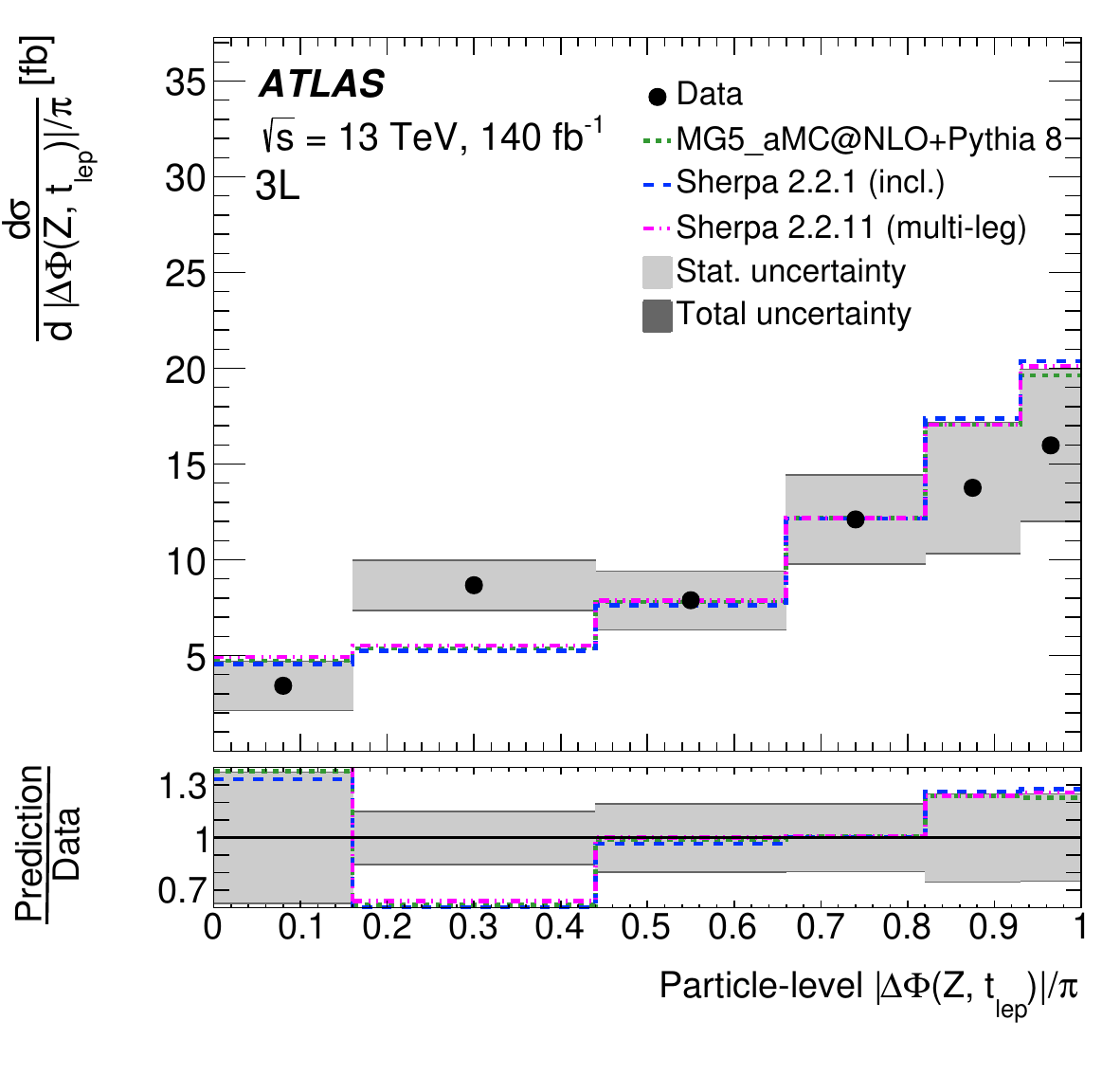}}
\hspace*{0.06\textwidth}
\subfloat[]{\includegraphics[width=0.46\textwidth]{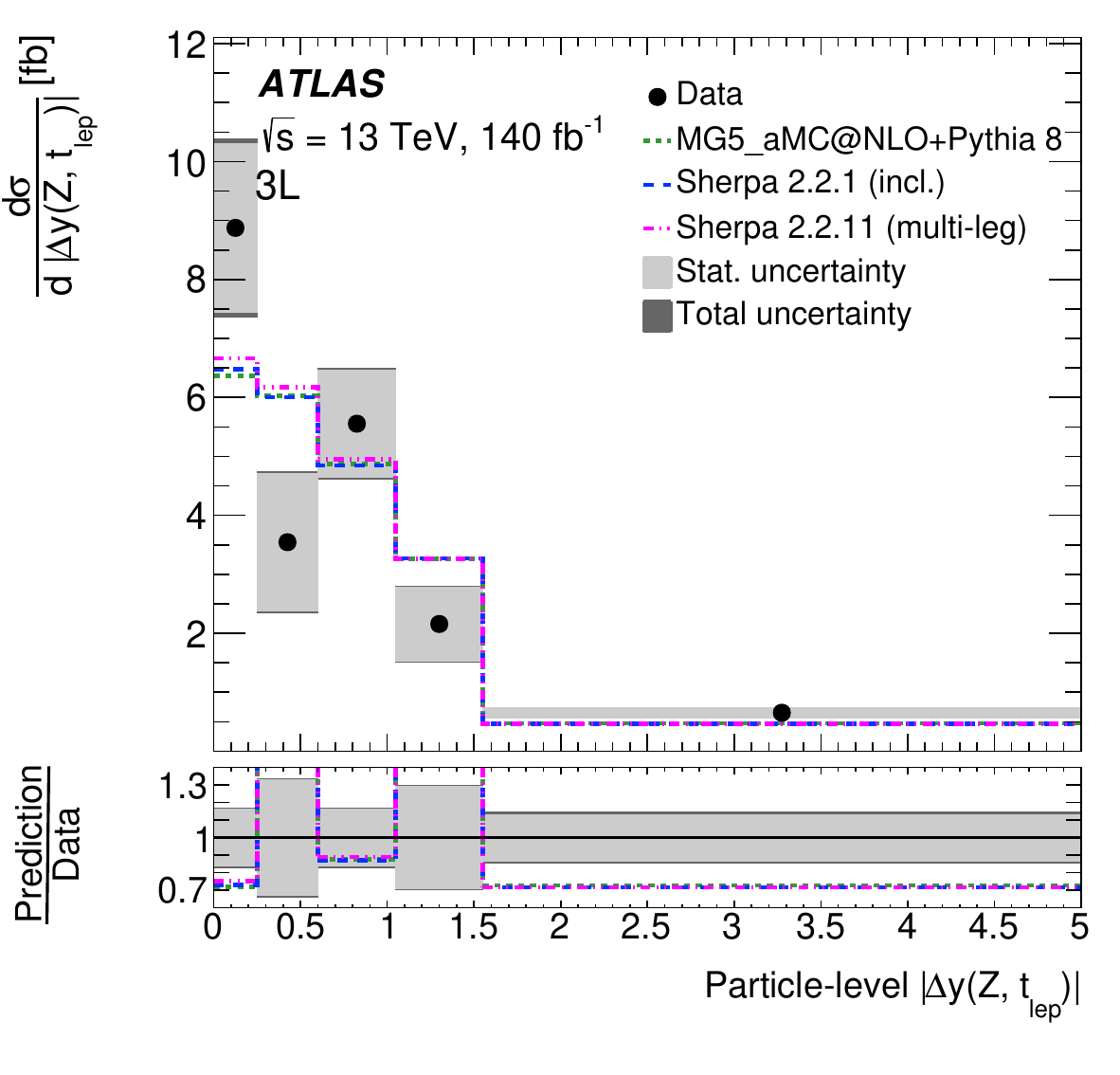}}
\caption{Absolute differential cross-section measurements, unfolded to particle level, as a function of (a) $\pT^{\ell,{\textrm{non-}}\Zboson}$ in the 3$\ell$ channel, (b) $|\Delta\Phi(\ell^{+}_{t}, \ell^{-}_{\bar{t}})|$ in the 4$\ell$ channel, (c) $|\Delta\Phi(\Zboson,t_{\mathrm{lep}})|/\pi$ and (d) $|\Delta y(\Zboson,t_{\mathrm{lep}})|$ in the 3$\ell$ channel. The dark grey band corresponds to the total uncertainty of the measurement; in some cases, it is almost fully covered by the light grey band, representing the dominant statistical uncertainty. Alternative generator predictions are overlaid as additional coloured lines.}
\label{fig:3l-4l-unfolded-results-2}
\end{figure}
 
\begin{figure}[!htb]
\centering
\subfloat[]{\includegraphics[width=0.46\textwidth]{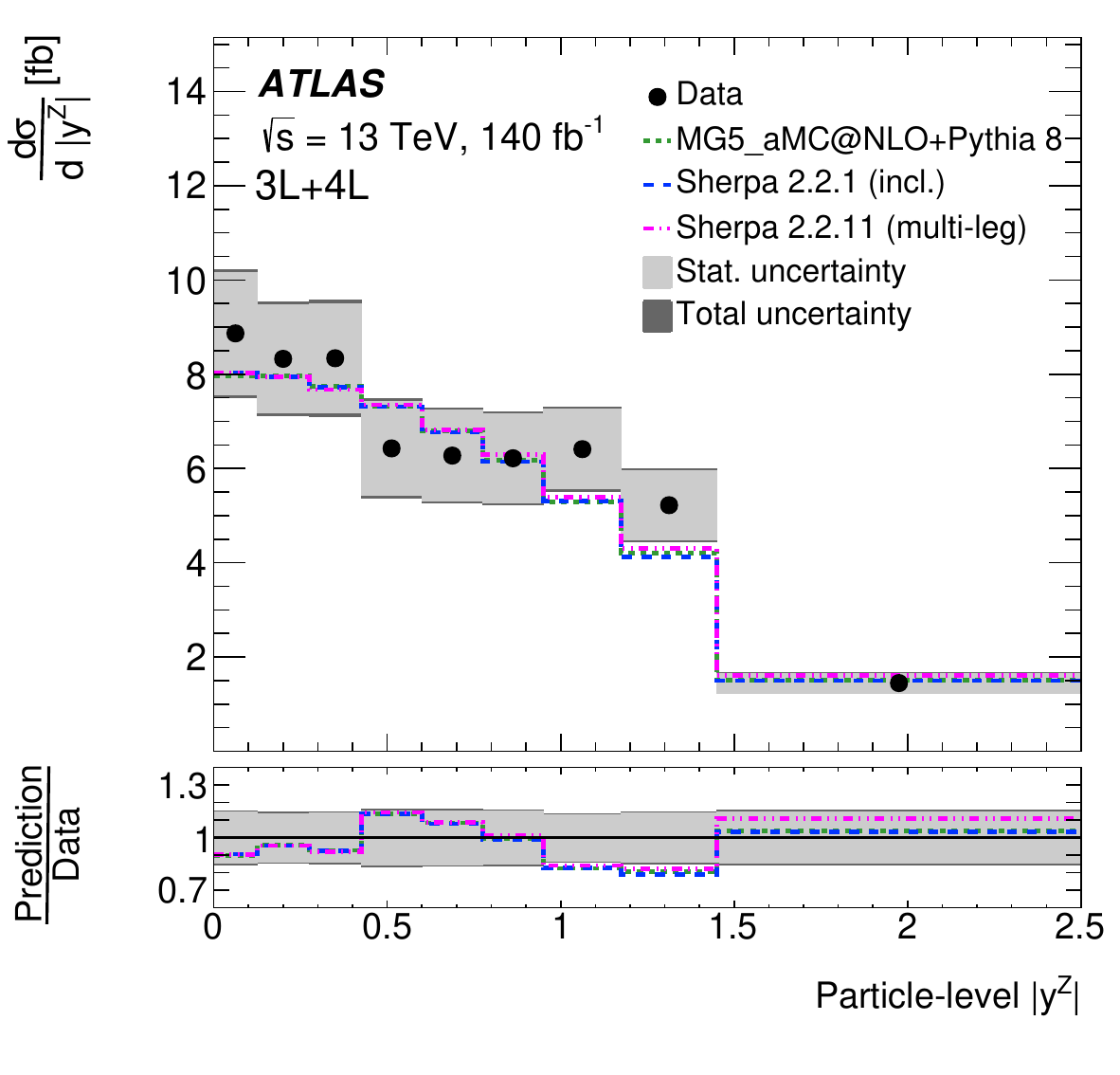}}
\hspace*{0.06\textwidth}
\subfloat[]{\includegraphics[width=0.46\textwidth]{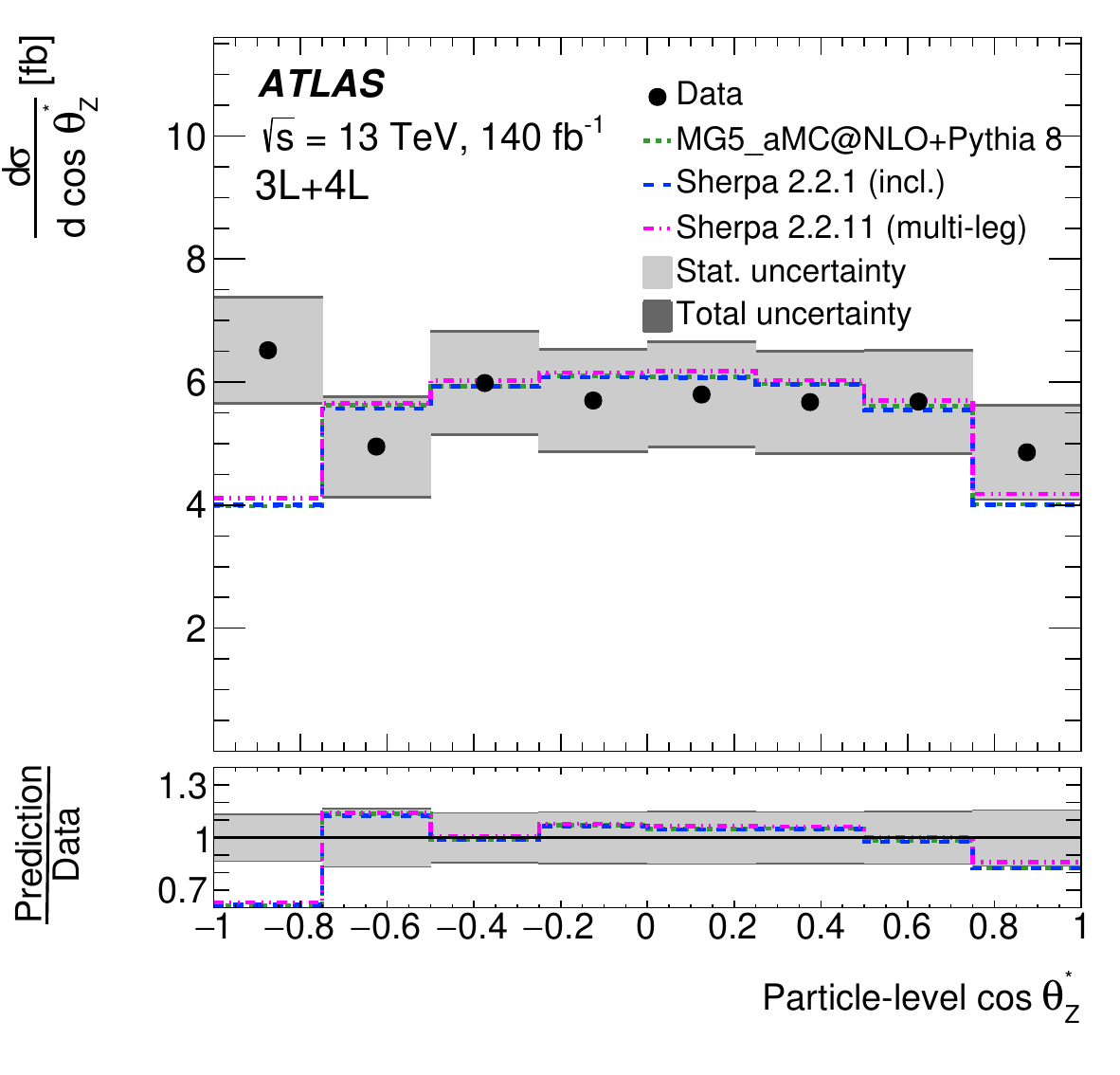}} \\
\subfloat[]{\includegraphics[width=0.46\textwidth]{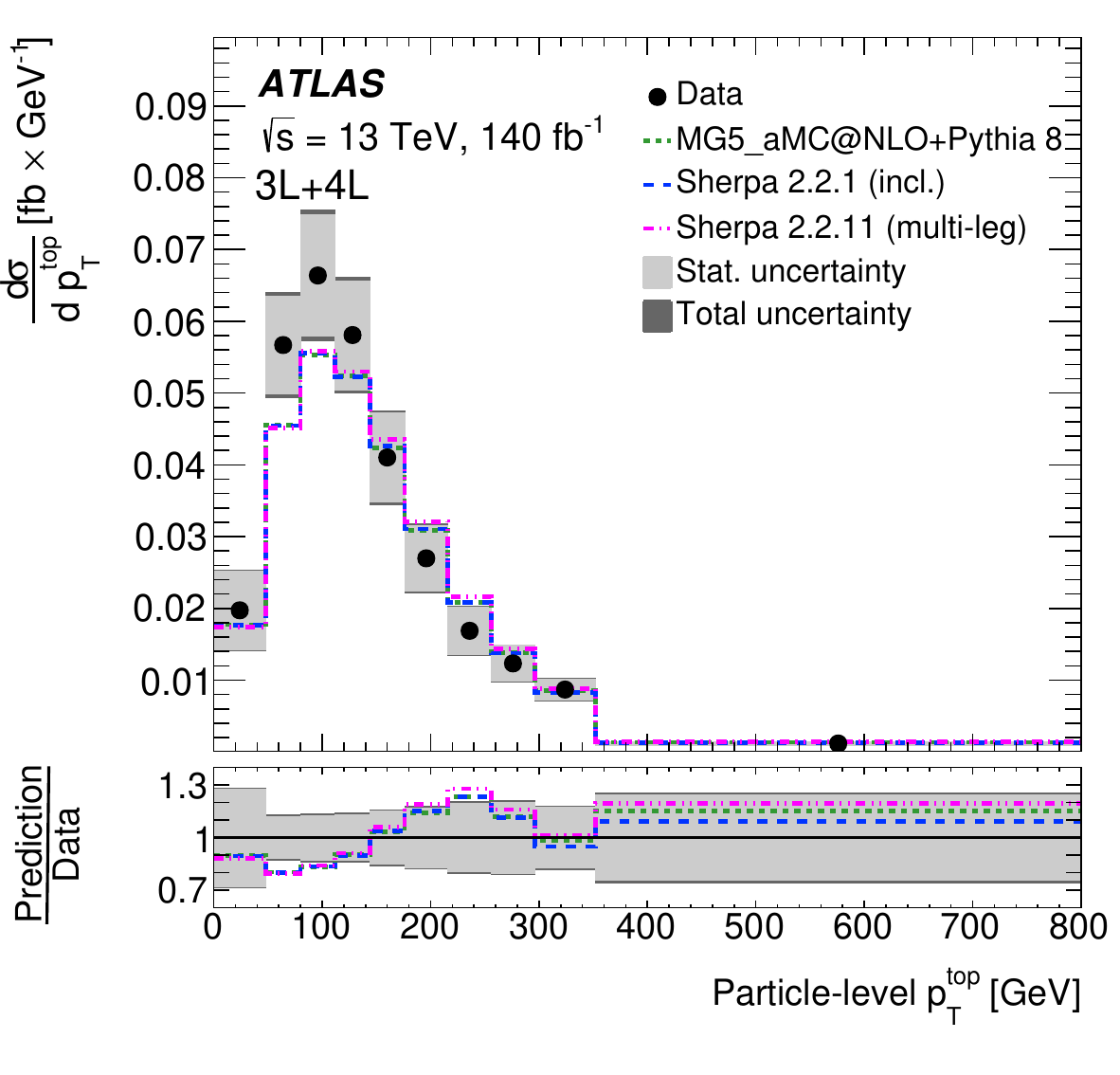}}
\hspace*{0.06\textwidth}
\subfloat[]{\includegraphics[width=0.46\textwidth]{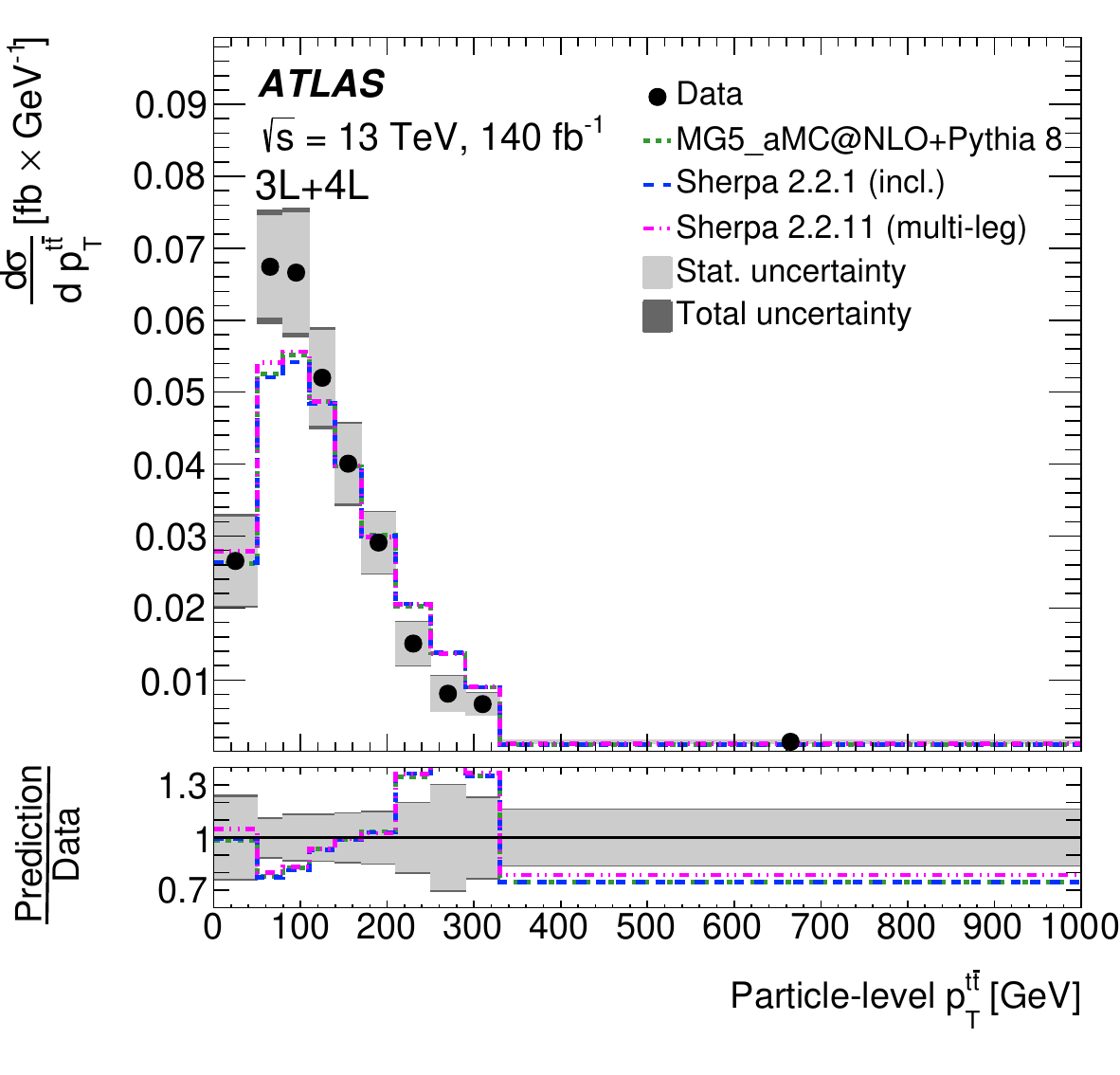}}
\caption{Absolute differential cross-section measurements, unfolded to particle level in the combined 3$\ell$ and 4$\ell$ channels, as a function of (a) $\lvert y^{Z}\rvert$, (b) $\cos{\theta^*_Z}$, (c) $\pT^{t}$, and (d) $\pT^{\ttbar}$. The dark grey band corresponds to the total uncertainty of the measurement; in some cases, it is almost fully covered by the light grey band, representing the dominant statistical uncertainty. Alternative generator predictions are overlaid as additional coloured lines.}
\label{fig:combination-unfolded-results-1}
\end{figure}
 
\begin{figure}[!htb]
\centering
\subfloat[]{\includegraphics[width=0.46\textwidth]{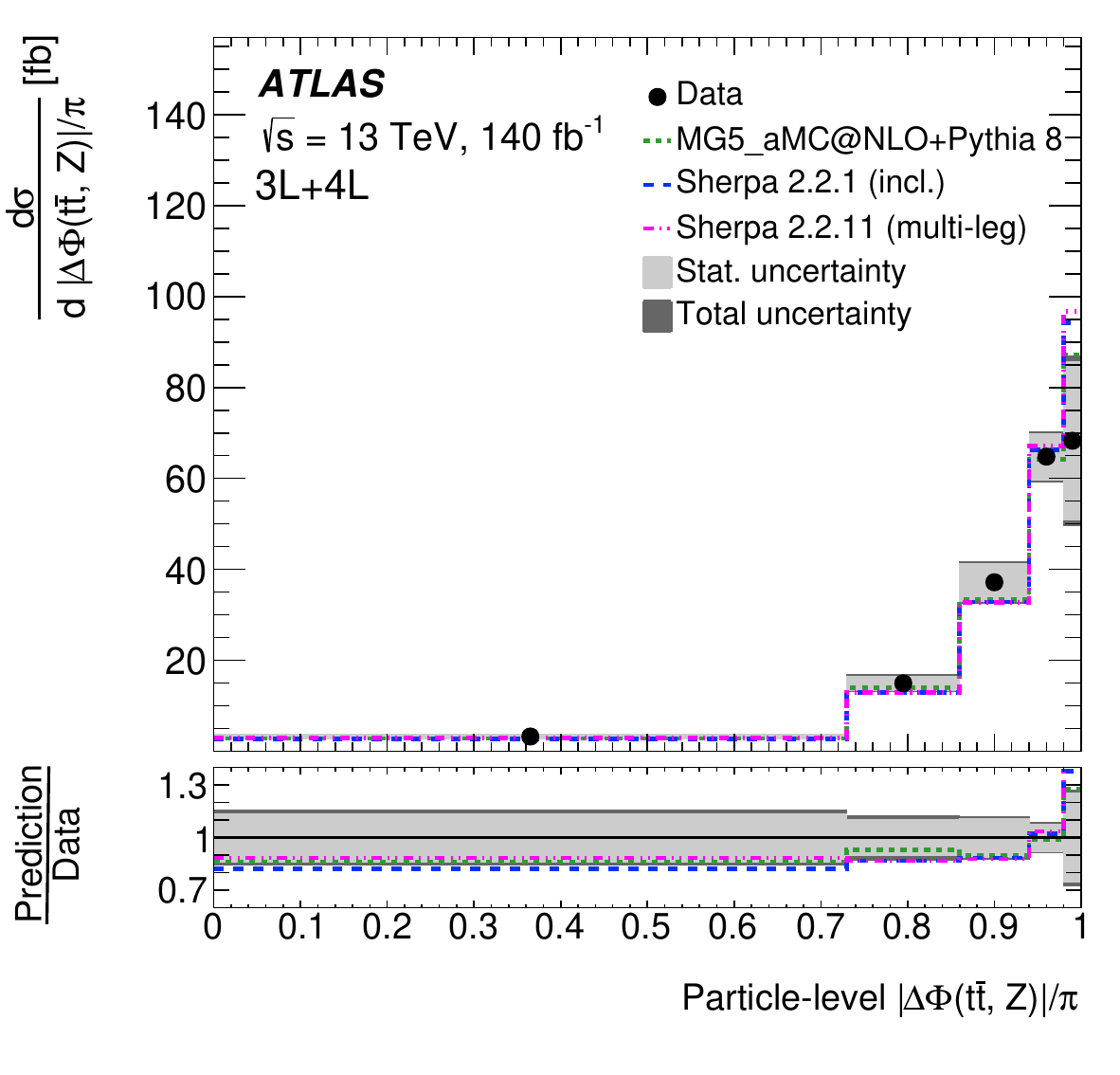}}
\hspace*{0.06\textwidth}
\subfloat[]{\includegraphics[width=0.46\textwidth]{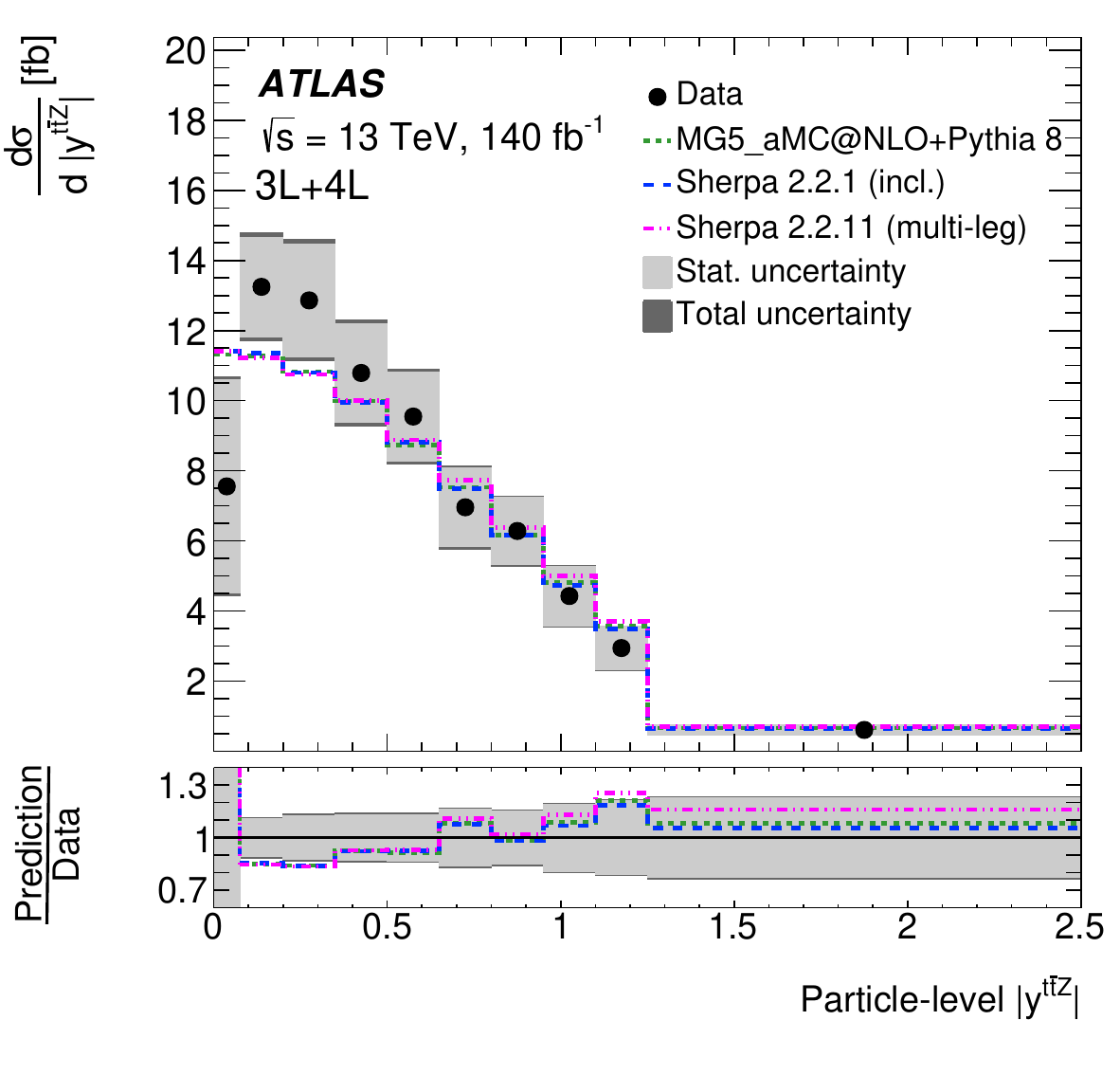}} \\
\subfloat[]{\includegraphics[width=0.46\textwidth]{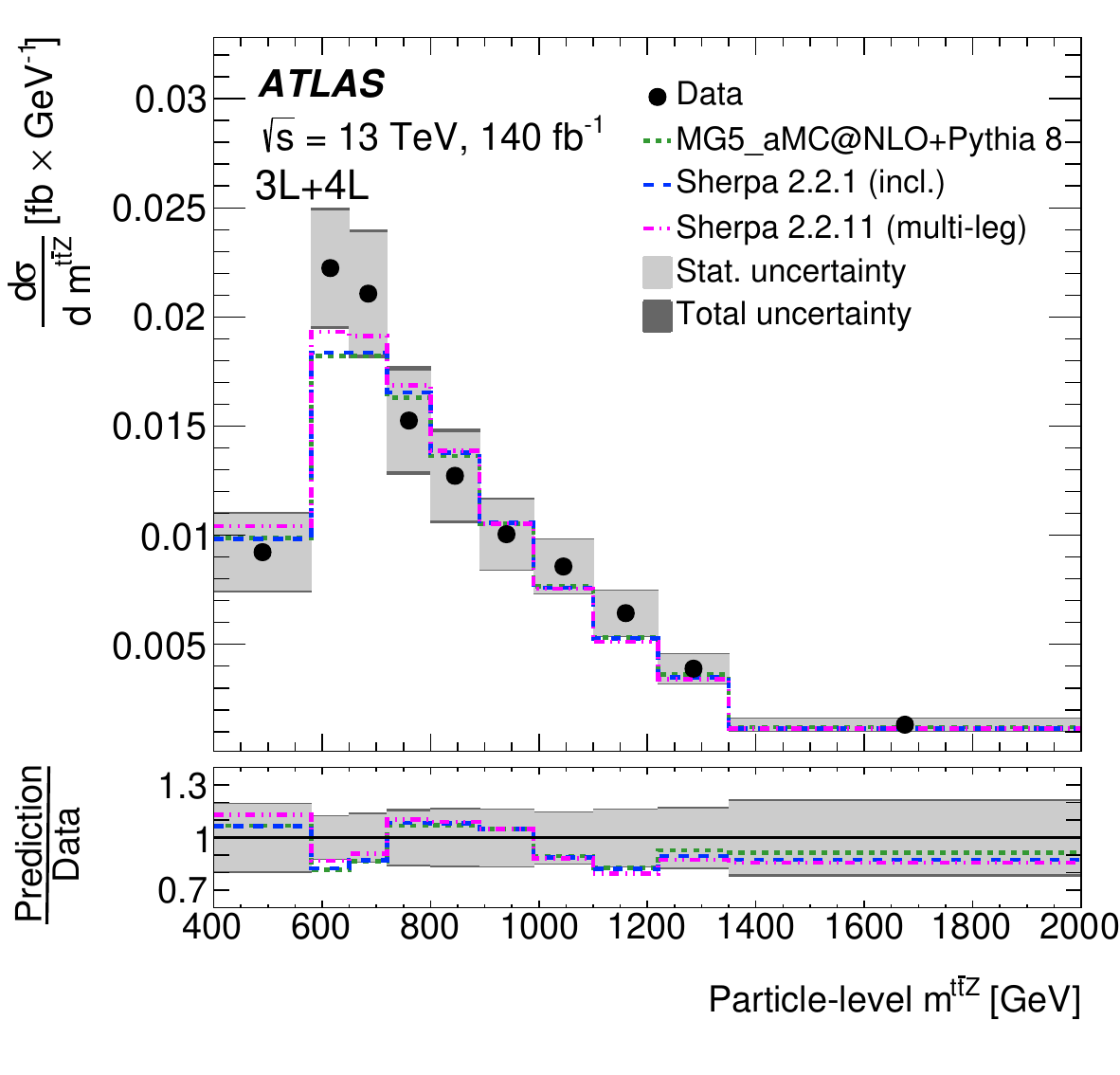}}
\hspace*{0.06\textwidth}
\subfloat[]{\includegraphics[width=0.46\textwidth]{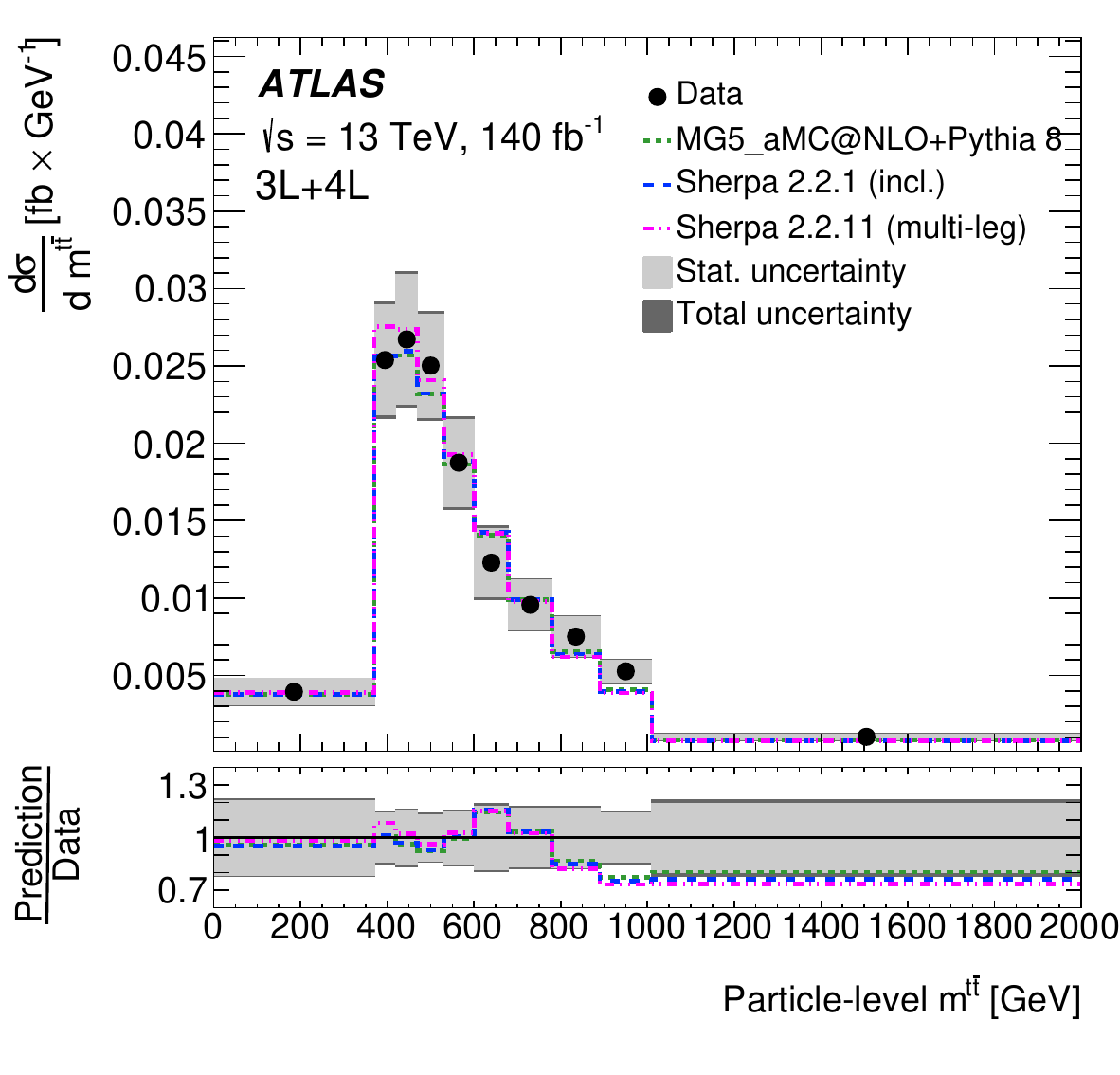}}
\caption{Absolute differential cross-section measurements, unfolded to particle level in the combined 3$\ell$ and 4$\ell$ channels, as a function of (a) $\lvert\Delta\Phi({\ttbar},\Zboson)\rvert/\pi$, (b) $\lvert y^{\ttZ}\rvert$, (c) $m^{\ttZ}$, and (d) $m^{\ttbar}$. The dark grey band corresponds to the total uncertainty of the measurement; in some cases, it is almost fully covered by the light grey band, representing the dominant statistical uncertainty. Alternative generator predictions are overlaid as additional coloured lines.}
\label{fig:combination-unfolded-results-2}
\end{figure}

\FloatBarrier


\section{Spin-correlation interpretation}

The first search for top-quark spin correlations in \ttZ events is presented, using detector-level distributions utilising the formalism in Ref.~\cite{Ravina:2021kpr}. Additional phenomenological studies probing anomalous couplings of the \ttZ process through spin correlations are presented in Ref.~\cite{Rahaman:2022dwp}.
Following the work in Ref.~\cite{Ravina:2021kpr}, the helicity ($k$) axis, transverse ($n$) axis and $r$ axis in the \ttbar rest frame are defined, and the polar angle of the charged lepton or down-type quark from the (anti-)top decay with respect to one of these axes, in the rest frame of its parent (anti-)top quark, is considered as a measure of (anti-)top polarisations and \ttbar spin correlations.
Six independent observables can thus be defined:
\begin{equation}\label{eq:spin-observables}
\coshelp,\quad \coshelm,\quad \costrap,\quad \costram,\quad \cosraxp,\quad \cosraxm,
\end{equation}
where the $\pm$ superscript indicates the sign of the charge of the lepton/quark.
As shown in Ref.~\cite{Ravina:2021kpr}, the coefficients of the spin density matrix can be extracted from the averages of the angular distributions corresponding to the observables listed in Eq.~\eqref{eq:spin-observables}.
These relations are summarised in Table~\ref{tab:spin-observables}; the $c_{ii}$ components are hereafter referred to as \enquote{spin correlations}, the $c_{ij} (i\neq j)$ and $c_{i}$ components as \enquote{spin cross-correlations}, and the $b_{i}^\pm$ components as \enquote{polarisations}.
The \ttZ process, differing from \ttbar production not only by the emission of an additional \Zboson~boson but also by different fractions of $q\bar{q}$- and $qg$-initiated production, leads to a different picture of top-quark spin correlations: the three coefficients $c_{rr}$, $c_{kk}$ and $c_{nn}$ adopt negative values and different magnitudes as in the \ttbar case, and a small longitudinal polarisation is induced by the emission of the \Zboson~boson, resulting in non-zero $b_r^\pm$ and $b_k^\pm$ coefficients~\cite{Ravina:2021kpr}.
Future measurements of the full spin density matrix in \ttZ production will be able to probe possible CP-violation effects and four-fermion interactions, with sensitivity complementary to \ttbar production~\cite{Brivio:2019ius}.
 
\begin{table}[!htb]
\footnotesize
\caption{Coefficients of the spin density matrix and their expressions as averages of angular distributions.}
\begin{center}
\begin{tabular}{ll}
\toprule
Coefficient & Expression \\
\midrule
$c_{rr}$    & $           -9 \langle \cosraxp\cdot\cosraxm \rangle$ \\
$c_{kk}$    & $           -9 \langle \coshelp\cdot\coshelm \rangle$ \\
$c_{nn}$    & $           -9 \langle \costrap\cdot\costram \rangle$ \\
\addlinespace[0.3em]
$c_{rk}$    & $           -9 \langle \cosraxp\cdot\coshelm + \cosraxm\cdot\coshelp \rangle$ \\
$c_{kn}$    & $           -9 \langle \coshelp\cdot\costram + \coshelm\cdot\costrap \rangle$ \\
$c_{rn}$    & $           -9 \langle \cosraxp\cdot\costram + \cosraxm\cdot\costrap \rangle$ \\
$c_{r}$     & $           -9 \langle \coshelp\cdot\costram - \coshelm\cdot\costrap \rangle$ \\
$c_{k}$     & $           -9 \langle \costrap\cdot\cosraxm - \costram\cdot\cosraxp \rangle$ \\
$c_{n}$     & $           -9 \langle \cosraxp\cdot\coshelm - \cosraxm\cdot\coshelp \rangle$ \\
\addlinespace[0.3em]
$b_{r}^+$   & $\hphantom{-}3 \langle \cosraxp \rangle$ \\
$b_{r}^-$   & $\hphantom{-}3 \langle \cosraxm \rangle$ \\
$b_{k}^+$   & $\hphantom{-}3 \langle \coshelp \rangle$ \\
$b_{k}^-$   & $\hphantom{-}3 \langle \coshelm \rangle$ \\
$b_{n}^+$   & $\hphantom{-}3 \langle \costrap \rangle$ \\
$b_{n}^-$   & $\hphantom{-}3 \langle \costram \rangle$ \\
\bottomrule
\end{tabular}
\label{tab:spin-observables}
\end{center}
\end{table}
 
In the $4\ell$ channel, all these observables can be defined in terms of leptons from the $\ttbar$ system, while in the $3\ell$ channel there is only one lepton that can be used to the define the polarisation observables.
In order to access the spin correlations in semileptonic $\ttbar$ events, the down-type quark from the hadronic \Wboson boson decay must be used.
Relying on the ${\sim}50\%$ branching ratio of hadronic \Wboson boson decays into $c\bar{s}/\bar{c}s$, $s$-candidates are selected by simultaneously applying $b$-tagging and $b$-vetoing to the jets selected by the top reconstruction algorithm as belonging to that \Wboson boson: if the jet is $b$-tagged at least at the $85\%$ WP but not at the $60\%$ WP, it is considered to be $c$-tagged, and its companion jet from the  \Wboson boson is the resulting $s$-jet.
The $3\ell$ events where the $s$-jet cannot be identified in this way are rejected.
The resulting identification efficiency is $42\pm 2\%$ in the simulated \ttZ MC samples.
 
A final observable of interest can be constructed in both the $3\ell$ and $4\ell$ channels: the opening angle between the two charged leptons (charged lepton and $s$-jet) from the dileptonic (semileptonic) \ttbar decay system, where each decay product is first boosted to the rest frame of its respective parent (anti-)top quark.
This angle $\varphi$ is particularly sensitive to spin correlations, and the following three relations hold~\cite{Ravina:2021kpr}:
\begin{equation*}
\dfrac{1}{\sigma}\dfrac{\dif\sigma}{\dif\cos\varphi}=\dfrac{1}{2}\left(1-D\cos\varphi\right), \quad D=-\dfrac{c_{rr}+c_{kk}+c_{nn}}{3}, \quad D=-3\langle\cos\varphi\rangle.
\end{equation*}
 
Since the angular observables listed in Table~\ref{tab:spin-observables} suffer from a small number of events in the $4\ell$ channel and, in general, highly non-diagonal migration matrices (since all the observables rely on the precise reconstruction of the top quarks), unfolding these distributions using the profile-likelihood method defined in Section~\ref{sec:results_differential} is not possible without a large amount of regularisation.
Instead, a detector-level template fit is preferred, comparing two \enquote{spin scenarios}:
\begin{equation*}
\mathcal{O} = f_\mathrm{SM}\cdot\mathcal{O}_\text{spin-on} + \left(1-f_\mathrm{SM}\right)\cdot\mathcal{O}_\text{spin-off}.
\end{equation*}
Each observable $\mathcal{O}$ (the 15 coefficients of the spin density matrix and the $\cos\varphi$ distribution) is thus fitted to a linear combination of a \enquote{SM} template (taken from the nominal \ttZ MC predictions) and a template whose values are those predicted in the absence of any spin correlation or top polarisation.
The latter predictions are identically zero for all spin coefficients, and therefore lead to a flat distribution of the $\cos\varphi$ observable.
These \enquote{spin-off} predictions at parton level are forward-folded through a migration matrix built from the nominal MC samples to produce corresponding detector-level templates, and are assigned the same uncertainties as the nominal templates.
The single parameter of interest, $f_\mathrm{SM}$, has a value of 0 in the absence of spin correlations and 1 in case of perfect agreement of the unfolded data with the SM.
 
Since some parameters of the \ttZ spin correlation matrix are zero within theoretical uncertainties \cite{Ravina:2021kpr}, they are excluded from the combined fit in order to improve its stability.
The only coefficients predicted to be significantly non-zero in the SM are the three spin correlations ($c_{rr}$, $c_{kk}$ and $c_{nn}$), one of the cross-correlations ($c_{rk}$), four of the polarisations ($b_r^\pm$ and $b_k^\pm$), and $D$.
Each observable is defined and measured in the combined $3\ell$ and $4\ell$ channels. The strategy outlined above results in the extraction of nine different values of $f_\mathrm{SM}$ at detector level, which are then combined in a profiled $\chi^2$-fit.
The fit fully takes into account the statistical overlap in the data, as well as the correlations between the NPs from the different measurements of $f_\mathrm{SM}$.
 
The expected and observed values of $f_\mathrm{SM}$ for each individual detector-level template fit to a single angular distribution are given in Table~\ref{tab:spin-results}.
For each individual measurement, it is checked that the fit does not exhibit any strong pull or constraint on the nuisance parameters, compared to the inclusive cross-section measurement presented in Section~\ref{sec:results_inclusive}, and the values are found to be consistent with the SM within uncertainties.
 
Using the covariance matrices, impacts and correlations of all NPs from each fit, these extracted values of $f_\mathrm{SM}$ are combined following the statistical prescriptions described above, and yield:
 
\begin{equation*}
f_\mathrm{SM}^\text{obs.} = 1.20 \pm 0.63\ \mathrm{(stat.)} \pm 0.25\ \mathrm{(syst.)} = 1.20 \pm 0.68\ \mathrm{(tot.)}.
\end{equation*}
 
The measured value agrees with the SM expectation, disfavouring the no-spin hypothesis with a significance of $1.8\sigma$, and is dominated by the statistical uncertainty. The main systematic uncertainties arise from the signal and background modelling, as well as the impact of \MET and flavour-tagging uncertainties on the reconstruction of the full \ttbar system.
 
\begin{table}[!htb]
\sisetup{round-mode = places, round-precision = 2}
\footnotesize
\caption{Values of the $f_\mathrm{SM}$ parameter extracted from the detector-level template fits to the angular distributions listed in Table~\ref{tab:spin-observables}. The \enquote{observed} values refer to measurements in data, whereas the \enquote{expected} values are obtained from an Asimov dataset (with $f_\mathrm{SM}=1$). The total errors quoted for each measurement include both the statistical uncertainty and the systematic uncertainties from all sources.}
\begin{center}
\begin{tabular}{llll}
\toprule
Distribution                                    & Channel & Expected values & Observed values \\
\midrule
$\cos\varphi$                                   & $3\ell + 4\ell$ & $1 ^{+\num{1.3931}} _{-\num{1.38374}}$ & $\num{-0.0926916} ^{+\num{1.34452}} _{-\num{1.28149}}$ \\
\addlinespace[0.6em]
$\cosraxp\cdot\cosraxm$                         & $3\ell + 4\ell$ & $1 ^{+\num{1.83302}} _{-\num{1.82424}}$ & $\hphantom{-}\num{1.17229} ^{+\num{1.79726}} _{-\num{1.76328}}$ \\
\addlinespace[0.6em]
$\coshelp\cdot\coshelm$                         & $3\ell + 4\ell$ & $1 ^{+\num{1.7833}} _{-\num{1.78129}}$ & $\hphantom{-}\num{1.38711} ^{+\num{1.72099}} _{-\num{1.72503}}$ \\
\addlinespace[0.6em]
$\costrap\cdot\costram$                         & $3\ell + 4\ell$ & $1 ^{+\num{1.86709}} _{-\num{1.85945}}$ & $\num{-1.05468} ^{+\num{2.05632}} _{-\num{1.95655}}$ \\
\addlinespace[0.6em]
$\cosraxp\cdot\coshelm + \cosraxm\cdot\coshelp$ & $3\ell + 4\ell$ & $1 ^{+\num{1.92679}} _{-\num{1.92631}}$ & $\hphantom{-}\num{0.361172} ^{+\num{1.9859}} _{-\num{1.92687}}$ \\
\addlinespace[0.6em]
$\cosraxp$                                      & $3\ell + 4\ell$ & $1 ^{+\num{1.80801}} _{-\num{1.80431}}$ & $\hphantom{-}\num{1.56105} ^{+\num{1.86016}} _{-\num{1.98397}}$ \\
\addlinespace[0.6em]
$\cosraxm$                                      & $3\ell + 4\ell$ & $1 ^{+\num{1.82061}} _{-\num{1.77952}}$ & $\hphantom{-}\num{1.81197} ^{+\num{1.62572}} _{-\num{1.68046}}$ \\
\addlinespace[0.6em]
$\coshelp$                                      & $3\ell + 4\ell$ & $1 ^{+\num{1.69371}} _{-\num{1.66637}}$ & $\hphantom{-}\num{2.00204} ^{+\num{1.6546}} _{-\num{1.69729}}$ \\
\addlinespace[0.6em]
$\coshelm$                                      & $3\ell + 4\ell$ & $1 ^{+\num{1.68189}} _{-\num{1.68098}}$ & $\hphantom{-}\num{2.30827} ^{+\num{1.67765}} _{-\num{1.67831}}$ \\
\bottomrule
\end{tabular}
\label{tab:spin-results}
\end{center}
\end{table}
 
\FloatBarrier


\section{SMEFT interpretation}
\label{sec:smeft}

The Standard Model effective field theory (SMEFT) \cite{Brivio:2017vri,Brivio:2020onw} provides a complete phenomenological extension of the SM beyond dimension-4 terms in the Lagrangian.
By employing the degrees of freedom and gauge symmetries of the SM, the SMEFT builds an infinite series of operators sorted by canonical dimension.
New physics effects from BSM theories characterised by a mass scale higher than the energies probed in a typical LHC collision can therefore be related to a set of operators $\left\{\mathcal{Q}\right\}$, together with their Wilson coefficients $\left\{C\right\}$.
Observed deviations from the SM can then be expressed in terms of $\left\{C\right\}$, without specifying a particular BSM model.
Similarly, LHC data found to be in agreement with the SM lead to constraints in the phenomenological space spanned by the Wilson coefficients $\left\{C\right\}$, which in turn can be translated into exclusions of various BSM scenarios.
The Lagrangian of the SMEFT thus reads:
\begin{equation*}
\mathcal{L}_\text{SMEFT} = \mathcal{L}_\text{SM} + \sum_{d>4}\mathcal{L}^{(d)},\quad \mathcal{L}^{(d)}=\sum_{i=1}^{n_d}\dfrac{C_i^{(d)}}{\Lambda^{d-4}}\mathcal{Q}_i^{(d)},
\end{equation*}
where $\Lambda$ is a suitable cut-off scale, chosen as the conventional $\Lambda=1~\TeV$ in this analysis.
The number of such operators $\mathcal{Q}^{(d)}$, $n_d$, is known up to $d=8$, but current computational tools only allow the study of SMEFT effects up to $d=6$ where generation of LHC-like collision events is required.
Operators contributing to $\mathcal{L}^{(5)}$ are known to include baryon- and lepton-number violating terms, and are therefore ignored in this analysis; similarly, higher-order operators are $\Lambda$-suppressed, so only $\mathcal{L}^{(6)}$ is considered in the following.
 
Using the EFT Monte Carlo samples described in Section~\ref{sec:samples} (produced with the SMEFTsim\,3.0 UFO model \cite{Brivio:2020onw}), the dependence of a generic observable $\mathcal{O}$ (e.g.\ an inclusive or differential cross section) on a set of Wilson coefficients $\left\{C\right\}$ can be parameterised as:
\begin{equation}\label{eq:smeft-parameterisation}
\mathcal{O} = \mathcal{O}_\text{SM} + \sum_i C_i A_i + \sum_{i,j} C_i C_j B_{ij}.
\end{equation}
The reweighting approach taken in the generation of these MC samples probes multiple values of the individual Wilson coefficients per event, as well as pairs of coefficients.
A simple quadratic fit to any observable of interest therefore yields the linear term $A_i$, which represents the interference between the SM and the SMEFT, and the quadratic term $B_{ij}$, which represents a pure SMEFT contribution.\footnote{In what follows, the \enquote{pure quadratic} and \enquote{cross} terms may be referred to, meaning $B_{ii}$ and $B_{ij, i\neq j}$ respectively.}
 
In order to remove possible discrepancies between the SMEFT prediction of the SM term $\mathcal{O}_\text{SM}$, obtained at LO in QCD ($\mathcal{O}_\text{SM}^\text{LO}$), and the observed \ttZ data, the SMEFT predictions for the observables are scaled by a factor $\mathcal{O}_\text{SM}^\text{NLO}/\mathcal{O}_\text{SM}^\text{LO}$, where $\mathcal{O}_\text{SM}^\text{NLO}$ is instead obtained from the modelling of the nominal \MGNLOPY[8] \ttZ sample.
 
The SMEFT predictions rely on the top-flavour structure of the SMEFTsim\,3.0 model, which is defined by the following assumptions:
\begin{itemize}
\item quarks of the first two generations and quarks of the third are described by independent fields, denoted by $(q,u,d)$ and $(Q,t,b)$ respectively;
\item a symmetry $U(2)^3=U(2)_q\times U(2)_u\times U(2)_d$ is imposed on the Lagrangian, under which only the light quarks transform (e.g. $q\mapsto\Omega_qq$, but $t\mapsto t$);
\item mixing effects in the quark sector are neglected (the CKM matrix is unity);
\item a symmetry $U(1)^3_{l+e}=U(1)_e\times U(1)_\mu\times U(1)_\tau$ is imposed on the Lagrangian, which corresponds to simple flavour diagonality in the lepton sector.
\end{itemize}
 
In this analysis, 20 dimension-6 SMEFT operators are considered, corresponding to 23 degrees of freedom (3 Wilson coefficients have a distinct imaginary part).
They are defined in Table~\ref{tab:smeft-operators}, and broadly fall into two classes: top--boson operators (classes 5 and 6 in the notation of Ref.~\cite{Brivio:2020onw}) and four-quark operators (classes 8a-c).
The \ttZ vertex is sensitive to particular combinations of some operators, which are made explicit in other models (such as dim6top \cite{Aguilar-Saavedra:2018ksv}):
\begin{align}\label{eq:redefinition-ttz-vertex-operators}
c_{tZ}           &= -\sin{\theta_W}\ctB + \cos{\theta_W}\ctW, \\
c_{\varphi Q}^-  &= \cHQ1 - \cHQ3, \nonumber 
\end{align}
 
where the operators on the left-hand side are in the notation of Ref.~\cite{Aguilar-Saavedra:2018ksv}, and $\theta_W$ is the Weinberg angle.
It is shown in the following that the above relations can be recovered by identifiying the directions in the EFT phase-space that are probed by this measurement.
 
\begin{table}[!htb]
\footnotesize
\caption{Definitions of the relevant dimension-6 SMEFT operators.
For the three top--boson operators indicated with a $\left(\star\right)$ the real and imaginary parts of the corresponding Wilson coefficients are considered separately.}
\label{tab:smeft-operators}
\def\arraystretch{1.7}
\begin{center}
\begin{tabular}{llll}
\toprule
& Operator & Definition & \\ 
\midrule
\multirow{6}{*}{\rotatebox[origin=c]{90}{\normalsize top--boson}}
& $\mathcal{Q}_{tW}$         & $(\bar Q \sigma^{\mu\nu} t) \sigma^i \tilde H W_{\mu\nu}^i$ & $\left(\star\right)$ \\ 
& $\mathcal{Q}_{tB}$         & $(\bar Q \sigma^{\mu\nu} t)  \tilde H B_{\mu\nu}$           & $\left(\star\right)$ \\ 
& $\mathcal{Q}_{tG}$         & $(\bar Q \sigma^{\mu\nu} T^a t) \tilde H G_{\mu\nu}^a$      & $\left(\star\right)$ \\ 
& $\mathcal{Q}_{HQ}^{(1)}$   & $(H^\dag i\overleftrightarrow{D}_\mu H) (\bar Q \gamma^\mu Q)$ &                   \\ 
& $\mathcal{Q}_{HQ}^{(3)}$   & $(H^\dag i\overleftrightarrow{D}^i_\mu H) (\bar Q \sigma^i\gamma^\mu Q)$ &         \\ 
& $\mathcal{Q}_{Ht}$         & $(H^\dag i\overleftrightarrow{D}_\mu H) (\bar t \gamma^\mu t)$ &                   \\ 
 
\midrule
\multirow{18}{*}{\rotatebox[origin=c]{90}{\normalsize four-quark}}
& $\mathcal{Q}_{tu}^{(1)}$   & $(\bar t \gamma_\mu t)(\bar u \gamma^\mu u)$ &                                     \\ 
& $\mathcal{Q}_{tu}^{(8)}$   & $(\bar t T^a\gamma_\mu t)(\bar u T^a\gamma^\mu u)$ &                               \\ 
& $\mathcal{Q}_{td}^{(1)}$   & $(\bar t \gamma_\mu t)(\bar d \gamma^\mu d)$ &                                     \\ 
& $\mathcal{Q}_{td}^{(8)}$   & $(\bar t T^a\gamma_\mu t)(\bar d T^a\gamma^\mu d)$ &                               \\ 
& $\mathcal{Q}_{qt}^{(1)}$   & $(\bar q \gamma_\mu q)(\bar t \gamma^\mu t)$ &                                     \\ 
& $\mathcal{Q}_{qt}^{(8)}$   & $(\bar q T^a\gamma_\mu q)(\bar t T^a\gamma^\mu t)$ &                               \\ 
& $\mathcal{Q}_{Qu}^{(1)}$   & $(\bar Q \gamma_\mu Q)(\bar u \gamma^\mu u)$ &                                     \\ 
& $\mathcal{Q}_{Qu}^{(8)}$   & $(\bar Q T^a\gamma_\mu Q)(\bar u T^a\gamma^\mu u)$ &                               \\ 
& $\mathcal{Q}_{Qd}^{(1)}$   & $(\bar Q \gamma_\mu Q)(\bar d \gamma^\mu d)$ &                                     \\ 
& $\mathcal{Q}_{Qd}^{(8)}$   & $(\bar Q T^a\gamma_\mu Q)(\bar d T^a\gamma^\mu d)$ &                               \\ 
& $\mathcal{Q}_{Qq}^{(1,1)}$ & $(\bar Q \gamma_\mu Q)(\bar q \gamma^\mu q)$ &                                     \\ 
& $\mathcal{Q}_{Qq}^{(3,1)}$ & $(\bar Q \sigma^i\gamma_\mu Q)(\bar q \sigma^i\gamma^\mu q)$ &                     \\ 
& $\mathcal{Q}_{Qq}^{(1,8)}$ & $(\bar Q T^a\gamma_\mu Q)(\bar q T^a\gamma^\mu q)$ &                               \\ 
& $\mathcal{Q}_{Qq}^{(3,8)}$ & $(\bar Q \sigma^iT^a\gamma_\mu Q)(\bar q \sigma^iT^a\gamma^\mu q)$ &               \\ 
 
\bottomrule
\end{tabular}
\end{center}
\end{table}
 
\FloatBarrier

Several different fitting scenarios are employed to extract as much information from the data as possible, in a self-consistent way.
Each fit is performed in the on-shell \ttZ fiducial region (combination of the $3\ell$ and $4\ell$ channels), separately for the top--boson and four-quark operators of Table~\ref{tab:smeft-operators}.
The differential distributions unfolded to particle level in Section~\ref{subsec:unfolded-results} are parameterised in terms of these EFT operators, and taken as joint input measurements for the EFT interpretation.
This detailed analysis of a single process provides a self-contained alternative to simultaneously fitting multiple $\ttbar+X$ processes, such as recently reported by CMS~\cite{CMS-TOP-22-006}.
 
For both the top--boson and four-quark categories of operators, a nominal full quadratic fit using both the linear and pure quadratic terms as well as the cross terms of the parameterisation shown in Eq.~\eqref{eq:smeft-parameterisation} is performed for all operators simultaneously.
Secondly, a fit where only the linear terms are present provides an approximate measure of the importance of the interference terms when compared with the full quadratic fit.
Full quadratic fits are also performed for each operator individually, where only a single Wilson coefficient is probed while all others are fixed to zero, in order to provide results which can be compared with those of other analyses.
 
Finally, following the work in Ref.~\cite{Brehmer:2016nyr}, the inverse covariance matrix of the unfolded measurements is rotated into the space of the Wilson coefficients $\left\{C\right\}$ (top--boson and four-quark operators together), considering only the linear SM/EFT interferences. The resulting matrix provides a lower bound on the Fisher information matrix. Its eigenvectors correspond to the directions in the space of Wilson coefficients probed by the measurements, with the corresponding eigenvalues providing a measure of the sensitivity achieved along these directions. The inverse of the square root of an eigenvalue can be understood as a predictor of the limit that would be obtained when fitting the linear combination of operators defined by the eigenvector, rather than the individual operators themselves. Only the eigenvectors whose eigenvalues are greater than 0.1 are considered, corresponding to a truncation of expected limits greater than ${\sim}\pi$, beyond the natural range of the Wilson coefficients of $\mathcal{O}\left(1\right)$ and the range of validity of the linear approximation.

A multimodal Gaussian likelihood function is implemented with the EFTfitter tool~\cite{EFTfitter2016}, relying on the BAT.jl package~\cite{Schulz:2021BAT}, to perform the EFT interpretation in a Bayesian statistical framework.
The overlap in data between the different measurements is suitably taken into account via correlation matrices.
The results of the various EFT fits described above are presented in Figure~\ref{fig:eft-limits-comparisons}\;for the two scenarios with either only top--boson operators or only four-quark operators, taking as input measurements the fiducial \ttZ cross section and the following normalised differential distributions defined in the combined $3\ell$ and $4\ell$ channels at particle level: $\pT^{Z}$, $\lvert y^{Z}\rvert$, $\cos{\theta^*_Z}$, $\pT^{t}$, $\lvert\Delta\Phi({\ttbar},\Zboson)\rvert$ and $\lvert y^{\ttZ}\rvert$.
Of these, $\pT^{Z}$ is typically the most sensitive to top--\Zboson operators, whereas $\pT^{t}$ is particularly relevant for constraining four-quark operators.
The corresponding numerical values are reported in Tables~\ref{tab:eft-limits-topboson} and~\ref{tab:eft-limits-fourquarks1}.
A uniform prior in the range $\left[-10,10\right]$ is used in every case.
The independent fits, where all Wilson coefficients other than the one considered are set to zero, offer the tightest constraints, typically $\vert C\vert/\Lambda^2 \lesssim 0.5~\TeV^{-2}$ in $95\%$ credible intervals.
 
The quadratic fits, capturing the full EFT picture at dimension 6, also show competitive constraints in the range $\vert C\vert/\Lambda^2 \lesssim 0.5{-}1~\TeV^{-2}$ for $95\%$ credible intervals.
Some of the operators are observed to feature slightly asymmetric limits or non-zero global modes (but are still compatible with the SM within their $1\sigma$ range); this pattern is a consequence of the interplay between the different operators in the global fit, as they do not all represent directions in the EFT space that can be individually probed by the various input measurements.
 
This pattern is also present in the linearised global fit, which can be used to gauge the relative importance of the SM/EFT interference terms.
For some operators (e.g.\ the imaginary and CP-violating parts of \ctG, \ctW and \ctB, or most of the colour-singlet four-quark operators), the results of the linear fits are not quoted since the corresponding interference term vanishes at LO in QCD.
As expected, \ctGRe and the top--\Zboson operators \cHQ1, \cHQ3 and \cHt have strong linear contributions; the inclusion of their pure EFT contributions in the quadratic fit, however, leads to constraints that are tighter by a factor of 2--4.
The real parts of the \ctW and \ctB operators, giving rise to weak dipole moments in the top quark, have comparatively smaller linear terms: the typically soft \Zboson-boson emission produces a momentum suppression, and there is an accidental cancellation between the $q\bar{q}$- and $gg$-initiated channels~\cite{Bylund:2016phk}.
On the other hand, while four-quark operators can interfere with SM \ttZ diagrams, the predominance of the $gg\to\ttZ$ production channel at the LHC drastically reduces the linearised limits when introducing pure EFT contributions in the quadratic fit: e.g.\ for \cQj31 and \cQj38, these effects produce improvements by factors of 6--30.
 
\begin{figure}[!htb]
\centering
\subfloat[]{\includegraphics[width=0.49\textwidth]{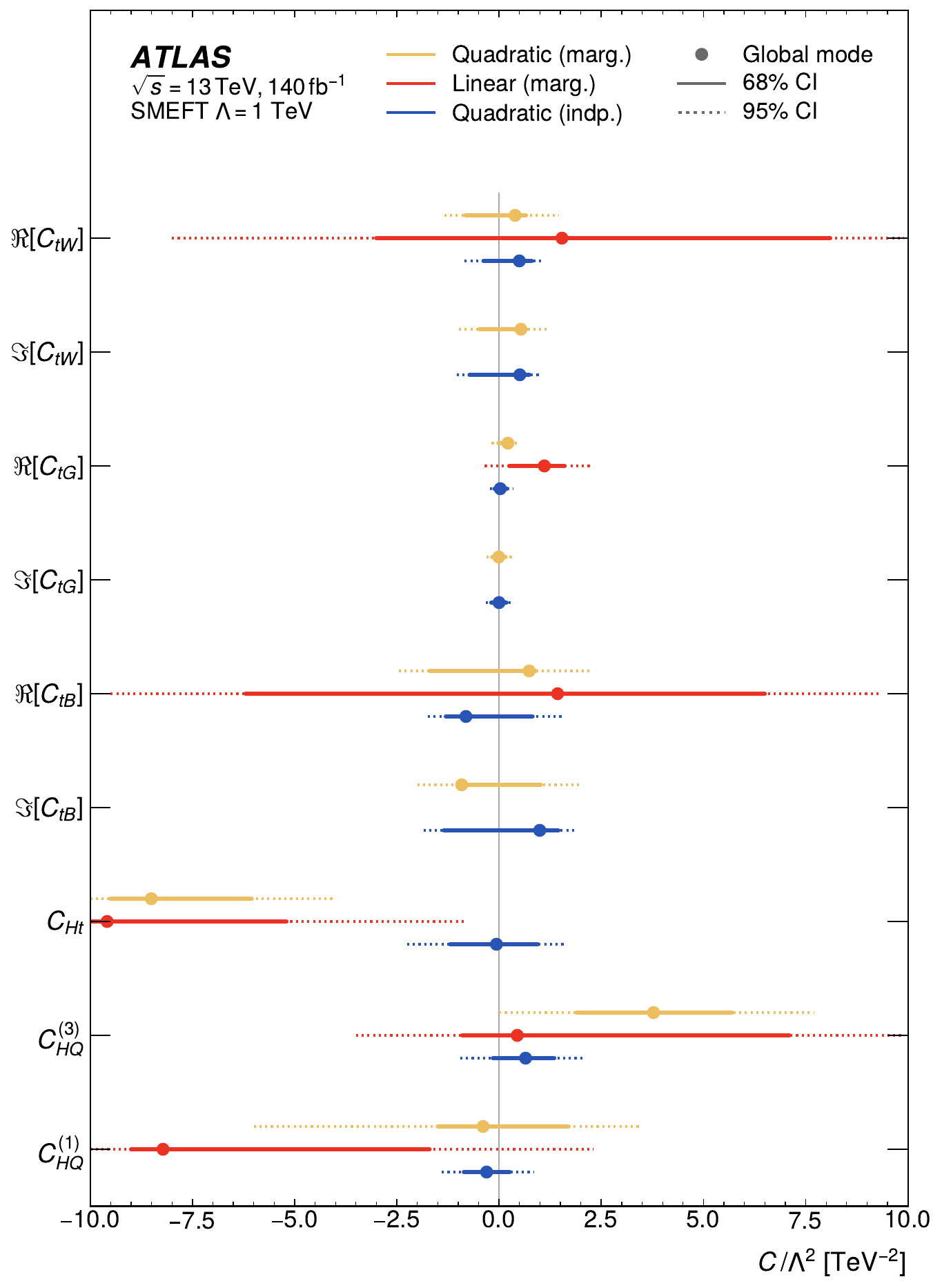}}
\subfloat[]{\includegraphics[width=0.49\textwidth]{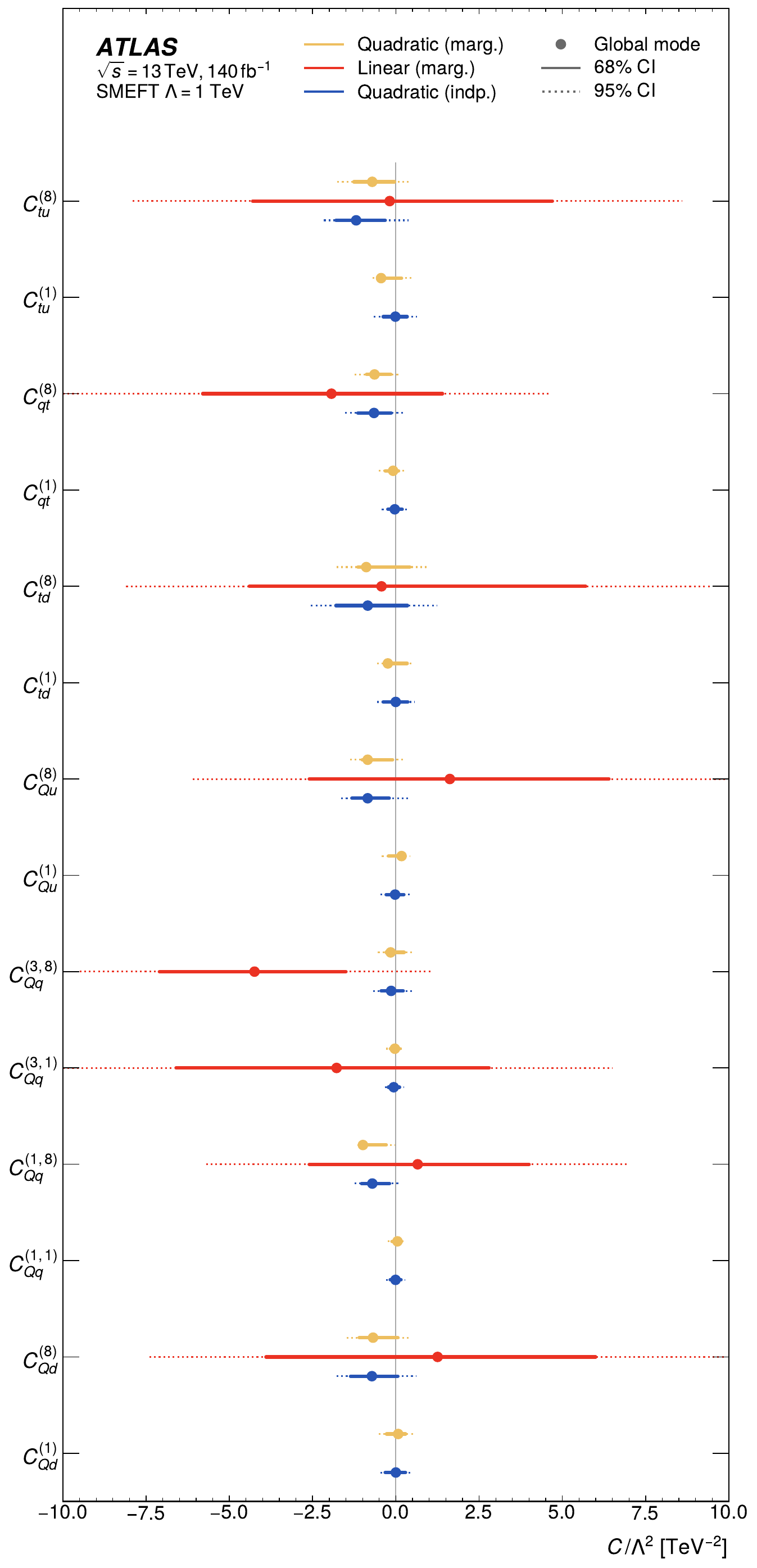}}
\caption{Comparison of the $68\%$ and $95\%$ credible intervals obtained in the (a) top--boson scenario and (b) four-quark scenario for the marginalised linear and quadratic fits, as well as the independent quadratic fit.
The imaginary (CP-violating) part of the \ctW and \ctB operators, as well as the colour-singlet four-quark operator, have no interference with the SM at leading order: no linear fit (red) is performed.
Also shown are the best-fit values (global mode) for each operator.}
\label{fig:eft-limits-comparisons}
\end{figure}
 
\begin{table}[!htb]
\footnotesize
\caption{Observed and expected $68\%$ and $95\%$ credible intervals for the top--boson operators,
comparing the results obtained from the marginalised linear and quadratic fits, as well as the independent quadratic fit.
Also shown are the best-fit values (global mode) for each operator.}
\label{tab:eft-limits-topboson}
\def\arraystretch{1.2}
\begin{scriptsize}
\begin{center}
\begin{tabular}{lllllll}
\toprule
\multicolumn{2}{l}{Wilson coefficient} & $68\%$ CI (exp.) & $95\%$ CI (exp.) & $68\%$ CI (obs.) & $95\%$ CI (obs.) & Best-fit \\
\midrule
 
\multirow{3}{*}{$C_{HQ}^{(1)}$} & $\mathcal{O}(\Lambda^{-2})$ (marg.) & $\left[-5.1,5.4\right]$ & $\left[-8.9,8.7\right]$ & $\left[-9.0,-1.7\right]$ & $\left[-10,2.3\right]$ & \hphantom{1}$-8$ \\
& $\mathcal{O}(\Lambda^{-4})$ (marg.) & $\left[-1.2,1.8\right]$ & $\left[-3.1,4.7\right]$ & $\left[-1.5,1.7\right]$ & $\left[-6.0,3.5\right]$ & \hphantom{1}$-0.4$ \\
& $\mathcal{O}(\Lambda^{-4})$ (indep.) & $\left[-0.58,0.56\right]$ & $\left[-1.1,1.1\right]$ & $\left[-0.86,0.26\right]$ & $\left[-1.4,0.84\right]$ & \hphantom{1}$-0.3$ \\
\addlinespace[0.5em]
 
\multirow{3}{*}{$C_{HQ}^{(3)}$} & $\mathcal{O}(\Lambda^{-2})$ (marg.) & $\left[-4.7,4.7\right]$ & $\left[-8.5,8.4\right]$ & $\left[-0.90,7.1\right]$ & $\left[-3.5,9.9\right]$ & \hphantom{$-1$}0.5 \\
& $\mathcal{O}(\Lambda^{-4})$ (marg.) & $\left[-1.1,2.6\right]$ & $\left[-2.8,4.4\right]$ & $\left[1.9,5.7\right]$ & $\left[0,7.7\right]$ & \hphantom{$-1$}4 \\
& $\mathcal{O}(\Lambda^{-4})$ (indep.) & $\left[-0.85,0.75\right]$ & $\left[-1.6,1.4\right]$ & $\left[-0.15,1.3\right]$ & $\left[-0.95,2.0\right]$ & \hphantom{$-1$}0.7 \\
\addlinespace[0.5em]
 
\multirow{3}{*}{$C_{Ht}$} & $\mathcal{O}(\Lambda^{-2})$ (marg.) & $\left[-4.3,4.2\right]$ & $\left[-7.9,8.1\right]$ & $\left[-10,-5.2\right]$ & $\left[-10,-0.80\right]$ & $-10$ \\
& $\mathcal{O}(\Lambda^{-4})$ (marg.) & $\left[-4.0,0.90\right]$ & $\left[-6.1,3.5\right]$ & $\left[-9.5,-6.0\right]$ & $\left[-10,-4.0\right]$ & \hphantom{1}$-9$ \\
& $\mathcal{O}(\Lambda^{-4})$ (indep.) & $\left[-1.0,0.95\right]$ & $\left[-2.0,1.7\right]$ & $\left[-1.2,0.95\right]$ & $\left[-2.2,1.6\right]$ & \hphantom{1}$-0.06$ \\
\addlinespace[0.5em]
 
\multirow{3}{*}{$\Im{\left[C_{tB}\right]}$} & $\mathcal{O}(\Lambda^{-2})$ (marg.) & --- & --- & --- & --- & --- \\
& $\mathcal{O}(\Lambda^{-4})$ (marg.) & $\left[-0.84,1.0\right]$ & $\left[-1.6,1.7\right]$ & $\left[-0.80,1.0\right]$ & $\left[-2.0,2.0\right]$ & \hphantom{1}$-0.9$ \\
& $\mathcal{O}(\Lambda^{-4})$ (indep.) & $\left[-1.0,1.0\right]$ & $\left[-1.6,1.6\right]$ & $\left[-1.4,1.5\right]$ & $\left[-1.9,1.9\right]$ & \hphantom{$-1$}1 \\
\addlinespace[0.5em]
 
\multirow{3}{*}{$\Re{\left[C_{tB}\right]}$} & $\mathcal{O}(\Lambda^{-2})$ (marg.) & $\left[-6.7,6.7\right]$ & $\left[-9.3,9.7\right]$ & $\left[-6.2,6.5\right]$ & $\left[-9.5,9.3\right]$ & \hphantom{$-1$}1 \\
& $\mathcal{O}(\Lambda^{-4})$ (marg.) & $\left[-1.3,0.90\right]$ & $\left[-2.3,2.0\right]$ & $\left[-1.7,0.90\right]$ & $\left[-2.5,2.3\right]$ & \hphantom{$-1$}0.7 \\
& $\mathcal{O}(\Lambda^{-4})$ (indep.) & $\left[-1.0,0.92\right]$ & $\left[-1.6,1.6\right]$ & $\left[-1.3,0.82\right]$ & $\left[-1.7,1.6\right]$ & \hphantom{1}$-0.8$ \\
\addlinespace[0.5em]
 
\multirow{3}{*}{$\Im{\left[C_{tG}\right]}$} & $\mathcal{O}(\Lambda^{-2})$ (marg.) & --- & --- & --- & --- & --- \\
& $\mathcal{O}(\Lambda^{-4})$ (marg.) & $\left[-0.19,0.17\right]$ & $\left[-0.32,0.32\right]$ & $\left[-0.16,0.16\right]$ & $\left[-0.30,0.31\right]$ & \hphantom{1}$-0.01$ \\
& $\mathcal{O}(\Lambda^{-4})$ (indep.) & $\left[-0.22,0.22\right]$ & $\left[-0.36,0.36\right]$ & $\left[-0.19,0.18\right]$ & $\left[-0.32,0.33\right]$ & \hphantom{$-1$}$0$ \\
\addlinespace[0.5em]
 
\multirow{3}{*}{$\Re{\left[C_{tG}\right]}$} & $\mathcal{O}(\Lambda^{-2})$ (marg.) & $\left[-0.70,0.70\right]$ & $\left[-1.4,1.3\right]$ & $\left[0.25,1.6\right]$ & $\left[-0.35,2.2\right]$ & \hphantom{$-1$}1 \\
& $\mathcal{O}(\Lambda^{-4})$ (marg.) & $\left[-0.11,0.23\right]$ & $\left[-0.27,0.38\right]$ & $\left[-0.015,0.32\right]$ & $\left[-0.18,0.43\right]$ & \hphantom{$-1$}0.2 \\
& $\mathcal{O}(\Lambda^{-4})$ (indep.) & $\left[-0.14,0.21\right]$ & $\left[-0.26,0.36\right]$ & $\left[-0.11,0.20\right]$ & $\left[-0.23,0.34\right]$ & \hphantom{$-1$}0.03 \\
\addlinespace[0.5em]
 
\multirow{3}{*}{$\Im{\left[C_{tW}\right]}$} & $\mathcal{O}(\Lambda^{-2})$ (marg.) & --- & --- & --- & --- & --- \\
& $\mathcal{O}(\Lambda^{-4})$ (marg.) & $\left[-0.56,0.56\right]$ & $\left[-1.1,1.1\right]$ & $\left[-0.48,0.62\right]$ & $\left[-0.98,1.2\right]$ & \hphantom{$-1$}0.5 \\
& $\mathcal{O}(\Lambda^{-4})$ (indep.) & $\left[-0.56,0.56\right]$ & $\left[-0.92,0.92\right]$ & $\left[-0.72,0.74\right]$ & $\left[-1.0,1.0\right]$ & \hphantom{$-1$}0.5 \\
\addlinespace[0.5em]
 
\multirow{3}{*}{$\Re{\left[C_{tW}\right]}$} & $\mathcal{O}(\Lambda^{-2})$ (marg.) & $\left[-5.8,5.9\right]$ & $\left[-9.4,9.7\right]$ & $\left[-3.0,8.1\right]$ & $\left[-8.0,9.9\right]$ & \hphantom{$-1$}2 \\
& $\mathcal{O}(\Lambda^{-4})$ (marg.) & $\left[-0.72,0.60\right]$ & $\left[-1.3,1.3\right]$ & $\left[-0.82,0.66\right]$ & $\left[-1.3,1.5\right]$ & \hphantom{$-1$}0.4 \\
& $\mathcal{O}(\Lambda^{-4})$ (indep.) & $\left[-0.52,0.60\right]$ & $\left[-0.88,0.92\right]$ & $\left[-0.38,0.80\right]$ & $\left[-0.84,1.0\right]$ & \hphantom{$-1$}0.5 \\
\addlinespace[0.5em]
 
\bottomrule
\end{tabular}
\end{center}
\end{scriptsize}
\end{table}
 
\begin{table}[!htb]
\footnotesize
\caption{Observed and expected $68\%$ and $95\%$ credible intervals for the four-quark operators,
comparing the results obtained from the marginalised linear and quadratic fits, as well as the independent quadratic fit.
Also shown are the best-fit values (global mode) for each operator.}
\label{tab:eft-limits-fourquarks1}
\def\arraystretch{1.2}
\begin{scriptsize}
\begin{center}
\begin{tabular}{lllllll}
\toprule
\multicolumn{2}{l}{Wilson coefficient} & $68\%$ CI (exp.) & $95\%$ CI (exp.) & $68\%$ CI (obs.) & $95\%$ CI (obs.) & Best-fit \\
\midrule
 
\multirow{3}{*}{$C_{Qd}^{(1)}$} & $\mathcal{O}(\Lambda^{-2})$ (marg.) & --- & --- & --- & --- & --- \\
& $\mathcal{O}(\Lambda^{-4})$ (marg.) & $\left[-0.31,0.32\right]$ & $\left[-0.54,0.55\right]$ & $\left[-0.28,0.29\right]$ & $\left[-0.52,0.53\right]$ & \hphantom{$-$}0.07 \\
& $\mathcal{O}(\Lambda^{-4})$ (indep.) & $\left[-0.39,0.37\right]$ & $\left[-0.56,0.56\right]$ & $\left[-0.32,0.29\right]$ & $\left[-0.47,0.46\right]$ & \hphantom{$-$}$0$ \\
\addlinespace[0.5em]
 
\multirow{3}{*}{$C_{Qd}^{(8)}$} & $\mathcal{O}(\Lambda^{-2})$ (marg.) & $\left[-4.8,5.0\right]$ & $\left[-8.7,8.6\right]$ & $\left[-3.9,6.0\right]$ & $\left[-7.4,9.9\right]$ & \hphantom{$-$}1 \\
& $\mathcal{O}(\Lambda^{-4})$ (marg.) & $\left[-0.90,0.34\right]$ & $\left[-1.5,0.82\right]$ & $\left[-1.1,0.060\right]$ & $\left[-1.5,0.46\right]$ & $-0.7$ \\
& $\mathcal{O}(\Lambda^{-4})$ (indep.) & $\left[-1.4,0.50\right]$ & $\left[-2.0,1.1\right]$ & $\left[-1.4,0.060\right]$ & $\left[-1.8,0.62\right]$ & $-0.7$ \\
\addlinespace[0.5em]
 
\multirow{3}{*}{$C_{Qq}^{(1,1)}$} & $\mathcal{O}(\Lambda^{-2})$ (marg.) & --- & --- & --- & --- & --- \\
& $\mathcal{O}(\Lambda^{-4})$ (marg.) & $\left[-0.14,0.17\right]$ & $\left[-0.29,0.30\right]$ & $\left[-0.11,0.17\right]$ & $\left[-0.24,0.30\right]$ & \hphantom{$-$}0.05 \\
& $\mathcal{O}(\Lambda^{-4})$ (indep.) & $\left[-0.21,0.21\right]$ & $\left[-0.34,0.34\right]$ & $\left[-0.17,0.16\right]$ & $\left[-0.29,0.29\right]$ & $-0.01$ \\
\addlinespace[0.5em]
 
\multirow{3}{*}{$C_{Qq}^{(1,8)}$} & $\mathcal{O}(\Lambda^{-2})$ (marg.) & $\left[-3.3,3.3\right]$ & $\left[-6.6,6.5\right]$ & $\left[-2.6,4.0\right]$ & $\left[-5.7,7.0\right]$ & \hphantom{$-$}0.7 \\
& $\mathcal{O}(\Lambda^{-4})$ (marg.) & $\left[-0.72,-0.090\right]$ & $\left[-1.0,0.19\right]$ & $\left[-0.89,-0.29\right]$ & $\left[-1.1,-0.030\right]$ & $-1$ \\
& $\mathcal{O}(\Lambda^{-4})$ (indep.) & $\left[-0.76,0.28\right]$ & $\left[-1.3,0.40\right]$ & $\left[-1.0,-0.19\right]$ & $\left[-1.2,0.080\right]$ & $-0.7$ \\
\addlinespace[0.5em]
 
\multirow{3}{*}{$C_{Qq}^{(3,1)}$} & $\mathcal{O}(\Lambda^{-2})$ (marg.) & $\left[-4.9,5.0\right]$ & $\left[-8.6,8.8\right]$ & $\left[-6.6,2.8\right]$ & $\left[-10,6.5\right]$ & $-2$ \\
& $\mathcal{O}(\Lambda^{-4})$ (marg.) & $\left[-0.18,0.13\right]$ & $\left[-0.32,0.26\right]$ & $\left[-0.17,0.12\right]$ & $\left[-0.29,0.24\right]$ & $-0.03$ \\
& $\mathcal{O}(\Lambda^{-4})$ (indep.) & $\left[-0.26,0.18\right]$ & $\left[-0.41,0.31\right]$ & $\left[-0.23,0.11\right]$ & $\left[-0.34,0.23\right]$ & $-0.06$ \\
\addlinespace[0.5em]
 
\multirow{3}{*}{$C_{Qq}^{(3,8)}$} & $\mathcal{O}(\Lambda^{-2})$ (marg.) & $\left[-2.9,2.9\right]$ & $\left[-5.7,5.8\right]$ & $\left[-7.1,-1.5\right]$ & $\left[-9.5,1.1\right]$ & $-4$ \\
& $\mathcal{O}(\Lambda^{-4})$ (marg.) & $\left[-0.31,0.29\right]$ & $\left[-0.57,0.55\right]$ & $\left[-0.29,0.25\right]$ & $\left[-0.54,0.48\right]$ & $-0.2 $\\
& $\mathcal{O}(\Lambda^{-4})$ (indep.) & $\left[-0.42,0.48\right]$ & $\left[-0.74,0.76\right]$ & $\left[-0.44,0.22\right]$ & $\left[-0.68,0.54\right]$ & $-0.1$ \\
\addlinespace[0.5em]
 
\multirow{3}{*}{$C_{Qu}^{(1)}$} & $\mathcal{O}(\Lambda^{-2})$ (marg.) & --- & --- & --- & --- & --- \\
& $\mathcal{O}(\Lambda^{-4})$ (marg.) & $\left[-0.26,0.24\right]$ & $\left[-0.46,0.44\right]$ & $\left[-0.22,0.24\right]$ & $\left[-0.42,0.43\right]$ & \hphantom{$-$}0.2 \\
& $\mathcal{O}(\Lambda^{-4})$ (indep.) & $\left[-0.37,0.35\right]$ & $\left[-0.59,0.56\right]$ & $\left[-0.29,0.24\right]$ & $\left[-0.47,0.43\right]$ & $-0.02$ \\
\addlinespace[0.5em]
 
\multirow{3}{*}{$C_{Qu}^{(8)}$} & $\mathcal{O}(\Lambda^{-2})$ (marg.) & $\left[-4.7,4.6\right]$ & $\left[-8.4,8.5\right]$ & $\left[-2.6,6.4\right]$ & $\left[-6.1,9.9\right]$ & \hphantom{$-$}2 \\
& $\mathcal{O}(\Lambda^{-4})$ (marg.) & $\left[-0.82,0.18\right]$ & $\left[-1.3,0.60\right]$ & $\left[-1.0,-0.10\right]$ & $\left[-1.4,0.27\right]$ & $-0.8$ \\
& $\mathcal{O}(\Lambda^{-4})$ (indep.) & $\left[-1.2,0.34\right]$ & $\left[-1.7,0.80\right]$ & $\left[-1.3,-0.20\right]$ & $\left[-1.6,0.36\right]$ & $-0.8$ \\
\addlinespace[0.5em]
 
\multirow{3}{*}{$C_{qt}^{(1)}$} & $\mathcal{O}(\Lambda^{-2})$ (marg.) & --- & --- & --- & --- & --- \\
& $\mathcal{O}(\Lambda^{-4})$ (marg.) & $\left[-0.28,0.16\right]$ & $\left[-0.47,0.34\right]$ & $\left[-0.32,0.080\right]$ & $\left[-0.51,0.27\right]$ & $-0.08$ \\
& $\mathcal{O}(\Lambda^{-4})$ (indep.) & $\left[-0.31,0.27\right]$ & $\left[-0.50,0.45\right]$ & $\left[-0.24,0.19\right]$ & $\left[-0.42,0.36\right]$ & $-0.03$ \\
\addlinespace[0.5em]
 
\multirow{3}{*}{$C_{qt}^{(8)}$} & $\mathcal{O}(\Lambda^{-2})$ (marg.) & $\left[-3.6,3.6\right]$ & $\left[-7.1,7.1\right]$ & $\left[-5.8,1.4\right]$ & $\left[-10,4.6\right]$ & $-2$ \\
& $\mathcal{O}(\Lambda^{-4})$ (marg.) & $\left[-0.68,0.080\right]$ & $\left[-1.1,0.36\right]$ & $\left[-0.88,-0.15\right]$ & $\left[-1.2,0.12\right]$ & $-0.6$ \\
& $\mathcal{O}(\Lambda^{-4})$ (indep.) & $\left[-0.90,0.36\right]$ & $\left[-1.6,0.54\right]$ & $\left[-1.1,-0.14\right]$ & $\left[-1.5,0.22\right]$ & $-0.7$ \\
\addlinespace[0.5em]
 
\multirow{3}{*}{$C_{td}^{(1)}$} & $\mathcal{O}(\Lambda^{-2})$ (marg.) & --- & --- & --- & --- & --- \\
& $\mathcal{O}(\Lambda^{-4})$ (marg.) & $\left[-0.32,0.38\right]$ & $\left[-0.53,0.57\right]$ & $\left[-0.35,0.34\right]$ & $\left[-0.56,0.54\right]$ & $-0.2$ \\
& $\mathcal{O}(\Lambda^{-4})$ (indep.) & $\left[-0.62,0.64\right]$ & $\left[-1.0,1.0\right]$ & $\left[-0.38,0.36\right]$ & $\left[-0.56,0.56\right]$ & \hphantom{$-$}$0$ \\
\addlinespace[0.5em]
 
\multirow{3}{*}{$C_{td}^{(8)}$} & $\mathcal{O}(\Lambda^{-2})$ (marg.) & $\left[-5.0,5.1\right]$ & $\left[-8.9,8.8\right]$ & $\left[-4.4,5.7\right]$ & $\left[-8.1,9.5\right]$ & $-0.4$ \\
& $\mathcal{O}(\Lambda^{-4})$ (marg.) & $\left[-1.0,0.68\right]$ & $\left[-1.8,1.3\right]$ & $\left[-1.1,0.42\right]$ & $\left[-1.8,0.94\right]$ & $-0.9$ \\
& $\mathcal{O}(\Lambda^{-4})$ (indep.) & $\left[-1.8,0.85\right]$ & $\left[-2.8,1.7\right]$ & $\left[-1.8,0.35\right]$ & $\left[-2.5,1.2\right]$ & $-0.8$ \\
\addlinespace[0.5em]
 
\multirow{3}{*}{$C_{tu}^{(1)}$} & $\mathcal{O}(\Lambda^{-2})$ (marg.) & --- & --- & --- & --- & --- \\
& $\mathcal{O}(\Lambda^{-4})$ (marg.) & $\left[-0.31,0.34\right]$ & $\left[-0.59,0.63\right]$ & $\left[-0.44,0.17\right]$ & $\left[-0.70,0.47\right]$ & $-0.4$ \\
& $\mathcal{O}(\Lambda^{-4})$ (indep.) & $\left[-0.46,0.46\right]$ & $\left[-0.76,0.76\right]$ & $\left[-0.38,0.34\right]$ & $\left[-0.66,0.64\right]$ & $-0.01$ \\
\addlinespace[0.5em]
 
\multirow{3}{*}{$C_{tu}^{(8)}$} & $\mathcal{O}(\Lambda^{-2})$ (marg.) & $\left[-4.6,4.4\right]$ & $\left[-8.4,8.2\right]$ & $\left[-4.3,4.7\right]$ & $\left[-7.9,8.6\right]$ & $-0.2$ \\
& $\mathcal{O}(\Lambda^{-4})$ (marg.) & $\left[-1.0,0.32\right]$ & $\left[-1.6,0.90\right]$ & $\left[-1.3,-0.040\right]$ & $\left[-1.8,0.46\right]$ & $-0.7$ \\
& $\mathcal{O}(\Lambda^{-4})$ (indep.) & $\left[-1.6,0.50\right]$ & $\left[-2.2,1.2\right]$ & $\left[-1.8,-0.32\right]$ & $\left[-2.2,0.38\right]$ & $-1$ \\
\addlinespace[0.5em]
 
\bottomrule
\end{tabular}
\end{center}
\end{scriptsize}
\end{table}
 
\FloatBarrier
 
The covariance matrix obtained in the linear fit is inverted to yield a lower bound on the underlying Fisher information matrix, and a new linear fit in the rotated EFT directions of sensitivity, as previously described, is performed.
The following eigenvectors are extracted, corresponding to eigenvalues larger than 0.1:
 
\begin{footnotesize}
\begin{align*}
\lambda_{1}= 40,\, \mathcal{F}_{1}: &+0.75\cdot\ctGRe +0.01\cdot\ctWRe +0.23\cdot\cHQ1 -0.15\cdot\cHQ3 -0.13\cdot\cHt -0.10\cdot\cQd8 \nonumber \\
&-0.42\cdot\cQj18 -0.15\cdot\cQj31 +0.06\cdot\cQj38 -0.13\cdot\cQu8 -0.05\cdot\ctd8 -0.33\cdot\ctj8 -0.08\cdot\ctu8,\\
\lambda_{2}= 8,\, \mathcal{F}_{2}: &-0.41\cdot\ctGRe +0.04\cdot\ctWRe -0.02\cdot\ctBRe -0.34\cdot\cHQ1 +0.27\cdot\cHQ3 +0.18\cdot\cHt \nonumber\\
&-0.12\cdot\cQd8 -0.57\cdot\cQj18 -0.22\cdot\cQj31 -0.05\cdot\cQj38 -0.22\cdot\cQu8 -0.05\cdot\ctd8 -0.39\cdot\ctj8 \nonumber\\
&-0.12\cdot\ctu8,\\
\lambda_{3}= 0.5,\, \mathcal{F}_{3}: &-0.07\cdot\ctGRe +0.06\cdot\ctWRe -0.01\cdot\ctBRe +0.18\cdot\cHQ1 +0.04\cdot\cHQ3 -0.33\cdot\cHt \nonumber\\
&-0.07\cdot\cQd8 +0.13\cdot\cQj18 +0.04\cdot\cQj31 -0.68\cdot\cQj38 -0.42\cdot\cQu8 -0.11\cdot\ctd8 +0.08\cdot\ctj8 \nonumber\\
&-0.41\cdot\ctu8.
\end{align*}
\end{footnotesize}
 
These relations are also visualised in Figure~\ref{fig:eft-fisher-matrix-full}.
Of particular interest is the combination $\mathcal{F}_1$, which clearly emerges as the most important direction of SMEFT constraints in the linear fit.
It features most prominently the \ctGRe operator together with the two four-quark operators \cQj18 and \ctj8 that couple the top-quark field to left-handed light-quark fields, but also involves a non-trivial combination of most of the other operators.
From the inverse square root of the eigenvalue $\lambda_1$, limits on $\mathcal{F}_1$ are expected to be set at ${\sim}0.15~\TeV^{-2}$.
 
While no clear picture of the four-quark sector appears, some of the expected relations between top--boson operators are recovered (see Eq.~\eqref{eq:redefinition-ttz-vertex-operators}).
The effective impact of \ctGRe in the linear combination $\mathcal{F}_{1}$ was gauged by separating it from the four-quark operators: it is much more constrained than the top--\Zboson operators, but surprisingly its constraints are similar to those of the leading combination of four-quark operators.
In other words, while the \ttZ process is mostly gluon-initiated and therefore sensitive to modifications of the top--gluon vertex by the operator \ctGRe, it is also highly affected by any non-SM structure in the quark-initated channel.
The linear combinations in the subspace of four-quark operators should be compared with those found elsewhere for the \ttbar process: the possibility of radiating the \Zboson boson in \ttZ production from an initial-state quark should provide novel directions of sensitivity.
 
The results of the linearised Fisher-rotated fit are presented in Figure~\ref{fig:eft-limits-comparisons-fisher} and Table~\ref{tab:eft-limits-fisher}.
The limit on $\mathcal{F}_1$ expected by construction is indeed recovered.
 
\begin{figure}[!htb]
\centering
\includegraphics[width=0.9\textwidth]{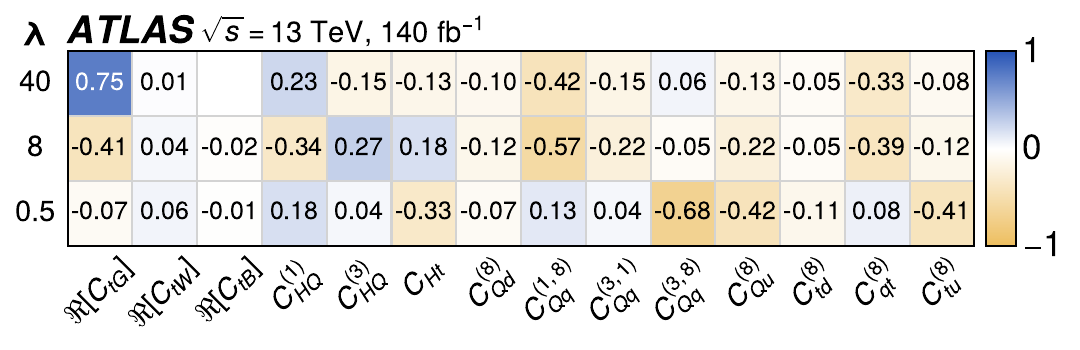}
\caption{Eigenvector decomposition of the Fisher information matrix obtained from the linear global EFT fit.
Each row represents a Fisher-rotated direction $\mathcal{F}$, for which the corresponding eigenvalue satisfies $\lambda>0.1$, expressed in terms of the underlying Wilson coefficients in the Warsaw basis (columns).}
\label{fig:eft-fisher-matrix-full}
\end{figure}
 
\begin{figure}[!htb]
\centering
\includegraphics[width=0.49\textwidth]{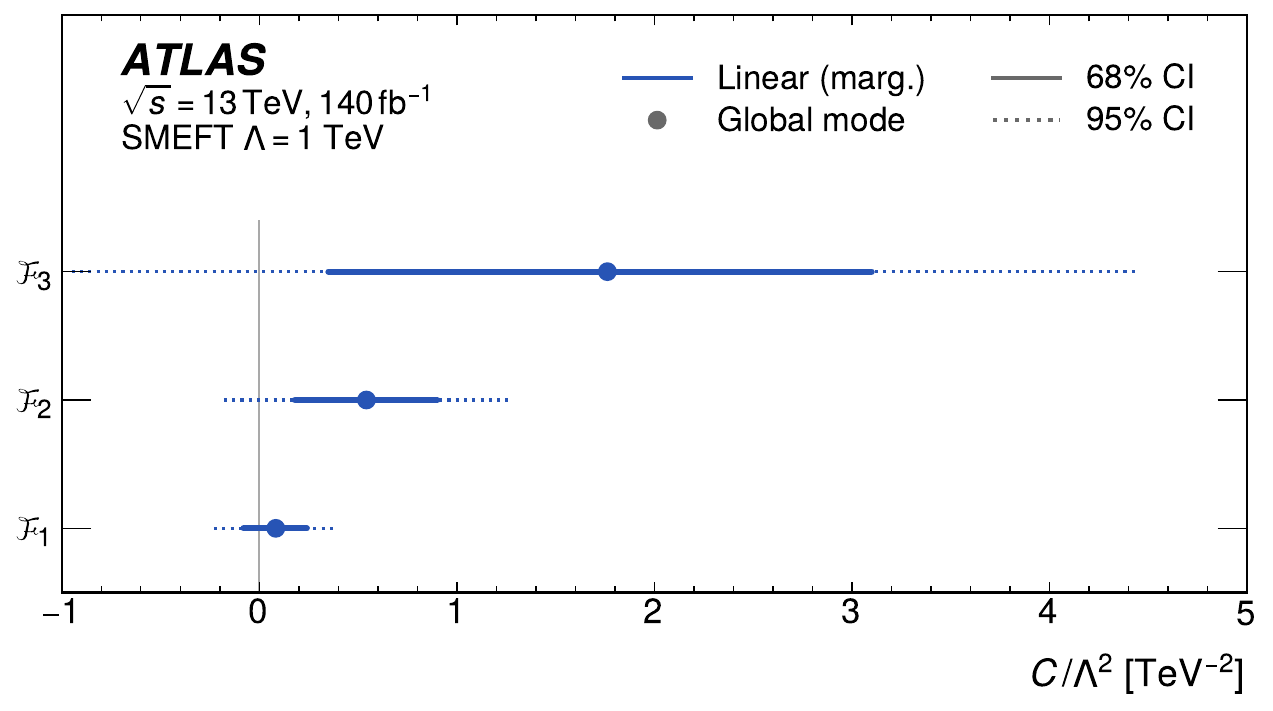}
\caption{Comparison of the $68\%$ and $95\%$ credible intervals obtained for the Fisher-rotated linear fit.
Also shown are the best-fit values (global mode) for each linear combination.}
\label{fig:eft-limits-comparisons-fisher}
\end{figure}
 
\begin{table}[!htb]
\footnotesize
\caption{Observed and expected $68\%$ and $95\%$ credible intervals for the main EFT directions of sensitivity, showing the results obtained from the Fisher-rotated linear fit. Also shown are the best-fit values (global mode) for each linear combination.}
\label{tab:eft-limits-fisher}
\def\arraystretch{1.7}
\begin{center}
\begin{tabular}{lllllll}
\toprule
\multicolumn{2}{l}{Wilson coefficient} & $68\%$ CI (exp.) & $95\%$ CI (exp.) & $68\%$ CI (obs.) & $95\%$ CI (obs.) & Best-fit \\
\midrule
 
\multirow{1}{*}{$\mathcal{F}_{1}$} & $\mathcal{O}(\Lambda^{-2})$ (marg.) & $\left[-0.15,0.16\right]$ & $\left[-0.30,0.31\right]$ & $\left[-0.080,0.24\right]$ & $\left[-0.23,0.39\right]$ & \hphantom{-}0.08 \\
\addlinespace[0.5em]
 
\multirow{1}{*}{$\mathcal{F}_{2}$} & $\mathcal{O}(\Lambda^{-2})$ (marg.) & $\left[-0.36,0.36\right]$ & $\left[-0.72,0.70\right]$ & $\left[0.18,0.90\right]$ & $\left[-0.18,1.3\right]$ & \hphantom{-}0.5 \\
\addlinespace[0.5em]
 
\multirow{1}{*}{$\mathcal{F}_{3}$} & $\mathcal{O}(\Lambda^{-2})$ (marg.) & $\left[-1.4,1.3\right]$ & $\left[-2.7,2.7\right]$ & $\left[0.35,3.1\right]$ & $\left[-0.95,4.5\right]$ & \hphantom{-}2 \\
\addlinespace[0.5em]
 
\bottomrule
\end{tabular}
\end{center}
\end{table}
 
\FloatBarrier


\FloatBarrier
 
\section{Conclusion}

This paper presents measurements of the inclusive and differential cross sections for production of a top-quark--top-antiquark pair in association with a \Zboson boson (\ttZ) in 13~\TeV proton--proton collisions.
The full \RunTwo dataset collected with the ATLAS experiment at the LHC between 2015 and 2018, amounting to an integrated luminosity of \lumi~\ifb, was used for this analysis.
The targeted final states feature two, three or four leptons, focusing on the decay of the \Zboson boson into a pair of electrons or muons but remaining inclusive in the decays of the \ttbar system.
This analysis supersedes, and largely improves upon, the previous \RunTwo ATLAS result.
 
The inclusive \ttZ cross section corresponding to the on-shell $Z$ phase-space region for difermion masses of $70<m_{f\kern-0.1em\bar{f}}<110~\GeV$ is measured to be $\sigma_{\ttZ} = 0.86\pm0.06\,\mathrm{pb}=0.86\pm0.04(\text{stat.})\pm0.04(\text{syst.})\,\mathrm{pb}$.
The observed result is in agreement with the SM prediction $\sigma_{\ttZ}^\text{NLO+NNLL}=0.86^{+0.08}_{-0.09}\,\mathrm{pb}$
and other calculations at NLO QCD and electroweak accuracy.
The result is limited by the systematic uncertainties associated with the modelling of background processes ($Z+$jets, non-prompt leptons), the renormalisation and factorisation scales of the signal process, and the determination of the integrated luminosity.
The total uncertainty in the \ttZ cross section is reduced by $38\%$ (a factor of two for the systematic uncertainties alone) compared to the previous analysis.
The improvements stem mainly from the multivariate techniques used throughout the analysis, and better modelling of both the signal and background processes, including data-driven approaches.
 
Detector-level observables sensitive to polarisation and spin correlations of the top quarks are statistically combined to search for such effects, for the first time in \ttZ production.
Templates from simulated \ttZ events, generated with and without the SM spin density matrix, are used to extract a fraction $f_\mathrm{SM}=1.20\pm 0.68$ reflecting the strength of spin correlations like those in the SM in \ttZ data events, with the uncertainty largely dominated by its statistical component.
This result is in agreement with the SM, and represents a $1.8\sigma$ departure from the null hypothesis of no spin correlations.
 
Furthermore, measurements of absolute and normalised differential production cross sections for the \ttZ process are presented, in a large number of observables sensitive to MC modelling and potential BSM effects.
Distributions at particle level and parton level, unfolded to specific fiducial volumes, are compared with simulations from various \ttZ MC generators and with fixed-order theoretical predictions.
Since the difference between the MC generators considered in this analysis is usually significantly smaller than the uncertainty of the measurement, it is not possible to decide which generator best describes the data.
 
Finally, the unfolded particle-level distributions are used to constrain potential BSM effects from dimension-6 operators in the framework of the SMEFT.
Various fitting scenarios are considered, in order to give a complete picture of EFT effects in \ttZ production.
A large number of operators are considered, affecting both the top--\Zboson coupling and \ttbar production, and also including four-quark operators.
The three most sensitive directions in the EFT space are identified through the Fisher information matrix of the measurements and lead to the tightest constraints.
Compared to recent EFT searches by CMS covering large regions of phase-space and including multiple $t\bar{t}+X$ processes,
the results presented in this paper leverage more information from the measurement of a single process.
Overall, no significant deviation from the SM is observed, and exclusion limits are placed on several top--electroweak and four-quark SMEFT operators.


\section*{Acknowledgements}


We thank CERN for the very successful operation of the LHC and its injectors, as well as the support staff at
CERN and at our institutions worldwide without whom ATLAS could not be operated efficiently.
 
The crucial computing support from all WLCG partners is acknowledged gratefully, in particular from CERN, the ATLAS Tier-1 facilities at TRIUMF/SFU (Canada), NDGF (Denmark, Norway, Sweden), CC-IN2P3 (France), KIT/GridKA (Germany), INFN-CNAF (Italy), NL-T1 (Netherlands), PIC (Spain), RAL (UK) and BNL (USA), the Tier-2 facilities worldwide and large non-WLCG resource providers. Major contributors of computing resources are listed in Ref.~\cite{ATL-SOFT-PUB-2023-001}.
 
We gratefully acknowledge the support of ANPCyT, Argentina; YerPhI, Armenia; ARC, Australia; BMWFW and FWF, Austria; ANAS, Azerbaijan; CNPq and FAPESP, Brazil; NSERC, NRC and CFI, Canada; CERN; ANID, Chile; CAS, MOST and NSFC, China; Minciencias, Colombia; MEYS CR, Czech Republic; DNRF and DNSRC, Denmark; IN2P3-CNRS and CEA-DRF/IRFU, France; SRNSFG, Georgia; BMBF, HGF and MPG, Germany; GSRI, Greece; RGC and Hong Kong SAR, China; ISF and Benoziyo Center, Israel; INFN, Italy; MEXT and JSPS, Japan; CNRST, Morocco; NWO, Netherlands; RCN, Norway; FCT, Portugal; MNE/IFA, Romania; MESTD, Serbia; MSSR, Slovakia; ARIS and MVZI, Slovenia; DSI/NRF, South Africa; MICINN, Spain; SRC and Wallenberg Foundation, Sweden; SERI, SNSF and Cantons of Bern and Geneva, Switzerland; NSTC, Taipei; TENMAK, T\"urkiye; STFC, United Kingdom; DOE and NSF, United States of America.
 
Individual groups and members have received support from BCKDF, CANARIE, CRC and DRAC, Canada; CERN-CZ, PRIMUS 21/SCI/017 and UNCE SCI/013, Czech Republic; COST, ERC, ERDF, Horizon 2020, ICSC-NextGenerationEU and Marie Sk{\l}odowska-Curie Actions, European Union; Investissements d'Avenir Labex, Investissements d'Avenir Idex and ANR, France; DFG and AvH Foundation, Germany; Herakleitos, Thales and Aristeia programmes co-financed by EU-ESF and the Greek NSRF, Greece; BSF-NSF and MINERVA, Israel; Norwegian Financial Mechanism 2014-2021, Norway; NCN and NAWA, Poland; La Caixa Banking Foundation, CERCA Programme Generalitat de Catalunya and PROMETEO and GenT Programmes Generalitat Valenciana, Spain; G\"{o}ran Gustafssons Stiftelse, Sweden; The Royal Society and Leverhulme Trust, United Kingdom.
 
In addition, individual members wish to acknowledge support from Chile: Agencia Nacional de Investigaci\'on y Desarrollo (FONDECYT 1190886, FONDECYT 1210400, FONDECYT 1230812, FONDECYT 1230987); China: National Natural Science Foundation of China (NSFC - 12175119, NSFC 12275265, NSFC-12075060); Czech Republic: PRIMUS Research Programme (PRIMUS/21/SCI/017); EU: H2020 European Research Council (ERC - 101002463); European Union: European Research Council (ERC - 948254), Horizon 2020 Framework Programme (MUCCA - CHIST-ERA-19-XAI-00), European Union, Future Artificial Intelligence Research (FAIR-NextGenerationEU PE00000013), Italian Center for High Performance Computing, Big Data and Quantum Computing (ICSC, NextGenerationEU), Marie Sklodowska-Curie Actions (EU H2020 MSC IF GRANT NO 101033496); France: Agence Nationale de la Recherche (ANR-20-CE31-0013, ANR-21-CE31-0013, ANR-21-CE31-0022, ANR-22-EDIR-0002), Investissements d'Avenir Idex (ANR-11-LABX-0012), Investissements d'Avenir Labex (ANR-11-LABX-0012); Germany: Baden-Württemberg Stiftung (BW Stiftung-Postdoc Eliteprogramme), Deutsche Forschungsgemeinschaft (DFG - 469666862, DFG - CR 312/5-1); Italy: Istituto Nazionale di Fisica Nucleare (FELLINI G.A. n. 754496, ICSC, NextGenerationEU); Japan: Japan Society for the Promotion of Science (JSPS KAKENHI JP21H05085, JSPS KAKENHI JP22H01227, JSPS KAKENHI JP22H04944, JSPS KAKENHI JP22KK0227); Netherlands: Netherlands Organisation for Scientific Research (NWO Veni 2020 - VI.Veni.202.179); Norway: Research Council of Norway (RCN-314472); Poland: Polish National Agency for Academic Exchange (PPN/PPO/2020/1/00002/U/00001), Polish National Science Centre (NCN 2021/42/E/ST2/00350, NCN OPUS nr 2022/47/B/ST2/03059, NCN UMO-2019/34/E/ST2/00393, UMO-2020/37/B/ST2/01043, UMO-2021/40/C/ST2/00187); Slovenia: Slovenian Research Agency (ARIS grant J1-3010); Spain: BBVA Foundation (LEO22-1-603), Generalitat Valenciana (Artemisa, FEDER, IDIFEDER/2018/048), La Caixa Banking Foundation (LCF/BQ/PI20/11760025), Ministry of Science and Innovation (MCIN \& NextGenEU PCI2022-135018-2, MICIN \& FEDER PID2021-125273NB, RYC2019-028510-I, RYC2020-030254-I, RYC2021-031273-I, RYC2022-038164-I), PROMETEO and GenT Programmes Generalitat Valenciana (CIDEGENT/2019/023, CIDEGENT/2019/027); Sweden: Swedish Research Council (VR 2018-00482, VR 2022-03845, VR 2022-04683, VR grant 2021-03651), Knut and Alice Wallenberg Foundation (KAW 2017.0100, KAW 2018.0157, KAW 2018.0458, KAW 2019.0447); Switzerland: Swiss National Science Foundation (SNSF - PCEFP2\_194658); United Kingdom: Leverhulme Trust (Leverhulme Trust RPG-2020-004); United States of America: U.S. Department of Energy (ECA DE-AC02-76SF00515), Neubauer Family Foundation.


\clearpage
\appendix
\part*{Appendix}
\addcontentsline{toc}{part}{Appendix}

This Appendix provides additional information about the MVA approaches used to measure the inclusive \ttZ cross section, as well as the full set of differential cross sections (absolute and normalised, unfolded to particle level and parton level) and further details of the unfolding procedure.
 
The pre-fit event yields in the fake lepton control regions are reported in Table~\ref{tab:3L_fake_yields_prefit}.
The corresponding pre-fit plots are shown in Figure~\ref{fig:3L-fake-cr-prefit}.
 
Tables~\ref{tab:2l_variable_definitions},~\ref{tab:3l_variable_definitions} and~\ref{tab:4l_variable_definitions} list the kinematic quantities used as inputs to the training of the DNNs in the 2$\ell$, 3$\ell$ and 4$\ell$ channels respectively.
Table~\ref{tab:diff_observables_binning} collects the specific binning used for each differential cross-section measurement; the same binning is used whether unfolding to particle level or parton level.
 
Figures~\ref{fig:combined-observed-unfolding-result-yz1} to~\ref{fig:combined-observed-unfolding-result-y_ttZ-particle} present the differential cross sections for observables defined in the combination of the 3$\ell$ and 4$\ell$ fiducial volumes.
Each figure contains four distributions: unfolded to particle level or parton level, absolute or normalised cross sections.
Figures~\ref{fig:trilepton-observed-unfolding-result-ht_leptons_3L-particle} to~\ref{fig:trilepton-observed-unfolding-result-n_jets_3L-particle} similarly correspond to observables defined in the 3$\ell$ channel, and Figures~\ref{fig:tetralepton-observed-unfolding-result-sum_pT_leptons} to~\ref{fig:tetralepton-observed-unfolding-result-nJets} to those defined in the 4$\ell$ channel.
 
Finally, Tables~\ref{tab:unfolding_p_values_absolute} and~\ref{tab:unfolding_p_values_normalised} summarise tests of the compatibility of the unfolded data and predictions from various MC generators, for the absolute and normalised differential spectra respectively.


 
\begin{table}[!htb]
\footnotesize
\caption{Pre-fit event yields in the fake-lepton control regions, obtained for an integrated luminosity of \lumi~\ifb.
The indicated errors include the Monte Carlo statistical uncertainty as well as all other systematic uncertainties discussed in Section~\ref{sec:systematics}.
Because of rounding, the values quoted for the total yield and its uncertainty may differ from the simple sum over all processes.
A dash (---) indicates event yields smaller than 0.1.}
\label{tab:3L_fake_yields_prefit}
\def\arraystretch{1.3}
\begin{center}
\begin{tabular}{l
r@{\(\,\pm\,\)}r
r@{\(\,\pm\,\)}r
r@{\(\,\pm\,\)}r}
\toprule
& \multicolumn{2}{c}{CR-\ttbar-e} & \multicolumn{2}{c}{CR-\ttbar-$\mu$} & \multicolumn{2}{c}{CR-Z-e} \\
\midrule
$t\bar{t}Z$               &  \numRF{2.67}{3} & \numRF{0.29}{2}          &  \numRF{0.72}{2} & \numRF{0.15}{2}          &  \numRF{33.4}{3} & \numRF{1.5}{2}\pho           \\
\ZZl                      &   \multicolumn{2}{c}{---}                   & \multicolumn{2}{c}{---}     &  \numRF{3.8}{2}\pho & \numRF{1.6}{2}\pho             \\
\ZZc                      &   \multicolumn{2}{c}{---}                   & \multicolumn{2}{c}{---}     &  \numRF{2.6}{2}\pho & \numRF{1.2}{2}\pho             \\
\ZZb                      &   \multicolumn{2}{c}{---}                   & \multicolumn{2}{c}{---}     &  \numRF{2.7}{2}\pho & \numRF{2.1}{2}\pho             \\
\WZl                      &  \numRF{0.3}{2} & \numRF{0.15}{2}           &  \multicolumn{2}{c}{---}                        &  \numRF{8.0}{2}\pho & \numRF{3.3}{2}\pho              \\
\WZc                      &  \numRF{0.39}{2} & \numRF{0.16}{2}          &  \multicolumn{2}{c}{---}                        &  \numRF{12}{2}\phdoo & \numRF{5}{1}\phdoo               \\
\WZb                      &  \numRF{0.18}{2} & \numRF{0.12}{2}          &  \multicolumn{2}{c}{---}                        &  \numRF{6.1}{2}\pho & \numRF{3.4}{2}\pho            \\
\tZq                      &  \numRF{0.2}{2} & \numRF{0.06}{1}           &  \multicolumn{2}{c}{---}                        &  \numRF{6.2}{2}\pho & \numRF{0.8}{1}\pho            \\
\tWZ                      &  \numRF{0.25}{2} & \numRF{0.05}{1}          &  \multicolumn{2}{c}{---}                        &  \numRF{3.6}{2}\pho & \numRF{0.4}{1}\pho            \\
\ttW                      &  \numRF{2.9}{2}\pho & \numRF{1.5}{2}\pho            &  \numRF{1.5}{2}\pho & \numRF{0.8}{1}\pho            &  \numRF{1.4}{2}\pho & \numRF{0.7}{1}\pho            \\
$t\bar{t}H$               &  \numRF{4.4}{2}\pho & \numRF{0.4}{1}\pho            &  \numRF{2.3}{2}\pho & \numRF{0.23}{2}           &  \numRF{3.15}{3} & \numRF{0.33}{2}          \\
Other                     &  \numRF{1.1}{2}\pho & \numRF{0.5}{1}\pho            &  \numRF{0.65}{2} & \numRF{0.33}{2}          &  \numRF{0.46}{2} & \numRF{0.24}{2}          \\
F-e-Other                 &  \numRF{147}{3}\phdoo & \numRF{9}{1}\phdoo              &  \multicolumn{2}{c}{---}                        &  \numRF{216}{3}\phdoo & \numRF{26}{2}\phdoo             \\
F-e-HF                    &  \numRF{830}{2}\phdoo & \numRF{70}{1}\phdoo             &  \multicolumn{2}{c}{---}                        &  \numRF{610}{2}\phdoo & \numRF{60}{1}\phdoo             \\
F-$\mu$-HF                &  \numRF{0.25}{2} & \numRF{0.09}{1}          &  \numRF{720}{2}\phdoo & \numRF{60}{1}\phdoo             &  \multicolumn{2}{c}{---}                        \\
F-Other                   &  \numRF{2.6}{2}\pho & \numRF{1.3}{2}\pho            &  \numRF{32}{2}\phdoo & \numRF{16}{2}\phdoo              &  \numRF{0.8}{1}\pho & \numRF{0.5}{1}\pho            \\
\midrule
Total                     &  \numRF{1000}{2}\phdoo & \numRF{80}{1}\phdoo            &  \numRF{760}{2}\phdoo & \numRF{60}{1}\phdoo             &  \numRF{910}{2}\phdoo & \numRF{90}{1}\phdoo             \\
\midrule
Data                      &  \multicolumn{2}{l}{\num{949}}                      &  \multicolumn{2}{l}{\num{786}}                      &  \multicolumn{2}{l}{\num{892}}                      \\
\bottomrule
\end{tabular}
\end{center}
\end{table}
 
\begin{figure}[!htb]
\centering
\subfloat[]{\includegraphics[width=0.32\textwidth]{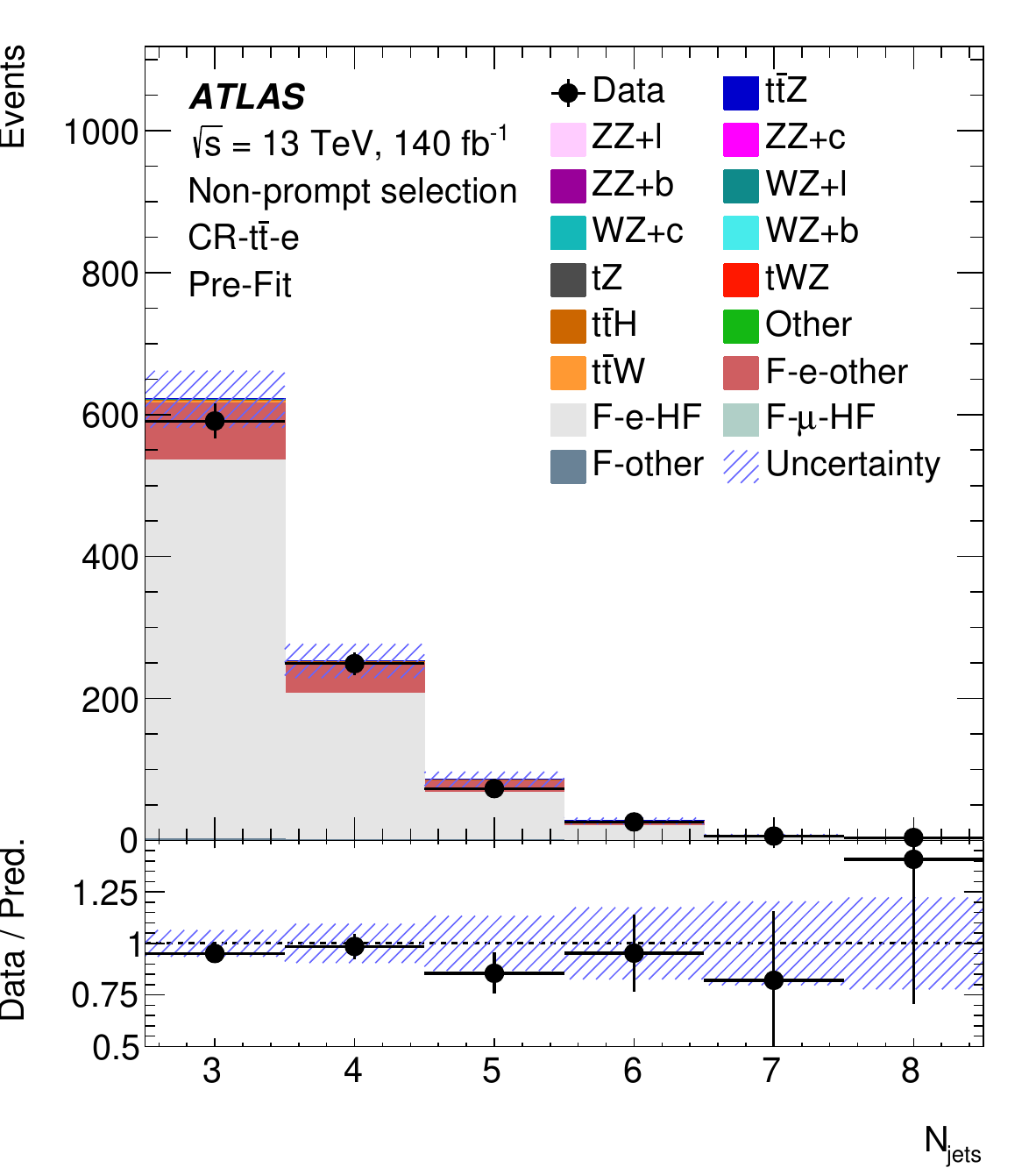}}
\subfloat[]{\includegraphics[width=0.32\textwidth]{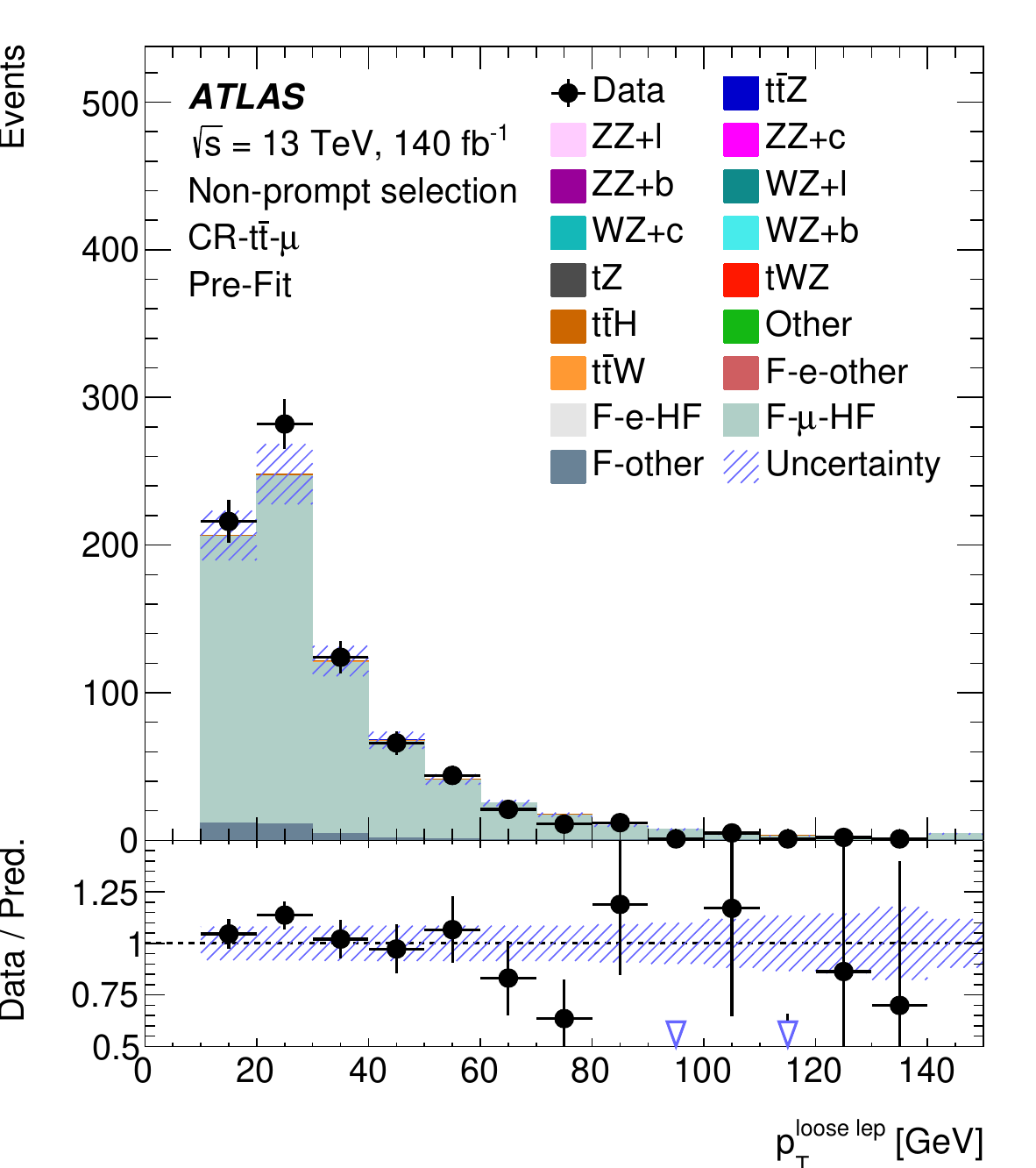}}
\subfloat[]{\includegraphics[width=0.32\textwidth]{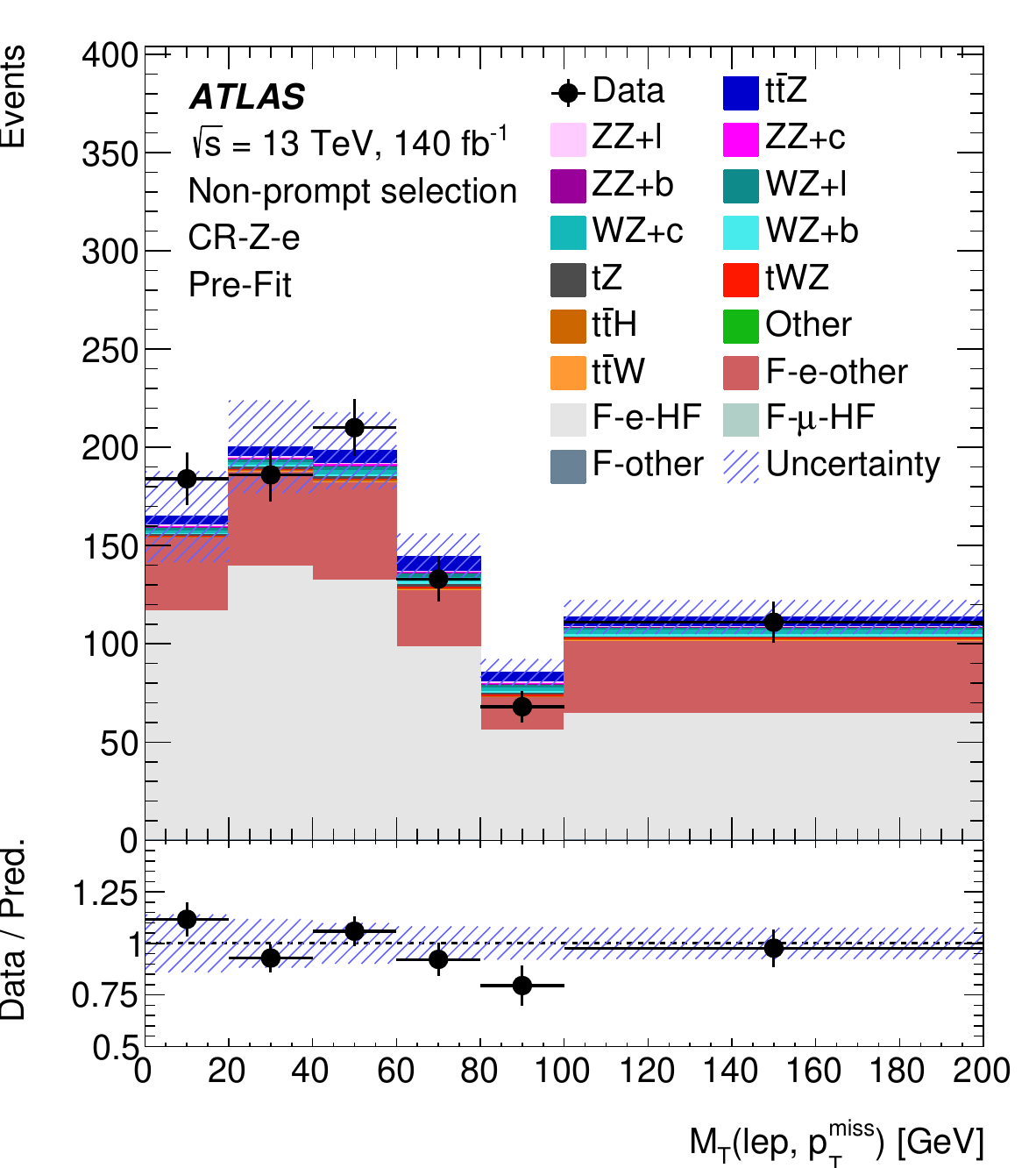}}
\caption{Pre-fit distributions of (a) number of jets in CR-\ttbar-e, (b) transverse momentum of the loose lepton in CR-\ttbar-$\mu$, and (c) of the \Wboson boson transverse mass in CR-Z-e. The shaded band corresponds to the total uncertainty (systematic and statistical) of the total SM prediction. The lower panel shows the ratio of the data to the SM prediction. The last bin includes also the overflow. The distributions (a) and (b) are not used in the fit, instead just an overall event yield in each of these regions is fitted.}
\label{fig:3L-fake-cr-prefit}
\end{figure}


\begin{table}[htbp]
\caption{Definition of the DNN input variables used in 2$\ell$OS signal regions.
Jets and leptons are ordered by their \pt starting with the largest.
To suppress the effect of mismodelling in events with high jet multiplicity,
only the first eight jets ordered by \pt are considered when calculating DNN input variables if an event has more than eight jets, otherwise all jets in the event are considered.}
\label{tab:2l_variable_definitions}
\def\arraystretch{1.3}
\begin{center}
\begin{tabular}{ll}
\toprule
\textbf{Variable} & \textbf{Definition}\\
\midrule
$\HT$ & Sum of $\pT$ of all objects (jets and leptons) in the event.\\ 
$H_{\mathrm{T}}^{\mathrm{jets}}$ & Sum of $\pT$ of all jets in the event.\\ 
$p_{\mathrm{T}}^{X.\mathrm{jet}}$ & $\pT$ of the $X$'th jet, where only the first eight jets are considered.\\ 
$p_{\mathrm{T}}^{X.\mathrm{lep}}$ & $\pT$ of the $X$'th lepton.\\ 
$W_{1t1W}$ & Weight for one-top hypothesis and 1$W$ from multi-hypothesis hadronic $t/W$ reconstruction.  \\
& It is the probability that the event contains all three jets from one of the top quarks and \\
& two light jets from the decay of the other top quark. More details are provided in Section~\ref{subsec:top_reco}.  \\
$W_{1t}$ & Weight for one-top hypothesis from multi-hypothesis hadronic $t/W$ reconstruction. \\
& The same as $W_{1t1W}$, with one top quark only.\\ 
$Centr_{\mathrm{jets}}$ & scalar sum of $\pT$ divided by sum of $E$ for all jets.\\ 
$\Delta R (b_1,b_2)$ & $\Delta R$ between two jets with highest $b$-tagging working point.  \\
&  The jets with the same working point are ordered by $p_{\mathrm{T}}$.  \\ 
$H_{1}^{\mathrm{jets}}$ & First Fox--Wolfram moment built from jets only. The first Fox--Wolfram moment is \\
&  defined as $H_1=\sum_{i,j}\frac{\vec{p_{i}}\cdot\vec{p_{j}}}{E_\text{vis}^{2}}$, where $\vec{p_{i}}$ and $\vec{p_{j}}$ are 3-momenta of $i$'th and $j$'th objects \\
&  (jet or lepton) and $E_\text{vis}$ is all visible energy in the event.\\ 
$N_{jj}^{m<50 \mathrm{GeV}}$ & Number of jet--jet combinations with mass lower than 50 GeV.\\ 
$m_Z$, $y^{Z}$, $p_{\mathrm{T}}^{Z}$  & Mass, rapidity and transverse momentum of the $Z$ boson.\\ 
$\mathrm{min}\left(M_{jj}^{\mathrm{ave}}\right)$ & Average (over the number of jets in event) minimum invariant mass of jet pairs. For each jet, \\
& the other jet which results in the minimum dijet invariant mass is found. \\
& The observable is the average of these masses over all jets in the event.\\%
$\Delta R(\ell,\ell)$ & $\Delta R$ between two leptons.\\ 
PCBT$_{Xj}$ & Discretised $b$-tagging efficiency (100--85--77--70--60\%) of the $X$-th jet.\\ 
$N_{\mathrm{lep}}^{\mathrm{top}}$ & Number of leptonic top candidates.\\
$N_{\mathrm{had}}^{\mathrm{top}}$ & Number of hadronic top candidates. \\
$N_{\mathrm{had}}^{W}$ & Number of hadronic $W$ candidates.\\ 
$E_{\mathrm{T}}^{\mathrm{miss}}$ & Missing transverse momentum in the event.\\ 
$H_1$ & First Fox--Wolfram momentent built from jets and leptons.\\ 
$p_{\mathrm{T}}^{t\bar{t}, \mathrm{spanet}}$ & Transverse momentum of the \ttbar system reconstructed from jets predicted by SPANet.\\ 
\addlinespace[0.5em]
\bottomrule
\end{tabular}
\end{center}
\end{table}
 
\begin{table}[htbp]
\caption{Definition of the DNN input variables used in 3$\ell$ signal regions. Jets and leptons are ordered by their \pt starting with the largest.}
\label{tab:3l_variable_definitions}
\def\arraystretch{1.3}
\begin{center}
\begin{tabular}{ll}
\toprule
\textbf{Variable} & \textbf{Definition} \\
\midrule
PCBT$_{b1}$ & Highest discretised $b$-tagging efficiency (100--85--77--70--60\%) of all jets in the event.\\ 
PCBT$_{b2}$ & Second-highest discretised $b$-tagging efficiency of all jets in the event.\\ 
Jet $p_{\textrm{T},i}$ & Transverse momentum of the $i$'th jet in the event where $i \in [1,4]$. \\ 
$E_{\textrm{T}}^{\textrm{miss}}$ & Missing transverse momentum in the event. \\ 
Lepton $p_{\textrm{T},i}$ & Transverse momentum of the $i$'th lepton in the event where $i \in [1,3]$. \\ 
$m_{t}^{\textrm{lep}}$ & Reconstructed mass of the leptonically decaying top quark. \\ 
$m_{t}^{\textrm{had}}$ & Reconstructed mass of the hadronically decaying top quark. \\ 
$N_{\textrm{jets}}$ & Jet multiplicity in the event. \\ 
Leading $b$-tagged jet $p_{\textrm{T}}$ & Transverse momentum of the jet with the highest discretised $b$-tagging efficiency. \\
& If two have the same bin the leading-\pt jet of the two is used. \\ 
$H_{\mathrm{T}}^{\mathrm{jets}}$ & Sum of the transverse momentum of all jets in the event. \\ 
$\Delta R(\ell_{i},b_{1})$ & Distance in $\Delta R$ between the $i$'th lepton and the $b$-tagged jet tagged with the \\
& highest working point in the event where $i \in [1,3]$. \\ 
$p_{\textrm{T},i}^{Z}$ & Transverse momentum of the first and second lepton $(i \in [1,2])$ assigned \\
& to the $Z$ boson based on their invariant mass being closest to the $Z$ mass. \\ 
$\eta_{i}^{Z}$ & Pseudorapidity of the first and second lepton $(i \in [1,2])$ assigned to the \\
& $Z$ boson based on their invariant mass being closest to the $Z$ mass. \\ 
Lepton $p_{\textrm{T}}^{\textrm{non-}Z}$ & Transverse momentum of the remaining lepton not assigned to the $Z$ boson. \\ 
\addlinespace[0.5em]
\bottomrule
\end{tabular}
\end{center}
\end{table}
 
\begin{table}[htbp]
\caption{Definition of the DNN input variables used in 4$\ell$ signal regions for the same-flavour (SF) and different-flavour (DF) trainings. Jets and leptons are ordered by their \pt starting with the largest.}
\label{tab:4l_variable_definitions}
\def\arraystretch{1.3}
\begin{center}
\begin{tabular}{llcc}
\toprule
\textbf{Variable} & \textbf{Definition} & \textbf{SF} & \textbf{DF}\\
\midrule
\MET & Missing transverse momentum in the event. & \checkmark & --- \\ 
$m^{\ell\ell, \textrm{non-}Z}$ & Invariant mass of two leptons which were not reconstructed & \checkmark & \checkmark \\ 
& as originating from the $Z$ boson.& & \\ 
2$\nu$SM weight & Output of the \textit{Two neutrino scanning method} for the event. & \checkmark & \checkmark \\ 
$\pT^{Z}$ & Transverse momentum of the OSSF lepton pair identified & \checkmark & \checkmark \\ 
& as $Z$ decay (invariant mass of lepton pair closest to $Z$ mass). & & \\ 
$m_{t}^{\ell b}$ & Invariant mass of lepton and $b$-tagged jet reconstructed as & \checkmark & \checkmark \\ 
& originating from the top quark by the \textit{Two neutrino scanning method}. & & \\ 
$m_{\bar{t}}^{\ell b}$ & Invariant mass of lepton and $b$-tagged jet reconstructed as & \checkmark & \checkmark \\ 
& originating from anti-top quark by \textit{Two neutrino scanning method}. & & \\ 
PCBT$_{b1}$ & Highest discretised $b$-tagging efficiency (100--85--77--70--60\%) of all  & \checkmark & ---\\
& jets in the event. & & \\ 
$\pT^{\textrm{lep}_{1}}$ & Transverse momentum of the leading lepton. & \checkmark & \checkmark \\ 
$\pT^{\textrm{jet}_{2}}$ & Transverse momentum of the sub-leading jet. & \checkmark & \checkmark \\ 
PCBT$_{b2}$ & Second-highest discretised $b$-tagging efficiency of all jets in the event. & & \\ 
$N_{\textrm{jets}}$ & Jet multiplicity in the event. & --- & \checkmark \\ 
$N_{b\textrm{-tagged jets}}$ & $b$-tagged (85\% efficiency) jet multiplicity in the event. & --- & \checkmark \\ 
\bottomrule
\end{tabular}
\end{center}
\end{table}
 
\FloatBarrier


 
\begin{table}[!htb]
\caption{Bin ranges for the differential observables defined in Table~\ref{tab:variable_definitions}.
The bin ranges are identical for particle- and parton-level measurements.}
\begin{center}
\begin{tabular}{llcl}
\toprule
Observable & Channels & Bins & Bin Ranges \\
\midrule
$\pT^{Z}$ [\GeV] & $3\ell + 4\ell$ & 8 & [0, 60, 100, 140, 180, 230, 280, 350, 1000] \\
\addlinespace[0.3em]
$\lvert y^{Z}\rvert$ & $3\ell + 4\ell$  & 9 & [0, 0.125, 0.275, 0.425, 0.6, 0.775, 0.95, 1.175, 1.45, 2.5] \\
\addlinespace[0.3em]
$\pT^{\ell,{\textrm{non-}}\Zboson}$ [\GeV] & $3\ell$ & 5 & [0, 35, 55, 80, 120, 500] \\
\addlinespace[0.3em]
$|\Delta y(\Zboson,t_{\mathrm{lep}})|$ & $3\ell$ & 5 & [0, 0.25, 0.6, 1.05, 1.55, 5] \\
\addlinespace[0.3em]
$|\Delta\Phi(\Zboson,t_{\mathrm{lep}})|/\pi$ & $3\ell$ & 6 & [0, 0.16, 0.44, 0.66, 0.82, 0.93, 1] \\
\addlinespace[0.3em]
$|\Delta\Phi(\ell^{+}_{t}, \ell^{-}_{\bar{t}})|/\pi$ & $4\ell$   & 7 & [0, 0.2, 0.37, 0.53, 0.67, 0.79, 0.89, 1] \\
\addlinespace[0.3em]
$\HT^{\ell}$ [\GeV] & $3\ell$ & 8 & [50, 130, 165, 195, 230, 275, 330, 405, 800] \\
\addlinespace[0.3em]
$\HT^{\ell}$ [\GeV] & $4\ell$ & 5 & [50, 195, 250, 315, 400, 800] \\
\addlinespace[0.3em]
$N_{\mathrm{jets}}$ & $3\ell$ & 4 & [2.5, 3.5, 4.5, 5.5, 10.5] \\
\addlinespace[0.3em]
$N_{\mathrm{jets}}$ & $4\ell$ & 3 & [1.5, 2.5, 3.5, 8.5] \\
\addlinespace[0.3em]
$\pT^{t}$ [GeV] & $3\ell + 4\ell$  & 10\phantom{1} & [0, 48, 80, 112, 144, 176, 216, 256, 296, 352, 800] \\
\addlinespace[0.3em]
$\pT^{\ttbar}$ [GeV] & $3\ell + 4\ell$ &  10\phantom{1} & [0, 50, 80, 110, 140, 170, 210, 250, 290, 330, 1000] \\
\addlinespace[0.3em]
$\lvert\Delta\Phi({\ttbar},\Zboson)\rvert/\pi$ & $3\ell + 4\ell$ & 5 & [0, 0.73, 0.86, 0.94, 0.98, 1] \\
\addlinespace[0.3em]
$m^{\ttbar}$ [GeV] & $3\ell + 4\ell$ & 10\phantom{1} &  [0, 370, 420, 470, 530, 600, 680, 780, 890, 1010, 2000]  \\
\addlinespace[0.3em]
$m^{\ttZ}$ [GeV] & $3\ell + 4\ell$ & 10\phantom{1} & [400, 580, 650, 720, 800, 890, 990, 1100, 1220, 1350, 2000]  \\
\addlinespace[0.3em]
$\lvert y^{\ttZ}\rvert$ & $3\ell + 4\ell$  & 10\phantom{1} & [0, 0.075, 0.2, 0.35, 0.5, 0.65, 0.8, 0.95, 1.1, 1.25, 2.5] \\
\addlinespace[0.3em]
$\cos{\theta^*_Z}$ & $3\ell + 4\ell$  & 8 & [$-1$, $-0.75$, $-0.5$, $-0.25$, 0, 0.25, 0.5, 0.75, 1] \\
\bottomrule
\end{tabular}
\label{tab:diff_observables_binning}
\end{center}
\end{table}
 
\FloatBarrier


 
\begin{figure}[!htb]
\centering
\subfloat[]{\includegraphics[width=0.46\textwidth]{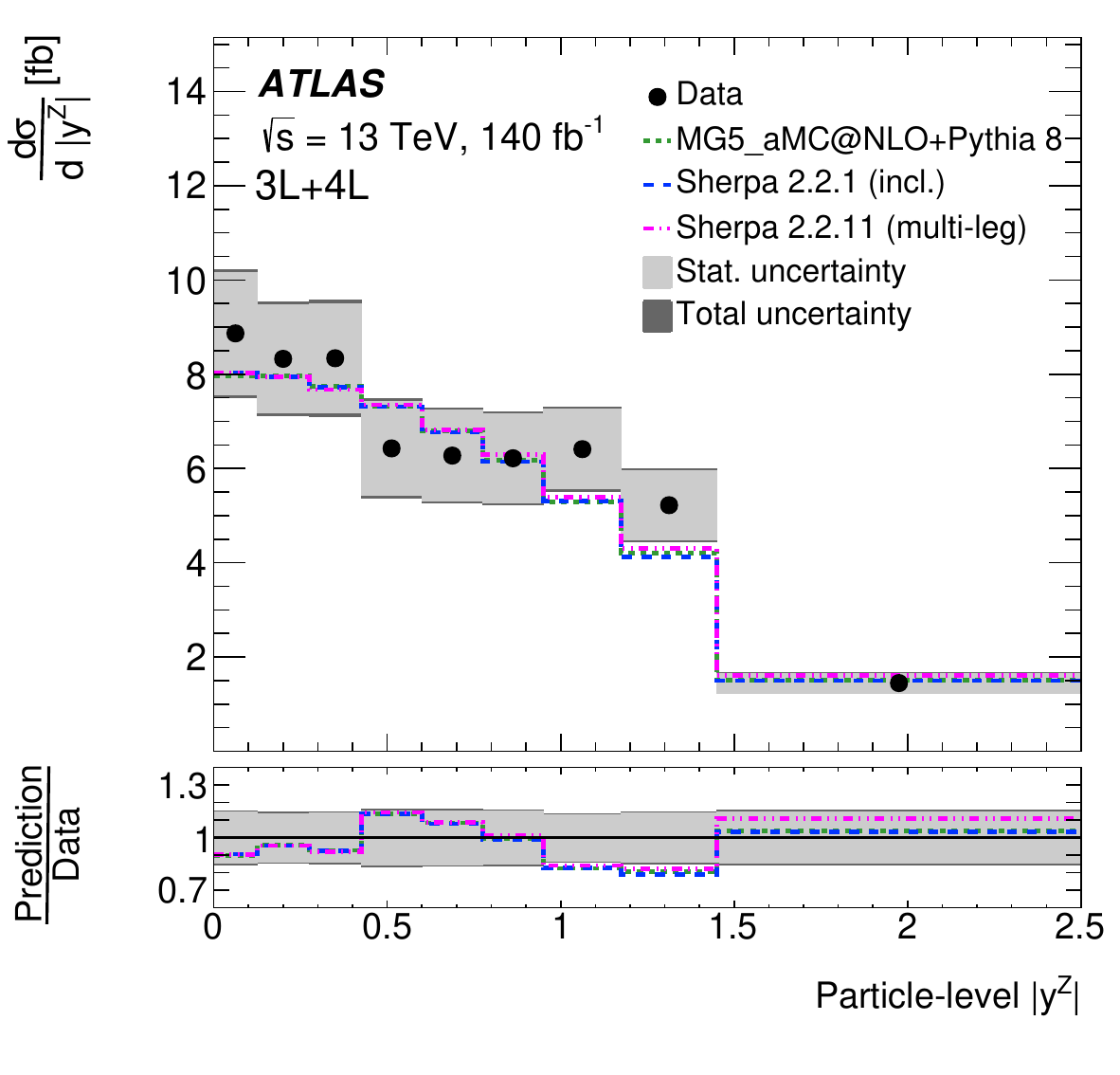}}
\hspace*{0.06\textwidth}
\subfloat[]{\includegraphics[width=0.46\textwidth]{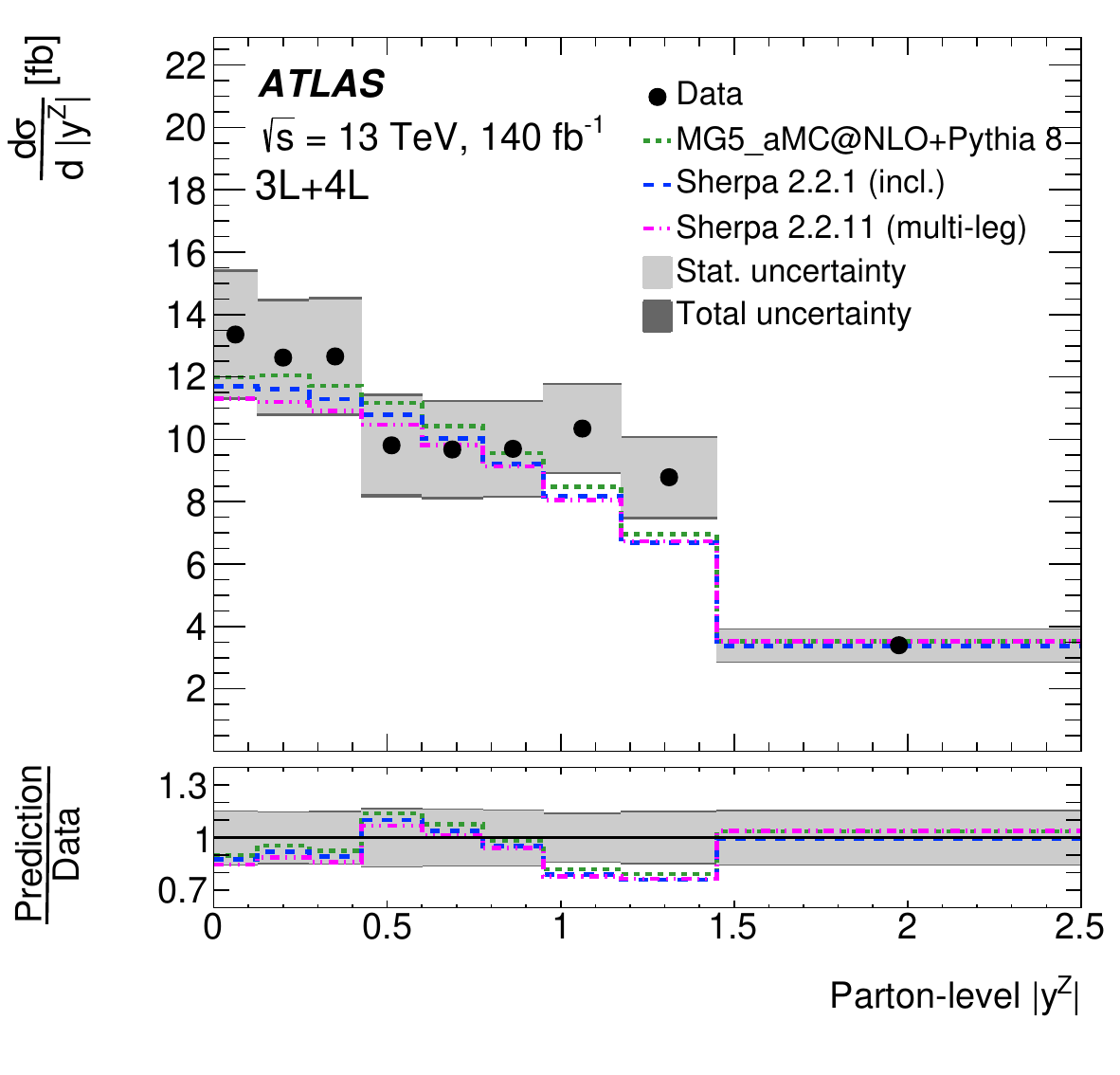}}\\
\subfloat[]{\includegraphics[width=0.46\textwidth]{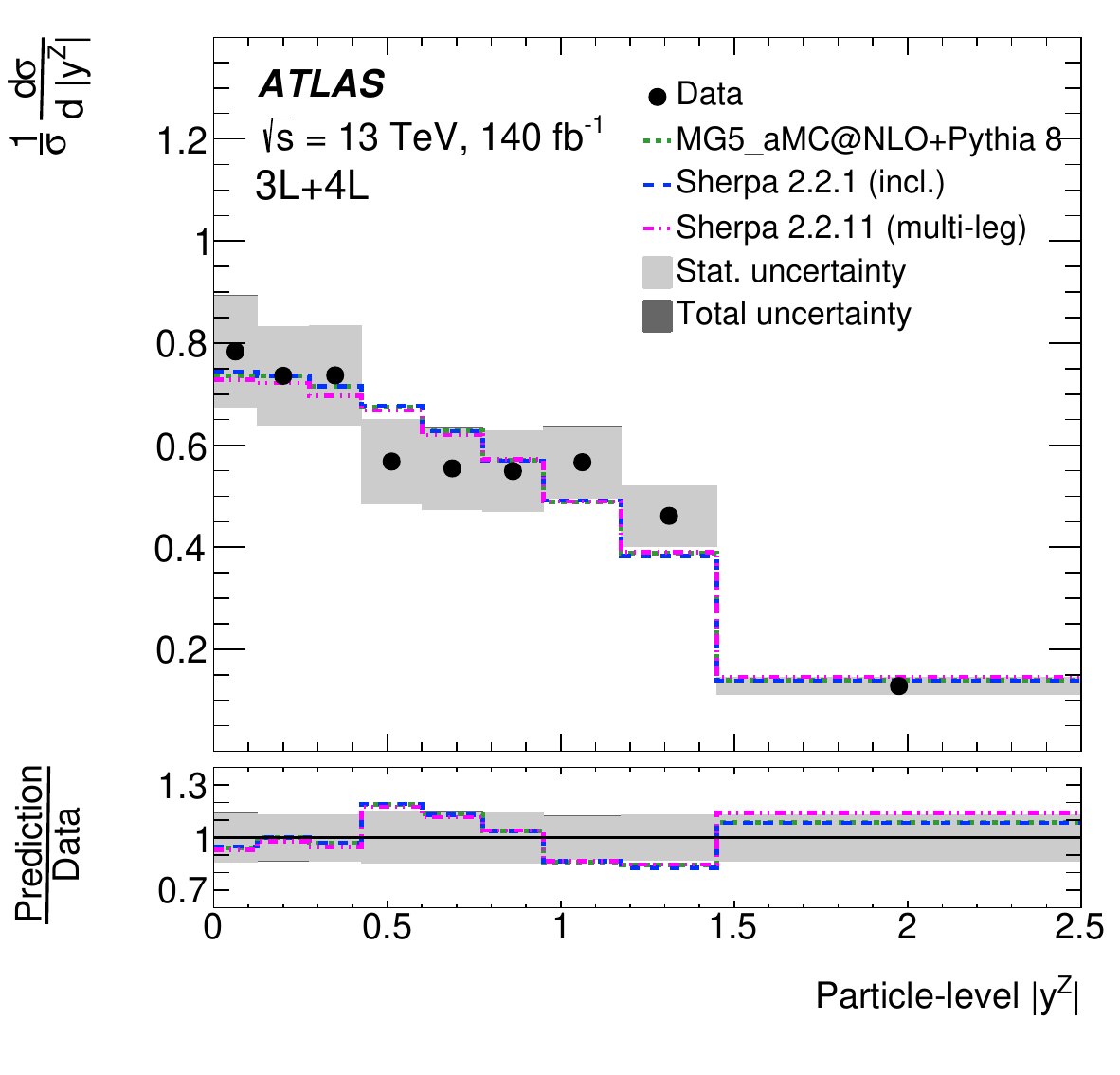}}
\hspace*{0.06\textwidth}
\subfloat[]{\includegraphics[width=0.46\textwidth]{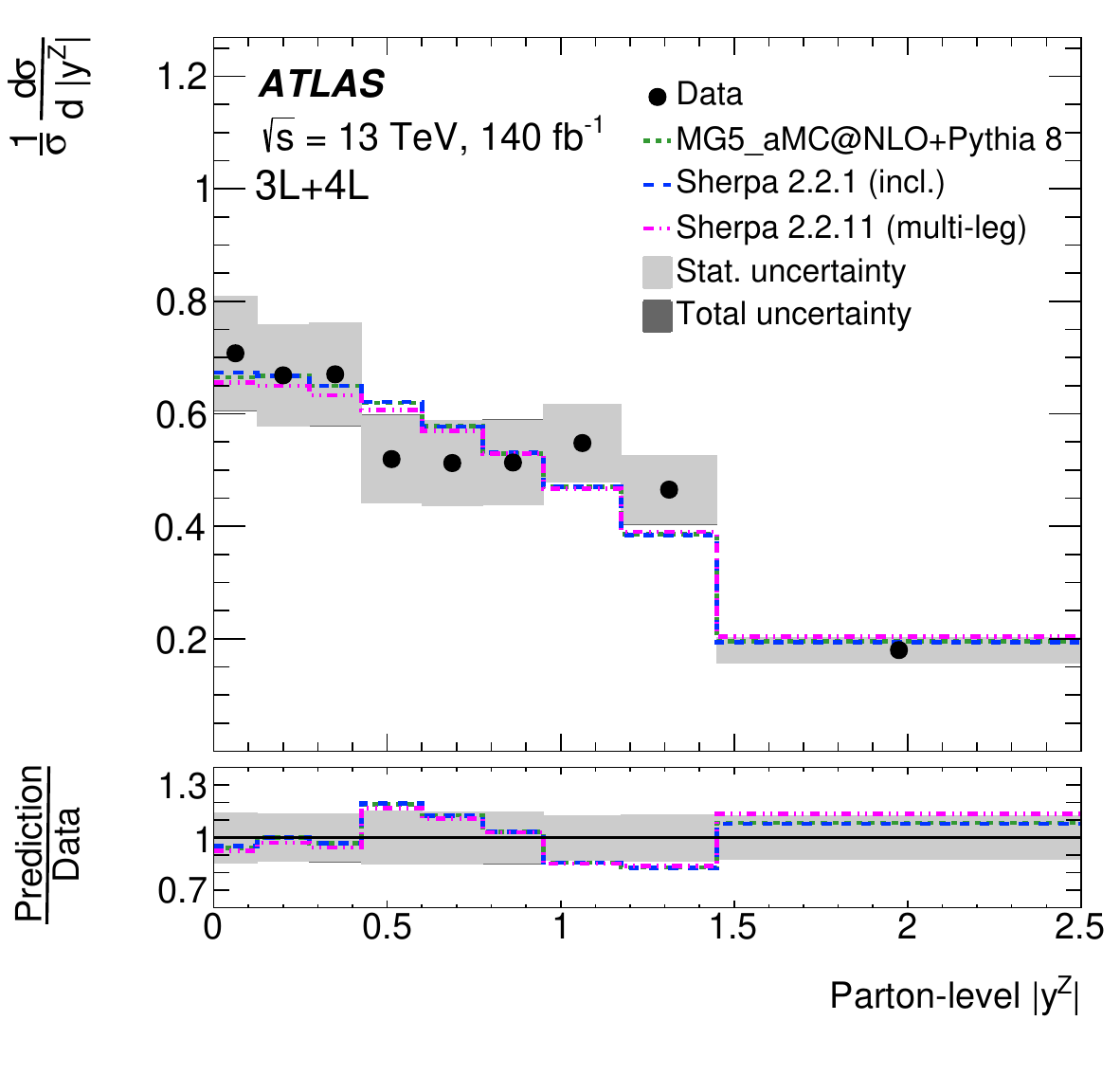}}
\caption{Cross-section measurement of the $\lvert y^{Z}\rvert$ observable in the combination of the $3\ell$ and $4\ell$ channels, absolute and normalised, unfolded to particle level (a,c) and parton level (b,d).}
\label{fig:combined-observed-unfolding-result-yz1}
\end{figure}
 
\begin{figure}[!htb]
\centering
\subfloat[]{\includegraphics[width=0.46\textwidth]{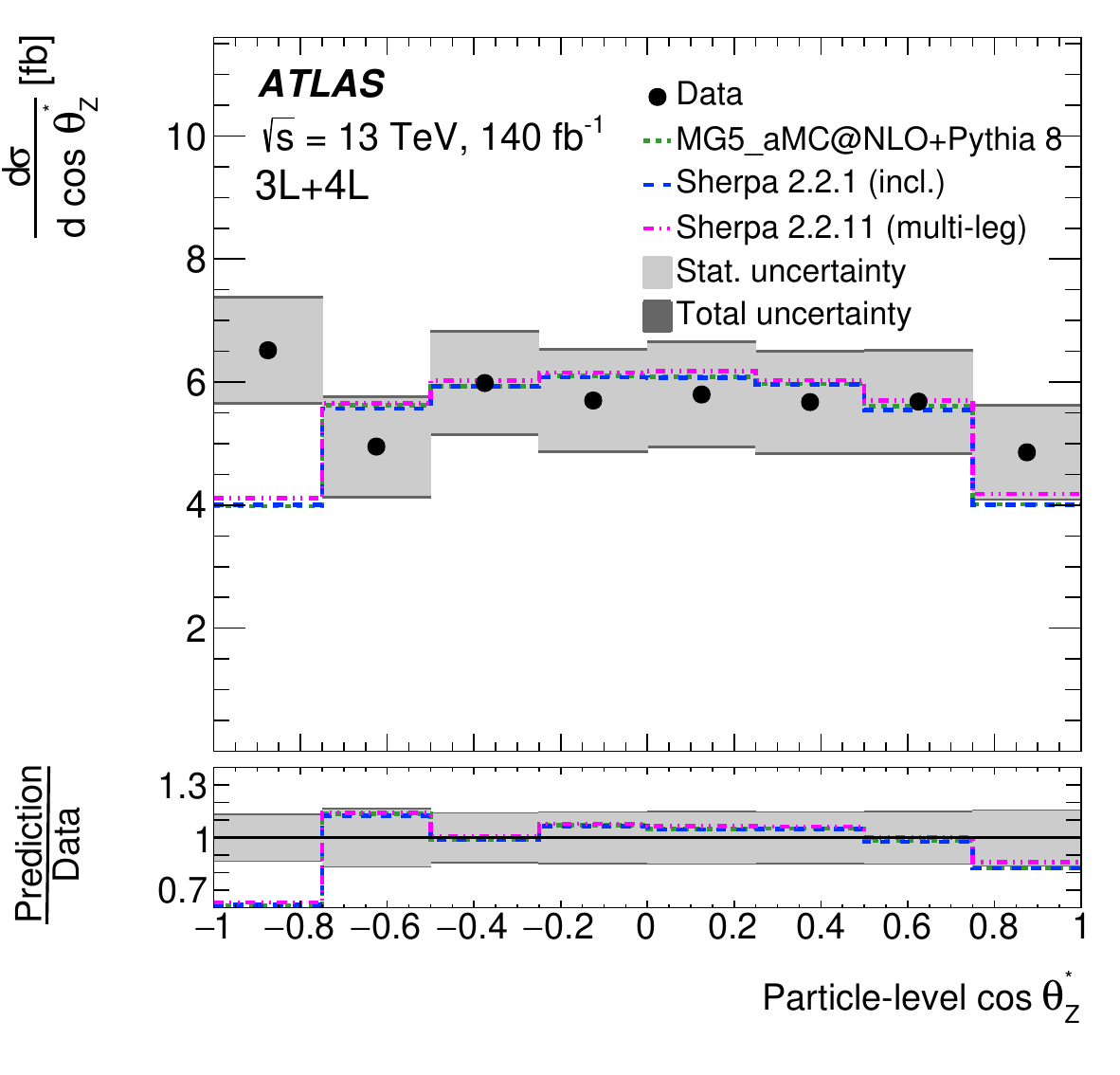}}
\hspace*{0.06\textwidth}
\subfloat[]{\includegraphics[width=0.46\textwidth]{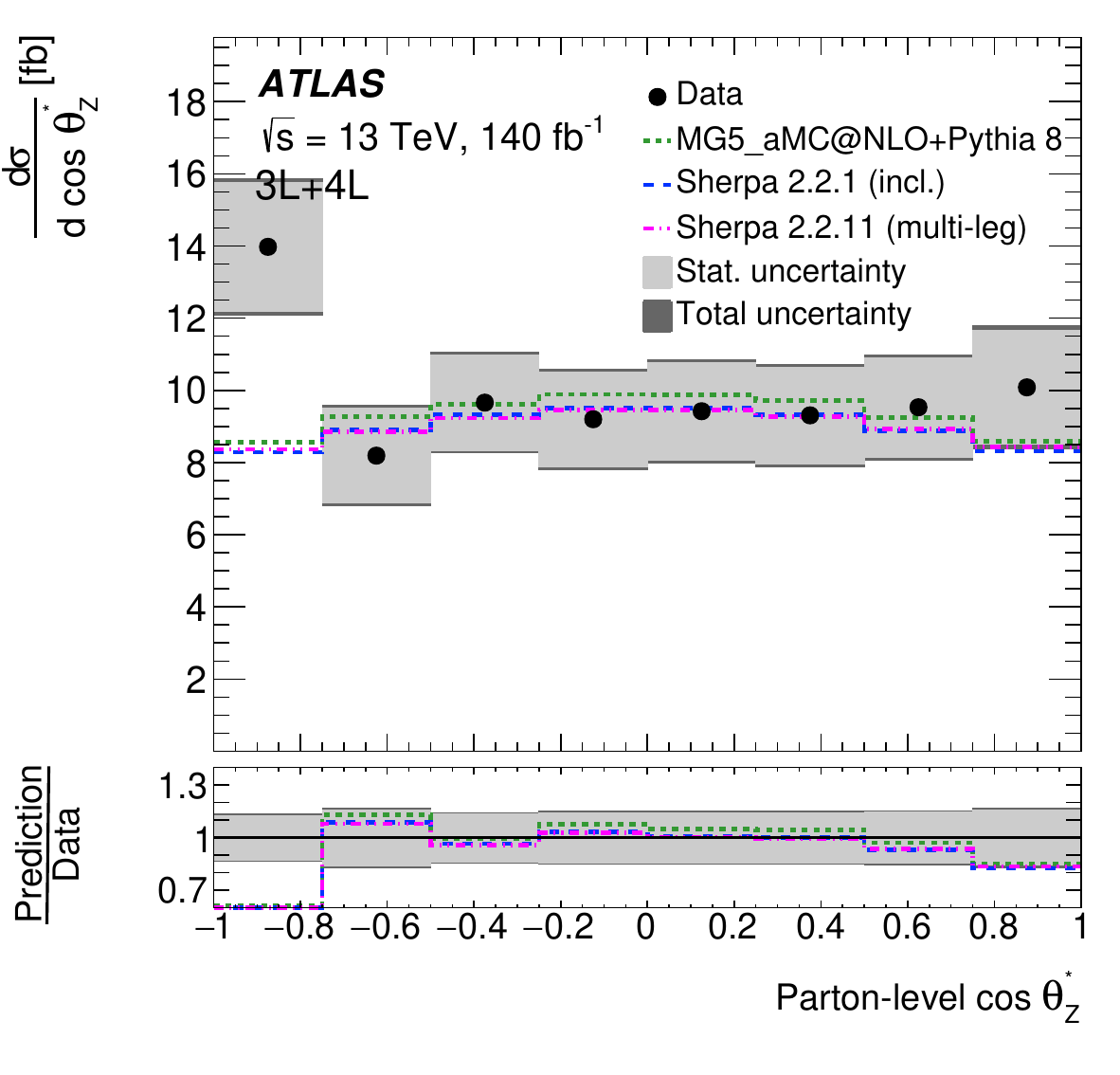}} \\
\subfloat[]{\includegraphics[width=0.46\textwidth]{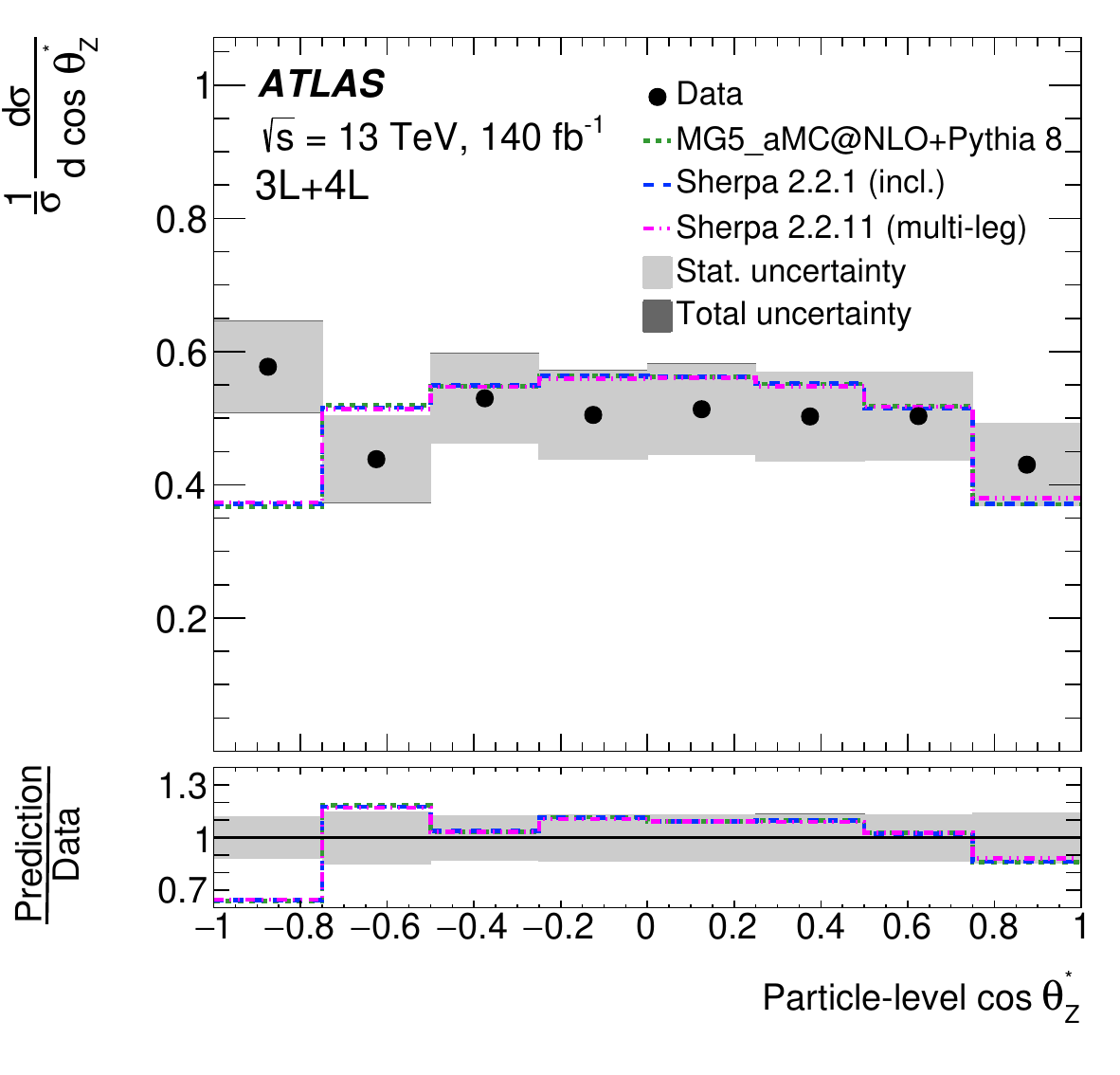}}
\hspace*{0.06\textwidth}
\subfloat[]{\includegraphics[width=0.46\textwidth]{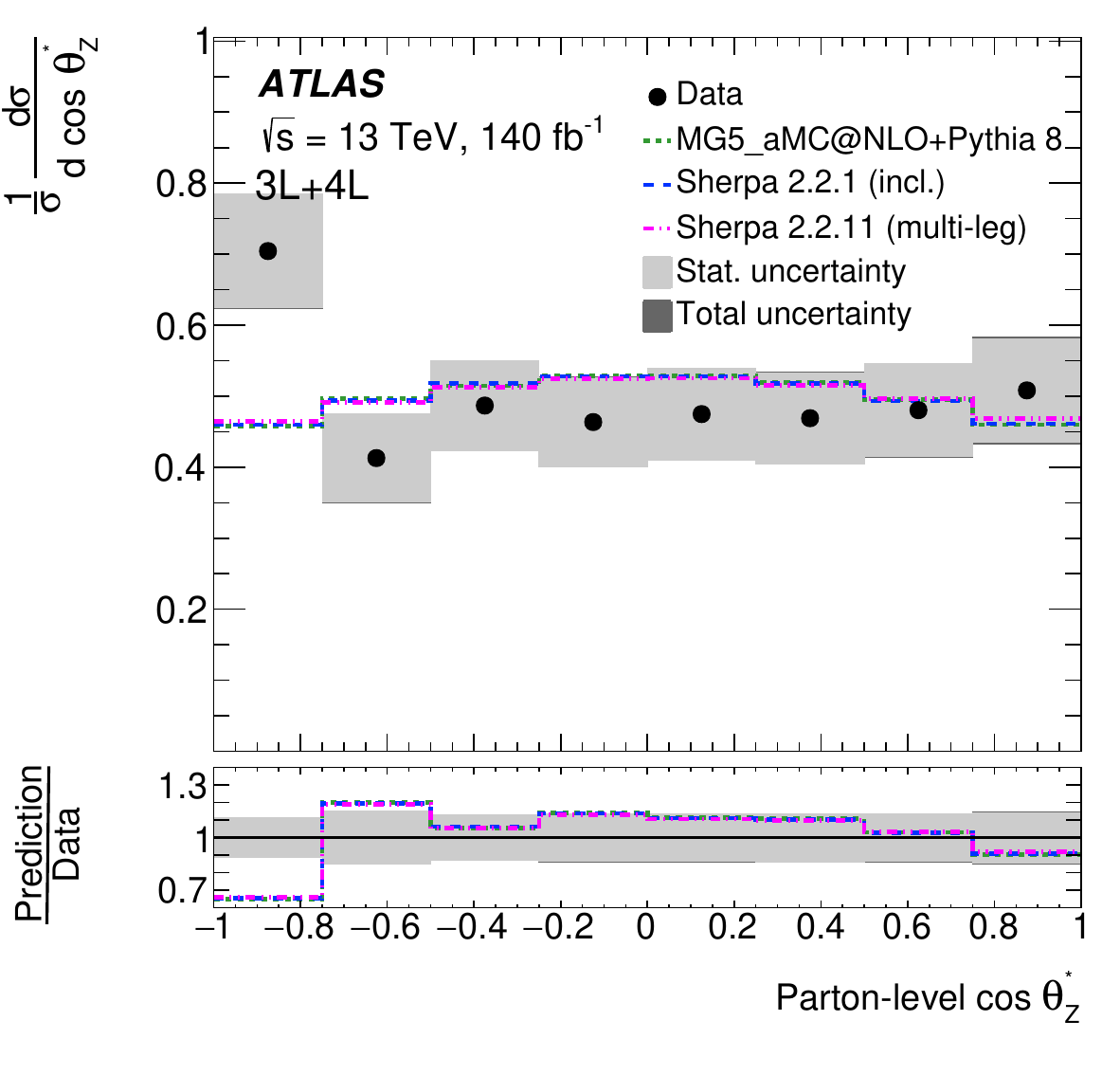}}
\caption{Cross-section measurement of the $\cos{\theta^*_Z}$ observable in the combination of the $3\ell$ and $4\ell$ channels, absolute and normalised, unfolded to particle level (a,c) and parton level (b,d).}
\label{fig:combined-observed-unfolding-result-cos-theta-starZ}
\end{figure}
 
\begin{figure}[!htb]
\centering
\subfloat[]{\includegraphics[width=0.46\textwidth]{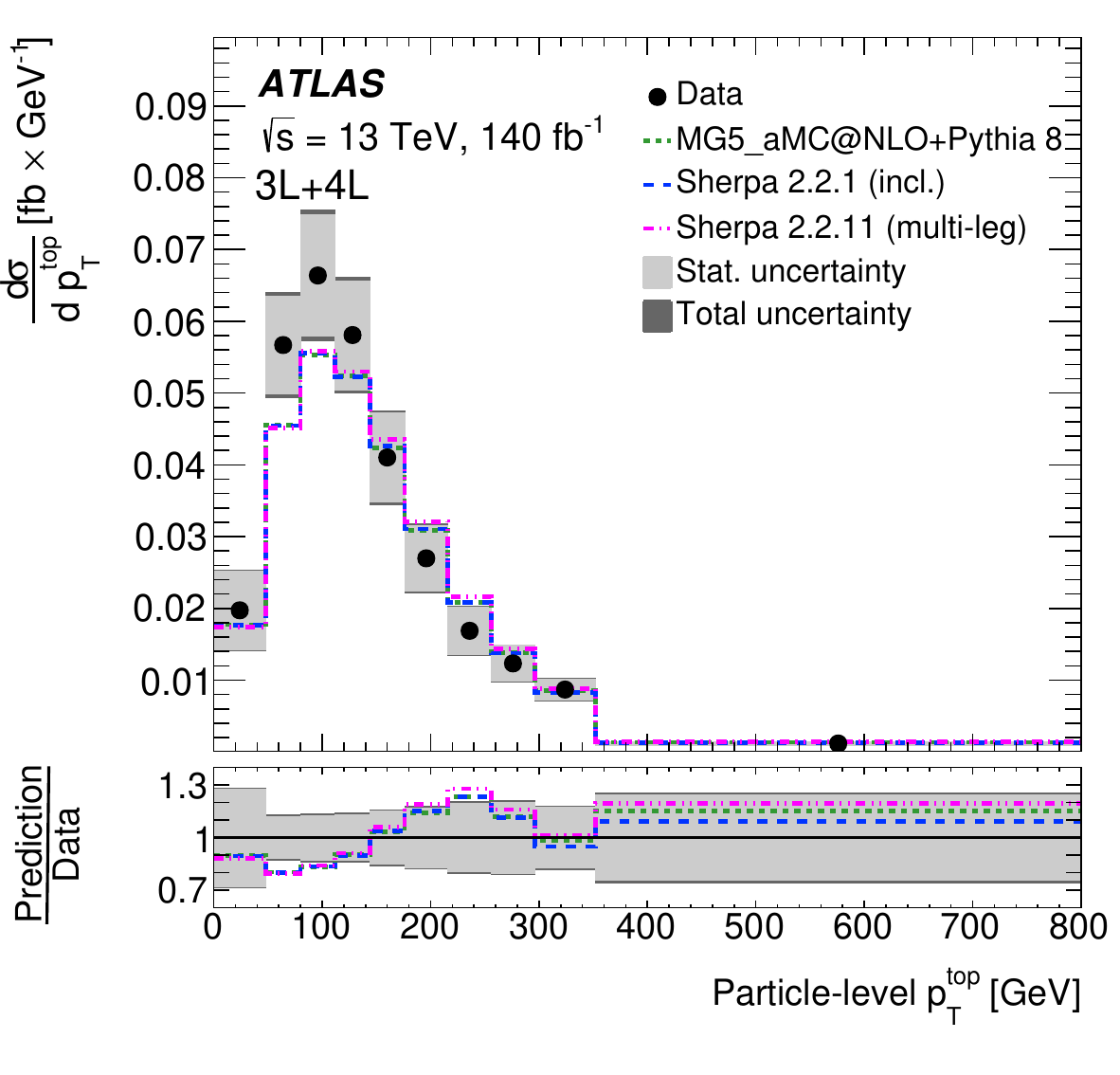}}
\hspace*{0.06\textwidth}
\subfloat[]{\includegraphics[width=0.46\textwidth]{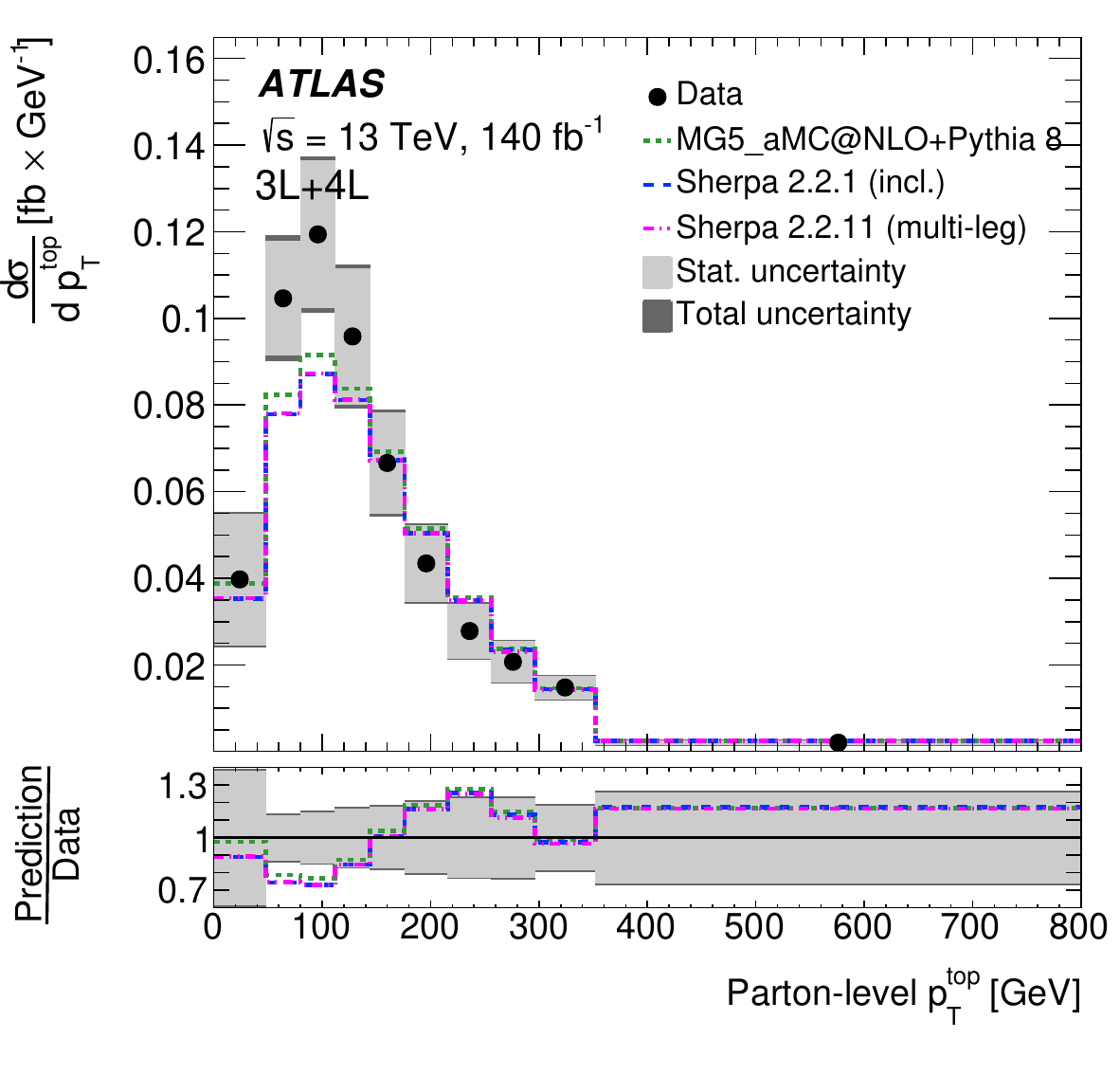}} \\
\subfloat[]{\includegraphics[width=0.46\textwidth]{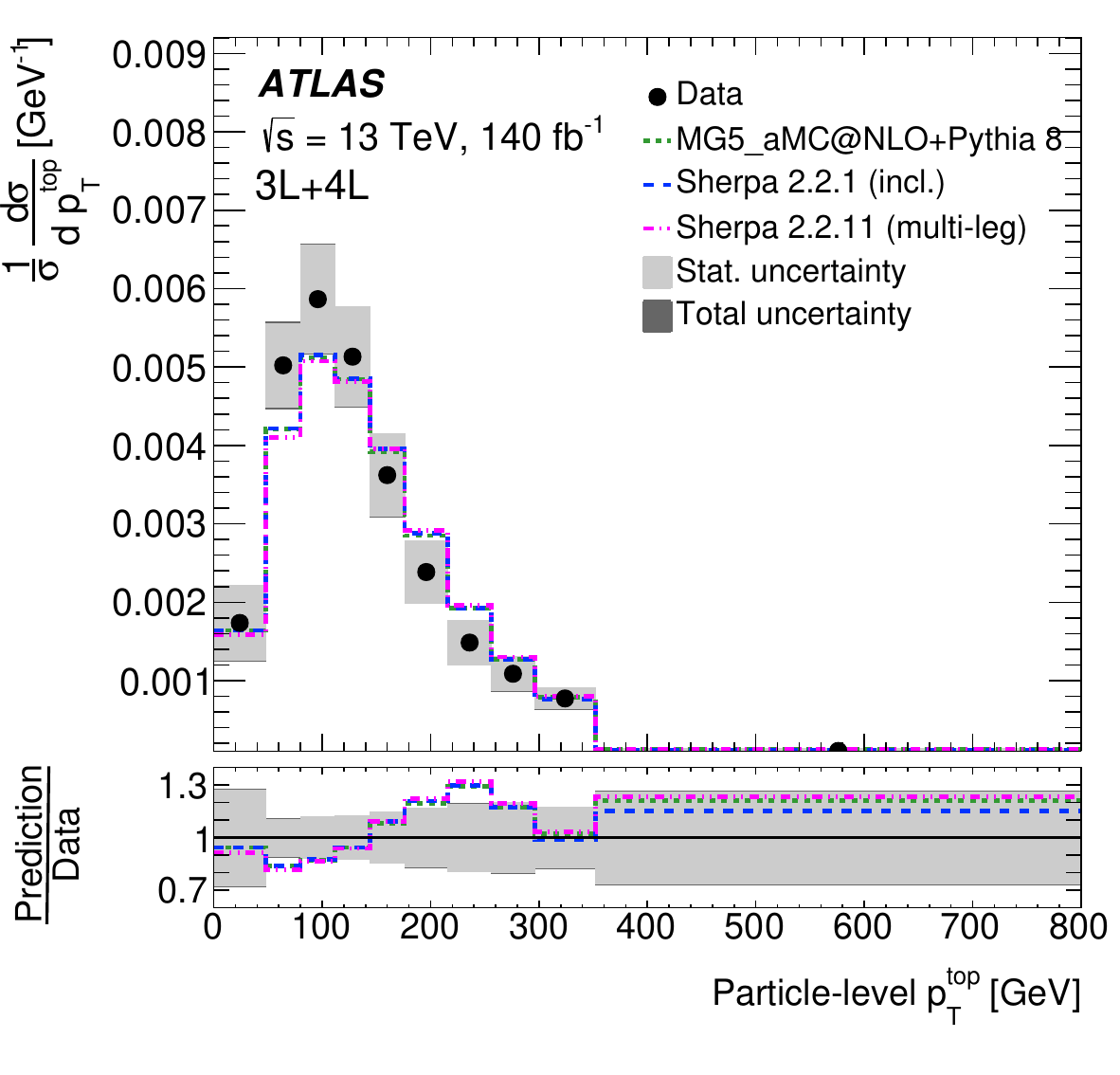}}
\hspace*{0.06\textwidth}
\subfloat[]{\includegraphics[width=0.46\textwidth]{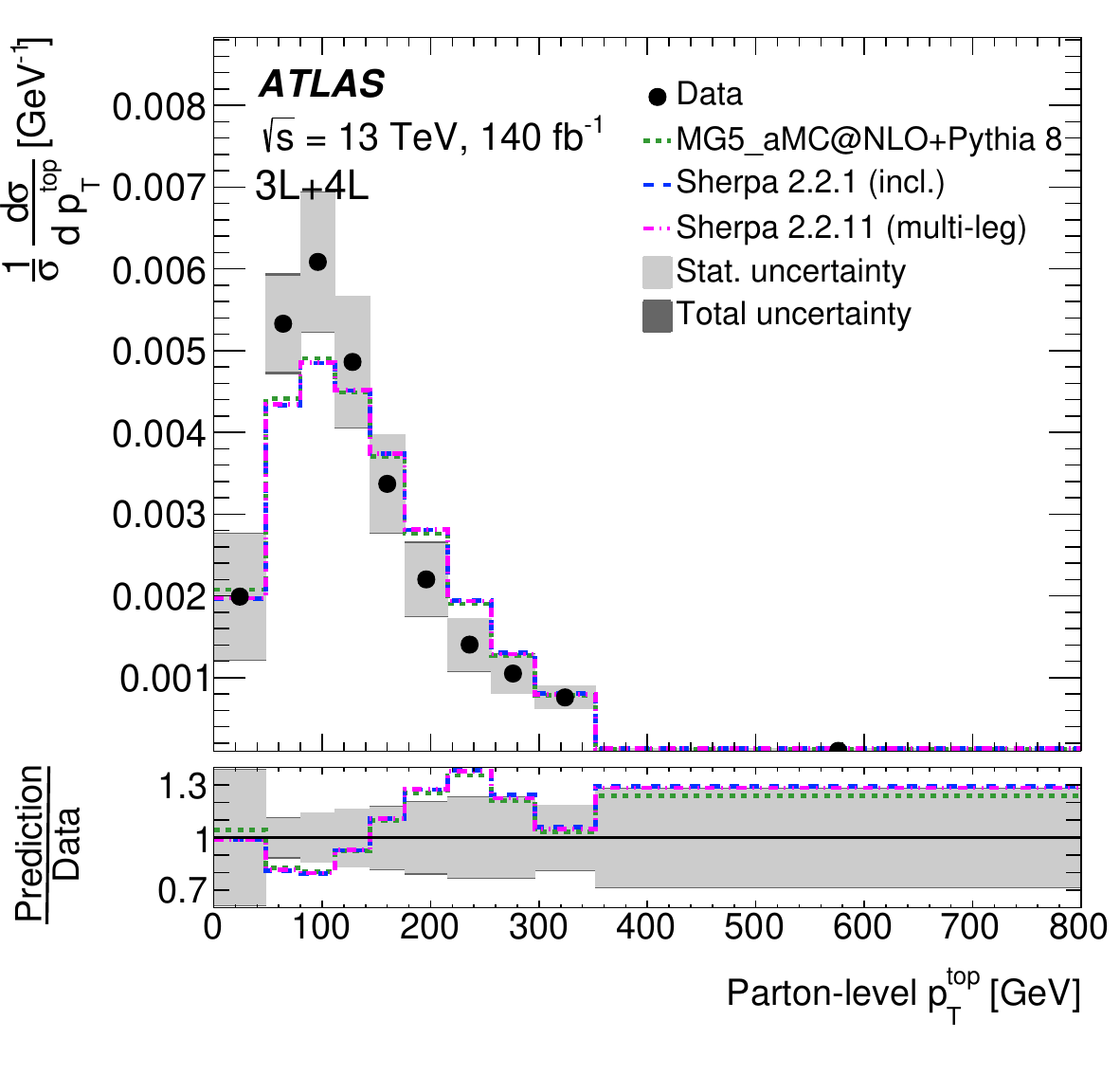}}
\caption{Cross-section measurement of the $\pT^{t}$  observable in the combination of the $3\ell$ and $4\ell$ channels, absolute and normalised, unfolded to particle level (a,c) and parton level (b,d).}
\label{fig:combined-observed-unfolding-result-pT_top-particle}
\end{figure}
 
\begin{figure}[!htb]
\centering
\subfloat[]{\includegraphics[width=0.46\textwidth]{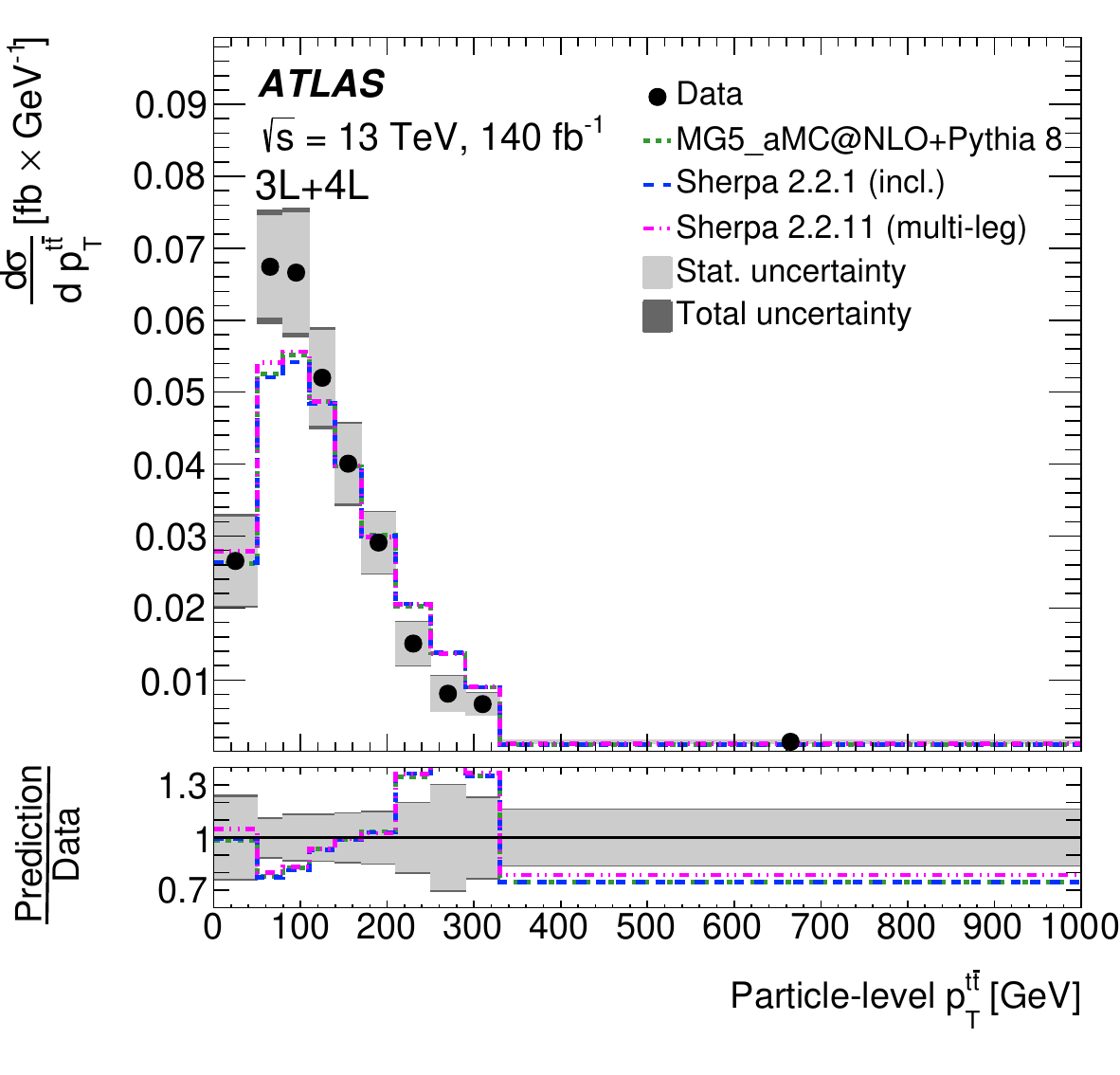}}
\hspace*{0.06\textwidth}
\subfloat[]{\includegraphics[width=0.46\textwidth]{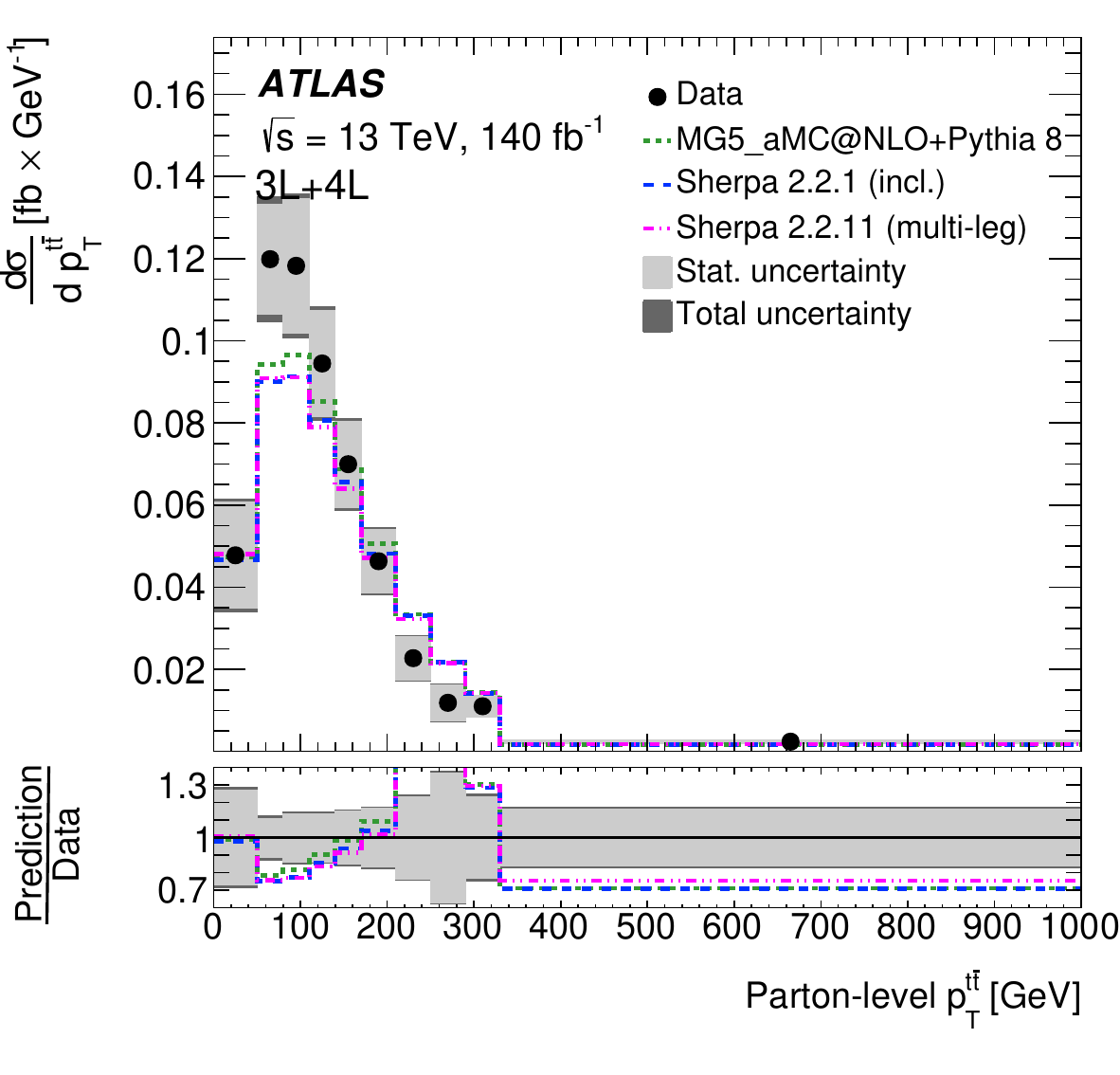}} \\
\subfloat[]{\includegraphics[width=0.46\textwidth]{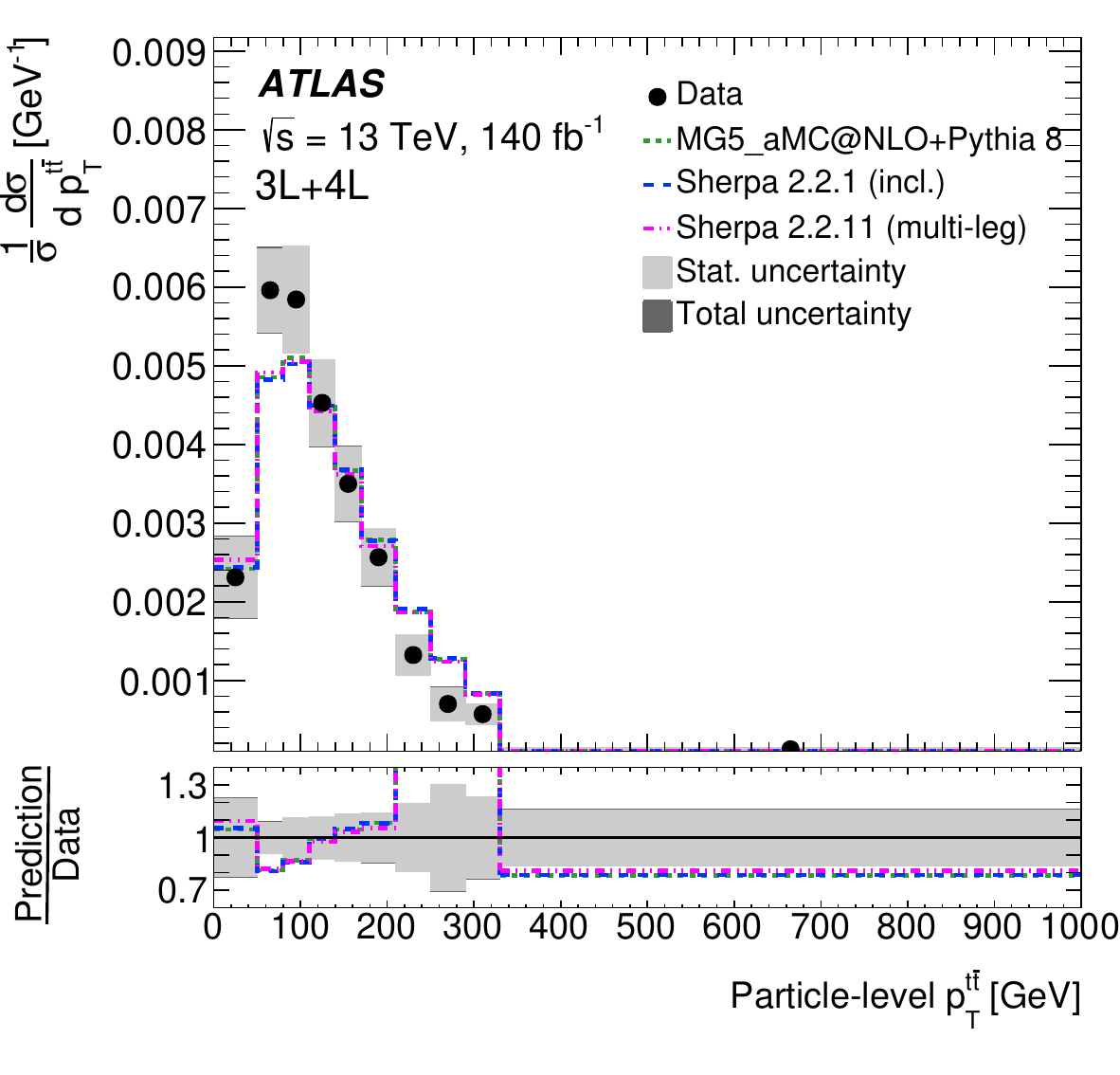}}
\hspace*{0.06\textwidth}
\subfloat[]{\includegraphics[width=0.46\textwidth]{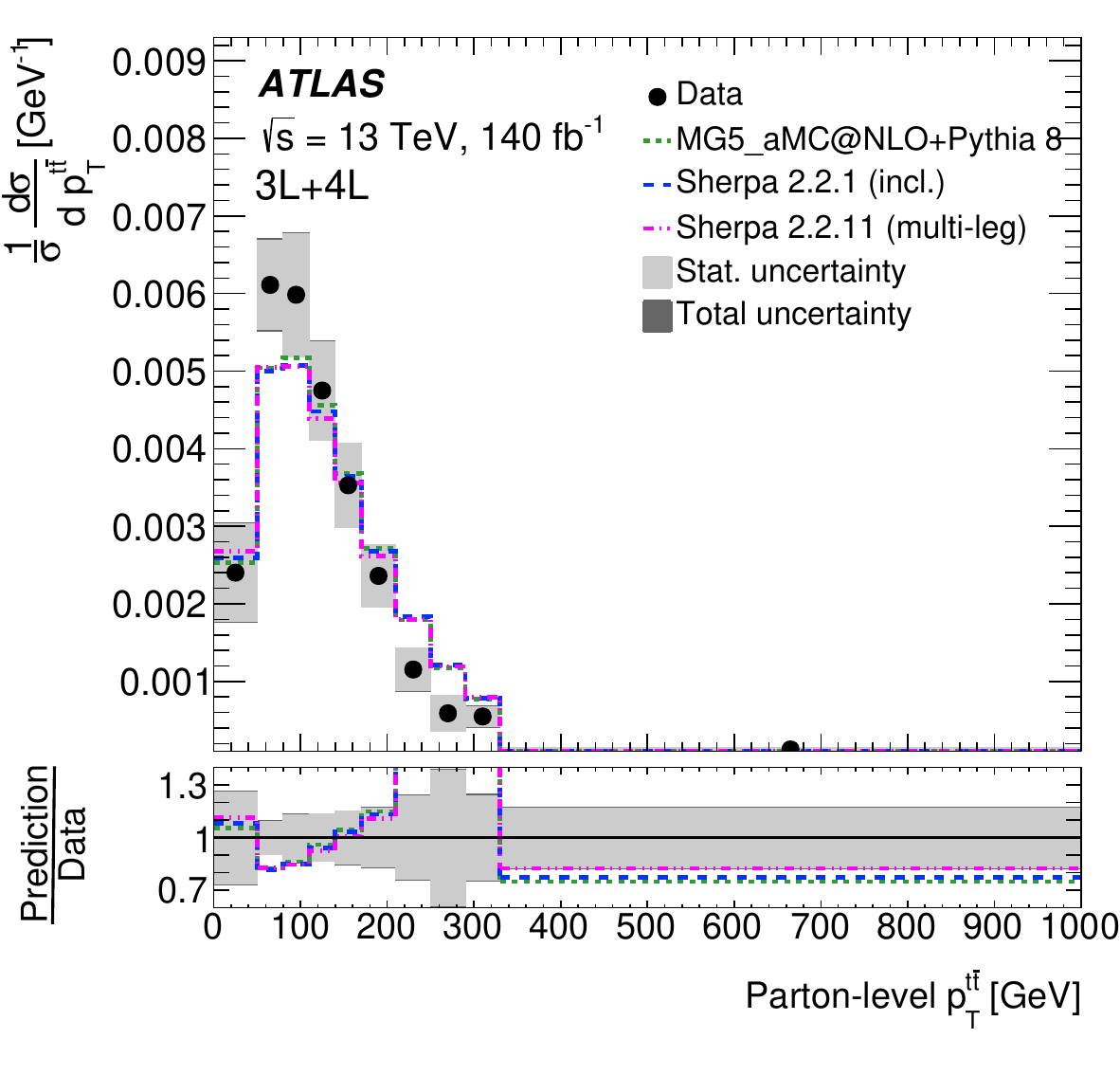}}
\caption{Cross-section measurement of the $\pT^{\ttbar}$  observable in the combination of the $3\ell$ and $4\ell$ channels, absolute and normalised, unfolded to particle level (a,c) and parton level (b,d).}
\end{figure}
 
\begin{figure}[!htb]
\centering
\subfloat[]{\includegraphics[width=0.46\textwidth]{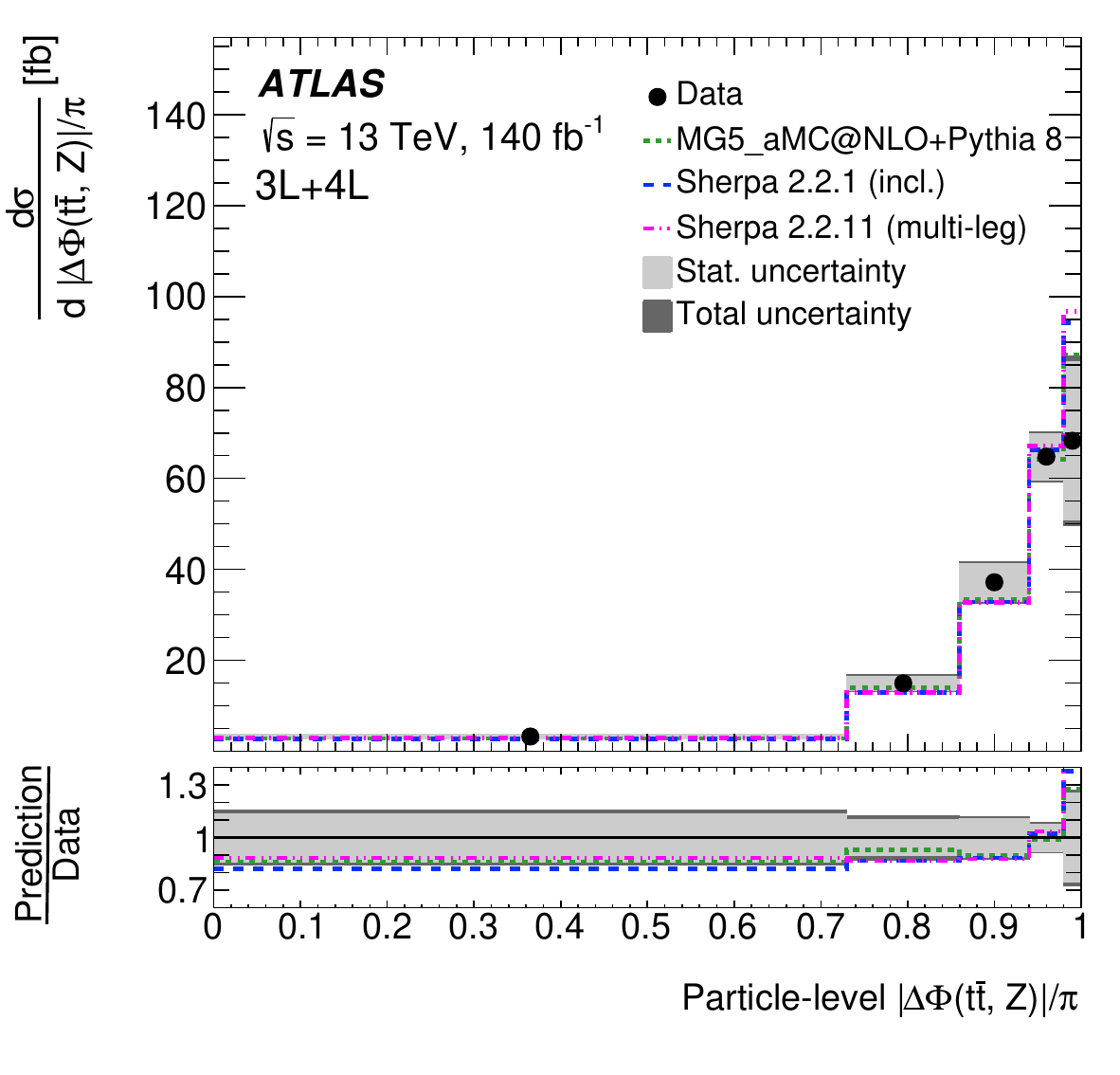}}
\hspace*{0.06\textwidth}
\subfloat[]{\includegraphics[width=0.46\textwidth]{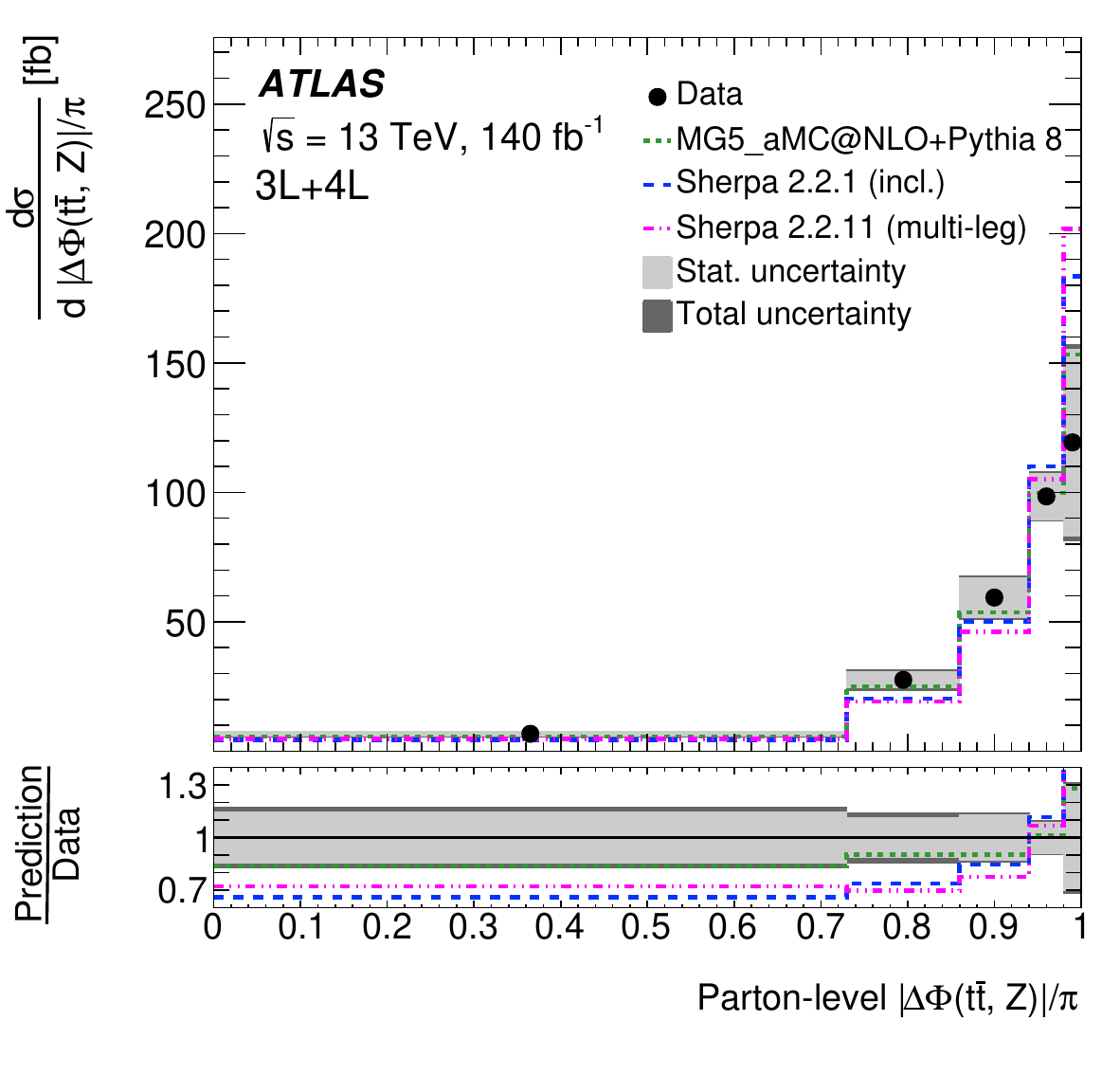}} \\
\subfloat[]{\includegraphics[width=0.46\textwidth]{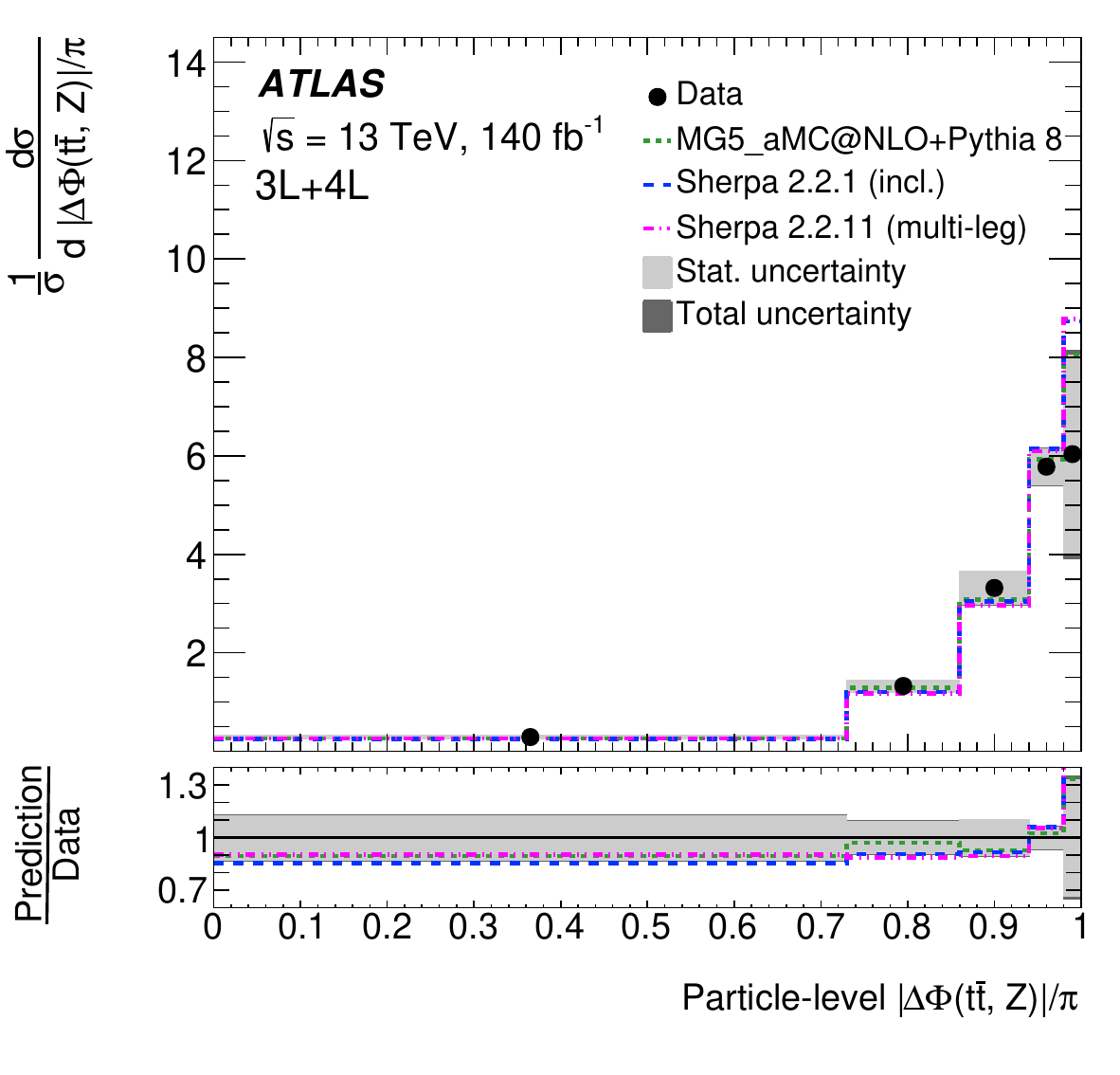}}
\hspace*{0.06\textwidth}
\subfloat[]{\includegraphics[width=0.46\textwidth]{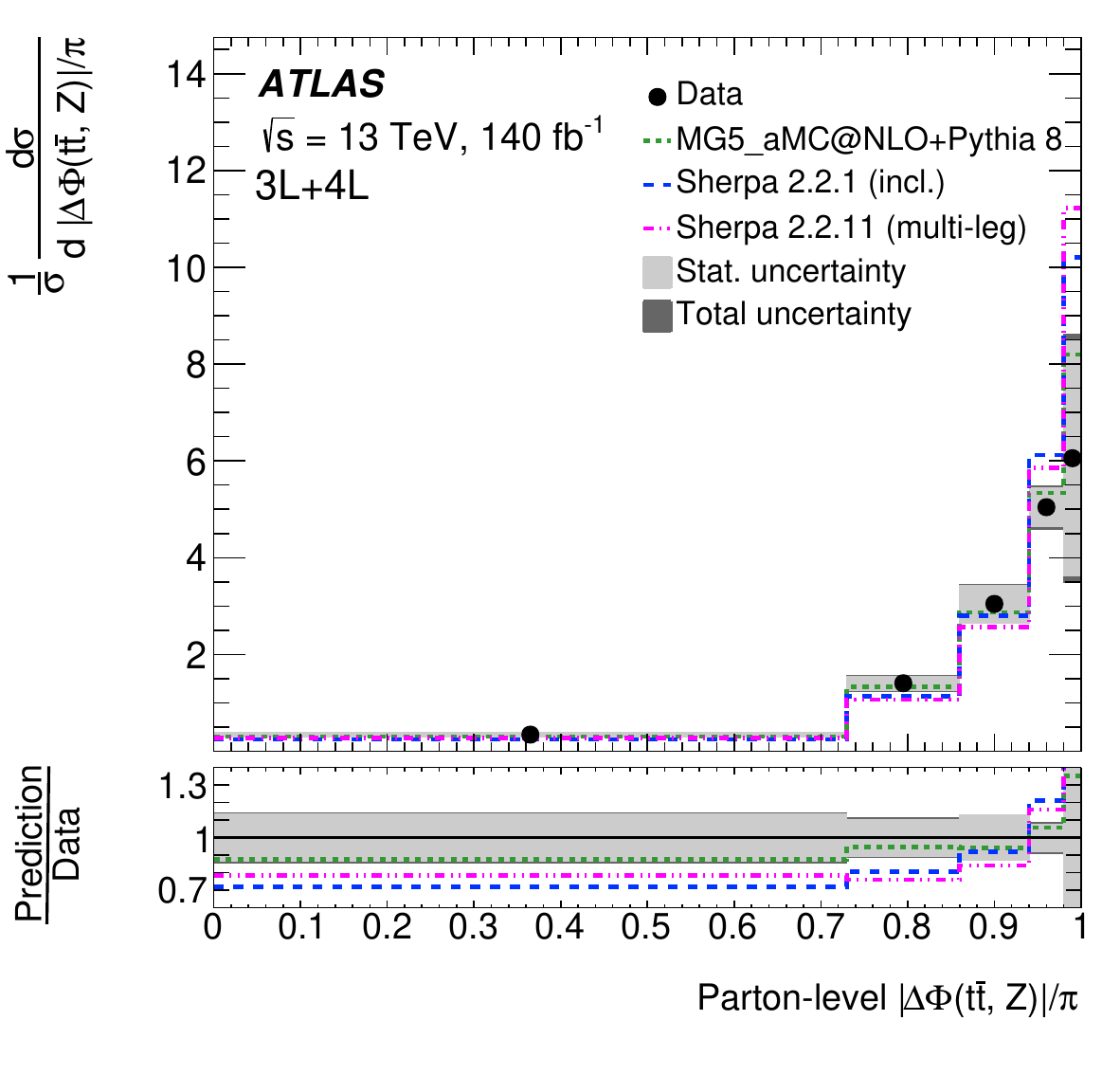}}
\caption{Cross-section measurement of the $\lvert\Delta\Phi({\ttbar},\Zboson)\rvert/\pi$  observable in the combination of the $3\ell$ and $4\ell$ channels, absolute and normalised, unfolded to particle level (a,c) and parton level (b,d).}
\end{figure}
 
\begin{figure}[!htb]
\centering
\subfloat[]{\includegraphics[width=0.46\textwidth]{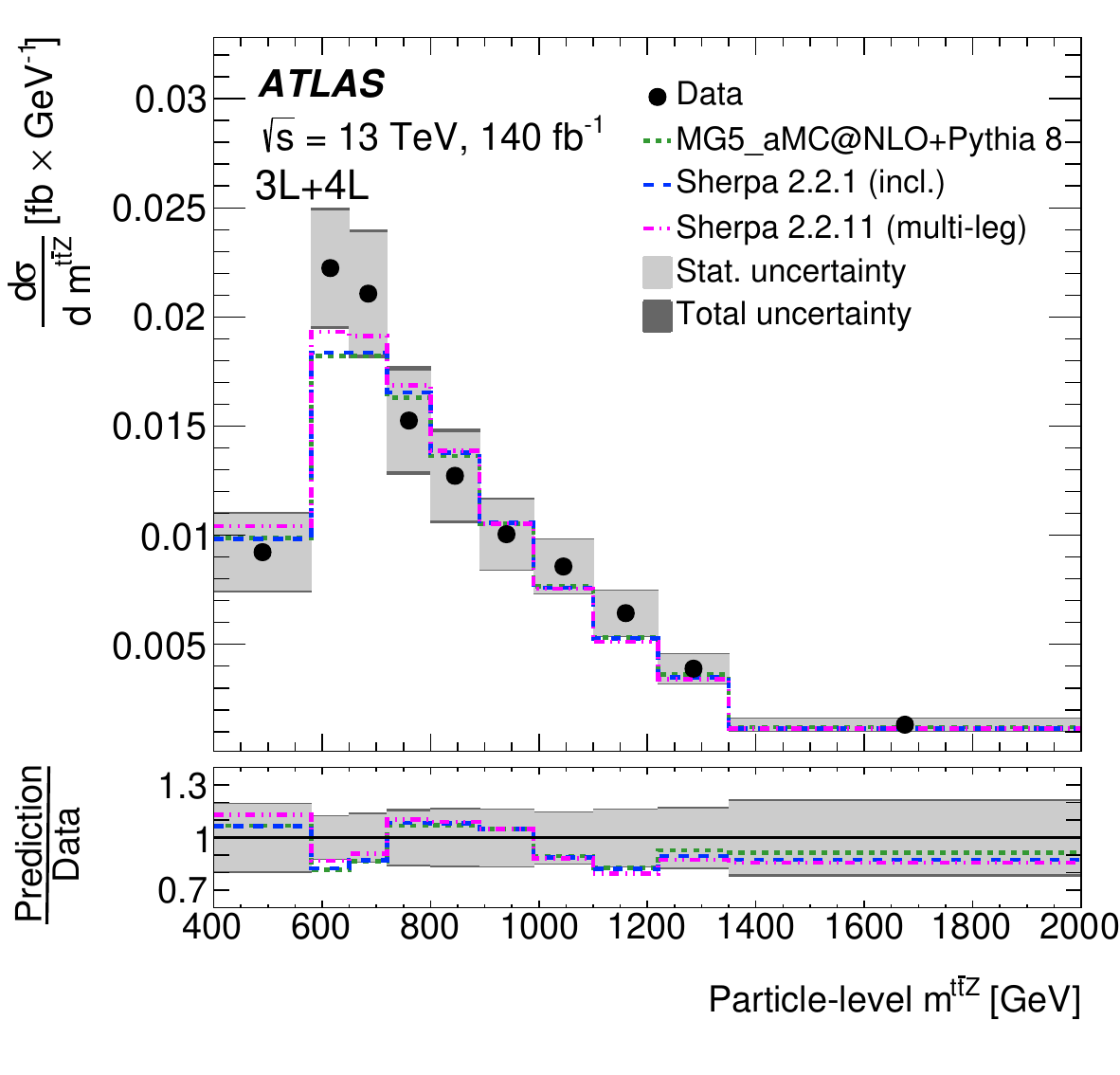}}
\hspace*{0.06\textwidth}
\subfloat[]{\includegraphics[width=0.46\textwidth]{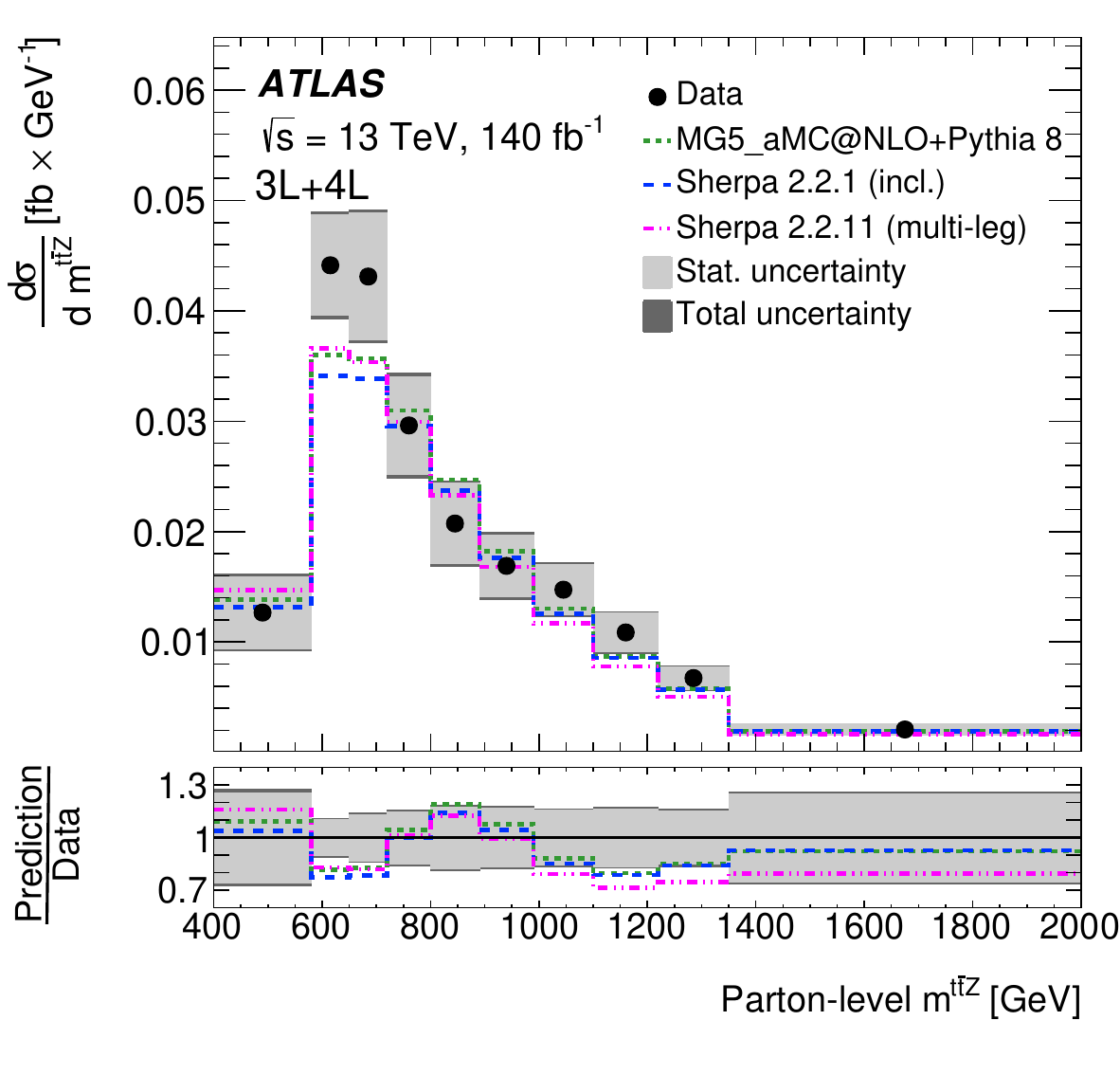}} \\
\subfloat[]{\includegraphics[width=0.46\textwidth]{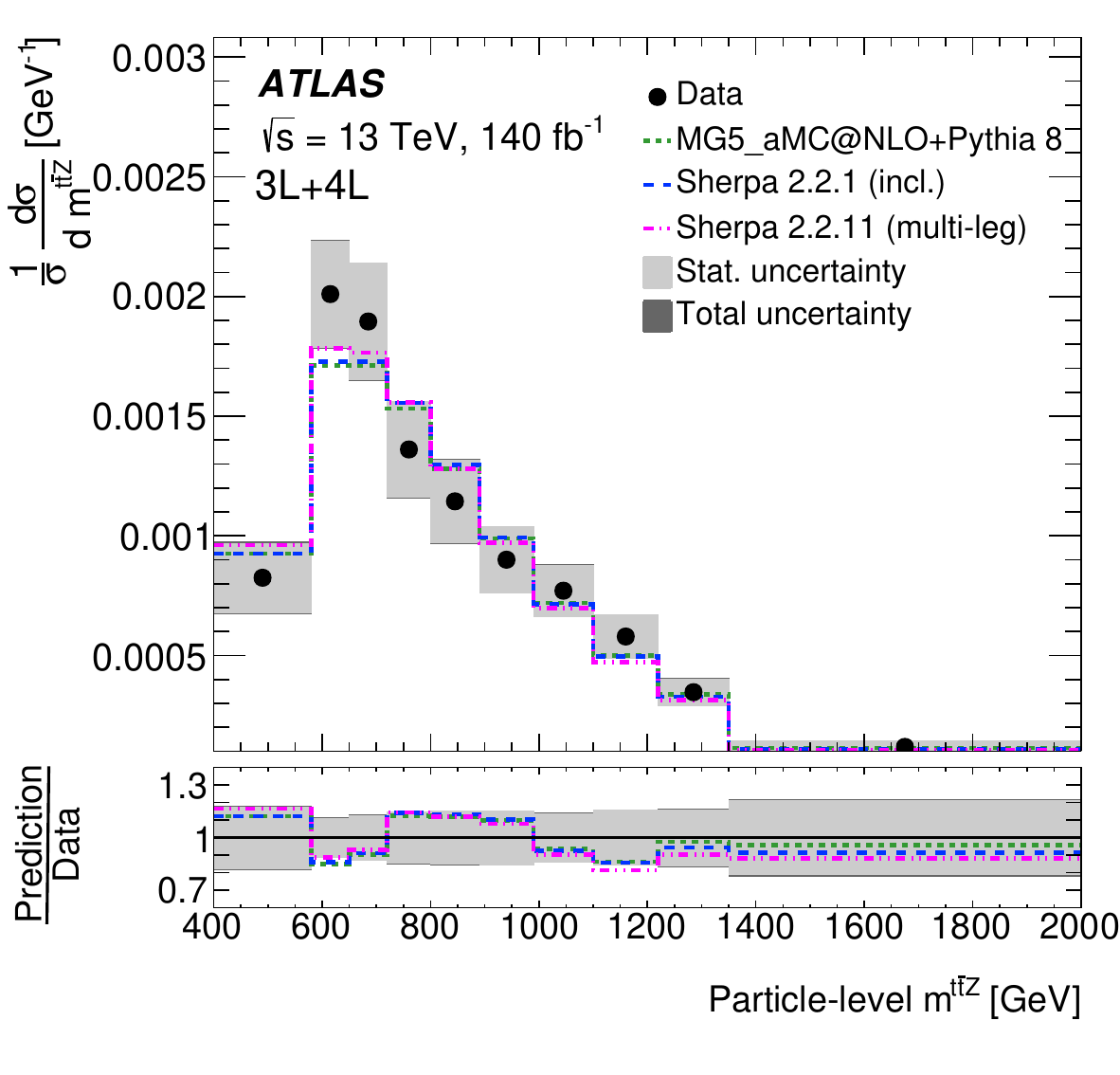}}
\hspace*{0.06\textwidth}
\subfloat[]{\includegraphics[width=0.46\textwidth]{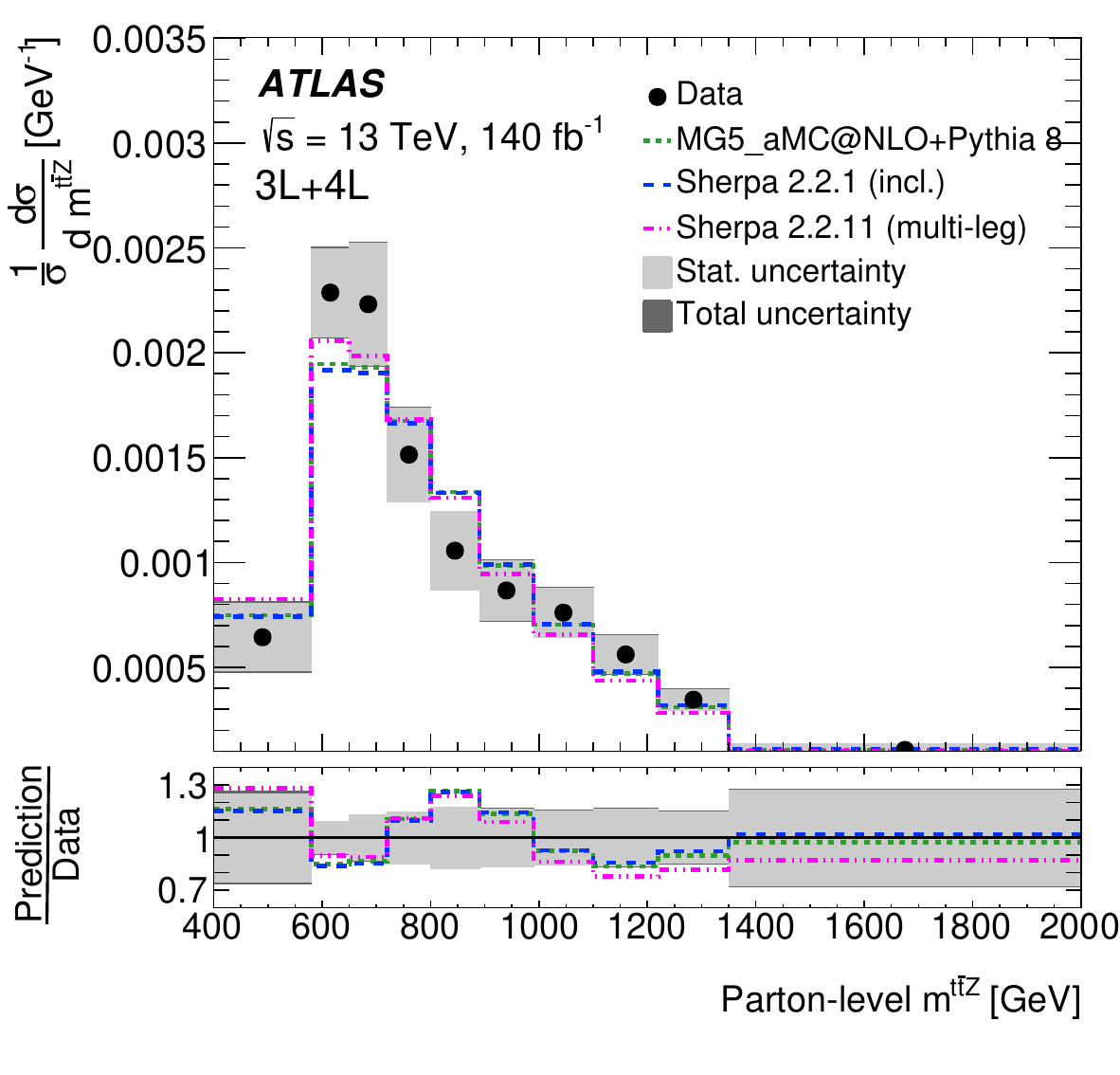}}
\caption{Cross-section measurement of the $m^{\ttZ}$  observable in the combination of the $3\ell$ and $4\ell$ channels, absolute and normalised, unfolded to particle level (a,c) and parton level (b,d).}
\end{figure}
 
\begin{figure}[!htb]
\centering
\subfloat[]{\includegraphics[width=0.46\textwidth]{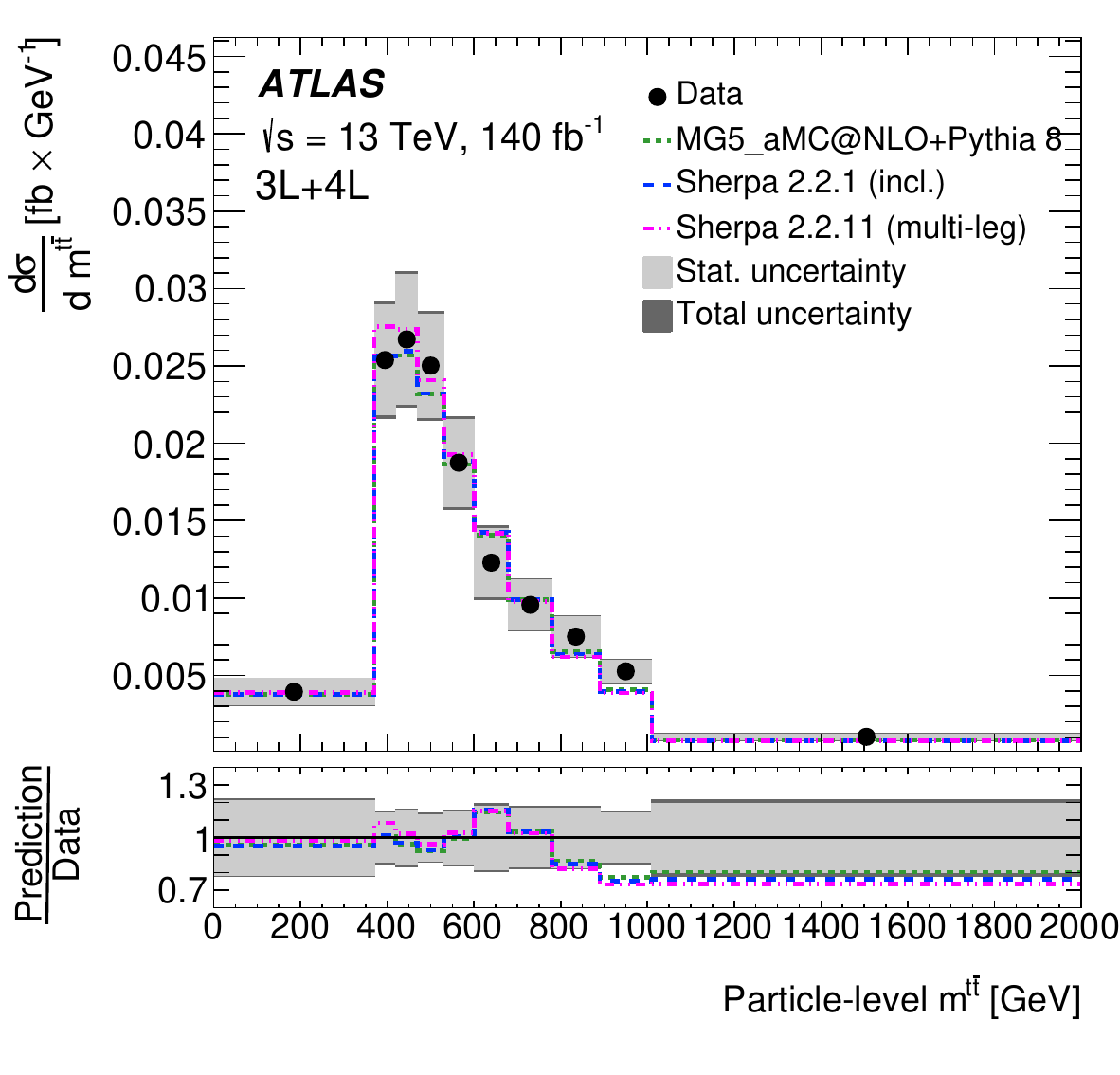}}
\hspace*{0.06\textwidth}
\subfloat[]{\includegraphics[width=0.46\textwidth]{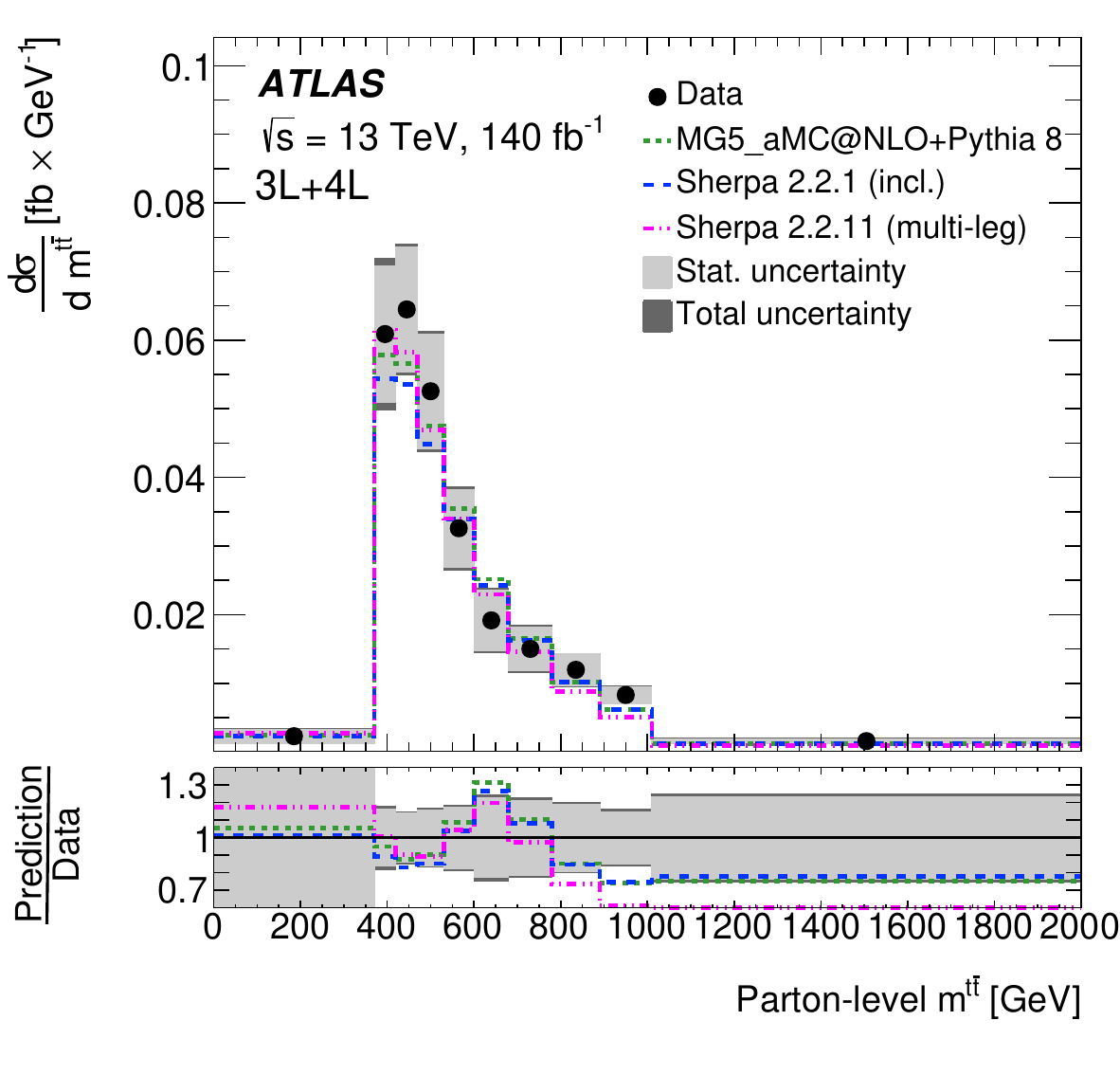}} \\
\subfloat[]{\includegraphics[width=0.46\textwidth]{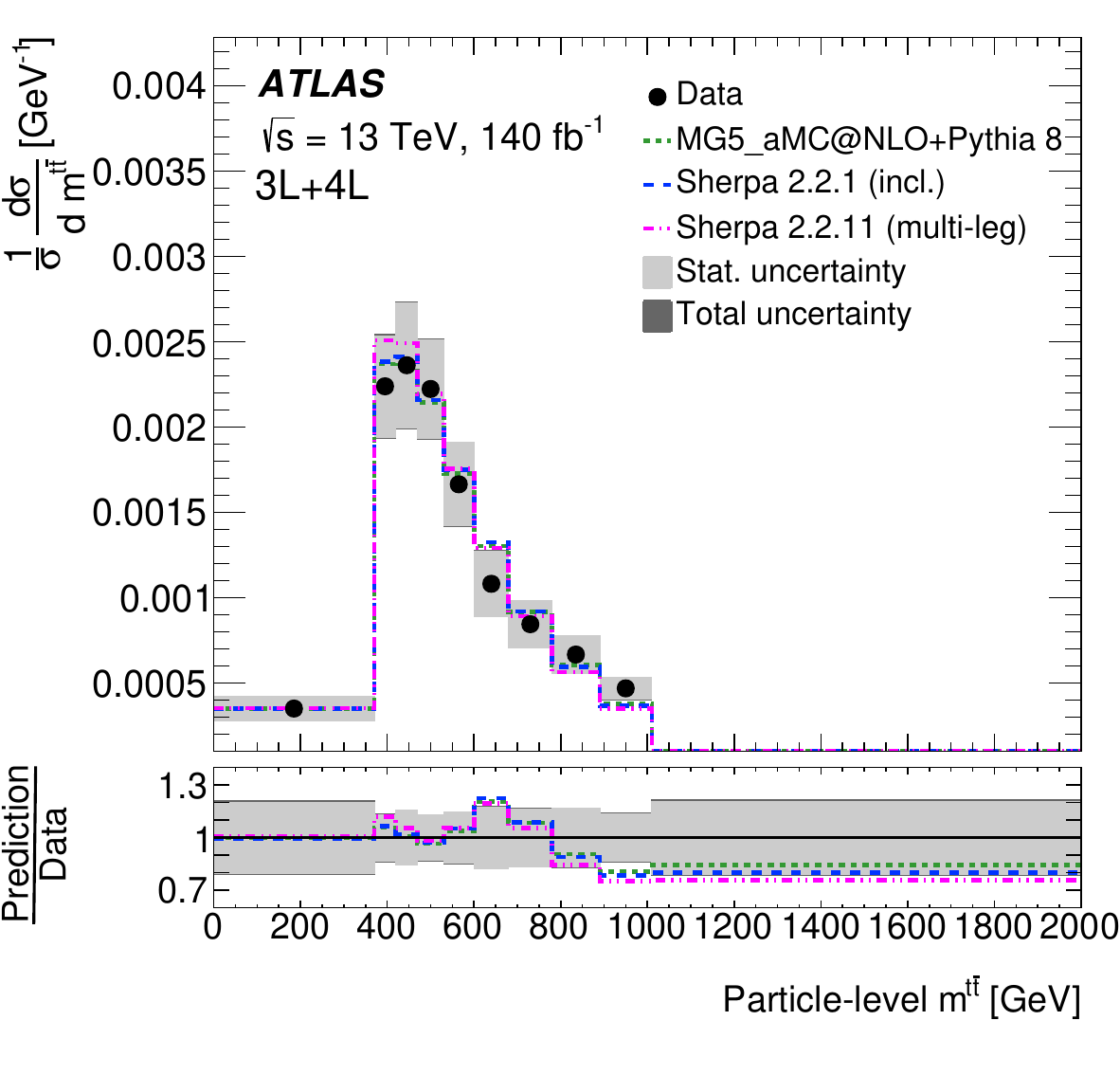}}
\hspace*{0.06\textwidth}
\subfloat[]{\includegraphics[width=0.46\textwidth]{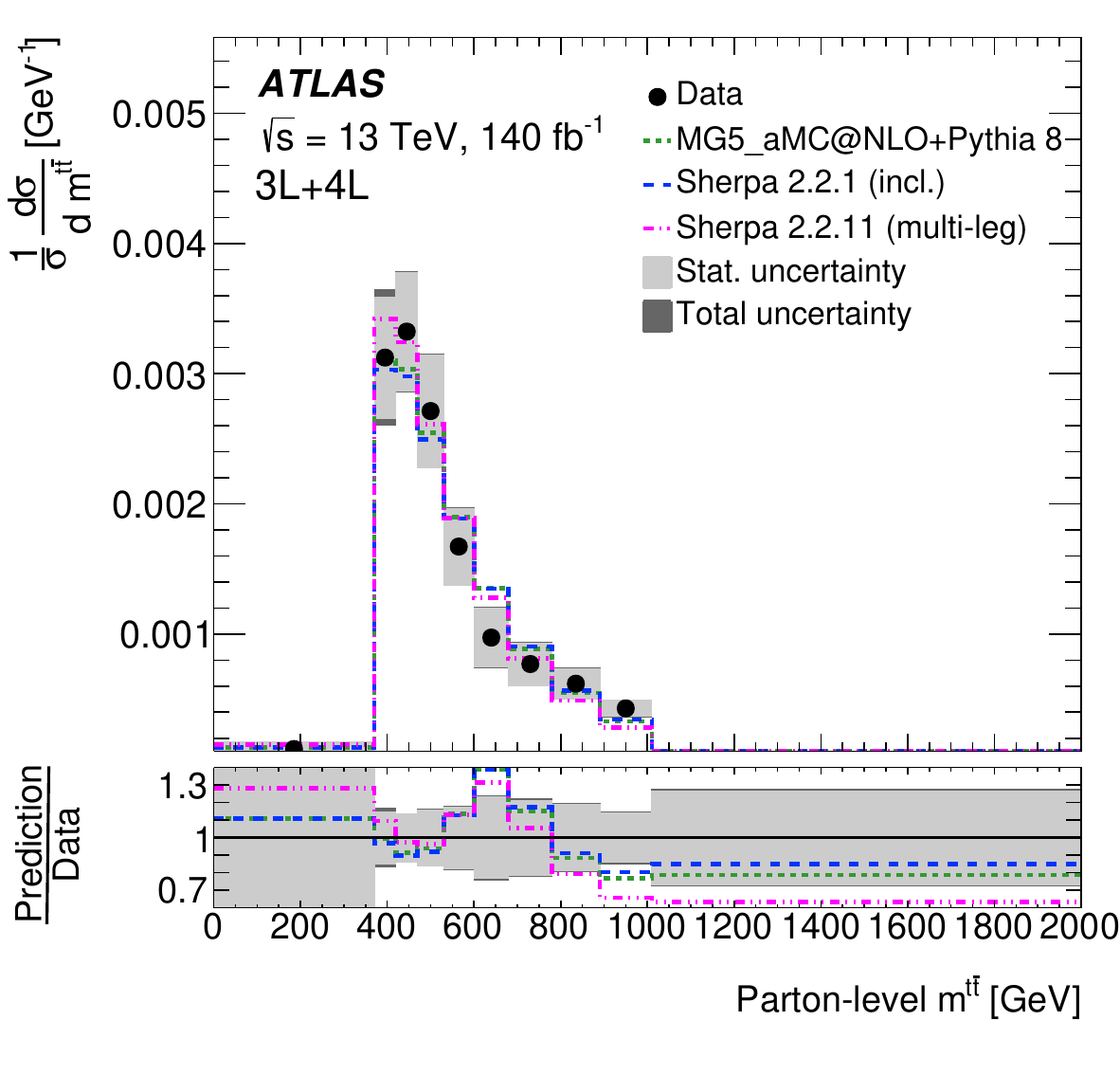}}
\caption{Cross-section measurement of the $m^{\ttbar}$  observable in the combination of the $3\ell$ and $4\ell$ channels, absolute and normalised, unfolded to particle level (a,c) and parton level (b,d).}
\end{figure}

\begin{figure}[!htb]
\centering
\subfloat[]{\includegraphics[width=0.46\textwidth]{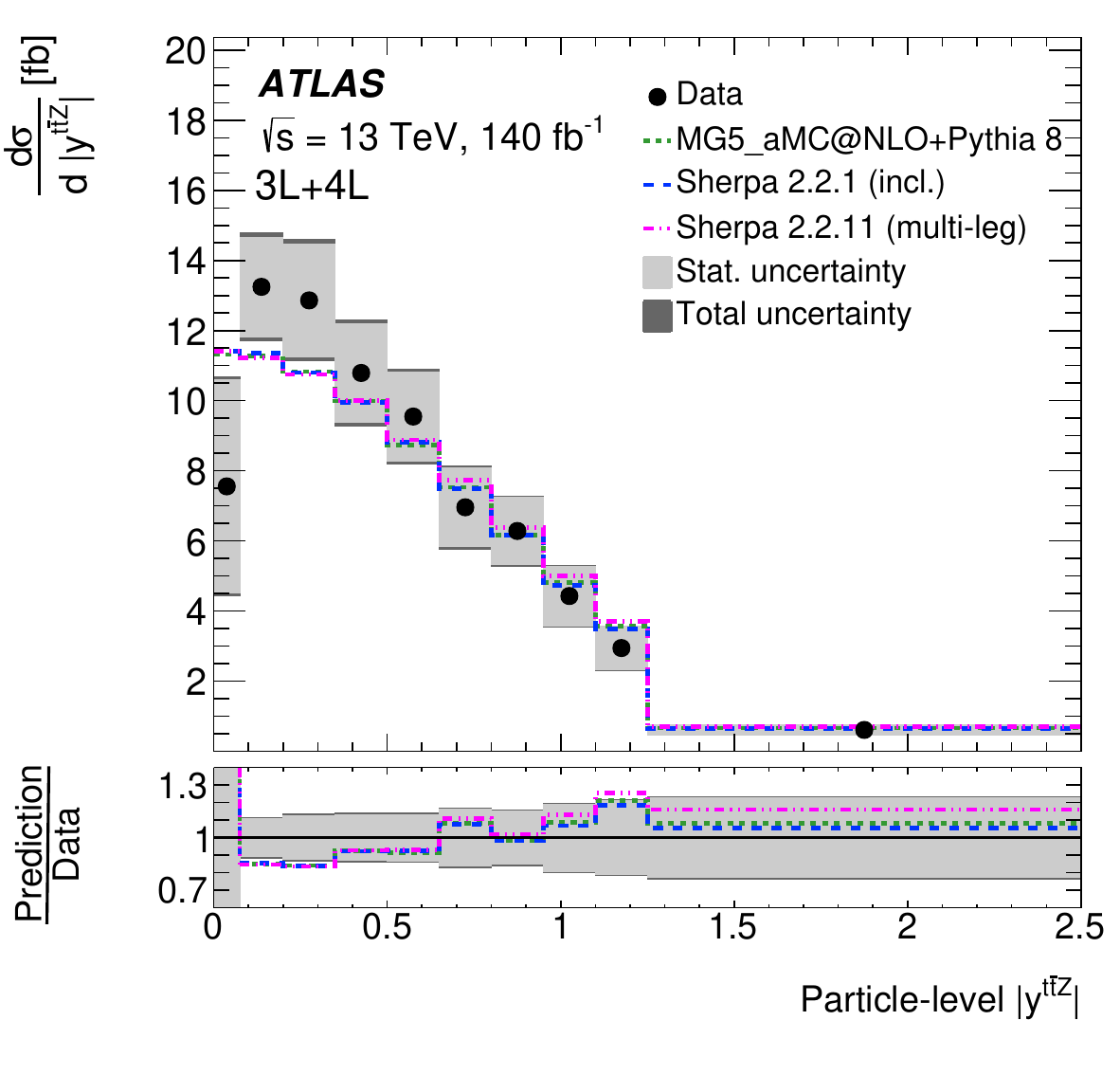}}
\hspace*{0.06\textwidth}
\subfloat[]{\includegraphics[width=0.46\textwidth]{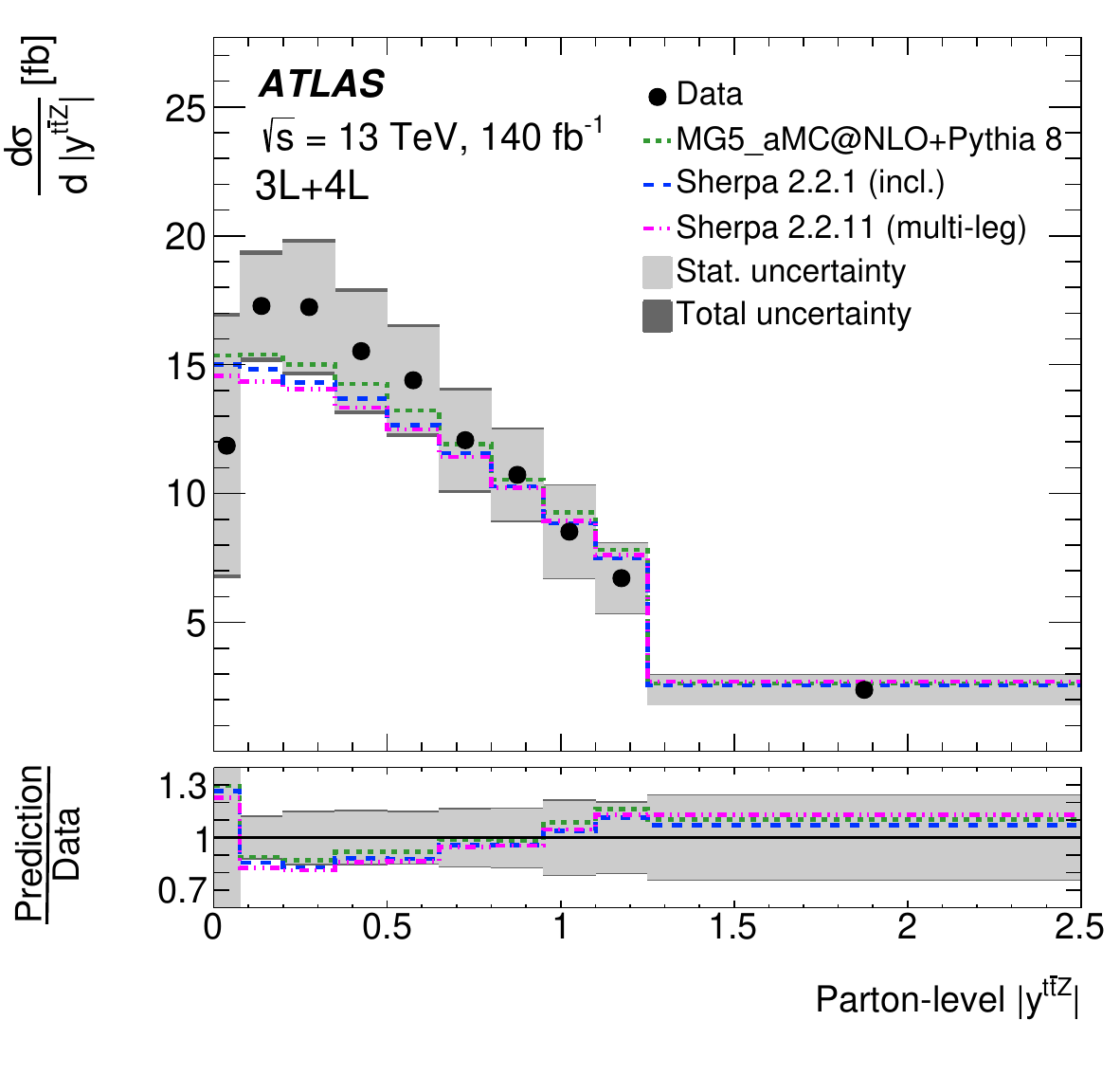}} \\
\subfloat[]{\includegraphics[width=0.46\textwidth]{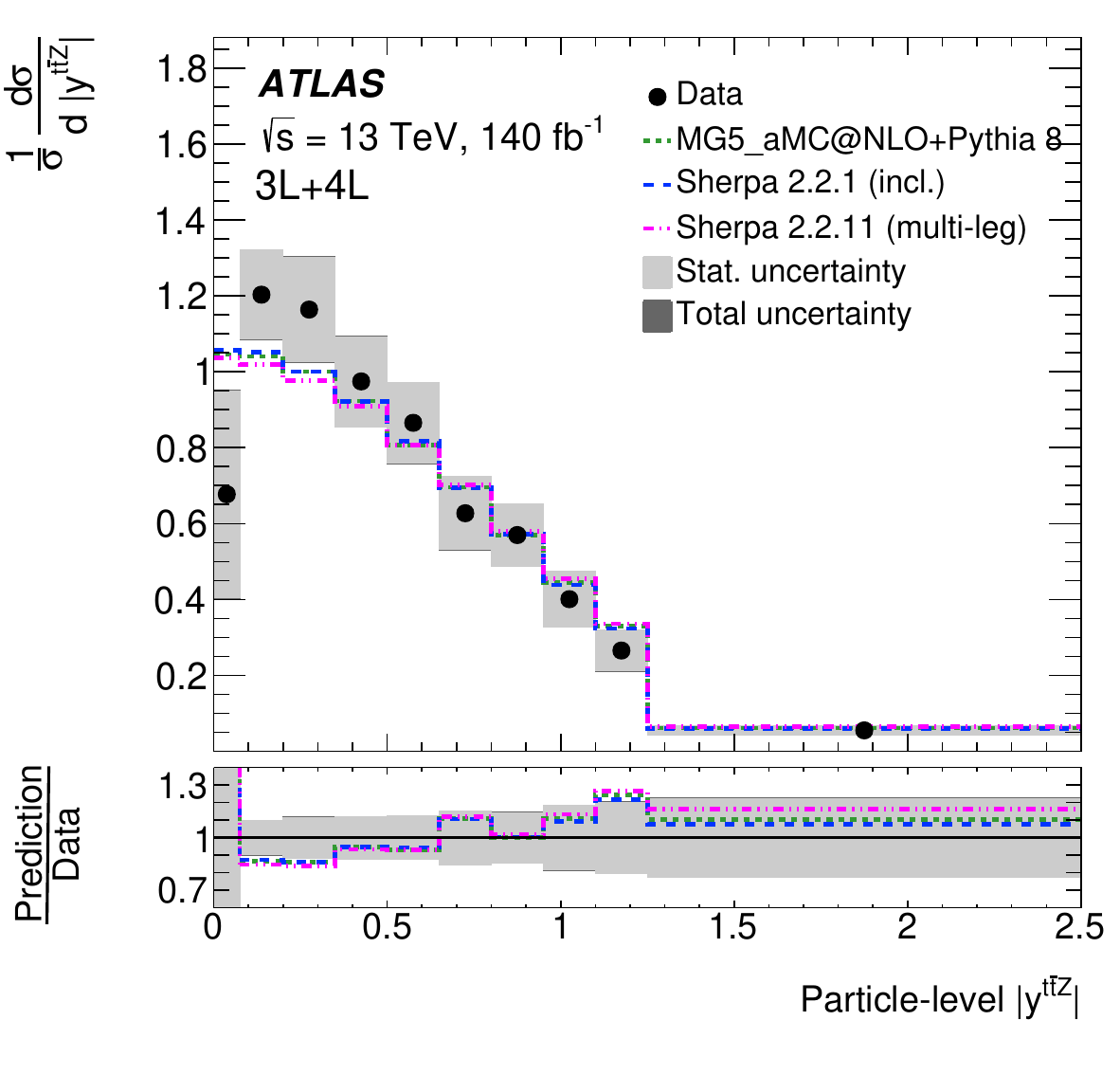}}
\hspace*{0.06\textwidth}
\subfloat[]{\includegraphics[width=0.46\textwidth]{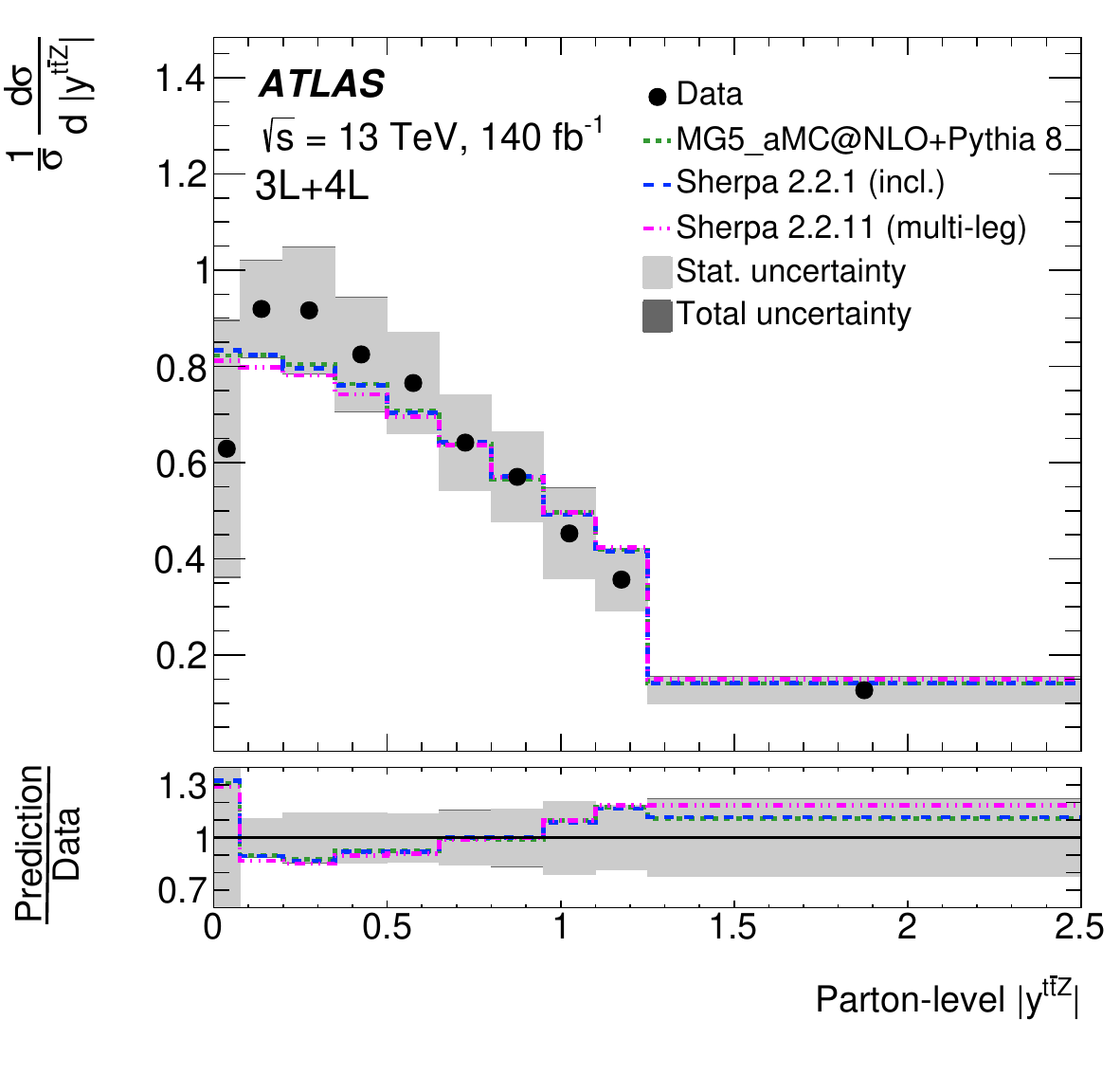}}
\caption{Cross-section measurement of the $\lvert y^{\ttZ}\rvert$  observable in the combination of the $3\ell$ and $4\ell$ channels, absolute and normalised, unfolded to particle level (a,c) and parton level (b,d).}
\label{fig:combined-observed-unfolding-result-y_ttZ-particle}
\end{figure}
 
\begin{figure}[!htb]
\centering
\subfloat[]{\includegraphics[width=0.46\textwidth]{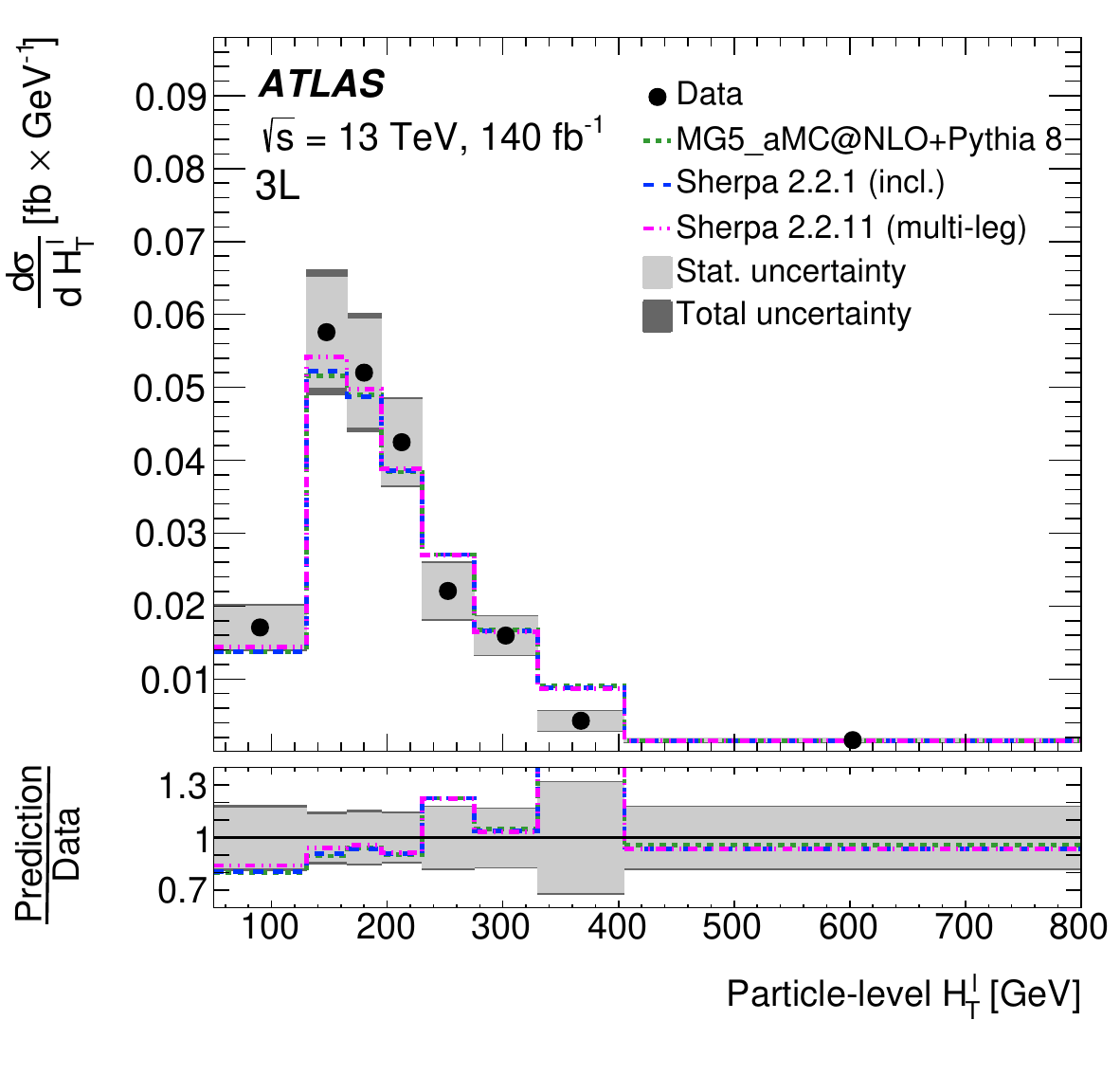}}
\hspace*{0.06\textwidth}
\subfloat[]{\includegraphics[width=0.46\textwidth]{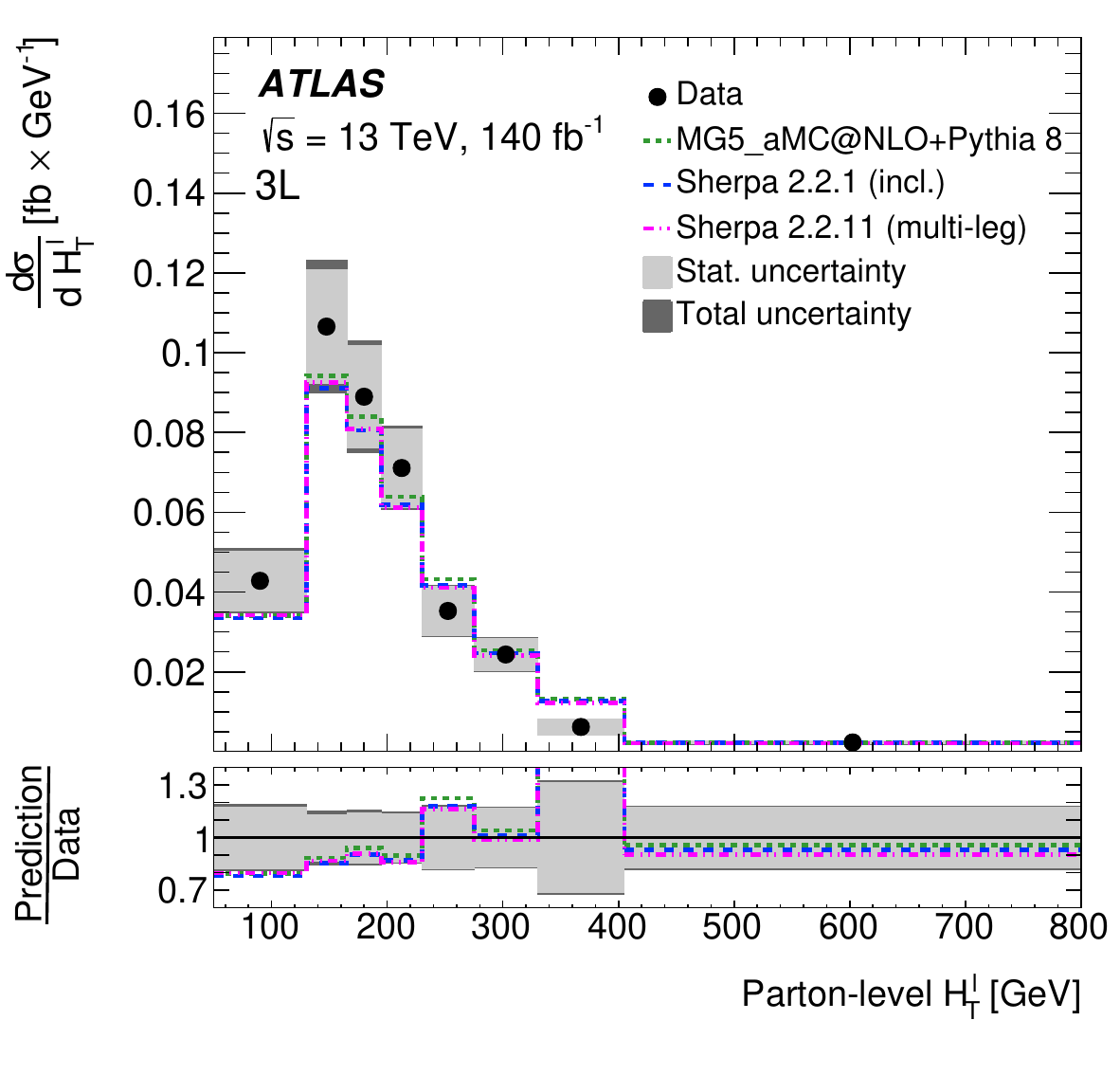}} \\
\subfloat[]{\includegraphics[width=0.46\textwidth]{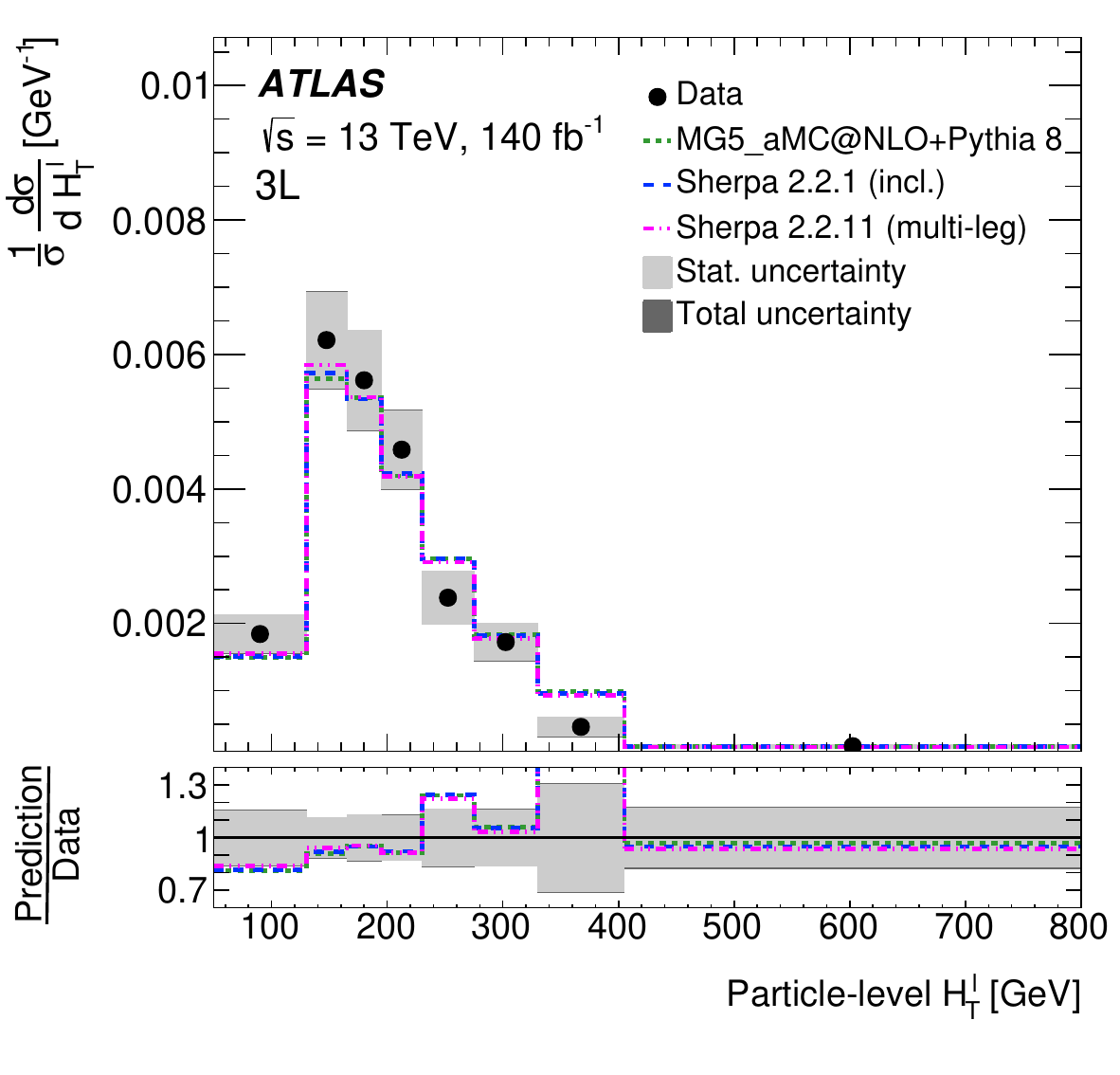}}
\hspace*{0.06\textwidth}
\subfloat[]{\includegraphics[width=0.46\textwidth]{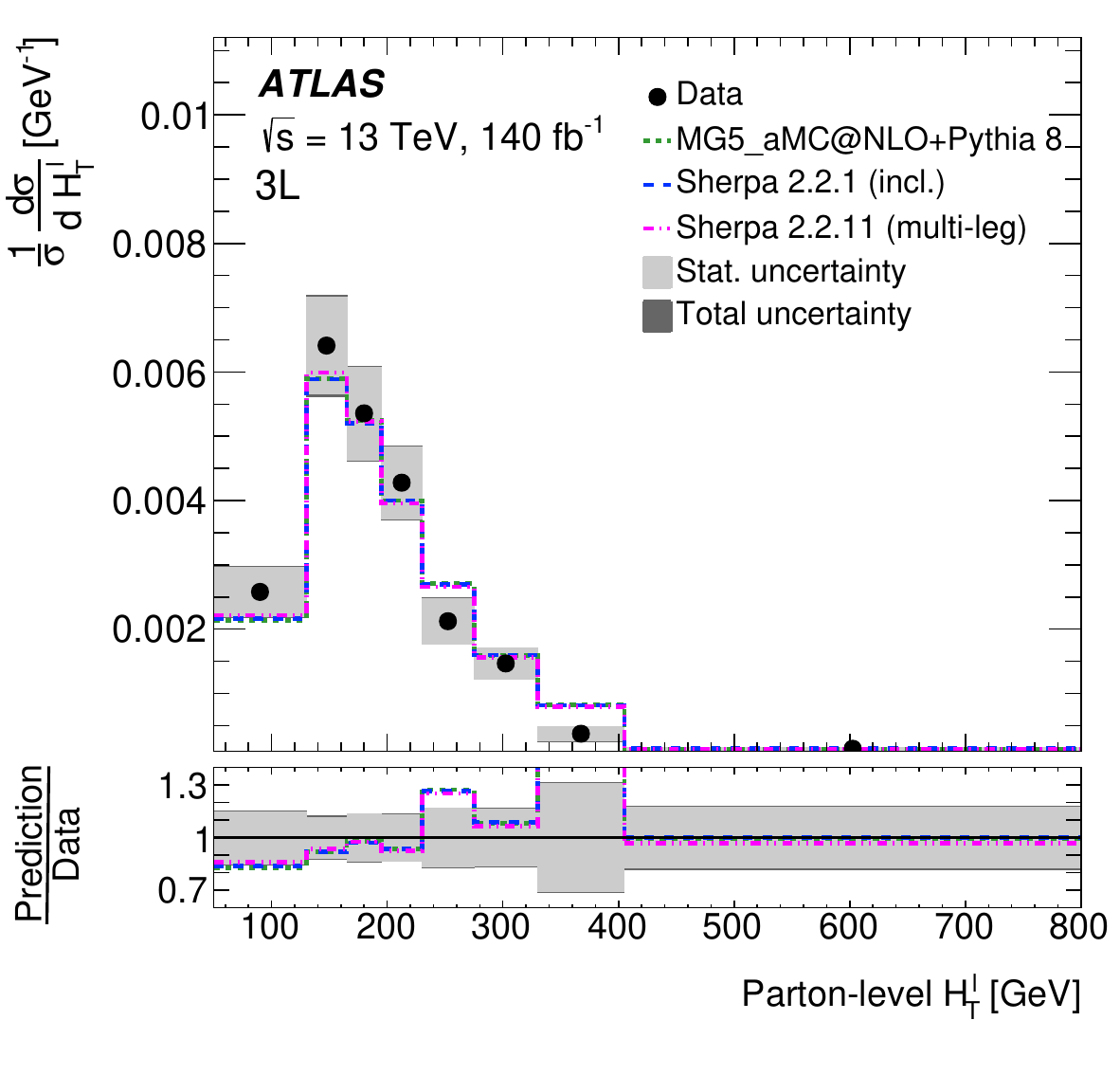}}
\caption{Cross-section measurement of the $H_{\text{T}}^{\ell}$  observable in the $3\ell$ channel, absolute and normalised, unfolded to particle level (a,c) and parton level (b,d).}
\label{fig:trilepton-observed-unfolding-result-ht_leptons_3L-particle}
\end{figure}
 
\begin{figure}[!htb]
\centering
\subfloat[]{\includegraphics[width=0.46\textwidth]{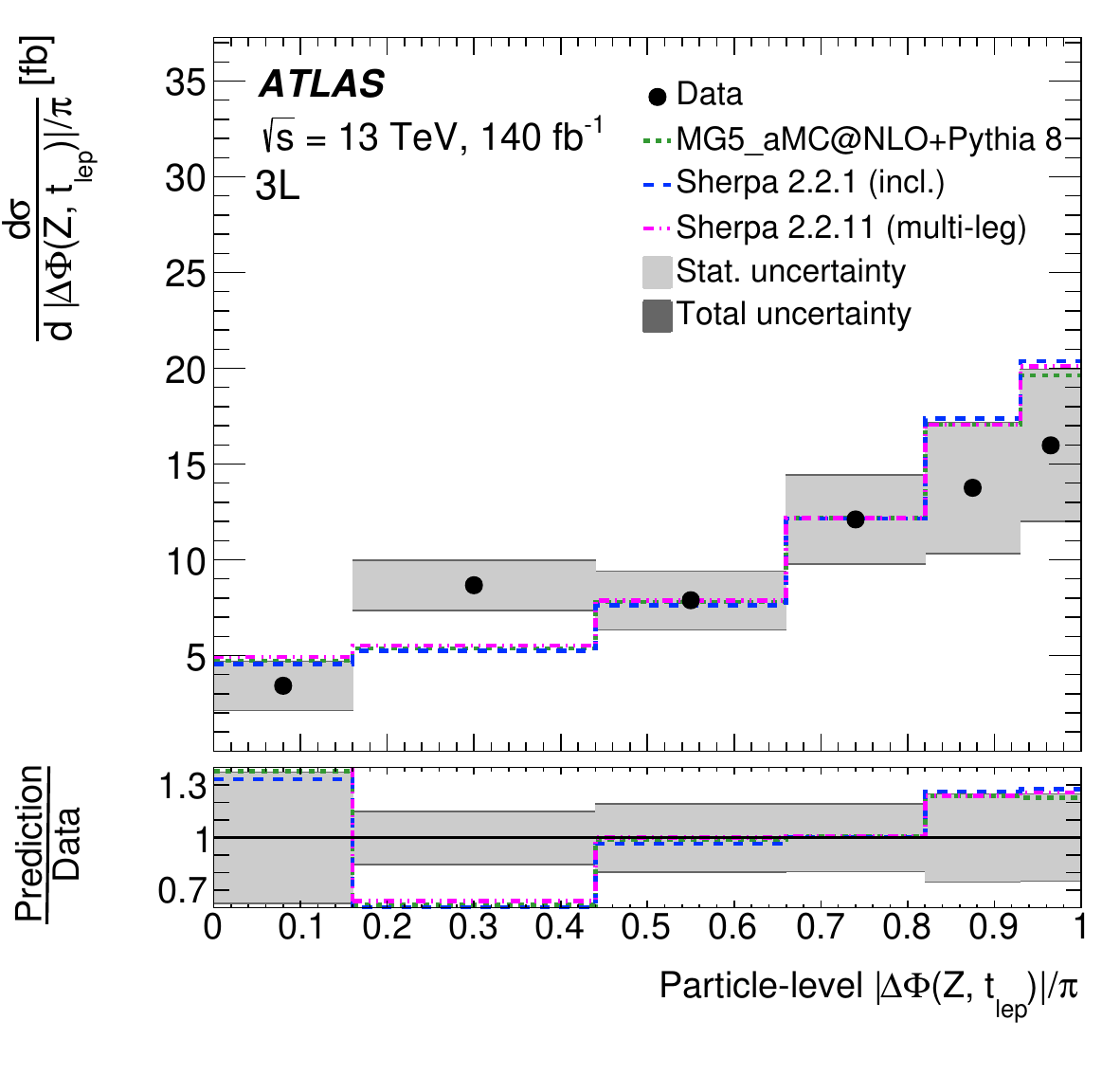}}
\hspace*{0.06\textwidth}
\subfloat[]{\includegraphics[width=0.46\textwidth]{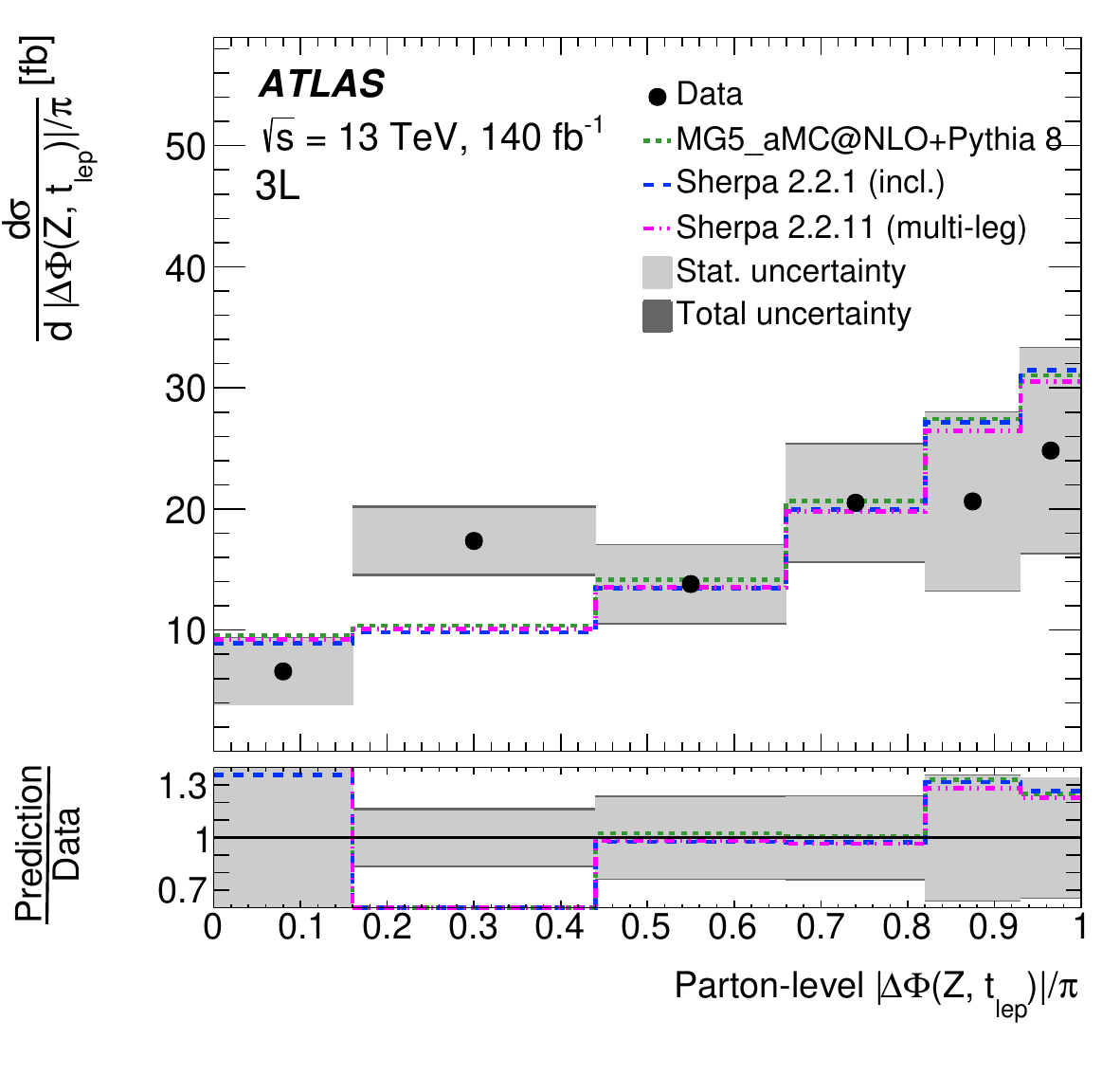}} \\
\subfloat[]{\includegraphics[width=0.46\textwidth]{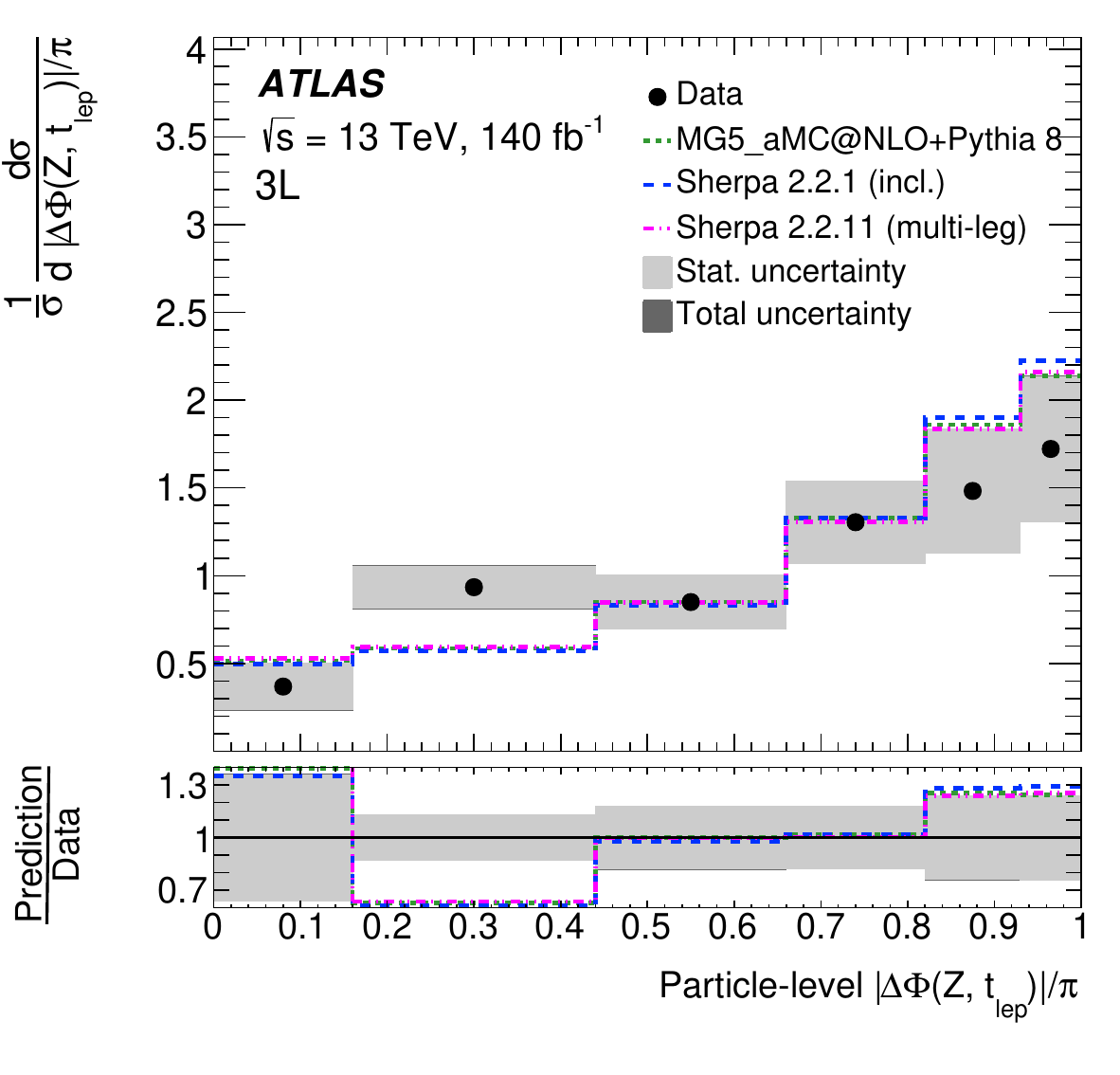}}
\hspace*{0.06\textwidth}
\subfloat[]{\includegraphics[width=0.46\textwidth]{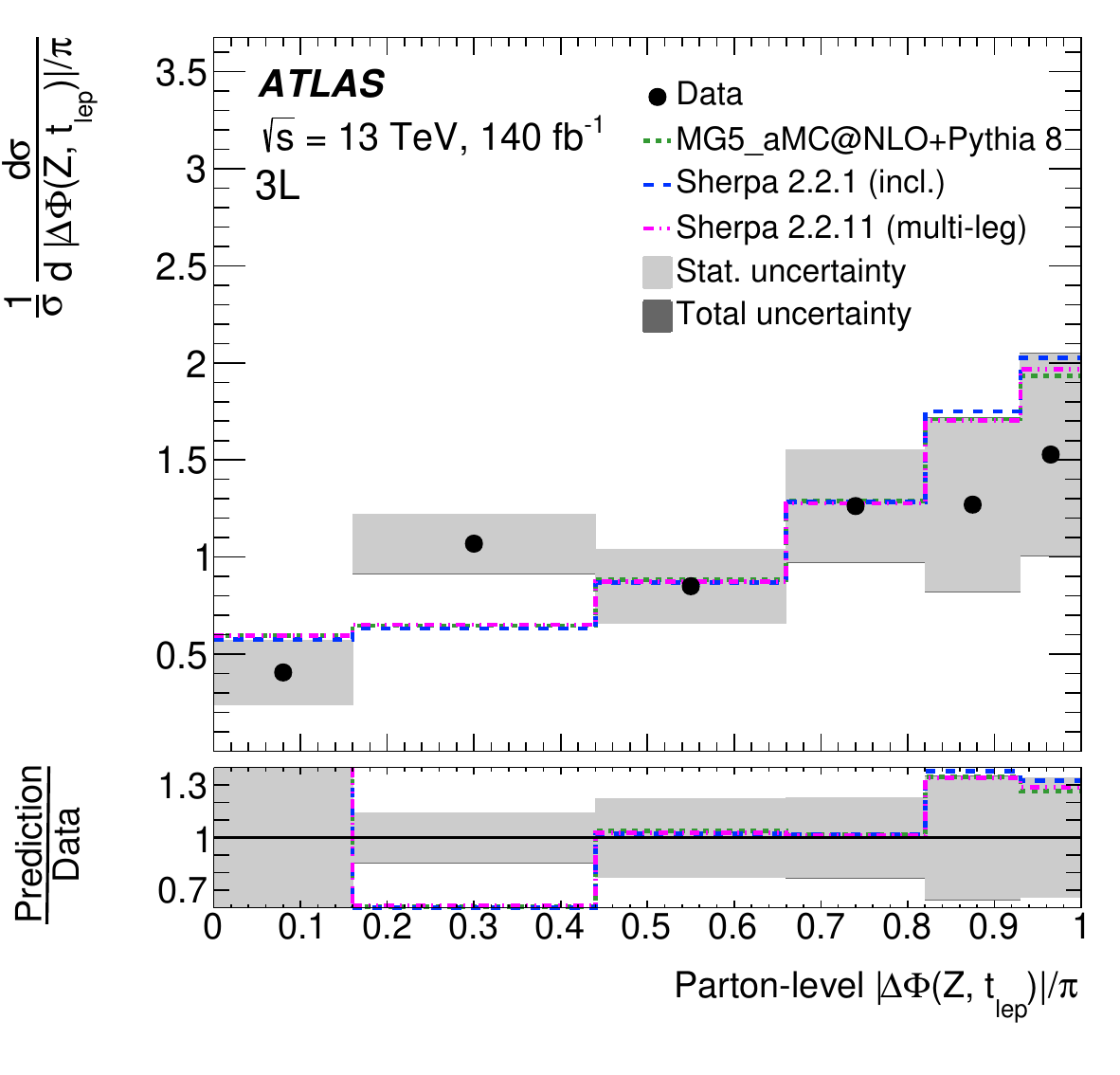}}
\caption{Cross-section measurement of the $|\Delta\Phi(\Zboson,t_{\mathrm{lep}})|/\pi$  observable in the $3\ell$ channel, absolute and normalised, unfolded to particle level (a,c) and parton level (b,d).}
\end{figure}
 
\begin{figure}[!htb]
\centering
\subfloat[]{\includegraphics[width=0.46\textwidth]{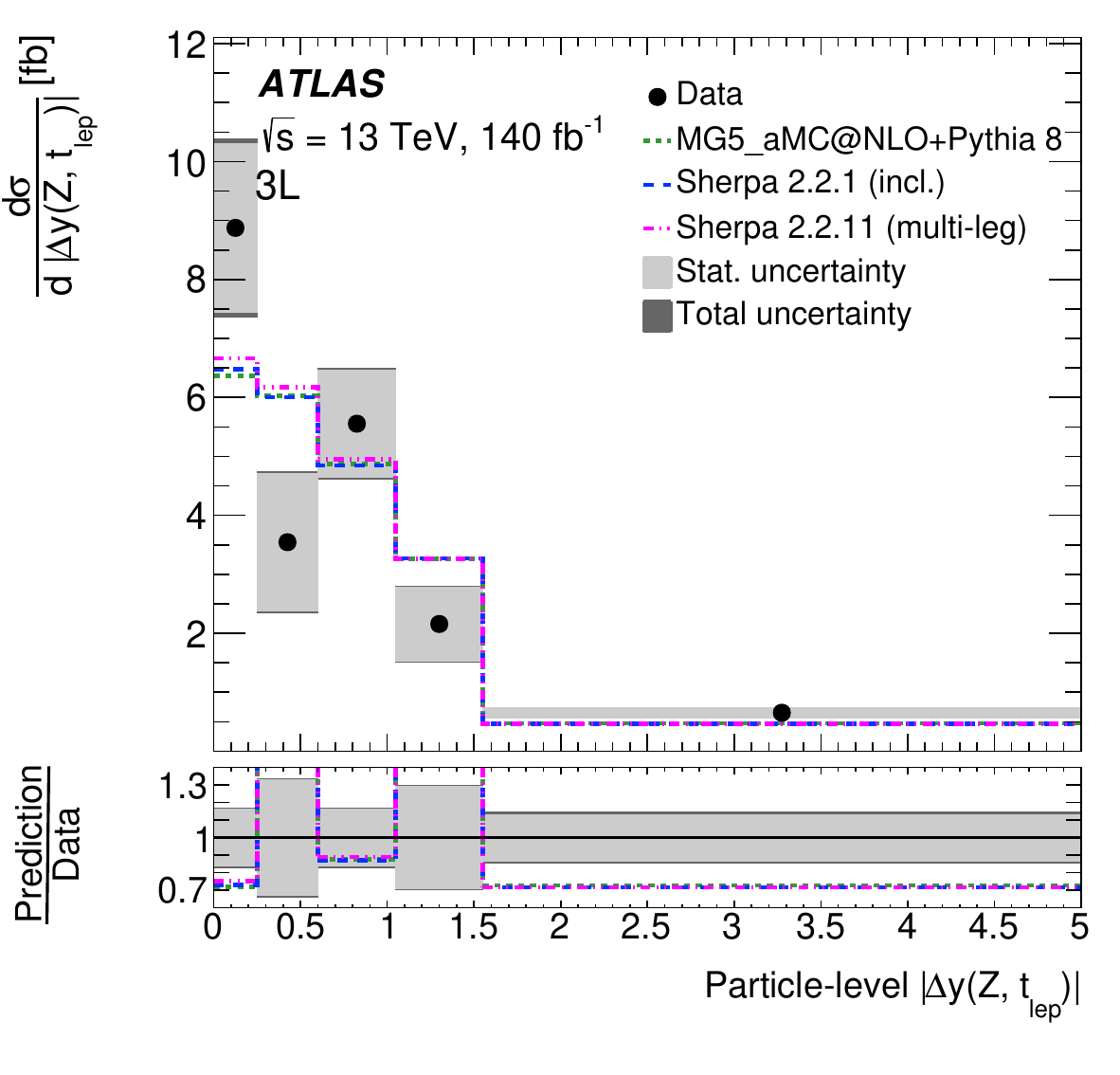}}
\hspace*{0.06\textwidth}
\subfloat[]{\includegraphics[width=0.46\textwidth]{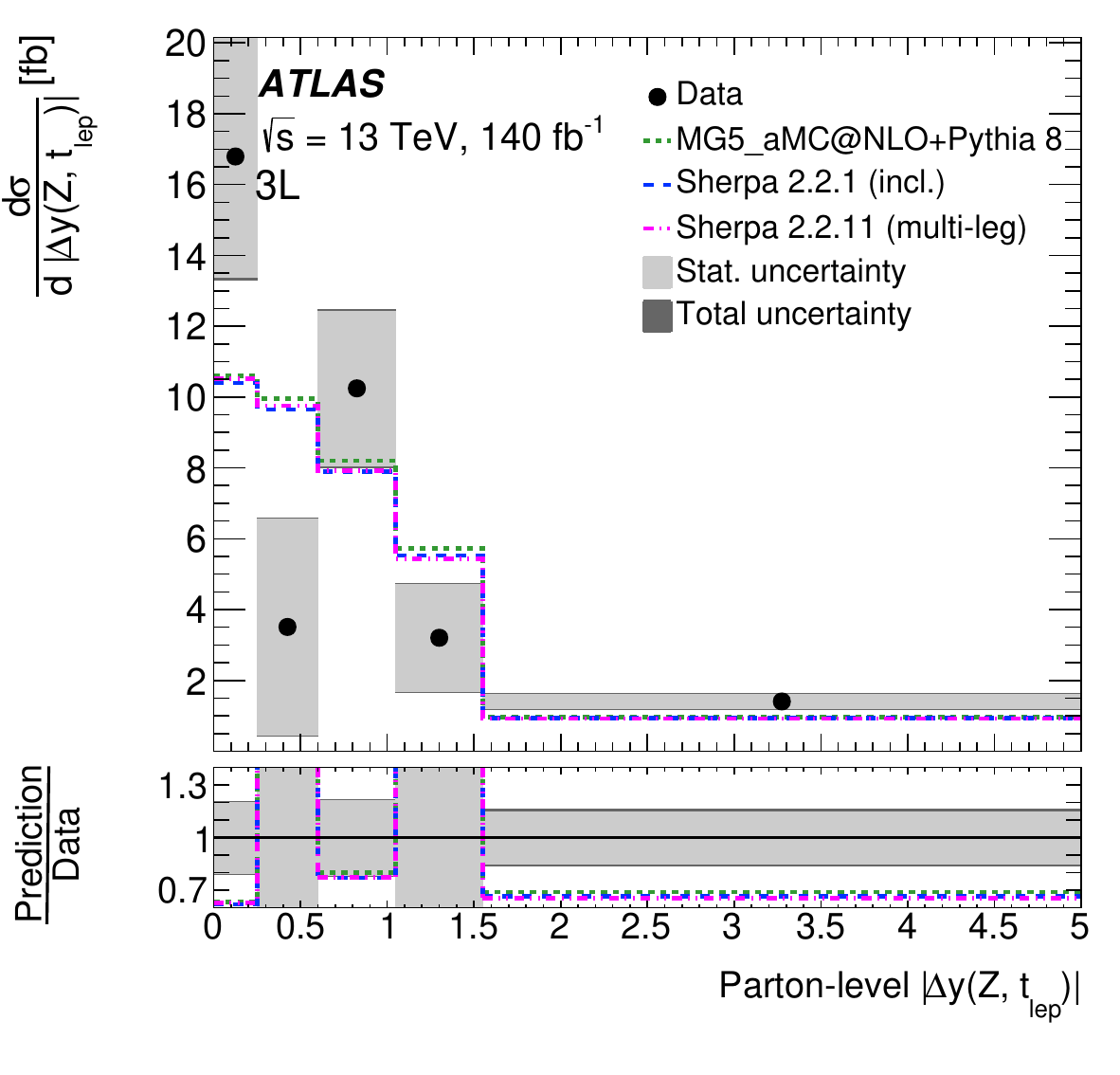}} \\
\subfloat[]{\includegraphics[width=0.46\textwidth]{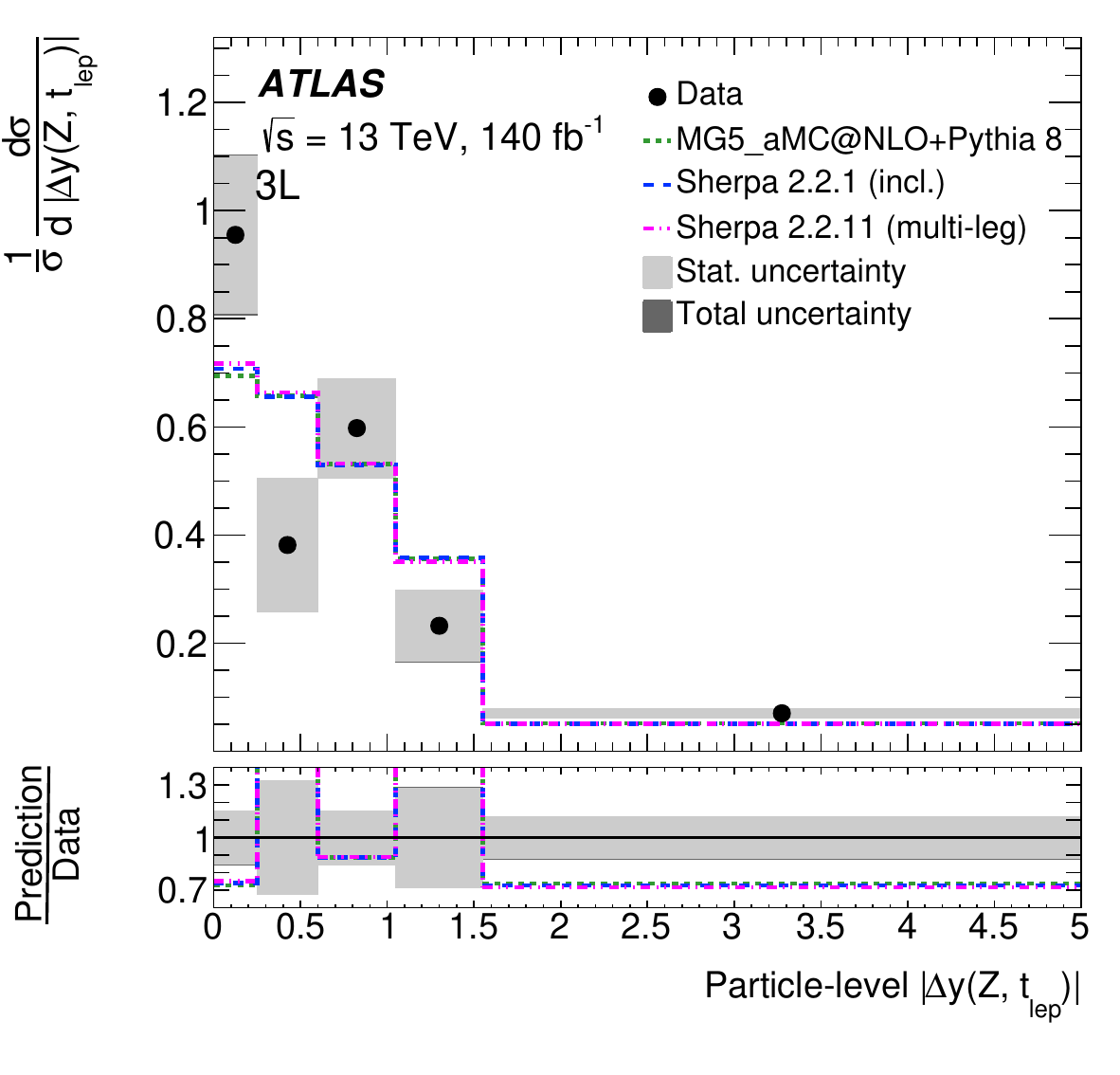}}
\hspace*{0.06\textwidth}
\subfloat[]{\includegraphics[width=0.46\textwidth]{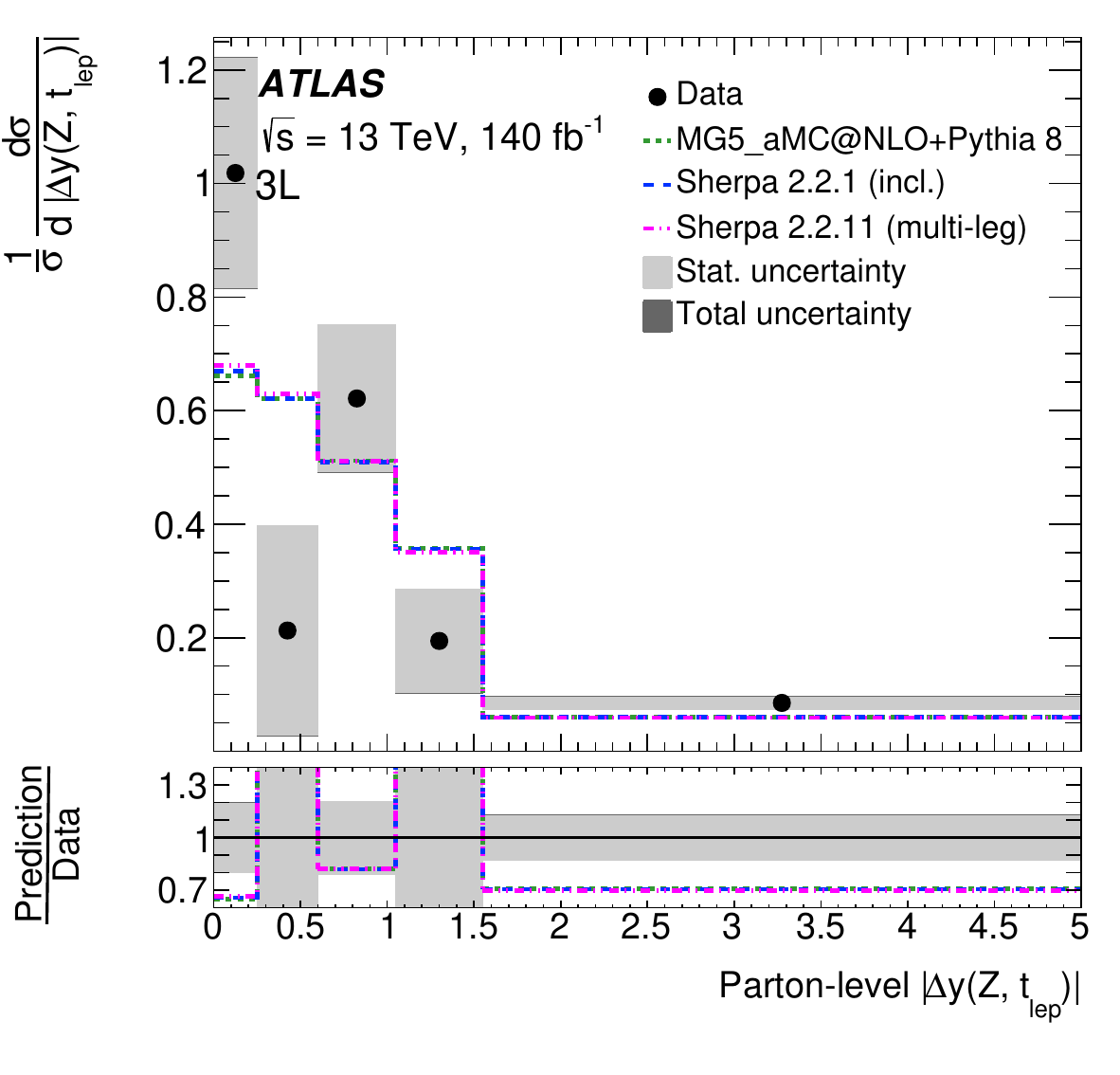}}
\caption{Cross-section measurement of the $|\Delta y(\Zboson,t_{\mathrm{lep}})|$  observable in the $3\ell$ channel, absolute and normalised, unfolded to particle level (a,c) and parton level (b,d).}
\end{figure}
 
\begin{figure}[!htb]
\centering
\subfloat[]{\includegraphics[width=0.46\textwidth]{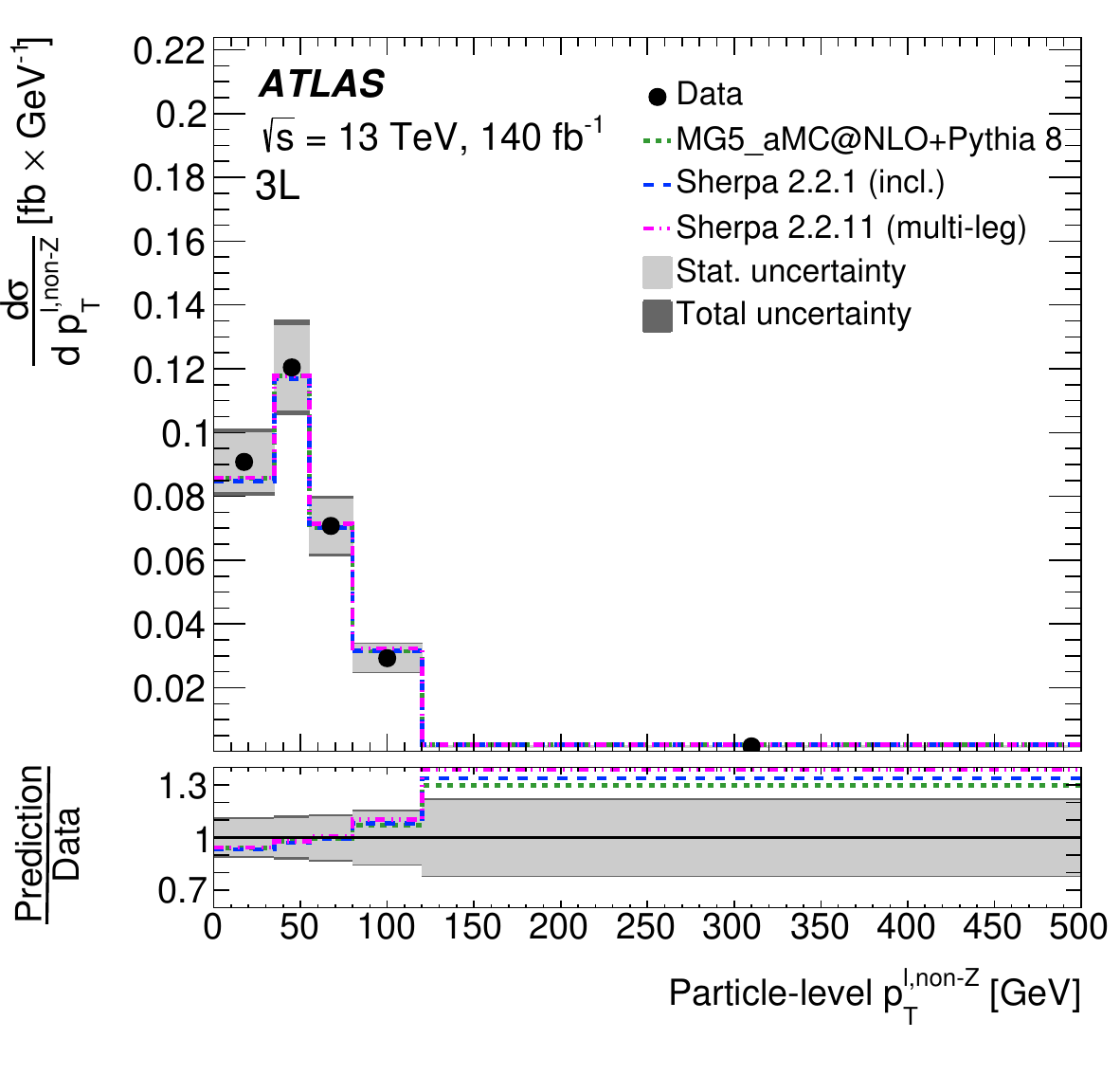}}
\hspace*{0.06\textwidth}
\subfloat[]{\includegraphics[width=0.46\textwidth]{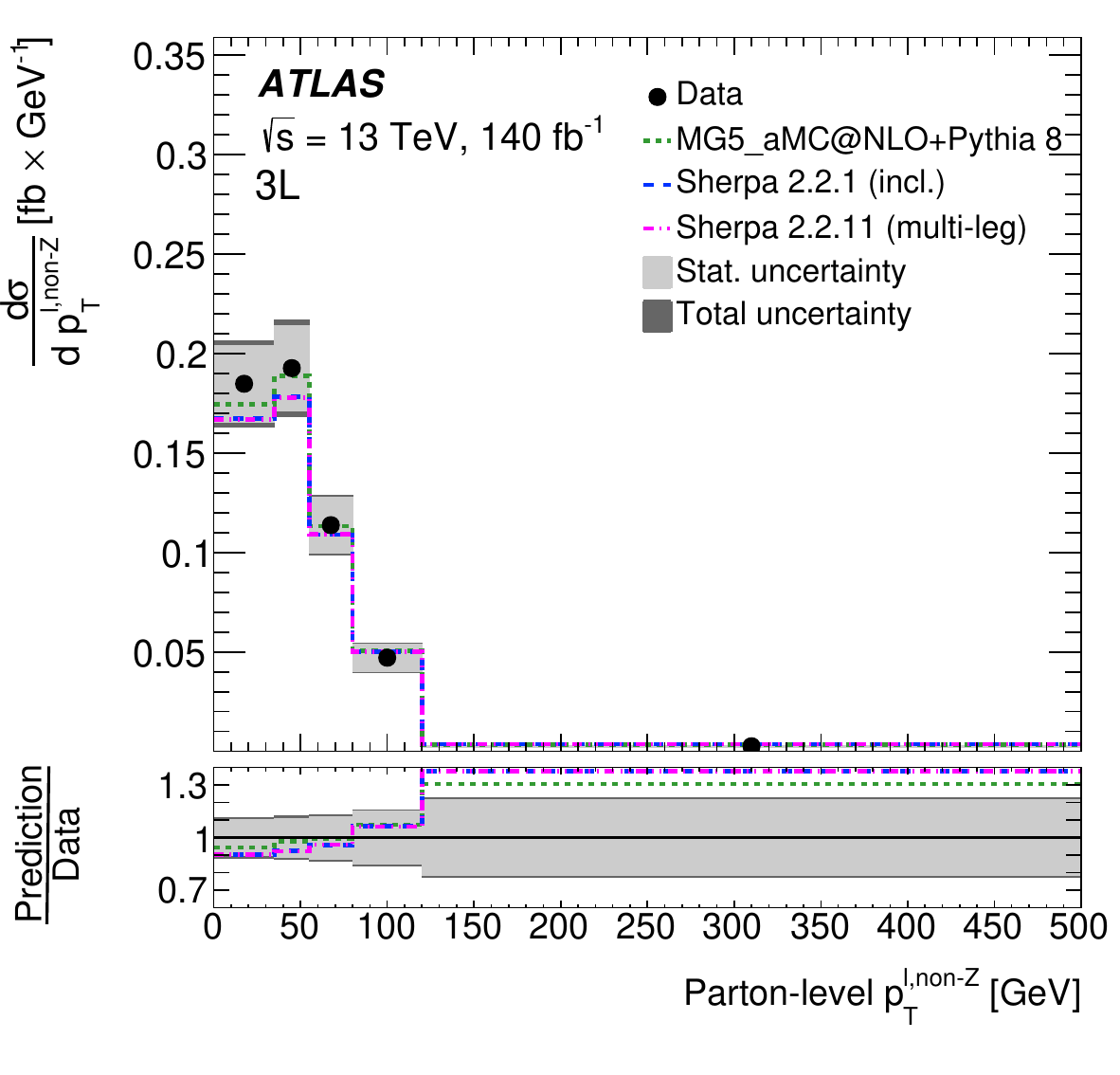}} \\
\subfloat[]{\includegraphics[width=0.46\textwidth]{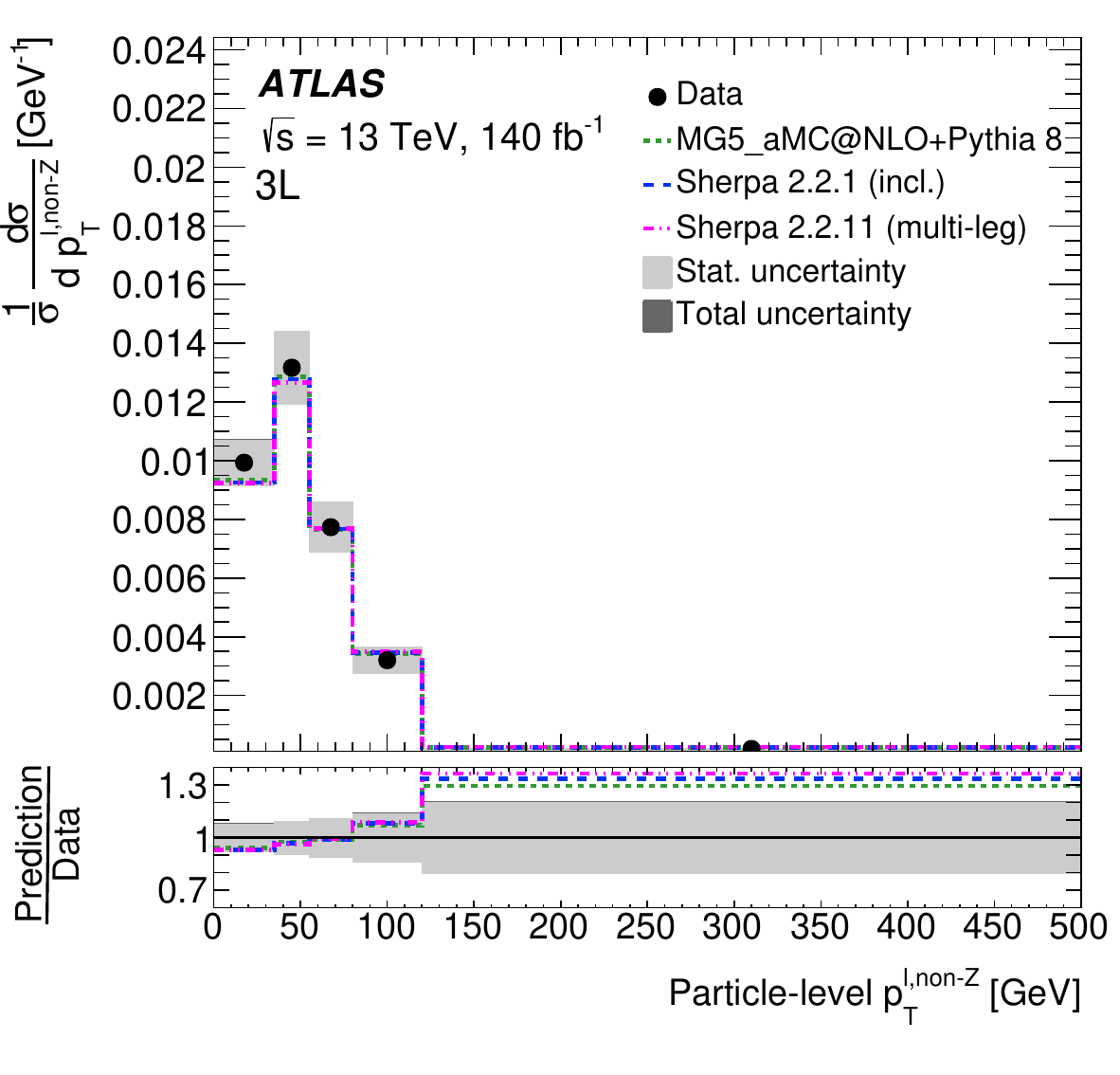}}
\hspace*{0.06\textwidth}
\subfloat[]{\includegraphics[width=0.46\textwidth]{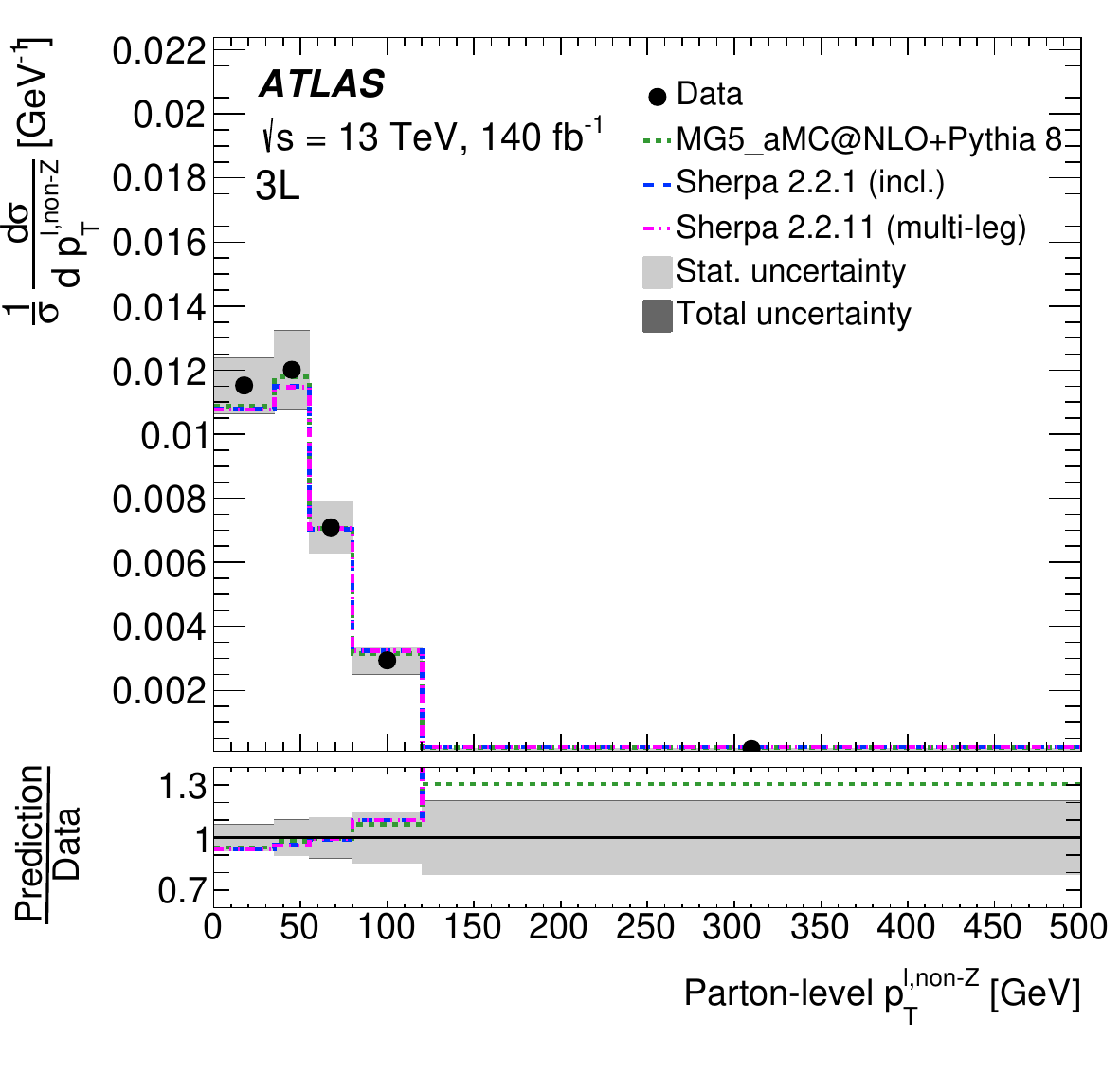}}
\caption{Cross-section measurement of the $\pT^{\ell,{\textrm{non-}}\Zboson}$  observable in the $3\ell$ channel, absolute and normalised, unfolded to particle level (a,c) and parton level (b,d).}
\end{figure}
 
\begin{figure}[!htb]
\centering
\subfloat[]{\includegraphics[width=0.46\textwidth]{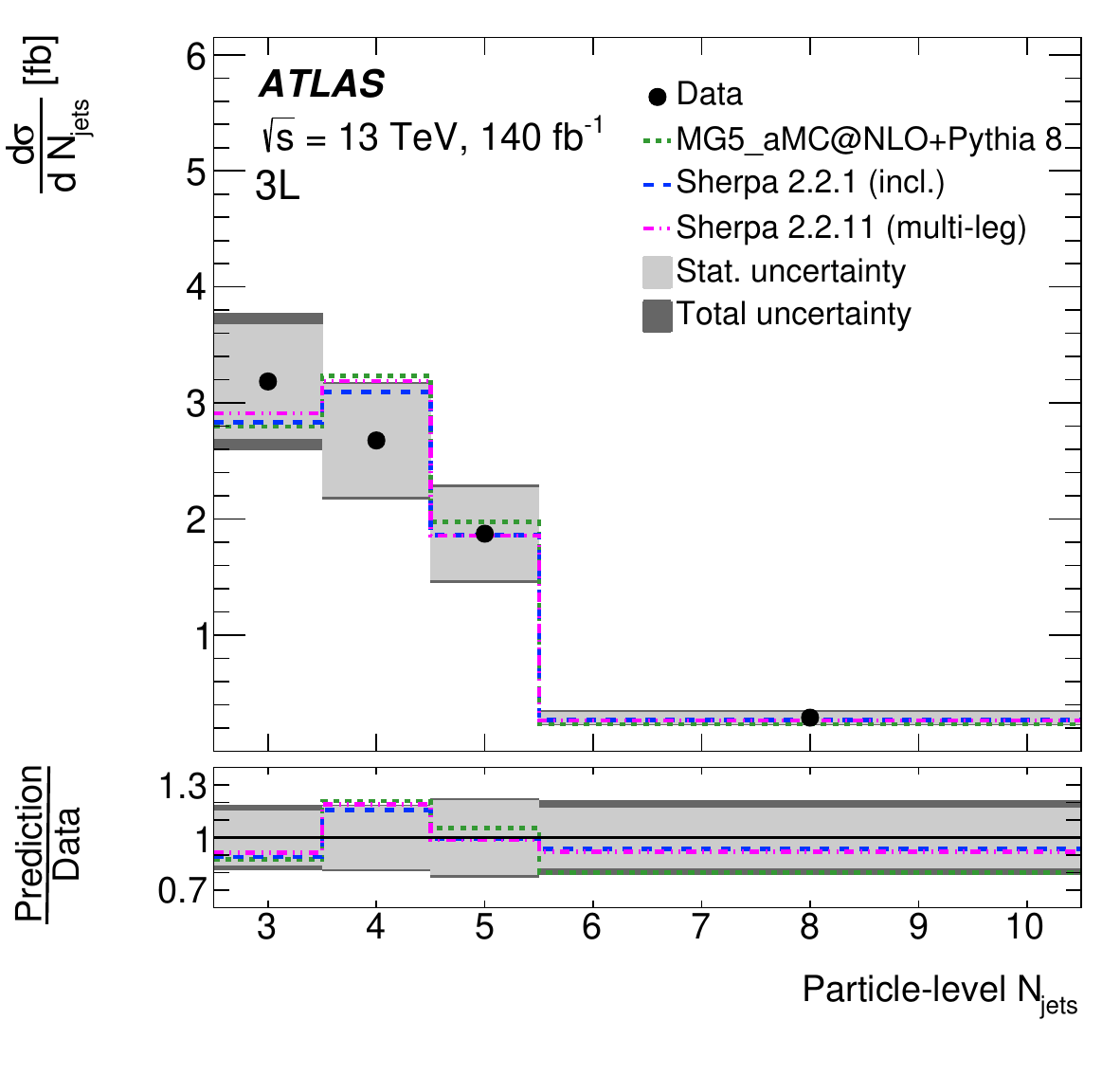}}
\hspace*{0.06\textwidth}
\subfloat[]{\includegraphics[width=0.46\textwidth]{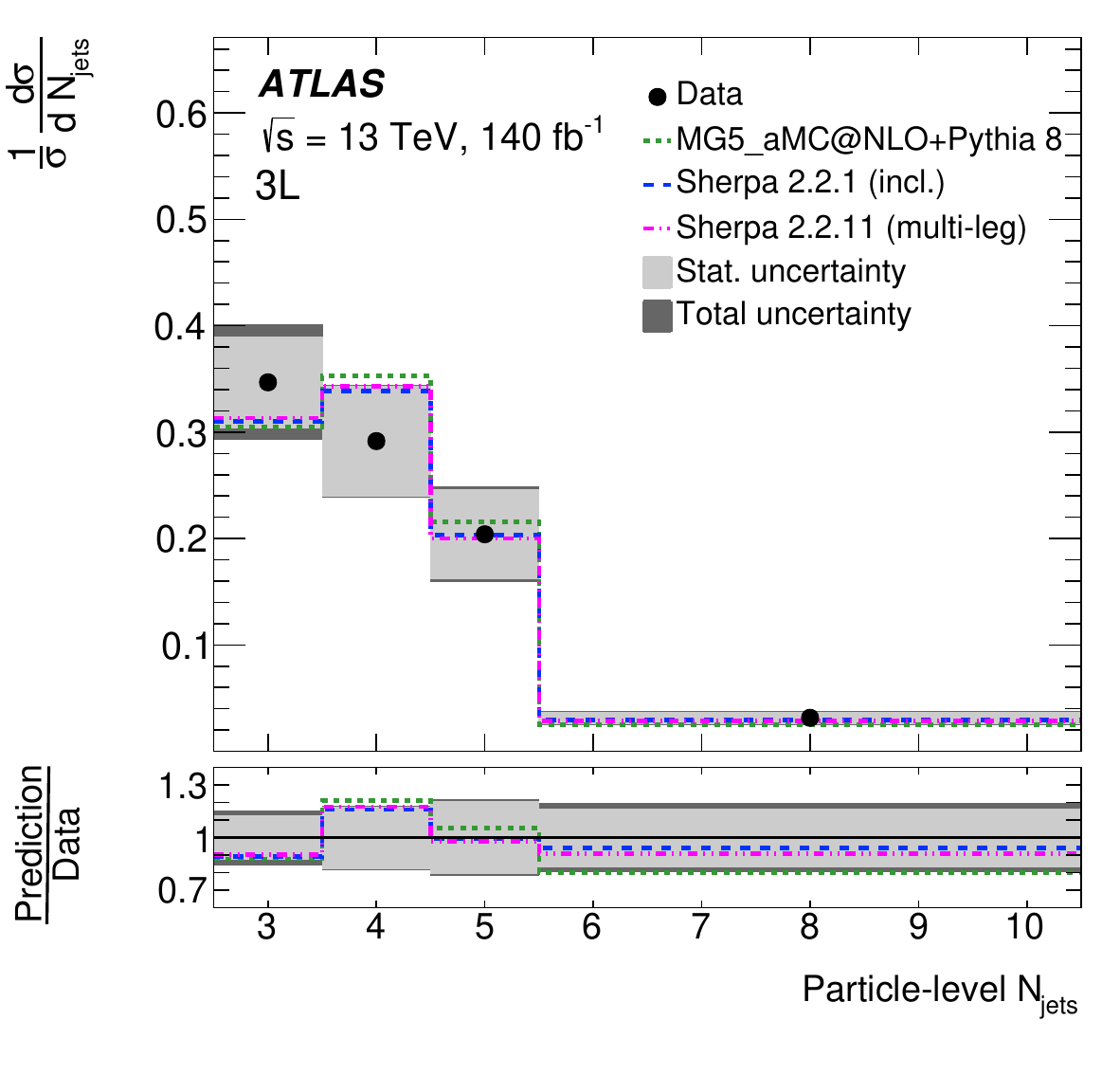}}
\caption{Cross-section measurement of the $N_{\mathrm{jets}}$  observable in the $3\ell$ channel, unfolded to particle level, absolute (a) and normalised (b).}
\label{fig:trilepton-observed-unfolding-result-n_jets_3L-particle}
\end{figure}
 
\begin{figure}[!htb]
\centering
\subfloat[]{\includegraphics[width=0.46\textwidth]{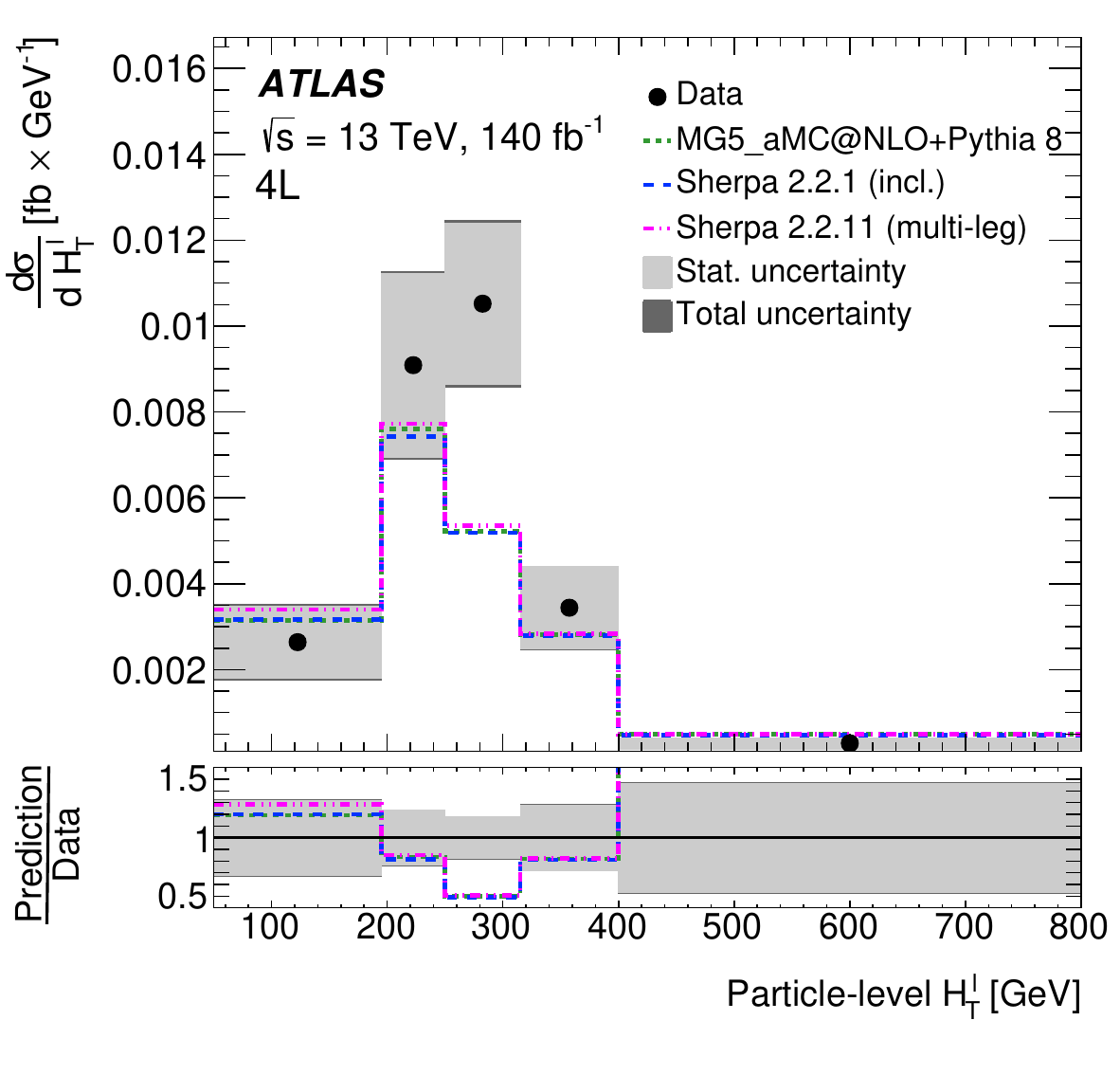}}
\hspace*{0.06\textwidth}
\subfloat[]{\includegraphics[width=0.46\textwidth]{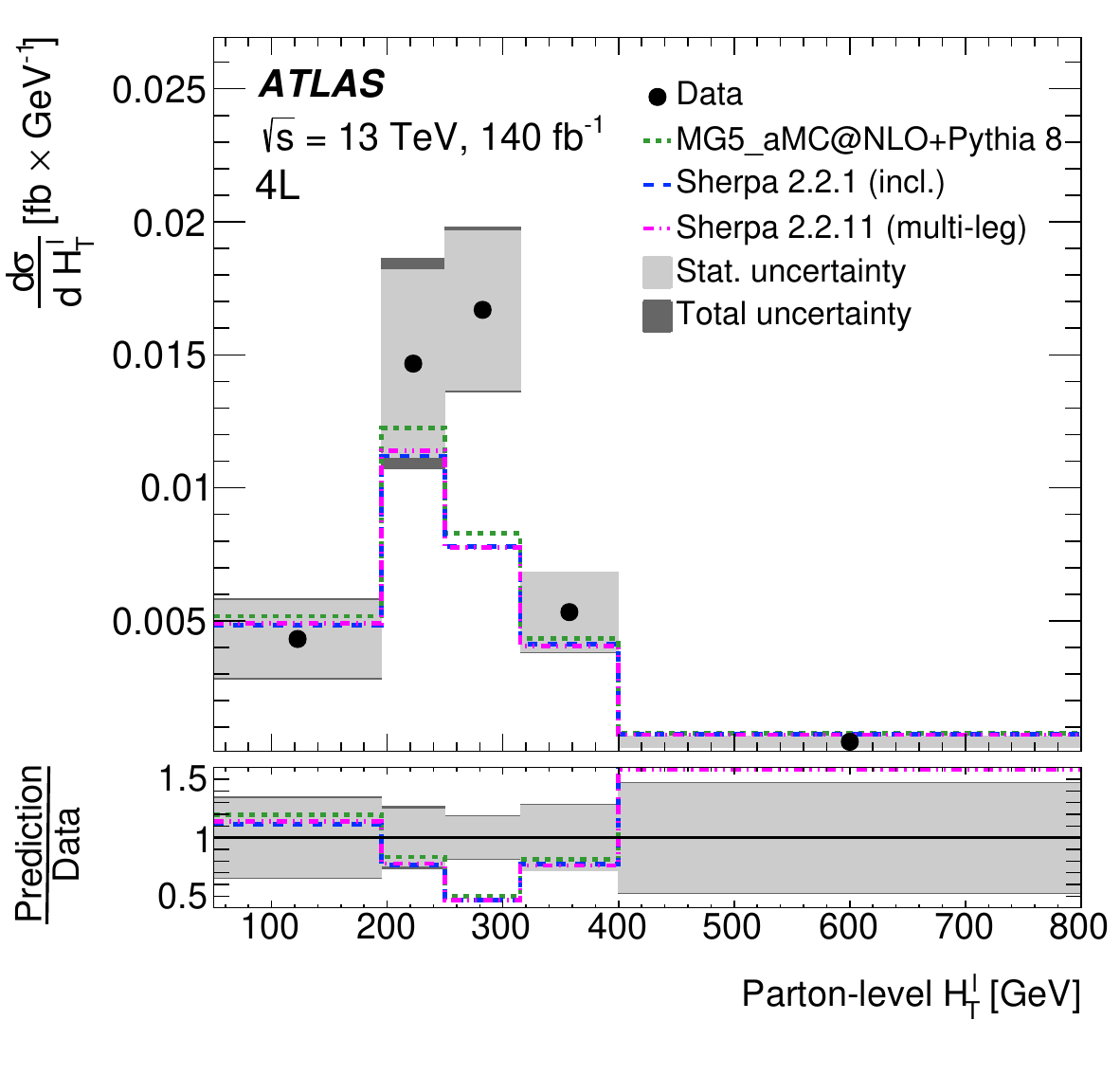}} \\
\subfloat[]{\includegraphics[width=0.46\textwidth]{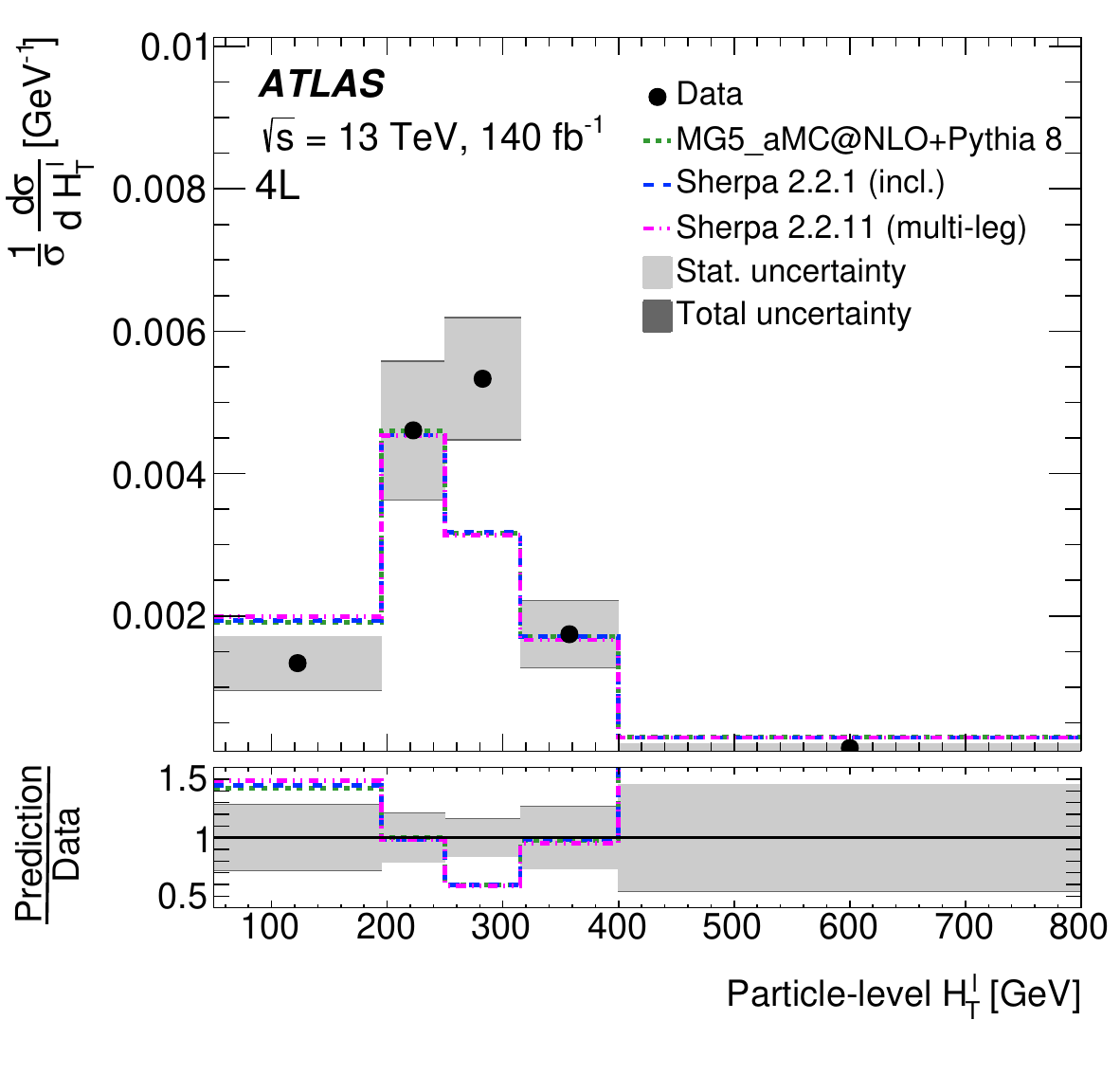}}
\hspace*{0.06\textwidth}
\subfloat[]{\includegraphics[width=0.46\textwidth]{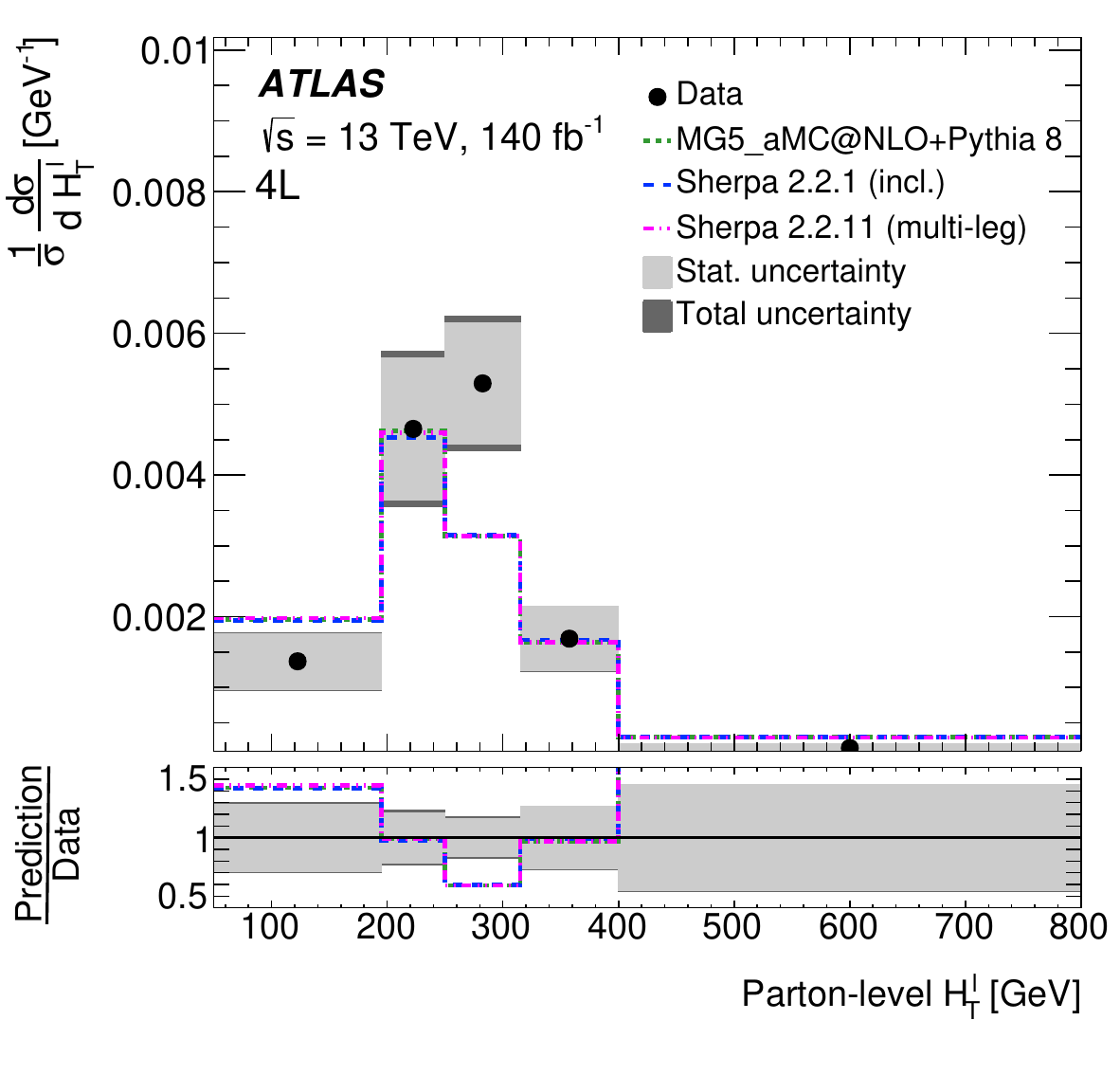}}
\caption{Cross-section measurement of the $\HT^{\ell}$ observable in the $4\ell$ channel, absolute and normalised, unfolded to particle level (a,c) and parton level (b,d).}
\label{fig:tetralepton-observed-unfolding-result-sum_pT_leptons}
\end{figure}
 
\begin{figure}[!htb]
\centering
\subfloat[]{\includegraphics[width=0.46\textwidth]{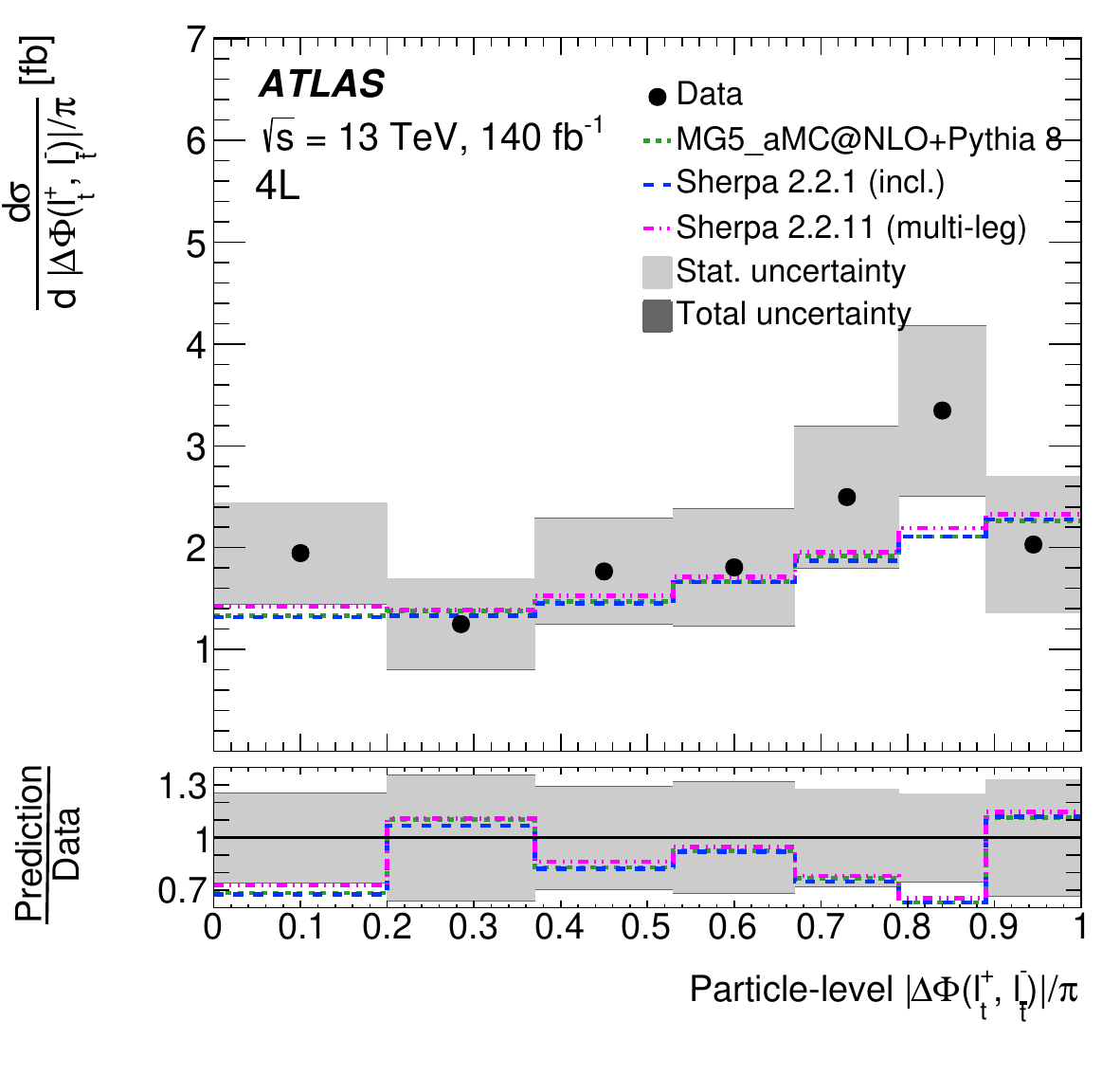}}
\hspace*{0.06\textwidth}
\subfloat[]{\includegraphics[width=0.46\textwidth]{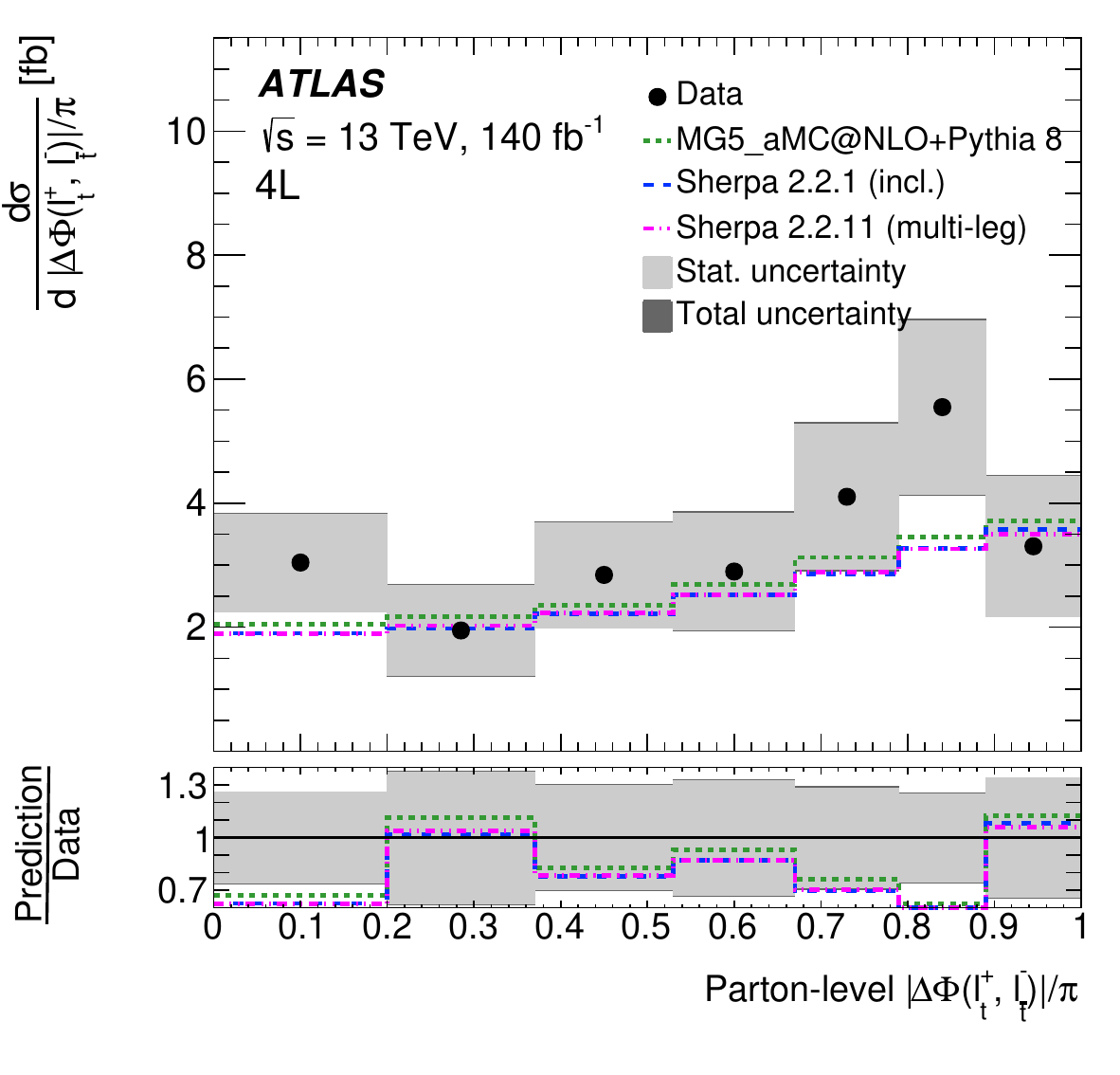}} \\
\subfloat[]{\includegraphics[width=0.46\textwidth]{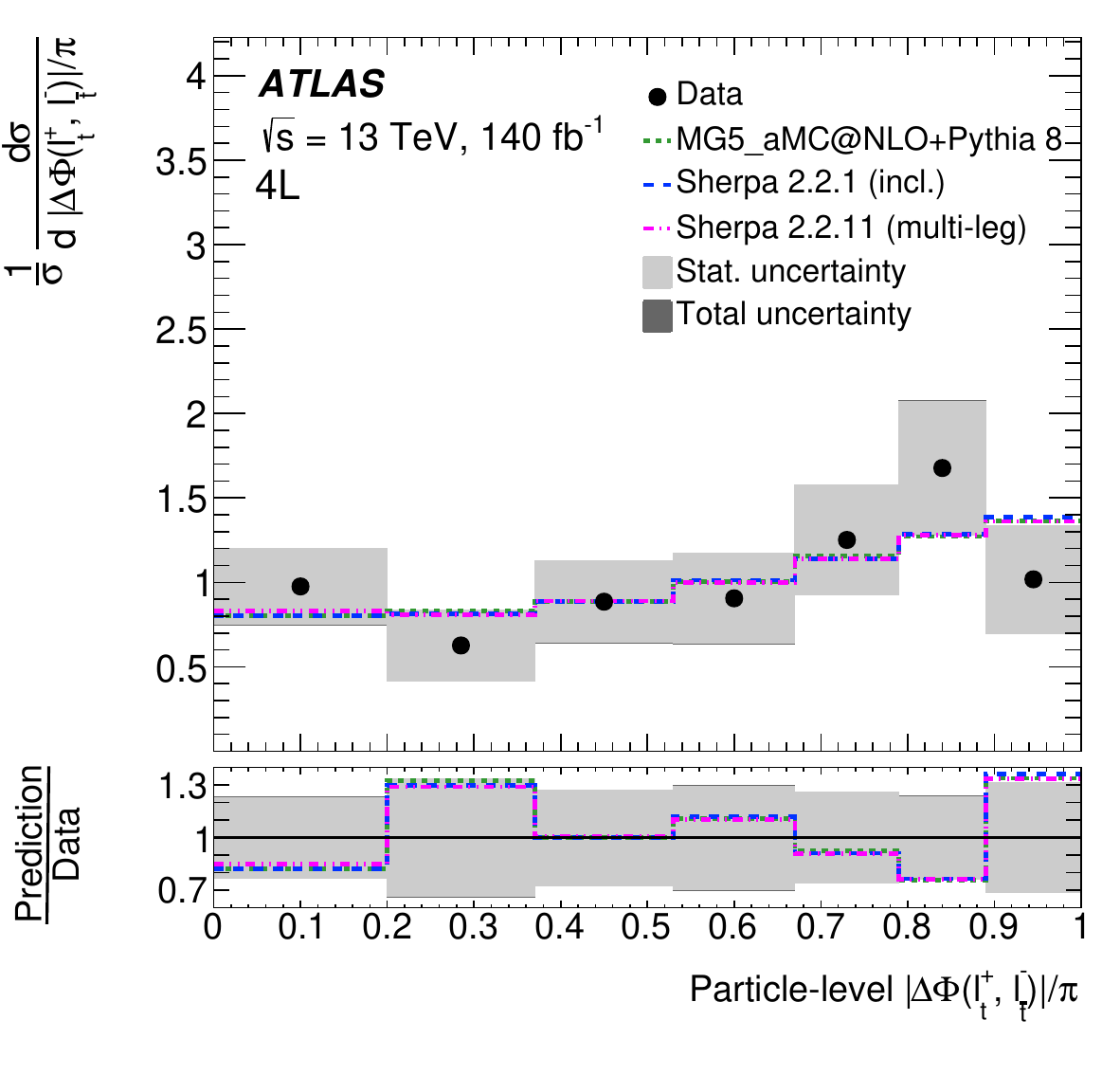}}
\hspace*{0.06\textwidth}
\subfloat[]{\includegraphics[width=0.46\textwidth]{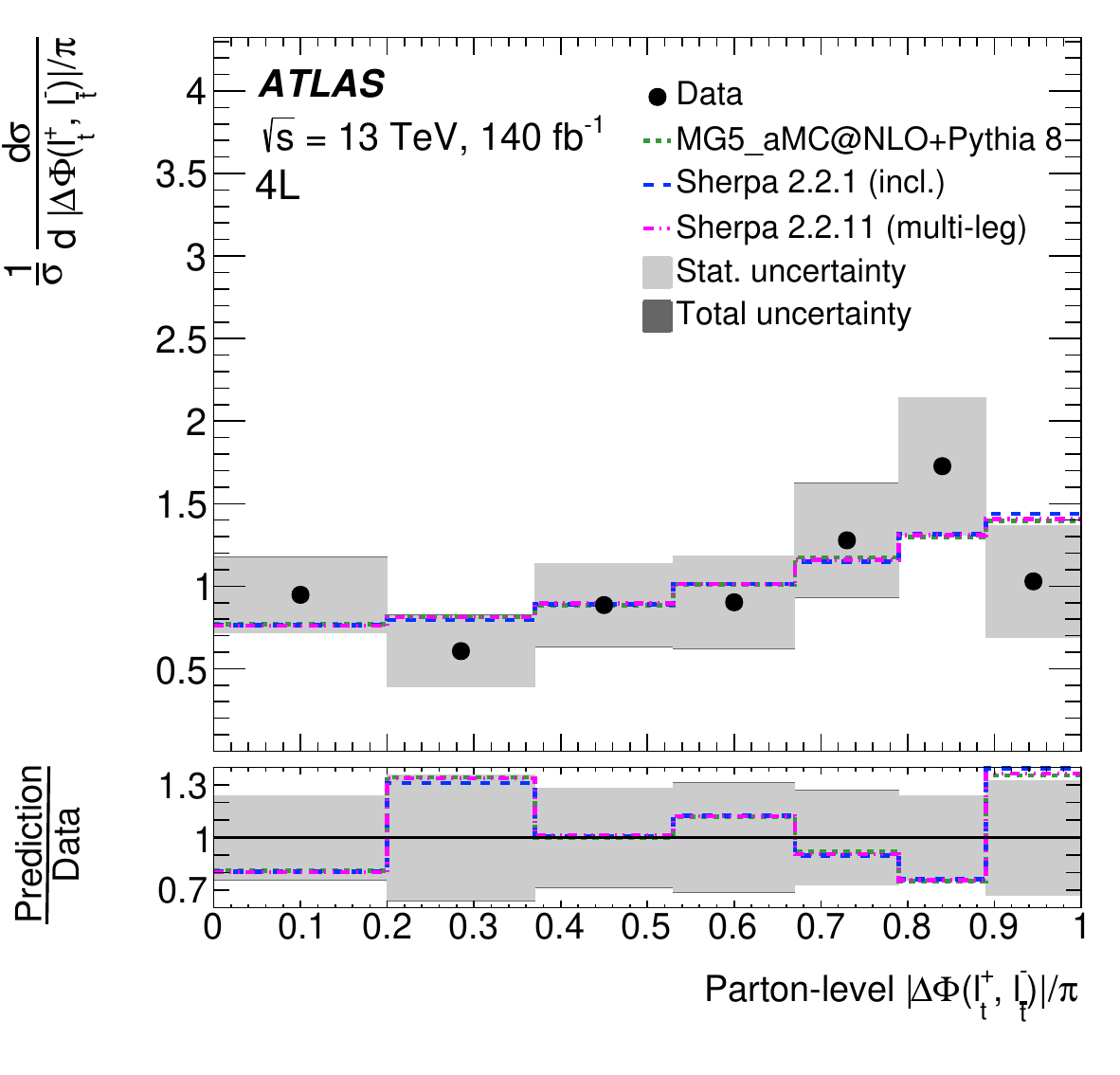}}
\caption{Cross-section measurement of the $|\Delta\Phi(\ell^{+}_{t}, \ell^{-}_{\bar{t}})|$ observable in the $4\ell$ channel, absolute and normalised, unfolded to particle level (a,c) and parton level (b,d).}
\end{figure}
 
\begin{figure}[!htb]
\centering
\subfloat[]{\includegraphics[width=0.46\textwidth]{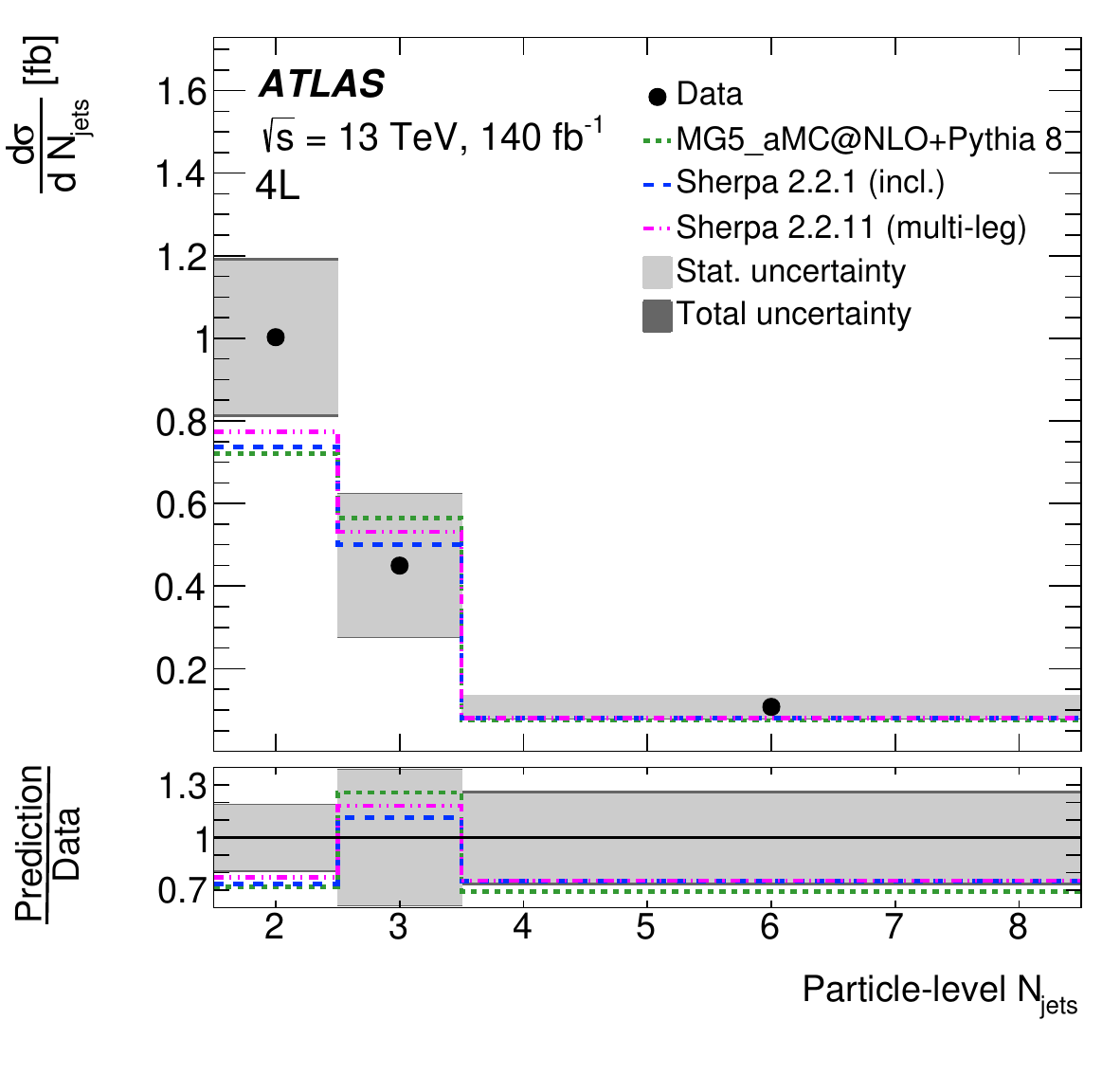}}
\hspace*{0.06\textwidth}
\subfloat[]{\includegraphics[width=0.46\textwidth]{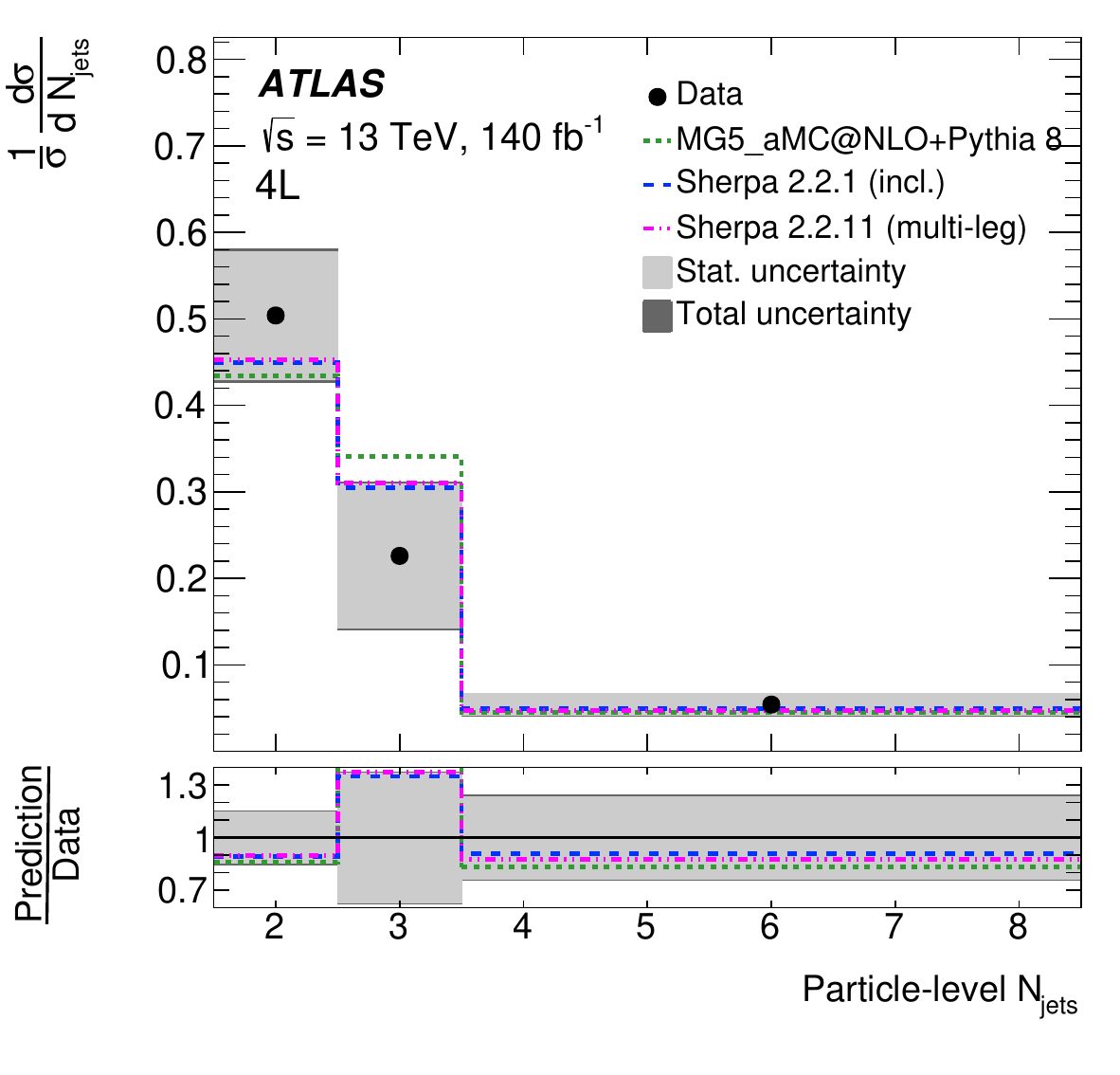}}
\caption{Cross-section measurement of the $N_{\mathrm{jets}}$ observable in the $4\ell$ channel, unfolded to particle level, absolute (a) and normalised (b).}
\label{fig:tetralepton-observed-unfolding-result-nJets}
\end{figure}
 
\FloatBarrier


 
\begin{table}[!htb]
\footnotesize
\caption{Summary of the compatibility tests, i.e.\ $p$-values, between the unfolded absolute differential spectra and the various predictions, in the $3\ell$ and $4\ell$ channels as well as in their combination, at particle-level and parton-level. }
\label{tab:unfolding_p_values_absolute}
\def\arraystretch{1.3}
\begin{center}
\begin{tabular}{cl ccc ccc cc} \\
\toprule
\multicolumn{2}{c}{} & \multicolumn{2}{c}{$\AMCatNLO{+}\PYTHIA[8]$} && \multicolumn{2}{c}{\SHERPA[2.2.1] (incl.)} && \multicolumn{2}{c}{\SHERPA[2.2.11] (multi-leg)}\\
\midrule
& Variable & parton & particle && parton & particle && parton & particle\\
\midrule
\multirow{5}{*}{\rotatebox{90}{$3\ell$}}      & $|\Delta\Phi(Z, t_\text{lep})|/\pi $ & 0.12 & 0.09 && 0.09 & 0.05 && 0.14 & 0.09 \\
& $|\Delta y(Z, t_\text{lep})| $ & 0.08  & 0.05 && 0.09    & 0.05 && 0.08  & 0.04 \\
& $\HT^{\ell} $            & 0.04  & 0.04 && 0.06    & 0.07 && 0.11  & 0.1\phantom{0} \\
& $\pT^{\ell, \textrm{non-}Z} $     & 0.75  & 0.75 && 0.44    & 0.63 && 0.44  & 0.51 \\
& $N_{\textrm{jets}} $    & -     & 0.55 && -       & 0.9\phantom{0}  && -     & 0.82 \\
\midrule
\multirow{3}{*}{\rotatebox{90}{$4\ell$}}      & $N_{\textrm{jets}}$  & - & 0.36 && - & 0.43 && - & 0.55   \\
& $|\Delta\Phi(\ell^{+}_{t}, \ell^{-}_{\bar{t}})|/\pi$   & 0.68    & 0.67 && 0.52    & 0.65 && 0.52    & 0.77 \\
& $\HT^{\ell}$               & 0.04 & 0.04 && 0.02 & 0.04 && 0.03 & 0.04 \\
\midrule
\multirow{9}{*}{\rotatebox{90}{$3\ell+4\ell$}}      & $|y^{Z}|$   & 0.77 & 0.78 && 0.70 & 0.77 && 0.64 & 0.77 \\
& $\pT^{Z}$                     & 0.09  & 0.08  &&  0.13  & 0.13 && 0.22  & 0.23 \\
& $\cos\theta^*_{Z}$               & 0.20  & 0.17  &&  0.21  & 0.19 && 0.24  & 0.22 \\
& $|\Delta\Phi(t\bar{t}, Z)|/\pi$ & 0.84  & 0.82  &&  0.08  & 0.53 && 0.07  & 0.56  \\
& $m^{t\bar{t}}$                  & 0.89  & 0.97  &&  0.8\phantom{0}   & 0.92 && 0.49  & 0.83  \\
& $m^{t\bar{t}Z}$                 & 0.86  & 0.93  &&  0.64  & 0.91 && 0.58  & 0.91  \\
& $\pT^{t}$          & 0.45  & 0.56  &&  0.2\phantom{0}   & 0.59 && 0.22 & 0.39  \\
& $\pT^{t\bar{t}}$              & 0.09  & 0.07  &&  0.05  & 0.05 && 0.07 & 0.1\phantom{0}  \\
& $|y^{t\bar{t}Z}|$               & 0.95  & 0.8\phantom{0}   &&  0.86  & 0.85 && 0.66  & 0.65  \\
\bottomrule
\end{tabular}
\end{center}
\end{table}

\begin{table}[!htb]
\footnotesize
\caption{Summary of the compatibility tests, i.e. $p$-values, between the unfolded normalised differential spectra and the various predictions, in the $3\ell$ and $4\ell$ channels as well as in their combination, at particle-level and parton-level. }
\label{tab:unfolding_p_values_normalised}
\def\arraystretch{1.3}
\begin{center}
\begin{tabular}{cl ccc ccc cc} \\
\toprule
\multicolumn{2}{c}{ } & \multicolumn{2}{c}{$\AMCatNLO{+}\PYTHIA[8]$} && \multicolumn{2}{c}{\SHERPA[2.2.1] (incl.)} && \multicolumn{2}{c}{\SHERPA[2.2.11] (multi-leg)}\\
\midrule
& Variable & parton & particle && parton & particle && parton & particle\\
\midrule
\multirow{5}{*}{\rotatebox{90}{$3\ell$}}      & $|\Delta\Phi(Z, t_\text{lep})|/\pi $ & 0.07 & 0.05 && 0.04 & 0.02 && 0.07 & 0.06 \\
& $|\Delta y(Z, t_\text{lep})| $   & 0.03 & 0.03 && 0.03  & 0.02  && 0.03 & 0.02 \\
& $\HT^{\ell} $              & 0.02 & 0.02 && 0.02  & 0.04  && 0.05 & 0.07 \\
& $\pT^{\ell, \textrm{non-}Z} $       & 0.63 & 0.63 && 0.29  & 0.5\phantom{0}   && 0.29 & 0.41 \\
&$N_{\textrm{jets}} $       & -    & 0.42 && -     & 0.81  && -    & 0.75 \\
\midrule
\multirow{3}{*}{\rotatebox{90}{$4\ell$}}      & $N_{\textrm{jets}}$  & - & 0.40 && - & 0.65 && - & 0.61   \\
& $|\Delta\Phi(\ell^{+}_{t}, \ell^{-}_{\bar{t}})|/\pi$  & 0.75 & 0.76 && 0.73  & 0.75 && 0.74  & 0.80 \\
& $\HT^{\ell}$              & 0.03 & 0.02 && 0.03  & 0.02 && 0.03  & 0.02 \\
\midrule
\multirow{9}{*}{\rotatebox{90}{$3\ell+4\ell$}}      & $|y^{Z}|$ & 0.71 & 0.72 && 0.71 & 0.71 && 0.70 & 0.69 \\
& $\pT^{Z}$                     & 0.04 & 0.03 && 0.06 & 0.06 && 0.13  & 0.14 \\
& $\cos\theta^*_{Z}$               & 0.11 & 0.11 && 0.12 & 0.12 && 0.15  & 0.15 \\
& $|\Delta\Phi(t\bar{t}, Z)|/\pi$ & 0.74 & 0.71 && 0.02 & 0.36 && 0.02  & 0.39  \\
& $m^{t\bar{t}}$                  & 0.79 & 0.88 && 0.87 & 0.73 && 0.27  & 0.43  \\
& $m^{t\bar{t}Z}$                 & 0.84 & 0.93 && 0.8\phantom{0}  & 0.91 && 0.81  & 0.89  \\
& $\pT^{t}$          & 0.03 & 0.23 && 0.01 & 0.29 && 0.01  & 0.1\phantom{0}  \\
& $\pT^{t\bar{t}}$              & 0.02 & 0.01 && 0.01 & 0.01 && 0.02  & 0.03  \\
& $|y^{t\bar{t}Z}|$               & 0.87 & 0.52 && 0.85 & 0.64 && 0.59  & 0.28  \\
 
\bottomrule
\end{tabular}
\end{center}
\end{table}
 
\FloatBarrier


\FloatBarrier
 
\printbibliography
 
\clearpage
 
\begin{flushleft}
\hypersetup{urlcolor=black}
{\Large The ATLAS Collaboration}

\bigskip

\AtlasOrcid[0000-0002-6665-4934]{G.~Aad}$^\textrm{\scriptsize 102}$,
\AtlasOrcid[0000-0002-5888-2734]{B.~Abbott}$^\textrm{\scriptsize 120}$,
\AtlasOrcid[0000-0002-1002-1652]{K.~Abeling}$^\textrm{\scriptsize 55}$,
\AtlasOrcid[0000-0001-5763-2760]{N.J.~Abicht}$^\textrm{\scriptsize 49}$,
\AtlasOrcid[0000-0002-8496-9294]{S.H.~Abidi}$^\textrm{\scriptsize 29}$,
\AtlasOrcid[0000-0002-9987-2292]{A.~Aboulhorma}$^\textrm{\scriptsize 35e}$,
\AtlasOrcid[0000-0001-5329-6640]{H.~Abramowicz}$^\textrm{\scriptsize 151}$,
\AtlasOrcid[0000-0002-1599-2896]{H.~Abreu}$^\textrm{\scriptsize 150}$,
\AtlasOrcid[0000-0003-0403-3697]{Y.~Abulaiti}$^\textrm{\scriptsize 117}$,
\AtlasOrcid[0000-0002-8588-9157]{B.S.~Acharya}$^\textrm{\scriptsize 69a,69b,m}$
% ...
\bigskip
\\

% ...
$^{29}$Physics Department, Brookhaven National Laboratory, Upton NY; United States of America.\\
% ...
$^{35}$$^{(a)}$Facult\'e des Sciences Ain Chock, Universit\'e Hassan II de Casablanca;$^{(b)}$Facult\'{e} des Sciences, Universit\'{e} Ibn-Tofail, K\'{e}nitra;$^{(c)}$Facult\'e des Sciences Semlalia, Universit\'e Cadi Ayyad, LPHEA-Marrakech;$^{(d)}$LPMR, Facult\'e des Sciences, Universit\'e Mohamed Premier, Oujda;$^{(e)}$Facult\'e des sciences, Universit\'e Mohammed V, Rabat;$^{(f)}$Institute of Applied Physics, Mohammed VI Polytechnic University, Ben Guerir; Morocco.\\
% ...
$^{49}$Fakult\"{a}t Physik , Technische Universit{\"a}t Dortmund, Dortmund; Germany.\\
% ...
$^{55}$II. Physikalisches Institut, Georg-August-Universit\"{a}t G\"ottingen, G\"ottingen; Germany.\\
% ...
$^{69}$$^{(a)}$INFN Gruppo Collegato di Udine, Sezione di Trieste, Udine;$^{(b)}$ICTP, Trieste;$^{(c)}$Dipartimento Politecnico di Ingegneria e Architettura, Universit\`a di Udine, Udine; Italy.\\
% ...
$^{102}$CPPM, Aix-Marseille Universit\'e, CNRS/IN2P3, Marseille; France.\\
% ...
$^{117}$Department of Physics, New York University, New York NY; United States of America.\\
% ...
$^{120}$Homer L. Dodge Department of Physics and Astronomy, University of Oklahoma, Norman OK; United States of America.\\
% ...
$^{150}$Department of Physics, Technion, Israel Institute of Technology, Haifa; Israel.\\
$^{151}$Raymond and Beverly Sackler School of Physics and Astronomy, Tel Aviv University, Tel Aviv; Israel.\\
% ...
$^{a}$ Also Affiliated with an institute covered by a cooperation agreement with CERN.\\
$^{b}$ Also at An-Najah National University, Nablus; Palestine.\\
$^{c}$ Also at Borough of Manhattan Community College, City University of New York, New York NY; United States of America.\\
$^{d}$ Also at Center for High Energy Physics, Peking University; China.\\
$^{e}$ Also at Center for Interdisciplinary Research and Innovation (CIRI-AUTH), Thessaloniki; Greece.\\
$^{f}$ Also at Centro Studi e Ricerche Enrico Fermi; Italy.\\
% ...
$^{m}$ Also at Department of Physics, King's College London, London; United Kingdom.\\
% ...

\end{flushleft}


\end{document}